\documentclass[fancy,11pt,twoside,mycap]{pers}
\ssp
\usepackage{graphics}
\usepackage{feynmf}

\begin{document}
\setlength{\unitlength}{1mm}  

\pagestyle{empty}
\topmargin-2pc
\begin{flushright}
SCIPP 99/25     \\
hep--ph/9906332
\end{flushright}
\vskip2cm

\begin{center}
{\Large\bf  Radiative corrections to the $Z b \bar{b}$ vertex \\[6pt]
 and constraints on extended Higgs sectors}\\[1cm]
{\large Heather E. Logan}\\[3pt]
{\it Santa Cruz Institute for Particle Physics  \\
   University of California, Santa Cruz, CA 95064, U.S.A.} \\[1.5cm]

{\bf Abstract}
\end{center}

We explore the radiative corrections to the process $Z \rightarrow b
\bar{b}$
in models with extended Higgs sectors.  The observables $R_b$ and $A_b$
are sensitive to these corrections.
$R_b$ is the hadronic branching fraction of $Z$ bosons to $b\bar{b}$,
$R_b = \Gamma(Z
\rightarrow b \bar{b})/\Gamma(Z \rightarrow \mathrm{hadrons})$.
$A_b$ is the $b$ quark asymmetry, $A_b =
(g_L^2 - g_R^2)/(g_L^2 + g_R^2)$ where $g_L$ and $g_R$ are the left and
right handed couplings of $Z$ to $b$ quarks.
We find that in models containing only doublets, singlets, or 
larger multiplets constrained by a custodial $SU(2)_{c}$ symmetry so that
$M_W = M_Z \cos\theta_W$ 
at tree level, the corrections involving
charged Higgs bosons always worsen agreement with experiment.  
The $R_{b}$ measurement can 
be used to set lower bounds on the charged Higgs masses in such models.
Corrections involving light
neutral Higgs bosons in models with enhanced $H^{0}b\bar{b}$ coupling
(large $\tan\beta$) can improve agreement with experiment over the
Standard Model.  
We present general formulas for the corrections to $R_b$ and $A_b$
in an arbitrary extended Higgs sector, and 
derive explicit results for a number of specific models.

\vskip2cm
\centerline{\it Presented in partial satisfaction of the requirements for
the degree of}
\centerline{\it Doctor of Philosophy in Physics at the University of 
California, Santa Cruz, June 1999}
\vfill
\clearpage
\pagestyle{persheader}

\makeatletter
\def\@cite#1#2{{[{#1}]\if@tempswa\typeout
{IJCGA warning: optional citation argument
ignored: `#2'} \fi}}


\newcount\@tempcntc
\def\@citex[#1]#2{\if@filesw\immediate\write\@auxout{\string\citation{#2}}\fi
  \@tempcnta\z@\@tempcntb\m@ne\def\@citea{}\@cite{\@for\@citeb:=#2\do
    {\@ifundefined
       {b@\@citeb}{\@citeo\@tempcntb\m@ne\@citea\def\@citea{,}{\bf ?}\@warning
       {Citation `\@citeb' on page \thepage \space undefined}}%
    {\setbox\z@\hbox{\global\@tempcntc0\csname b@\@citeb\endcsname\relax}%
     \ifnum\@tempcntc=\z@ \@citeo\@tempcntb\m@ne
       \@citea\def\@citea{,}\hbox{\csname b@\@citeb\endcsname}%
     \else
      \advance\@tempcntb\@ne
      \ifnum\@tempcntb=\@tempcntc
      \else\advance\@tempcntb\m@ne\@citeo
      \@tempcnta\@tempcntc\@tempcntb\@tempcntc\fi\fi}}\@citeo}{#1}}
\def\@citeo{\ifnum\@tempcnta>\@tempcntb\else\@citea\def\@citea{,}%
  \ifnum\@tempcnta=\@tempcntb\the\@tempcnta\else
   {\advance\@tempcnta\@ne\ifnum\@tempcnta=\@tempcntb \else \def\@citea{--}\fi
    \advance\@tempcnta\m@ne\the\@tempcnta\@citea\the\@tempcntb}\fi\fi}
\makeatother

\def\nicefrac#1#2{\hbox{${#1\over #2}$}}



\title{Radiative corrections to the {$Z \lowercase{b \bar{b}}$} vertex and 
constraints on extended Higgs sectors}
\author{Heather Erin Logan}
\degreeyear{1999}
\degreemonth{June}
\degree{Doctor of Philosophy}
\field{Physics}

\chair{Professor Howard Haber}
\committeememberone{Professor Michael Dine}
\committeemembertwo{Professor Bruce Schumm}
\numberofmembers{3}

\campus{Santa Cruz}


\begin{frontmatter}

\tableofcontents

\end{frontmatter}



\chapter{Introduction}

The Standard Model (SM) of electroweak interactions 
\cite{Glashow70,Weinberg67,Salam68} 
has been 
tested and confirmed to unprecedented precision in the past several 
years at the $e^{+}e^{-}$ colliders LEP at CERN and SLC at SLAC
(for recent data, see \cite{Clare99}).
Precision measurements of many electroweak observables
have confirmed that the electroweak interactions are well 
described by a spontaneously broken SU(2) $\times$ U(1) gauge symmetry.  
However, these measurements have not allowed us to determine the 
dynamics of the symmetry breaking.

The couplings of quarks and leptons to the 
$Z$ boson and the mass of the $Z$ and $W$ have been measured.  
At tree level, the couplings of quarks and leptons to the $Z$ are entirely
determined by the gauge structure of the theory.  
These couplings depend only on the
SU(2) $\times$ U(1) quantum numbers of the quarks and leptons, and the 
electric charge and weak mixing angle $\sin^2 \theta_W$.
  
The measurement of the $Z$ and $W$ masses provides us with one more piece of 
information about electroweak symmetry breaking (EWSB).  In the SM, the $Z$ and
$W$ masses $M_Z$ and $M_W$ are related at tree-level by
\begin{equation}
M_W = M_Z \cos \theta_W.
\end{equation}
This relation is satisfied experimentally to better than 1\% 
\cite{Altarelli98}.  In the SM
this relation is a consequence of the presence of an unbroken global 
SU(2) symmetry of the EWSB sector, often called ``custodial SU(2)
symmetry'' \cite{Sikivie80}.  The three Goldstone bosons and the three
SU(2) gauge currents transform as triplets under the custodial symmetry.
(For a pedagogical discussion see \cite{Haber91}.)

At the one--loop level the situation is different.  The electroweak
measurements are precise enough to begin to probe the effects of one--loop
corrections in the couplings of quarks and leptons to the $Z$ and the 
$W$ and $Z$ masses.  By measuring the one--loop corrections we can gain 
more information about the EWSB sector.

In the SM, the electroweak symmetry is broken by the Higgs mechanism
\cite{Higgs64,Englert64,Guralnik64}.
A set of scalar (Higgs) fields are introduced, with a potential which is
symmetric under SU(2) $\times$ U(1).  The potential has a continuous set 
of degenerate minima at nonzero field values; the symmetry is spontaneously
broken by the ground state choosing one of the degenerate minima.

The minimal SM Higgs sector consists of one complex SU(2) doublet of scalar 
fields.  After EWSB, three of the 
degrees of freedom are ``eaten'' by the $W$ and $Z$ bosons, giving them 
mass, and leaving one CP--even neutral Higgs boson $H^0$ in the physical
spectrum.  (For a review of the properties of the SM Higgs boson, 
see \cite{HHG}.)

Since the couplings of Higgs bosons to fermions and gauge bosons are 
proportional to the fermion or gauge boson mass, one--loop corrections 
involving Higgs bosons coupled to $W$, $Z$ or third--generation quarks 
can be significant.  In the SM, loop corrections involving $H^0$ coupling
to gauge bosons depend logarithmically on the $H^0$ mass.  A 
fit to the electroweak data gives an upper bound on the SM Higgs mass of
$M_H < 220$ GeV at the 95\% confidence level \cite{Clare99}.
In the SM the Higgs couplings to third--generation quarks do not give us
additional information about the Higgs sector.  Such corrections would 
contribute to the decay $Z \rightarrow b \bar{b}$; the decays
$Z \rightarrow t \bar{t}$ and $W^+ \rightarrow t \bar{b}$ and its 
complex conjugate are kinematically forbidden for on--shell $W$ and $Z$.
The coupling of $H^0$ to $b$ quarks is too small to make an 
observable contribution to $Z \rightarrow b \bar{b}$.  The coupling of the
charged Goldstone bosons $G^{\pm}$ to $t \bar{b}$ is large enough to 
make an observable contribution to $Z \rightarrow b \bar{b}$, but 
the contribution
is fixed by electroweak symmetry; it depends only on the $W$ and $t$--quark
masses, the electric charge and $\sin^2 \theta_W$
\cite{Akhundov86,Beenakker88,Bernabeu88,Hollik90,Lynn90}. 

Many extensions to the minimal SM Higgs sector are possible.  (For a 
review and references see \cite{HHG}.)  As in the SM, extended models 
typically must 
contain at least one SU(2) doublet in order to give mass
to the fermions.  They can also contain additional SU(2) doublets,
singlets, and/or larger multiplets.  In general, extended Higgs sectors 
will yield charged Higgs bosons and additional neutral Higgs bosons in the 
physical spectrum.
Extended Higgs sectors contribute to the radiative corrections to 
the process $Z \rightarrow b \bar{b}$
through the charged Higgs couplings to $ t \bar{b} $ and the 
neutral Higgs couplings to $ b \bar{b} $.  

The process $Z \rightarrow b \bar{b}$ yields two 
observable quantities, $R_b$ and $A_b$.
$R_{b}$ is the hadronic 
branching ratio of $Z$ to $b$ quarks,
\begin{equation}
	R_{b} \equiv \frac{\Gamma(Z \rightarrow b \bar{b})}
	{\Gamma(Z \rightarrow {\rm hadrons})}
\end{equation}
and $A_b$ is the $b$ asymmetry,
\begin{equation}
A_b = \frac{\sigma(e^-_L \to b_F) - \sigma(e^-_L \to b_B)
	+ \sigma(e^-_R \to b_B) - \sigma(e^-_R \to b_F)}
	{\sigma(e^-_L \to b_F) + \sigma(e^-_L \to b_B)
	+ \sigma(e^-_R \to b_B) + \sigma(e^-_R \to b_F)},
\end{equation}
where $e^-_{L,R}$ are left and right handed initial--state electrons
and $b_{F,B}$ are final--state $b$ quarks moving in the forward and 
backward directions.  The forward direction is defined as the direction
of the initial--state electrons.
In terms of the $b$ quark couplings to $Z$,
\begin{equation}
	A_{b} = \frac{(g^{L}_{Zb\bar{b}})^{2} - (g^{R}_{Zb\bar{b}})^{2}}
		{(g^{L}_{Zb\bar{b}})^{2} + (g^{R}_{Zb\bar{b}})^{2}}.
\end{equation}

Until 1996 the $R_b$ measurement was significantly higher than 
the SM prediction \cite{LEPandSLD}, and 
a number of models were introduced to bring the prediction into better 
agreement with experiment 
\cite[and references therein]{Rb}.  These include models with a 
modified tree-level $Zb\bar{b}$ coupling \cite{Field98}, 
a significantly lower value for the top mass \cite{fourgenSUSY}, 
or extra particles which contribute to $R_{b}$ through loop corrections 
\cite{Grant,Comelli,Drees,Cline,Chankowski}.
The last two approaches take advantage of the fact that the discrepancy 
in $R_{b}$ was the same size as the top-mass-dependent loop corrections to
$Z \rightarrow b \bar{b}$, which arise from the exchange of longitudinally
polarized $W^{\pm}$ bosons (\textit{i.e.}, the SM Goldstone bosons 
$G^{\pm}$).  
Since 1996 the $R_b$ measurement has 
come closer to the SM prediction but is still slightly high.  
It is best used to 
constrain models that would predict a lower $R_b$ than the SM.

In this thesis we introduce a parameterization for a general extended Higgs 
sector and calculate the contribution to $Z \rightarrow b \bar{b}$ from 
one-loop radiative corrections involving singly charged and neutral Higgs
bosons.  
We obtain general expressions for the corrections to the left- and 
right-handed $ Z b \bar{b} $ couplings, and then use the measurements of 
$R_b$ and $A_b$ to constrain specific models.
This approach has the advantage of yielding general formulas for the 
corrections in terms of the couplings and masses of the Higgs bosons.
The formulas can then be specialized to any extended Higgs model by 
inserting the appropriate couplings.  Kundu and Mukhopadhyaya 
\cite{Kundu96} have taken the same approach and calculated the 
charged Higgs boson contributions to $Z \rightarrow b \bar{b}$ in a general
extended Higgs sector.  However, the neutral Higgs boson contributions in a 
general extended Higgs sector
do not appear in the literature.
Previously, the corrections to $Z \rightarrow b \bar{b}$ in extended
Higgs sectors had only been computed for the two Higgs doublet model
(2HDM) \cite{Denner91,Djouadi91,Boulware91,Grant}.

This thesis does not take into account corrections to $Z \rightarrow 
b \bar{b}$ coming from loops involving supersymmetric particles.
However, in the limit of large superpartner masses, the supersymmetric
contributions decouple \cite{Haber93,Herrero98,Herrero99}.  
In this limit, our formulas are relevant in
supersymmetric models with extended Higgs sectors.

The method of parameterizing a general extended Higgs 
sector developed in this thesis 
can also be used to calculate Higgs boson corrections 
to other processes, such as $b \rightarrow s \gamma$ and $b \rightarrow
c \tau^- \bar{\nu}_{\tau}$.  

This thesis is organized as follows.
In chapter \ref{sec:data_constr} we discuss the measurements of 
$R_b$ and $A_b$ 
and the constraints that they put on the $ Z b \bar{b} $ couplings.
In chapter \ref{sec:extendedHiggs} we
introduce the two Higgs doublet model and then generalize to an 
arbitrary extended Higgs sector, as well as describing some of the 
features that such models contain.  
In chapters \ref{sec-H+} and 
\ref{sec-H0} we calculate the radiative corrections to the $Zb\bar{b}$
coupling.  In chapter \ref{sec-H+} we consider the loops involving charged
Higgs bosons while in chapter \ref{sec-H0} we consider the loops involving
neutral Higgs bosons.  
In chapter
\ref{sec:models} we apply the general formulas
for loop corrections to a number of specific models and exhibit constraints
on the charged Higgs sector.  We first consider extended Higgs sectors 
containing only doublets and singlets, and then extend the
analysis to Higgs sectors containing larger multiplets as well.  We 
summarize our conclusions in chapter \ref{sec:conclusions}.
  
In the appendices we summarize a number of extended Higgs models, and
list formulas which we have used in our analysis.
In appendix \ref{sec:Zqqcouplings} we list the tree--level $Z$ couplings
to quarks in the SM.  In appendix \ref{app:HVcouplings} we list the
Higgs couplings to vector bosons for an arbitrary extended Higgs sector.
In appendices \ref{app:2HDM}, \ref{app:2doub1trip}, and 
\ref{app:SU2c} we describe the details of some of the extended Higgs 
models which are considered in our analysis.  In appendix \ref{app:2HDM}
we list the couplings in the two Higgs doublet model.  In appendix
\ref{app:2doub1trip} we describe the models containing two Higgs doublets 
and one triplet.  In appendix \ref{app:SU2c} we describe a class of 
models in which the Higgs sector preserves a ``custodial'' SU(2)
symmetry.
In appendix \ref{app:loopints} we give details of the calculation of
one--loop integrals.  
Finally, in appendices \ref{app:Abderiv}, \ref{app:SMinputs}, and
\ref{app:dirsearches}, we describe the experimental data used in this
thesis.  In appendix \ref{app:Abderiv} we describe how the $A_b$
measurement was extracted from the data.  In appendix \ref{app:SMinputs}
we list the SM parameters used in our numerical calculations.
Lastly, in appendix \ref{app:dirsearches}, we describe the lower
bounds on Higgs masses from direct searches.




\chapter{Constraints from the data}
\label{sec:data_constr}

The radiative corrections to $Z\rightarrow b \bar{b}$
modify the $ Z b \bar{b} $ couplings from their tree-level values.
In this section we show how the experimental constraints on $R_b$ and 
$A_b$ constrain the possible values of the effective $Zb\bar{b}$ couplings.
These constraints will provide limits on the radiative corrections.
The effective couplings are
\begin{equation}
	\bar{g}_b^{L,R} = g_{Zb\bar{b}}^{L,R} + \delta g^{L,R}
\end{equation}
where $\bar{g}_b^{L,R}$ are the radiatively--corrected effective 
couplings, $g_{Zb\bar{b}}^{L,R}$ are the tree--level couplings, and 
$\delta g^{L,R}$ contain the radiative corrections.
Our notation and the tree--level couplings are listed in appendix 
\ref{sec:Zqqcouplings}.

The effective couplings $\bar{g}_b^{L,R}$ are extracted from the 
measured values of $R_b$ and $A_b$ in the next section.  In order to 
use these to constrain new physics, the SM prediction for
$\bar{g}_b^{L,R}$ must be known precisely.  This requires an accurate 
value of $\sin^2\theta^{lept}_{eff}$, which can be affected by oblique
corrections from new physics.  This issue is addressed in section
\ref{sec:STU}.

\section{Extracting the effective $Zb\bar{b}$ couplings from $R_b$ 
and $A_b$}
\label{sec:gfromRbAb}

Following the discussion by Field \cite{Field98} and using his notation,
the effective couplings $\bar{g}_b^{L,R}$ are related to $R_b$ and $A_b$
as follows.
\begin{eqnarray}
R_b &=& \left[ 1 + \frac{S_b}{\bar{s}_b C_b^{QCD} C_b^{QED}} \right]^{-1} \\
	\nonumber \\
A_b &=& \frac{2 \bar{r}_b (1-4\mu_b)^{1/2}}
	{1 - 4\mu_b + (1+2\mu_b)\bar{r}_b^2} 
\end{eqnarray}
where
\begin{eqnarray}
\bar{r}_b &=& \frac{\bar{v}_b}{\bar{a}_b}  \\
\bar{s}_b &=& (\bar{a}_b)^2(1-6\mu_b) + (\bar{v}_b)^2 \\
\bar{v}_b &=& \bar{g}_b^L + \bar{g}_b^R  \\
\bar{a}_b &=& \bar{g}_b^L - \bar{g}_b^R \\
S_b &=& \sum_{q \neq b} (\bar{a}_q)^2 + (\bar{v}_q)^2.
\end{eqnarray}
$\mu_b$ is a correction factor coming from the nonzero mass of the $b$ quark.
Using the running $b$ quark mass in the $\overline{\rm MS}$
scheme 
evaluated at $M_Z$, $\bar{m}_b(M_Z) = 3.0$
GeV \cite{Fusaoka98}, this correction factor is 
$\mu_b = (\bar{m}_b(M_Z)/M_Z)^2 
\simeq 1.0 \times 10^{-3}$.

$C_b^{QED}$ and $C_b^{QCD}$ are QED and QCD correction factors,
\begin{equation}
C_b^{QED} = 1 + \delta^{QED}_b - \langle \delta^{QED}_{q \neq b} \rangle
\end{equation}
\begin{equation}
C_b^{QCD} = 1 + \delta^{QCD}_b - \delta^{QCD}_{q \neq b},
\end{equation}
where $\langle \delta^{QED}_{q \neq b} \rangle$ denotes the average 
of $\delta^{QED}_{q \neq b}$ over $u$, $d$, $c$, and $s$ quarks, and,
\begin{eqnarray}
\delta_q^{QED} &=& \frac{3 (e_q)^2}{4\pi} \alpha(M_Z) \\
\delta_{q \neq b}^{QCD} &=& 1.00 \left( \frac{\alpha_s(M_Z)}{\pi} \right)
	+ 1.42 \left( \frac{\alpha_s(M_Z)}{\pi} \right)^2 \\
\delta_b^{QCD} &=& 0.99 \left( \frac{\alpha_s(M_Z)}{\pi} \right)
	- 1.55 \left( \frac{\alpha_s(M_Z)}{\pi} \right)^2,
\end{eqnarray}
where $e_q$ is the quark electric charge in units of the positron charge.
Numerically, with
$\alpha_s(M_Z) = 0.12$ and $\alpha^{-1}(M_Z) = 128.9$,
\begin{eqnarray}
C_b^{QED} &=& 0.99975  \\
C_b^{QCD} &=& 0.9953.
\end{eqnarray}
The non-$b$ quark couplings are written as,
\begin{eqnarray}
\bar{a}_q &=& \sqrt{\rho_q} T_3^q \\
\bar{v}_q &=& \sqrt{\rho_q} (T_3^q - 2 e_q (\bar{s}^q_W)^2)
\end{eqnarray}
where, assuming non-$b$ quark universality, for all $q \neq b$,
\begin{eqnarray}
\sqrt{\rho_q} &=& \sqrt{\rho_l} = 2 |\bar{a}_l|   \\
(\bar{s}_W^q)^2 &=& \frac{1}{4} (1-\bar{r}_l),
\end{eqnarray}
and $T^q_3$ is the third component of weak isospin of quark $q$.
The SM prediction for the effective couplings is then \cite{Field98},
\begin{eqnarray}
\bar{g}^L_b &=& -0.4208  \\
\bar{g}^R_b &=& 0.0774.
\end{eqnarray}
Together with the SM predictions for the leptonic couplings,
\begin{eqnarray}
\bar{a}_l &=& -0.50124 \\
\bar{r}_l &=& 0.07332,
\end{eqnarray}
these yield the SM predictions for $R_b$ and $A_b$,
\begin{eqnarray}
	R_b^{SM} &=& 0.21587 \\
	A_b^{SM} &=& 0.935.
\end{eqnarray}
These are the SM predictions quoted in \cite{Clare99}.  (For a discussion
of how $R_b^{SM}$ was obtained, see appendix \ref{app:SMinputs}.)
	
The measured values are \cite{Clare99},
\begin{eqnarray}
R_b &=& 0.21680 \pm 0.00073
	\label{eqn:Rbmeasured} \\
A_b &=& 0.895 \pm 0.016
	\label{eqn:Abmeasured}
\end{eqnarray}
$R_b$ is measured directly at LEP and SLD.
$A_b$ is measured directly at SLD from the left-right forward-backward 
asymmetry, and indirectly at LEP from 
the measured value of $A_e$ and
the forward-backward asymmetry 
$A_{FB}^{0,b} = \frac{3}{4} A_{e} A_{b}$. 
Details of the calculation of $A_b$, and the inputs used,
are listed in
appendix \ref{app:Abderiv}.   
The $R_b$ measurement is $1.3 \sigma$ above the SM prediction, and the 
$A_b$ measurement is $2.5 \sigma$ below the SM prediction.

\begin{figure}
\resizebox{\textwidth}{!}{\rotatebox{270}{\includegraphics{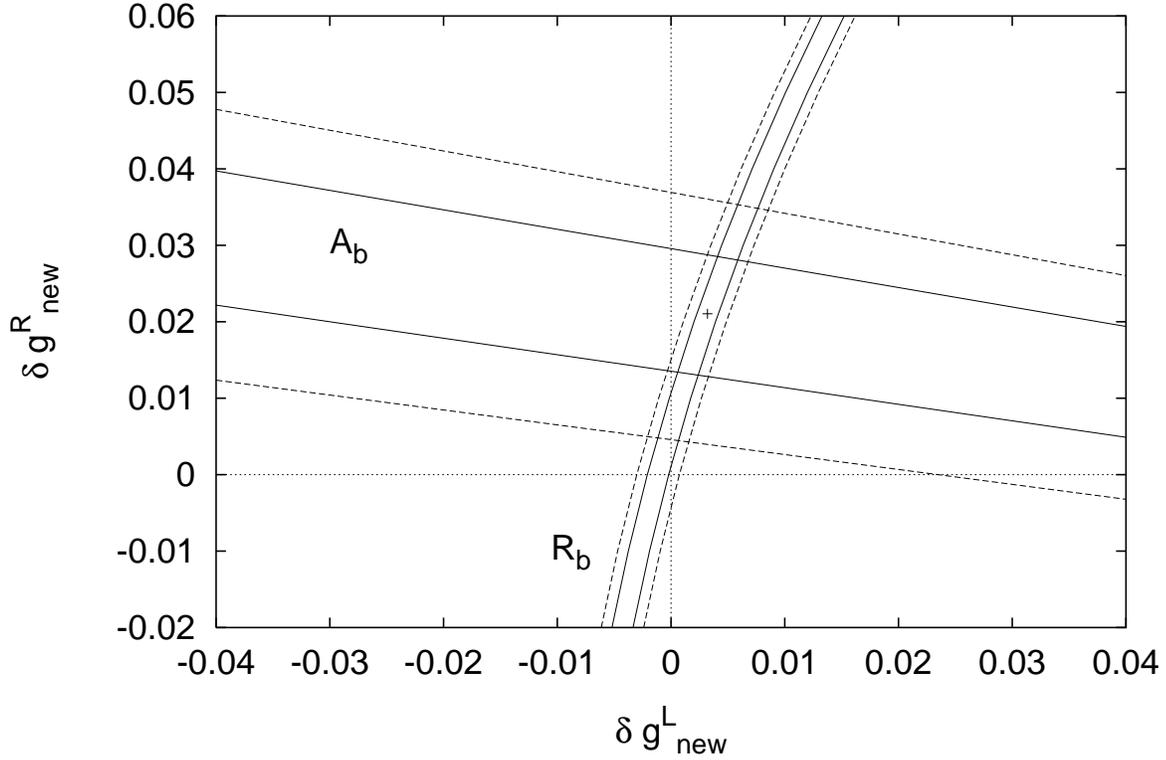}}}
\caption[Constraints on $\delta g^{R,L}$ from $R_b$ and $A_b$]
{The constraints from $R_b$ and $A_b$ on the right-- and   
left--handed $Z b \bar{b}$ couplings.  Plotted are the allowed deviations
$\delta g^{R,L}_{\rm new}$ of
the couplings from their SM values.  The 1 $\sigma$
errors are shown as solid lines and the 2 $\sigma$ errors as dashed lines.
The central value, at 
$\delta g^L_{\rm new} = 0.0032$ and 
$\delta g^R_{\rm new} = 0.0210$,
is marked by the cross.
}
\label{fig:RbAb}
\end{figure}

Defining
\begin{equation}
(\bar{g}_b^{L,R})_{\rm expt} = (\bar{g}_b^{L,R})_{SM} 
	+ \delta g^{L,R}_{\rm new}
\end{equation}
we plot the experimental constraints from $R_b$ and $A_b$ on 
$\delta g^{L,R}_{\rm new}$ in figure \ref{fig:RbAb}.
The central value is at $\delta g^L = 0.0032$ and $\delta g^R = 0.0210$.
Comparing these to the SM predictions,
we see that $\delta g^L$ is a $1\%$ correction while $\delta g^R$
is close to a $30\%$ correction.
This is in approximate agreement with the results of Field,
\cite{Field98} who found that a model independent fit of the $Z$ pole 
data yielded a right-handed $b$ quark coupling $42\%$
above the SM prediction.

It is also useful to expand $R_b$ and $A_b$ about their SM values, to
first order in $\delta g^{R,L}_{\rm new}$.  Using the SM parameters given 
above, we find
\begin{eqnarray}
\delta R_b &=& -0.7788 \delta g^L_{\rm new} + 0.1410 \delta g^R_{\rm new}  
\label{eqn:Rbsigns} \\
\delta A_b &=& -0.2984 \delta g^L_{\rm new} - 1.623 \delta g^R_{\rm new}.
\end{eqnarray}
Note that a positive $\delta g^L_{\rm new}$ decreases both $R_b$ and $A_b$, 
while a positive $\delta g^R_{\rm new}$ increases $R_b$ and decreases $A_b$.

\section{Tree--level $Zb\bar{b}$ couplings: The effect of oblique 
corrections}
\label{sec:STU}

In the SM, all electroweak observables are fixed by the measurement of 
three quantities, commonly chosen to be the electromagnetic fine structure
constant $\alpha$, the muon decay constant $G_{\mu}$, and the $Z$ mass.
In particular, by measuring these quantities, one can predict the value of 
$\sin^2\theta^{lept}_{eff}$.  In practice, many more electroweak observables
are measured and a fit is made to the SM parameters 
(see e.g., \cite{CERN97}).

However, the dependence of $\sin^2\theta^{\rm lept}_{\rm eff}$ on other
electroweak observables can be modified in models with new physics that
contributes to oblique corrections.  These modifications are parameterized
by the Peskin--Takeuchi parameters $S$, $T$, and $U$ \cite{Peskin92}.
In particular \cite{Grant98},
\begin{equation}
\sin^2\theta^{\rm lept}_{\rm eff} - [\sin^2\theta^{\rm lept}_{\rm eff}]_{SM}
	\equiv \delta s^2_W
	= \frac{\alpha}{c^2_W - s^2_W}
	\left[ \frac{1}{4} S - s^2_W c^2_W T \right].
\end{equation}
Nonzero values of the $S$ and $T$ parameters therefore modify
the prediction for the tree--level $Zb\bar{b}$ couplings 
$g_{Zb\bar{b}}^{L,R}$.

The $S$, $T$, and $U$ parameters are defined relative to a reference SM,
with a fixed SM Higgs mass.  In the reference SM they are all zero.
For $M_H^{SM} = M_Z$, a fit of the electroweak data gives \cite{Caso98}
\begin{eqnarray}
S &=& -0.16 \pm 0.14  \\
T &=& -0.21 \pm 0.16  \\
U &=& 0.25 \pm 0.24.
\end{eqnarray}
In order to understand the significance of oblique corrections of this 
size, we compute the corrections to the SM predictions for 
$R_b$ and $A_b$ due to $S$, $T$, and $U$.  To first order in $\delta s^2_W$, 
\begin{eqnarray}
\delta R_b &=& \frac{R_b}{\bar{s}_b} \left[ 4 \bar{v}_b g^R_b (1-R_b)
	+ \frac{32 R_b \bar{a}_l^2}{C_b^{QCD} C_b^{QED}} 
	\left(\frac{1}{2 s^2_W} - \frac{10}{9} \right) \right]
	\frac{\delta s^2_W}{s^2_W}  \\
	&=& 0.0373 \delta s^2_W 
	= 1.35 \times 10^{-4} S  - 9.62 \times 10^{-5} T  \\
\delta A_b &=& \left[ A_b \left(1 - A_b \frac{(1+2\mu_b)}{\sqrt{1-4\mu_b}}
	\bar{r}_b \right) \frac{2 g^R_b}{\bar{v}_b} \right]
	\frac{\delta s^2_W}{s^2_W}  \\
	&=& -0.642 \delta s^2_W  
	= -2.32 \times 10^{-3} S  + 1.66 \times 10^{-3} T  
\end{eqnarray}
where we have used the SM values all parameters as given in section 
\ref{sec:gfromRbAb} and $[\sin^2\theta^{\rm lept}_{\rm eff}]_{SM} \equiv 
(s^2_W)_{SM} = 0.23157$.  The oblique corrections to the SM predictions 
for $R_b$ and $A_b$ do not depend on $U$.

Inserting the measured central values of $S$ and $T$, we find
\begin{eqnarray}
\delta R_b &=& -1.4 \times 10^{-6}  \\
\delta A_b &=& 2.3 \times 10^{-5}.
\end{eqnarray}
Comparing these corrections to the measured values in equations 
\ref{eqn:Rbmeasured} and \ref{eqn:Abmeasured}, we see that the correction
due to nonzero values of $S$ and $T$ is less than $1\%$ of the experimental
error on both $R_b$ and $A_b$.  We can safely neglect these
corrections.




\chapter{Models with extended Higgs sectors}
\label{sec:extendedHiggs}

A wide variety of extensions to the minimal SM Higgs sector are 
possible.  (For a review and references see \cite{HHG}.) 
In this section we discuss some of the interesting 
features of models with extended Higgs sectors.  
We also list some important formulas for the couplings and 
Goldstone bosons in the two Higgs doublet model and a general
extended Higgs sector.  These formulas will be used later in
the corrections to the process $Z \rightarrow b \bar{b}$.

\subsubsection{The $\rho$ parameter in an extended Higgs sector}

In an extended Higgs sector which contains one or more multiplets
larger than doublets,
there is the possibility of having $\rho \neq 1$ at tree level.
It is well known that $\rho=1$ at tree--level in a
Higgs sector containing only doublets and singlets 
\cite{Lee1,Ross1}.  In a general Higgs sector, however, the 
tree--level $\rho$ parameter is given by \cite{HHG,Tsao1}
\begin{equation}
\rho \equiv \frac{m_{W}^{2}}{m_{Z}^{2} c_{W}^{2}} = 
\frac{\sum_{k} 2(T_{k}(T_{k}+1) - Y_{k}^{2}/4) v_{k}^{2}
	+ \sum_{i} 2T_{i}(T_{i}+1) v_{i}^{2}  }
	{\sum_{k} Y_{k}^{2} v_{k}^{2}},
\end{equation}
where $k$ runs over the complex multiplets and 
$i$ runs over the real multiplets in the Higgs sector, and 
$c_{W} = \cos\theta_W$.  The Higgs vacuum expectation values (vevs)
$v_k$ and $v_i$ for each multiplet are defined as,
\begin{equation}
\langle \phi^0_k \rangle = v_k/\sqrt{2}
\label{eqn:3complexvevnorm}
\end{equation}
for complex representations, and
\begin{equation}
\langle \eta^0_i \rangle = v_i
\label{eqn:3realvevnorm}
\end {equation}
for real representations.

Experimentally, $\rho$ has been shown to be very close to one;
in particular, $\Delta \rho \equiv \rho - 1 = (3.9 \pm 1.2) \times 10^{-3}$
(\cite{Altarelli98}, in which $\Delta \rho = \epsilon_1$).
Certain multiplets automatically satisfy $\rho = 1$.  These are multiplets
for which \cite{HHG},
\begin{equation}
(2 T + 1)^2 - 3 Y^2 = 1.
\label{eqn:rhois1}
\end{equation}
This equation is satisfied by
the singlet $(T,Y)=(0,0)$, the familiar doublet $(\frac{1}{2},1)$,
and a series of complicated larger multiplets, $(3,4)$, 
$(\frac{25}{2},15)$, etc.
Higgs sectors that contain only multiplets of this type yield $\rho = 1$ 
without any fine--tuning of the parameters of the Higgs potential.

However, problems arise when one attempts to construct a Higgs sector
in which the only multiplets larger than doublets satisfy 
equation \ref{eqn:rhois1}.  First, the Higgs sector must contain
at least one doublet in order to give mass to the fermions.
Then the Higgs potential will be forced by 
${\rm SU}(2)_{L} \times {\rm U}(1)_{Y}$--invariance to have an accidental 
continuous global symmetry (e.g., a separate U(1) rotation of each of
the multiplets larger than doublets).  
The global symmetry is spontaneously broken when the larger
multiplets get vevs, resulting in massless Goldstone bosons in the physical 
spectrum.  As described in reference \cite{Nir89}, the physical
spectrum must also contain a light CP--even Higgs boson $H^0$ with mass on
the order of the vev of the larger multiplet.  This is required 
because the mass splitting between the massless Goldstone boson
 and $H^0$ is on the order
of the vev that breaks the accidental global symmetry.
Such a massless Goldstone boson 
is then ruled out by the experimental limits on 
$Z \rightarrow a^0 H^{0}$, where $a^0$ is the massless Goldstone boson
\cite{Nir89}.  

The accidental global symmetries can be eliminated by 
introducing a set of new Higgs multiplets to couple the 
U(1) rotations of the larger multiplets to those of the doublet,
so that the model has only one U(1) symmetry, that of hypercharge.
However, these new multiplets will in general spoil $\rho=1$.  
There are two ways that $\rho \approx 1$ can be maintained.
First, we can require that the
vevs of the new multiplets must be small enough to satisfy the 
experimental constraints on 
$\rho$.  This requires an unnatural fine--tuning.
A completely general Higgs sector can be made to 
agree with $\rho \approx 1$ by fine--tuning the parameters of the model
so that the multiplets that would contribute to $\rho \neq 1$ have 
very small vevs.
Second, we can introduce a set of new Higgs multiplets such that
the entire Higgs sector preserves a custodial ${\rm SU}(2)_c$ symmetry.
This can be done for each of the multiplets that satisfy equation
\ref{eqn:rhois1}.  The new multiplets eliminate the accidental
U(1) symmetry, while at the same time the ${\rm SU}(2)_c$ 
symmetry ensures that 
$\rho = 1$ by ensuring that equal masses are given to the
$W^{\pm}$ and $W^3$ gauge bosons.  Models with ${\rm SU}(2)_c$
symmetry are discussed in detail in appendix \ref{app:SU2c}.

In general, the ${\rm SU}(2)_c$ symmetry is preserved 
through a conspiracy of the vevs and electroweak quantum  
numbers of the Higgs multiplets in the model.  This
conspiracy can be made exact to all orders in the Higgs self--couplings
by requiring that the Higgs potential be invariant under the custodial 
${\rm SU}(2)_{c}$ symmetry.  This is only possible when the Higgs sector
consists of certain sets of multiplets, which transform together under an
${\rm SU}(2)_L \times {\rm SU}(2)_R$ symmetry.
Such a model involving triplet Higgs fields
has been constructed by Georgi and Machacek \cite{Georgi1}.
It was considered in greater depth by Chanowitz and Golden 
\cite{Chanowitz1}, who showed that a Higgs potential for the model could be
constructed that was invariant under the full ${\rm SU}(2)_L \times 
{\rm SU}(2)_R$.  
This ensured that radiative corrections from Higgs self--interactions 
preserved ${\rm SU}(2)_c$.  A more detailed study of the phenomenology of the
model \cite{Gunion1} and naturalness problems from one--loop effects 
\cite{Gunion2} was made by Gunion, Vega, and Wudka.  
The fields in this model consist of one $Y=1$ complex doublet, one real ($Y=0$)
triplet, and one $Y=2$ complex triplet.  The ${\rm SU}(2)_c$ 
symmetry ensures that the vevs of the neutral members of the two triplets
are equal, yielding $\rho=1$.  This model must be fine tuned because
${\rm SU}(2)_c$--breaking terms arise in the Higgs potential
at the one--loop level \cite{Gunion2} from corrections involving the
hypercharge interactions.  For more details, see appendix \ref{app:SU2c}.

\subsubsection{Higgs couplings in an extended Higgs sector}

An extended Higgs sector that contains
multiplets larger than doublets can have Higgs couplings 
which differ from the analogous couplings in the SM and in models
containing only doublets and singlets.  Some 
couplings can be enhanced relative to their values in models 
containing only doublets and singlets, and other couplings exist that 
are not present in simpler models.
For simplicity, we assume that the Higgs sector is CP--conserving.
We denote the CP--even neutral Higgs bosons by $H^0_i$, the 
CP--odd neutral Higgs bosons by $A^0_i$, and the charged Higgs
bosons by $H^+_i$, $H^{++}_i$, etc.

The Higgs couplings to $Z$ and $W^{\pm}$ are affected 
because of their dependence
on the isospin of the electroweak eigenstates involved.  For 
example [see equations \ref{eq:ZHA} -- \ref{eq:ZZH}],
the couplings $g_{ZH^{0}_{i}A^{0}_{j}}$, 
$g_{ZH^{+}_{i}H^{-}_{i}}$, $g_{W^{+}W^{-}H_i^{0}}$, and 
$g_{ZZH_i^0}$
can be enhanced in a model containing multiplets larger than doublets.
This can lead
to an enhancement of the production cross section for certain Higgs bosons 
through the processes $Z^{*} \rightarrow H^{0}_{i}A^{0}_{j}$,
$Z^{*} \rightarrow H^{+}_{i}H^{-}_{i}$, $W^{*} \rightarrow WH_i^{0}$,
and $Z^* \to ZH_i^0$,
as well as enhancement of the loop correction to $Z \rightarrow b \bar{b}$
from the diagram of figure~\ref{fig:h+loops}(a).

Certain couplings exist in models with multiplets larger than 
doublets that are zero in
models containing only doublets and singlets.  For example, off--diagonal
charged Higgs couplings to $Z$ (equation \ref{eq:Z+-}),
and the $H^{\pm}_{i}W^{\mp}Z$ vertex, described in 
\cite{HHG} and references therein,
are generally nonzero in Higgs sectors that include multiplets larger than
doublets.  The off--diagonal charged Higgs couplings to $Z$ can lead to 
loop corrections to $Z \rightarrow b \bar{b}$ involving two charged 
Higgs bosons of different mass in the loop.  
The Feynman diagram for this process
is given in figure \ref{fig:h+loops}(a).


\section{The two Higgs doublet model}

In this section we briefly review some properties of the two Higgs 
doublet model (2HDM).  A working knowledge of this model will be useful when
we consider a general extended Higgs sector in the next section.
For a more complete treatment see reference \cite{HHG,Grant}.  The 
discussion below follows \cite{Grant}.  The complete Higgs couplings 
to fermions and the $Z$--Higgs--Higgs couplings in this model
are listed in appendix \ref{app:2HDM}.

The 2HDM is the usual ${\rm SU}(2)_L \times {\rm U}(1)_Y$
SM with an extended Higgs sector consisting of two complex
doublets of scalar fields, $\Phi_{1}$ and $\Phi_{2}$, with hypercharge 
$Y= 1$.  
Note that any model that contains a complex Higgs multiplet 
$\Phi$ with hypercharge $Y$ can be rewritten in terms of the conjugate
multiplet $i\sigma_2 \Phi^*$ with hypercharge $-Y$.
The electroweak gauge symmetry ${\rm SU}(2)_L \times {\rm U}(1)_Y$ 
is broken down
to electromagnetic ${\rm U}(1)_{EM}$ by choosing a Higgs
potential such that the two real neutral Higgs 
fields acquire the vevs $v_{1}$ and $v_{2}$.  The Goldstone 
bosons are then
\begin{equation}
G^{0} = \cos \beta \phi_{1}^{0,i} + \sin \beta \phi_{2}^{0,i}
\end{equation}
\begin{equation}
G^{+} = \cos \beta \phi_{1}^+ + \sin \beta \phi_{2}^+
\end{equation}
where the Higgs doublets are $\Phi_{k} = (\phi_{k}^{+}, \phi_{k}^{0})$,
the neutral component is
$\phi_{k}^{0} = \frac{1}{\sqrt{2}}(v_k + \phi_{k}^{0,r} + i\phi_{k}^{0,i})$,
and the ratio of the vevs is parameterized by $\tan \beta = v_{2}/v_{1}$.
The $W^{\pm}$ and $Z$ bosons acquire mass through the Higgs mechanism 
and the fermions acquire mass through their Yukawa couplings to 
the Higgs bosons.  The vevs of the two Higgs doublets are 
constrained by the $W$ mass,
$M_{W}^{2} = g^{2} (v_{1}^{2} + v_{2}^{2}) / 4 = g^2 v_{SM}^2 / 4$,
where $v_{SM} = 246$ GeV.

In addition to the Goldstone bosons, the 2HDM contains one 
charged Higgs boson $H^{+}$, one CP--odd neutral Higgs boson $A^{0}$, and two 
CP--even neutral Higgs bosons $h^{0}$ and $H^{0}$.  The Higgs 
mass eigenstates are 
\begin{equation}
H^{+} = -\sin \beta \phi_{1}^{+} + \cos \beta \phi_{2}^{+}
\end{equation}
\begin{equation}
A^{0} = -\sin \beta \phi_{1}^{0,i} + \cos \beta \phi_{2}^{0,i}
\label{eq:A02HDM}
\end{equation}
\begin{equation}
h^{0} = - \sin\alpha \phi_{1}^{0,r} + \cos\alpha \phi_{2}^{0,r}
\label{eq:H102HDM}
\end{equation}
\begin{equation}
H^{0} = \cos\alpha \phi_{1}^{0,r} + \sin\alpha \phi_{2}^{0,r}.
\label{eq:H202HDM}
\end{equation}
The two CP--even neutral states are defined so that $h^0$ is 
lighter than $H^0$.  $\alpha$ is a mixing angle determined by the 
Higgs potential.

In the SM, the diagonalization of the quark mass matrix automatically
diagonalizes the Yukawa couplings of the neutral Higgs boson to quarks.
Thus in the SM, there are no tree--level 
flavor--changing neutral Higgs interactions.
In the 2HDM with the most 
general Higgs Yukawa couplings, however, 
flavor--changing neutral Higgs interactions can arise.  These interactions
are severely constrained by the measurements of 
$K^0 - \bar{K}^0$ and $B^0 - \bar{B}^0$ mixing, which arise at the
one--loop level in the SM.  Because the constraints on flavor--changing
neutral Higgs couplings involving first--generation quarks are the 
strongest, it has been suggested that the flavor--changing couplings 
should be proportional to the masses of the quarks involved in the
coupling, so that the couplings are of order $\sqrt{m_im_j}/v_{SM}$,
where $v_{SM} = 246$ GeV is the SM Higgs vev.  Even with the
Yukawa couplings suppressed by first--generation quark masses, however,
the measurements of $K^0 - \bar{K}^0$ and $B^0 - \bar{B}^0$ mixing
require that $M_{A^0}$ is above 2 TeV \cite{Atwood97}.  The constraint
on $M_{A^0}$ from flavor--changing neutral Higgs interactions 
is significantly stronger than the constraints on Higgs masses 
from $R_b$ that we will present.

The severe constraints on flavor--changing neutral Higgs interactions 
led Glashow and Weinberg \cite{Glashow77} and Paschos \cite{Paschos77}
to introduce a discrete symmetry in order to forbid tree--level 
flavor--changing neutral Higgs interactions in models with more than
one doublet.  They showed that a sufficient condition to 
eliminate flavor--changing neutral Higgs interactions in a model containing
more than one Higgs doublet is that the fermions
of each charge receive their mass from couplings to exactly one
neutral Higgs field.  (Note that mass terms
for quarks that conserve baryon number and are 
${\rm SU}(2)_L \times {\rm U}(1)_Y$--invariant can only arise from couplings
to Higgs doublets with hypercharge $Y= \pm 1$.)

With the discrete symmetry of references \cite{Glashow77,Paschos77}, 
there are two possible configurations for the 
quark Yukawa couplings in the 2HDM, referred to as the Type I and 
Type II models.  In the Type I model,
all the quarks couple to $\Phi_1$, and not to $\Phi_2$.  
In the Type II model, the 
down--type quarks couple to $\Phi_1$ and the up--type quarks
couple to $\Phi_2$.  The Higgs sector in the minimal supersymmetric model
is a Type II 2HDM.

In a Type I model, one Higgs doublet $\Phi_{1}$ gives mass to both 
$t$ and $b$ quarks.  The Yukawa couplings are,
\begin{equation}
\label{eqn:lambdatI}
\lambda_t = \frac{\sqrt{2} m_t}{v_1}
\end{equation}
\begin{equation}
\label{eqn:lambdabI}
\lambda_b = \frac{\sqrt{2} m_b}{v_1}.
\end{equation}
Note that in a Type I model, $\lambda_b / \lambda_t = m_b/m_t$,
so $\lambda_b \ll \lambda_t$ for all values of $v_1$.

In a Type II model, 
$\Phi_{1}$ couples to $b$ quarks and 
$\Phi_{2}$ couples to $t$ quarks.  The quark Yukawa couplings are then,
\begin{equation}
\label{eqn:lambdatII}
\lambda_t = \frac{\sqrt{2} m_t}{v_2}
\end{equation}
\begin{equation}
\label{eqn:lambdabII}
\lambda_b = \frac{\sqrt{2} m_b}{v_1}
\end{equation}
Note that in a Type II model, $\lambda_b/ \lambda_t = (m_b/m_t) \tan\beta$,
so $\lambda_b$ can be enhanced relative to $\lambda_t$ 
by choosing $v_{1}$ much less than $v_{2}$ (i.e., choosing $\tan\beta$
to be large).  The Yukawa couplings
for the 2HDM Higgs mass eigenstates are listed in appendix \ref{app:2HDM}.


\section{A general extended Higgs sector}

In this thesis we consider the effects of a general extended Higgs
sector on the $Z \rightarrow b \bar{b}$ decay rate.  
We remind the reader that an extended 
Higgs sector consists of a number of scalars organized into 
multiplets according to their transformation properties under 
${\rm SU}(2)_L \times {\rm U}(1)_Y$.  
The Higgs sector must contain at least one
${\rm SU}(2)_L$ doublet to give mass to the SM fermions.  The Higgs
sector is divided into complex representations, denoted by $\phi_{k}$,
and real 
representations, denoted by $\eta_{i}$.  We define a real representation as 
consisting of a real multiplet of fields with integer weak isospin
and hypercharge $Y=0$, as in reference \cite{HHG}.  
We also assume that the Higgs
sector is CP--conserving, so that the neutral Higgs mass eigenstates are
either CP--even or CP--odd.  We will denote a CP--even state
by $H^0_i$ and a CP--odd state by $A^0_j$.  A Higgs
potential is chosen to break ${\rm SU}(2)_L \times {\rm U}(1)_Y$ down to 
${\rm U}(1)_{EM}$ such that the neutral member of each of the 
Higgs multiplets acquires a vev.  We denote the vevs of complex
representations by $v_{k}$ and the vevs of real representations by
$v_{i}$.  The vevs are normalized as in equations \ref{eqn:3complexvevnorm}
and \ref{eqn:3realvevnorm}.  These vevs are constrained by the $W$ 
mass, which for a general extended Higgs sector is given by,
\begin{equation}
M_{W}^2 = \frac{g^2}{4} \left\{ \sum_{k}2v_{k}^{2}(T_{k}(T_{k}+1)-
	\frac{Y_{k}^{2}}{4})
                + \sum_{i}2v_{i}^{2}T_{i}(T_{i}+1) \right\}
	= \frac{g^2}{4} v_{SM}^2,
\end{equation}
where $v_{SM} = 246$ GeV.
The Goldstone bosons are given by,
\begin{equation}
G^{0} = \frac{\sum_{k}Y_{k}v_{k}\phi_{k}^{0,i}}
{\sqrt{\sum_{k}Y_{k}^{2}v_{k}^{2}}}
\end{equation}
\begin{eqnarray}
G^{+} &=&  \left\{ \sum_{k}\left[ \left[T_{k}(T_{k}+1) - Y_{k}
                (Y_{k}-2)/4 \right]^{1/2} v_{k}\phi_{k}^{+} 
		\right. \right. \nonumber \\
	& &	- \left. \left. \left[T_{k}(T_{k}+1) - Y_{k}
                (Y_{k}+2)/4 \right]^{1/2} v_{k}(\phi_{k}^{-})^{*} \right]
	+ \sum_{i} \left[ 2T_{i}(T_{i}+1) \right]^{1/2} v_{i}\eta_{i}^{+} 
		\right\} 
			\nonumber \\ 
  & & \times \left\{ \sum_{k}2v_{k}^{2}(T_{k}(T_{k}+1)-Y_{k}^{2}/4)
		+ \sum_{i}2v_{i}^{2}T_{i}(T_{i}+1) \right\}^{-1/2},
\label{eq:G+}
\end{eqnarray}
and we use the phase convention,
\begin{equation}
G^- = - (G^+)^*.
\end{equation}
Note that for a Higgs boson in a complex representation, $(\phi^Q)^*$ is a 
state with charge $-Q$ but is not the same as $\phi^{-Q}$.
For a Higgs boson in a
real representation, we use the phase convention $(\eta^+)^* = - \eta^-$.

In a Higgs sector that contains only multiplets for which $\rho = 1$ 
automatically (see equation \ref{eqn:rhois1}),
the $W$ mass and the formula for
$G^+$ simplify to,
\begin{equation}
M_{W}^2 = \frac{g^2}{4} \sum_{k} Y_{k}^{2} v_{k}^{2}
\end{equation}
\begin{equation}
G^{+} = \frac{\sum_{k} \left[ \left[ (Y_k^2 + Y_k)/2 \right]^{1/2}
				v_{k} \phi_{k}^{+}
			- \left[ (Y_k^2 - Y_k)/2 \right]^{1/2}
                                v_{k} (\phi_{k}^{-})^{*} \right]}
	{\sqrt{\sum_{k}Y_{k}^{2}v_{k}^{2}}}.
\end{equation}

The Yukawa couplings of a general extended Higgs sector take the form
of either a Type I or Type II model, and are defined in the same
way as in the 2HDM.  If the extended model contains only one Higgs
doublet, $\Phi_1$, it is necessarily a Type I model, with the
Yukawa couplings given in equations \ref{eqn:lambdatI} and 
\ref{eqn:lambdabI}.  If the extended model contains two or more
Higgs doublets, then it can be either a Type I model or a Type II
model.  If all the quarks couple only to one Higgs doublet $\Phi_1$,
the model is Type I, with the Yukawa couplings given in equations 
\ref{eqn:lambdatI} and \ref{eqn:lambdabI}.  Alternatively,
if the down--type quarks couple to one Higgs doublet $\Phi_1$ and 
the up--type quarks couple to a different Higgs doublet
$\Phi_2$, then the model is Type II, with the 
Yukawa couplings given in equations 
\ref{eqn:lambdatII} and \ref{eqn:lambdabII}.

When the Higgs mass--squared matrix is diagonalized, 
the electroweak eigenstates 
mix to form mass eigenstates.  Recall that we have assumed that the
Higgs sector is CP--conserving.  We denote the CP--even neutral Higgs
bosons by $H^0_i$ and the CP--odd neutral Higgs bosons by $A^0_i$.
The couplings of the Higgs mass eigenstates to quarks take the form,
\begin{equation}
i (g^{L}_{H\bar{q}q} P_{L} + g^{R}_{H\bar{q}q} P_{R})
	= i (g^{V}_{H\bar{q}q} + g^{A}_{H\bar{q}q} \gamma_{5}).
\end{equation}
The individual couplings to $b\bar{b}$ and $b\bar{t}$ in a Type II
model are given by,
\begin{equation}
g_{H_{i}^{0}b\bar{b}}^{V} = - \frac{1}{\sqrt{2}} \lambda_b
	\langle H_{i}^{0} | \phi_{1}^{0,r} \rangle   \label{eq:gV}
\end{equation}
\begin{equation}
g_{A_{i}^{0}b\bar{b}}^{A} = - \frac{i}{\sqrt{2}} \lambda_b
        \langle A_{i}^{0} | \phi_{1}^{0,i} \rangle   \label{eq:gA}
\end{equation}
\begin{equation}
g_{H_{i}^{+}\bar{t}b}^{R} = - \lambda_b 
	\langle H_{i}^{+} | \phi_{1}^{+} \rangle   \label{eq:gR}
\end{equation}
\begin{equation}
\label{eq:gLtb}
g_{H_{i}^{+}\bar{t}b}^{L} = + \lambda_t
        \langle H_{i}^{+} | \phi_{2}^{+} \rangle.   \label{eq:gL}
\end{equation}
The couplings for a Type I model are obtained by 
replacing $\phi_2^+$ with $\phi_1^+$ in equation \ref{eq:gL}; the 
other couplings remain the same.

The $Z$--Higgs--Higgs couplings take the form given in equation
\ref{eqn:HHVdiagram}.
The $Z$--Higgs--Higgs couplings involving neutral and 
singly--charged Higgs bosons are,
\begin{equation}
g_{ZH_{i}^{0}A_{j}^{0}} = \frac{ie}{s_{W}c_{W}} \sum_{k=1}^{N}
	\langle H_{i}^{0} | \phi_{k}^{0,r} \rangle
	\langle A_{j}^{0} | \phi_{k}^{0,i} \rangle
	T^{3}_{\phi_{k}^{0}} 			 \label{eq:ZHA}
\end{equation}
\begin{equation}
g_{ZH_{i}^{+}H_{j}^{-}} = - \frac{e}{s_{W}c_{W}} \left\{ \sum_{k=1}^{N}
	\langle H_{i}^{+} | \phi_{k}^{+} \rangle
	\langle H_{j}^{+} | \phi_{k}^{+} \rangle T^{3}_{\phi_{k}^{+}}
	- s_{W}^{2} \delta_{ij}  \right\}.
\label{eq:Z+-}
\end{equation}
For completeness, we also give the $W^+W^-H_i^0$ and $ZZH_i^0$ couplings.
The $V$--Higgs--Higgs ($V=W$,$Z$) take the form given in equation
\ref{eqn:HVVdiagram}.  The $W^+W^-H_i^0$ coupling is,
\begin{equation}
g_{W^{+}W^{-}H_i^{0}}
= g^{2} \sum_{k}\langle H_i^{0}|\phi_{k}^{0,r} \rangle v_{k}
\left(T_{k}(T_{k}+1) - \frac{Y_{k}^{2}}{4}\right),
\label{eq:WWH}
\end{equation}
and the $ZZH_i^0$ coupling is,
\begin{equation}
g_{ZZH_i^0} = \frac{g^2}{2c^2_W} \sum_k 
	\langle H_i^{0}|\phi_{k}^{0,r} \rangle v_{k} Y^2_k
\label{eq:ZZH}
\end{equation}
where $s_W = \sin\theta_W$, $c_W = \cos\theta_W$, 
and $T^{3}_{\phi}$ is the third component of the weak isospin of 
$\phi$.  
The complete Higgs--Vector boson couplings can be found in appendix
\ref{app:HVcouplings}.

Although the $Z$--Higgs--Higgs couplings are diagonal in the electroweak 
basis, they are not necessarily diagonal in the mass eigenstate basis.
In addition, the $ZH^{+}H^{-}$ couplings can differ from the SM
$ZG^{+}G^{-}$ coupling.  This can happen in a general model
if $H^+$ has some admixture of a multiplet larger than a doublet.
In the SM, the $ZG^+G^-$ coupling is,
\begin{equation}
g_{ZG^{+}G^{-}} =  - \frac{e}{s_{W}c_{W}} 
	\left(\nicefrac{1}{2} - s_{W}^{2} \right).
\end{equation}




\chapter{Charged Higgs corrections to $Z \rightarrow b \bar{b}$}
\label{sec-H+}

\begin{figure}
\begin{center}
\resizebox{8cm}{!}
        {\includegraphics*[100pt,430pt][370pt,700pt]{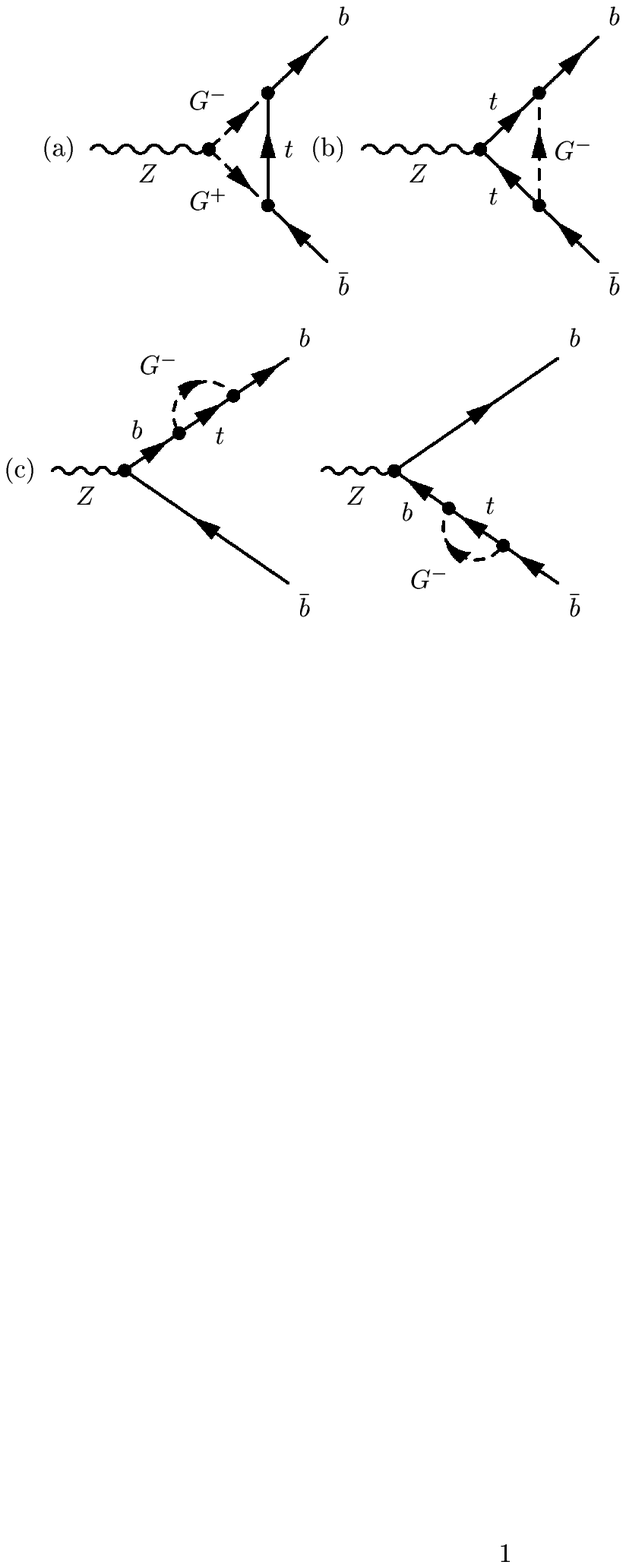}}
\end{center}
\caption
{Feynman diagrams of the leading $m_t^2$ contributions to the 
electroweak corrections to
$Z\rightarrow b \bar{b}$ in the SM.}
\label{fig:gloops}
\end{figure}

In the SM, the $Zb\bar{b}$ couplings
receive a correction from the exchange of the longitudinal components
of the $W^{\pm}$ and $Z$ bosons.  The
Feynman diagrams for these corrections are shown in 
figure~\ref{fig:gloops}.
\begin{figure}
\begin{center}
\resizebox{8cm}{!}
        {\includegraphics*[100pt,295pt][370pt,700pt]{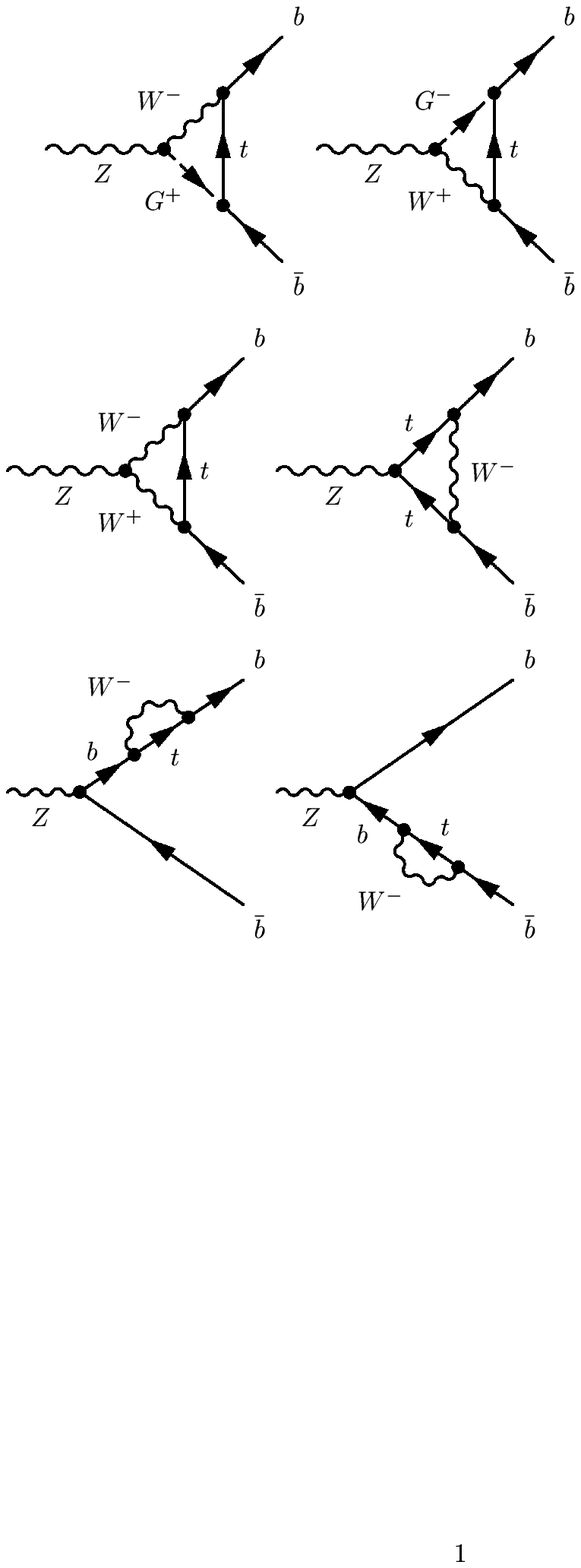}}
\end{center}
\caption
{Feynman diagrams of the subleading electroweak corrections to
$Z\rightarrow b \bar{b}$ in the SM.}
\label{fig:gloops2}
\end{figure}
We work in the 't Hooft-Feynman gauge, in which the longitudinal
components of $W^{\pm}$ and $Z$ are just the Goldstone bosons
$G^{\pm}$ and $G^0$.  In this gauge the
Goldstone bosons are physical degrees of freedom and have 
masses $M_{W}$ and $M_{Z}$,
respectively.  The diagrams in figure~\ref{fig:gloops} give the leading 
$m_t^2$ contribution to $\delta g^{L,R}$ in the SM.  
A detailed review of the calculation of these diagrams is given in
reference \cite{Hollik2}.  
The diagrams in figure \ref{fig:gloops2} also contribute to $\delta g^{L,R}$
but their contributions are suppressed by a factor of $m_Z^2/m_t^2$ compared 
to the diagrams of figure \ref{fig:gloops}.

\begin{figure}
\begin{center}
\resizebox{8cm}{!}
        {\includegraphics*[100pt,295pt][370pt,700pt]{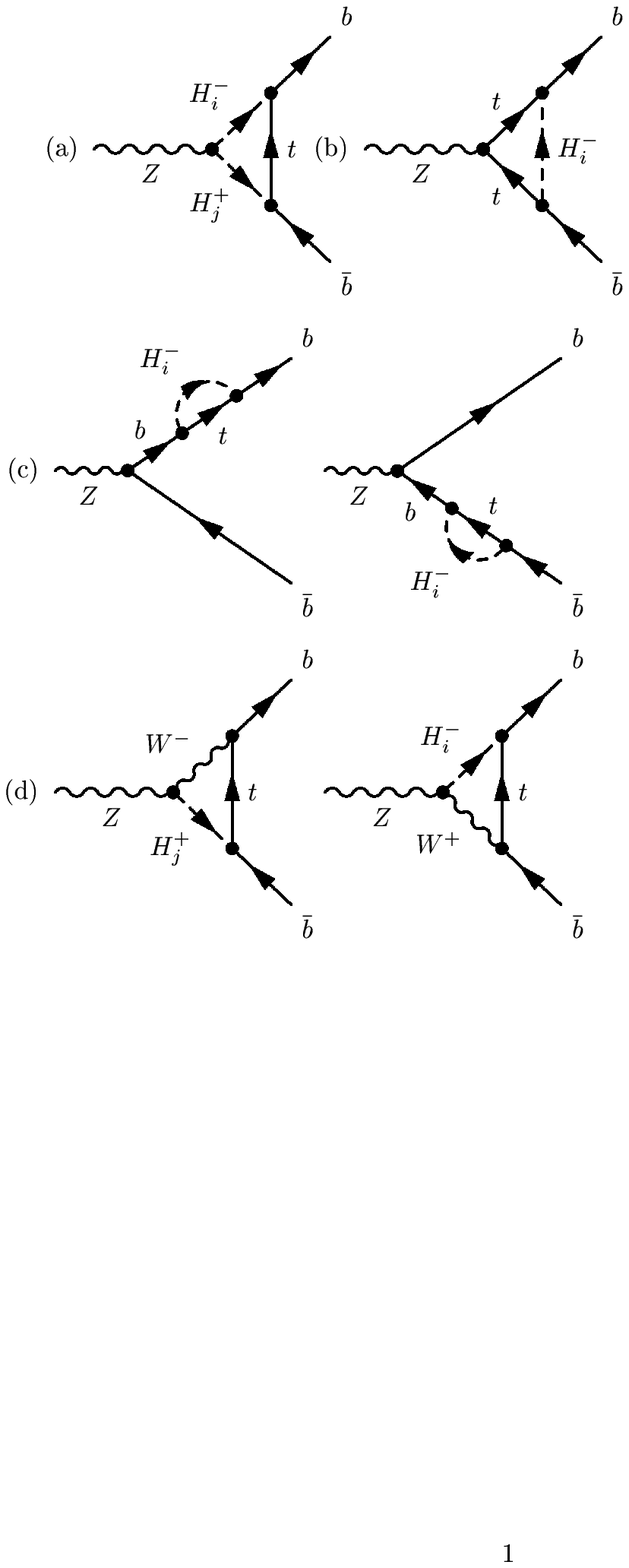}}
\end{center}
\caption
{Feynman diagrams of the electroweak corrections to
$Z\rightarrow b \bar{b}$ in a model
with an extended Higgs sector. 
}
\label{fig:h+loops}
\end{figure}

In an extended Higgs sector which contains singly charged Higgs states
$H_{i}^{\pm}$, the corrections to $\delta g^{L,R}$
arise from the diagrams of figure~\ref{fig:h+loops}, where
$H_i^{\pm}$ runs over all the singly charged states in the Higgs sector,
including $G^{\pm}$.

In calculating the corrections shown in figure \ref{fig:h+loops}
we make the following approximation.  We keep only the leading term
in powers of $m_t^2 / M_Z^2$.  In $\delta g^L$ this leading term 
is proportional to $m_t^2$, where the two powers of $m_t$ come from
the left--handed Higgs--quark couplings $g^L_{H^+_i \bar{t} b}$.
In $\delta g^R$ the right--handed Higgs--quark couplings are proportional
to $m_b^2 \tan^2\beta$, so the leading term in $\delta g^R$ 
does not grow with increasing $m_t$.
This approximation has been used in calculating the
large $m_t^2$--dependent
corrections to $R_b$ in the SM in the classic papers
\cite{Beenakker88,Bernabeu88,Hollik90,Lynn90}, 
and in calculating the 
corrections in extended Higgs sectors in references 
\cite{Denner91,Djouadi91,Boulware91,Grant,Kundu96}.

In figure \ref{fig:h+loops}(d), the $W^{\pm}$ are longitudinally 
polarized.
The two diagrams in figure \ref{fig:h+loops}(d) 
involving a $ZW^+H^-_i$ vertex can be nonzero in 
models containing Higgs multiplets 
larger than doublets.  However, their contribution to $R_b$ and $A_b$
is suppressed by a factor of $m_Z^2/m_t^2$ compared to diagrams 
\ref{fig:h+loops}(a), (b)
and (c), and we will neglect them.
Evaluating diagrams \ref{fig:h+loops}(a), (b), and (c), 
we get
\begin{equation}
\delta g^{L,R}(a) = \frac{1}{16 \pi^{2}} \sum_{i,j} 
g^{L,R}_{H^{+}_{i}\bar{t}b} g^{L,R}_{H^{+}_{j}\bar{t}b} 
g_{ZH^{+}_{i}H^{-}_{j}} \times
2 C_{24}(m_b^2, M_Z^2, m_b^2; m_{t}^2, M_{i}^2, M_{j}^2)
\end{equation}
\begin{eqnarray}
\delta g^{L,R}(b) = - \frac{1}{16 \pi^{2}} \sum_{i}
&(g^{L,R}_{H^{+}_{i}\bar{t}b})^{2} &
\{ -2g^{R,L}_{Zt\bar{t}}C_{24} + \frac{1}{2}g^{R,L}_{Zt\bar{t}} + 
g^{L,R}_{Zt\bar{t}}m_{t}^{2}C_{0} \}  \nonumber \\
	& & (m_b^2, M_Z^2, m_b^2; M_{i}^2,m_{t}^2,m_{t}^2) 
\end{eqnarray}
\begin{equation}
\delta g^{L,R}(c) = \frac{1}{16 \pi^{2}} \sum_{i}
(g^{L,R}_{H^{+}_{i}\bar{t}b})^{2} g^{L,R}_{Zb\bar{b}}
B_{1}(m_{b}^{2}; m_{t}^2, M_{i}^2).
\end{equation}
The two-- and three--point integrals $C_{24}$, $C_{0}$, and 
$B_{1}$ are defined in appendix \ref{app:loopints}.
The sums over $i$ and $j$ run over all the singly
charged Higgs mass eigenstates $H_{i}^{+}$ as well as the Goldstone 
boson $G^{+}$.  Where no ambiguity is involved, we have given the 
arguments of groups of tensor integrals that depend on the same 
variables only once at the end of the group.  These expressions for 
$\delta g^{L}$ agree with those of Kundu \textit{et al.} \cite{Kundu96}.
For compactness we will drop
the first three arguments of the three--point integrals, 
$(m_b^2, M_Z^2, m_b^2)$, because these arguments are the same in all the
expressions.  The first three arguments of the three--point integrals
depend only on the masses of the on--shell external particles.

Collecting the results, and expressing the corrections in terms 
of the quark Yukawa couplings, we obtain
for a Type II model,
\begin{eqnarray}
\label{eqn:dgLH+}
\delta g^L &=& -\frac{1}{16 \pi^2} \lambda_t^2 \frac{e}{s_Wc_W}
	\sum_{i,j} 
	\langle H^+_i | \phi^+_2 \rangle \langle H^+_j | \phi^+_2 \rangle
	\left\{ \sum_{k=1}^{N}
	\langle H_{i}^{+} | \phi_{k}^{+} \rangle
	\langle H_{j}^{+} | \phi_{k}^{+} \rangle T^{3}_{\phi_{k}^{+}}
	- s_{W}^{2} \delta_{ij}  \right\}  \nonumber \\
& &	\times 2 C_{24} (m_t^2, M_i^2, M_j^2)  \nonumber \\
& & - \frac{1}{16 \pi^2} \lambda_t^2 \sum_i 
	\langle H^+_i | \phi^+_2 \rangle^2 
	\left\{ -2 g^R_{Zt\bar{t}} C_{24} + \frac{1}{2}g^R_{Zt\bar{t}}
	+ g^L_{Zt\bar{t}} C_0 \right\} (M_i^2, m_t^2, m_t^2)  \nonumber \\
& & + \frac{1}{16 \pi^2} \lambda_t^2 g^L_{Zb\bar{b}} \sum_i 
	\langle H^+_i | \phi^+_2 \rangle^2
	B_1 (m_b^2; m_t^2, M_i^2)
\end{eqnarray}
\begin{eqnarray}
\label{eqn:dgRH+}
\delta g^R  &=& -\frac{1}{16 \pi^2} \lambda_b^2 \frac{e}{s_Wc_W}
	\sum_{i,j}
	\langle H^+_i | \phi^+_1 \rangle \langle H^+_j | \phi^+_1 \rangle
	\left\{ \sum_{k=1}^{N}
	\langle H_{i}^{+} | \phi_{k}^{+} \rangle
	\langle H_{j}^{+} | \phi_{k}^{+} \rangle T^{3}_{\phi_{k}^{+}}
	- s_{W}^{2} \delta_{ij}  \right\}  \nonumber \\
& &	\times 2 C_{24} (m_t^2, M_i^2, M_j^2)  \nonumber \\
& & - \frac{1}{16 \pi^2} \lambda_b^2 \sum_i 
	\langle H^+_i | \phi^+_1 \rangle^2 
	\left\{ -2 g^L_{Zt\bar{t}} C_{24} + \frac{1}{2}g^L_{Zt\bar{t}}
	+ g^R_{Zt\bar{t}} C_0 \right\} (M_i^2, m_t^2, m_t^2)  \nonumber \\
& & + \frac{1}{16 \pi^2} \lambda_b^2 g^R_{Zb\bar{b}} \sum_i
	\langle H^+_i | \phi^+_1 \rangle^2
	B_1 (m_b^2; m_t^2, M_i^2).
\end{eqnarray}
The corrections for a Type I model are obtained by replacing 
$\phi_2^+$ with $\phi_1^+$ in $\delta g^L$.

We see that $\delta g^L$ is proportional to $\lambda_t^2$ and 
$\delta g^R$ is proportional to $\lambda_b^2$.  Clearly,
$\delta g^R$ is negligible 
compared to $\delta g^L$, except in a Type II model when $\lambda_b$ 
is enhanced by small $v_1$.
In this situation there is also a significant contribution to 
$\delta g^{L,R}$ coming from loops involving the neutral
Higgs bosons, as described in the next section.

In the Type II 2HDM, $\delta g^R$ is proportional to 
$(m_b \tan \beta)^2$, while 
$\delta g^L$ is proportional to $(m_t \cot \beta)^2$. 
At large $\tan\beta$,
$\delta g^R$ is enhanced and $\delta g^L$
is suppressed.  $\lambda_t$ and $\lambda_b$ are the same size
when $\tan\beta = m_t/m_b \simeq 50$.
However, because of their different dependence on the 
$Zq\bar{q}$ couplings,
$\delta g^L$ and $\delta g^R$ are the same size 
when $\tan \beta \simeq 10$. 

The formulas in equations \ref{eqn:dgLH+} -- \ref{eqn:dgRH+} 
can be simplified a great deal.  Electromagnetic gauge invariance requires
that the terms proportional to $s^{2}_{W}$ (from the $Zq\bar{q}$ and 
$ZH^{+}H^{-}$ couplings) add to zero in the limit $M_{Z}^{2} \rightarrow 0$.  
This provides a check on our calculations.  In our approximation
we neglect terms of order $M_{Z}^{2}/m_{t}^{2}$.
Using the expansions for
the two-- and three--point integrals given in appendix \ref{app:loopints}
and neglecting terms of order 
$M_{Z}^{2}/m_{t}^{2}$ in the three--point integrals, 
we find that the terms proportional to 
$s^2_W$ cancel.  The corrections can then be written as
%
\begin{eqnarray}
\delta g^{L,R} = && \mp \frac{1}{16\pi^{2}} \frac{e}{s_{W}c_{W}}
		\sum_{i} 
		(g_{H^{+}_{i}\bar{t}b}^{L,R})^{2} 
		\frac{1}{2} m_{t}^{2} C_{0}(M_{i}^2,m_{t}^2,m_{t}^2)	
		\nonumber \\
	      && - \frac{1}{16\pi^{2}} \frac{e}{s_{W}c_{W}}
		\sum_{i}
		(g_{H^{+}_{i}\bar{t}b}^{L,R})^{2}
		\sum_{k} \langle H_{i}^{+} | \phi_{k}^{+} \rangle^{2}
		(T^{3}_{\phi_{k}^{+}} - \frac{1}{2}) 
		2 C_{24}(m_{t}^2,M_{i}^2,M_{i}^2)
			\nonumber \\
	      && - \frac{1}{16\pi^{2}} \frac{e}{s_{W}c_{W}}
		\sum_{i} \sum_{j \neq i} 
		(g_{H^{+}_{i}\bar{t}b}^{L,R})(g_{H^{+}_{j}\bar{t}b}^{L,R})
		 \sum_{k}
		\langle H_{i}^{+} | \phi_{k}^{+} \rangle
		\langle H_{j}^{+} | \phi_{k}^{+} \rangle T^{3}_{\phi_{k}^{+}}
			\nonumber \\
		& & \times 2 C_{24}(m_{t}^2,M_{i}^2,M_{j}^2).
\label{eq:h+loops2}
\end{eqnarray}
The third term in equation \ref{eq:h+loops2} is the sum of the diagrams
\ref{fig:h+loops}(a) for two different charged Higgs bosons $H^{+}_{i}$ and
$H^{+}_{j}$ in the loop.  It is only nonzero when there are
nonzero off--diagonal $ZH^+_iH^-_j$ couplings ($i\neq j$).
The second term describes the contribution to diagrams \ref{fig:h+loops}(a)
from diagonal $ZH^{+}_{i}H^{-}_{i}$ couplings when
$T^{3}_{\phi_{k}^{+}}$ is different from $1/2$.  
This term is only nonzero when
the Higgs sector contains multiplets larger than doublets.
The first term comes from the sum of diagrams \ref{fig:h+loops}(b) and (c),
plus the remaining part of diagram \ref{fig:h+loops}(a) with 
$T^{3}_{\phi_{k}^{+}} = 1/2$.  This part of diagram
\ref{fig:h+loops}(a) is what we would get if we replaced all of the 
$ZH^{+}H^{-}$ couplings with the SM $ZG^{+}G^{-}$ coupling, 
$g_{ZG^{+}G^{-}} = \frac{-e}{s_{W}c_{W}} (\frac{1}{2} - s^{2}_{W})$.
Note that for $m_t \gg M_Z$, $C_0(M_i^2,m_t^2,m_t^2)$ is negative
(see appendix \ref{app:loopints}).
Therefore the first term of $\delta g^{L}$ ($\delta g^{R}$) is always
positive (negative) definite, which decreases the prediction for $R_{b}$.

We can learn much from equation \ref{eq:h+loops2}
about the contributions to 
$\delta g^{L,R}$ from charged Higgs boson exchange in various types of Higgs 
sectors.  First, if the Higgs sector contains only doublets and
singlets, $T^{3}_{\phi_{k}^{+}} = 1/2$ and there are no 
off-diagonal $ZH^{+}H^{-}$ couplings.  Then the second and third terms 
of equation \ref{eq:h+loops2} are
zero.  We are left with the first term,
\begin{eqnarray}
\label{eqn:H+doubletsonly}
\delta g^{L,R} & = & \mp \frac{1}{16\pi^{2}} \frac{e}{2s_{W}c_{W}}
		\sum_{i}
                (g_{H^{+}_{i}\bar{t}b}^{L,R})^{2} m_{t}^{2} 
		C_{0}(M_{i}^2,m_{t}^2,m_{t}^2)
			\nonumber \\
	& = & \delta g^{L,R}_{SM}  \pm \frac{1}{16\pi^{2}}
		  \frac{e}{2s_{W}c_{W}}
		\sum_{i \neq G^{+}} (g_{H^{+}_{i}\bar{t}b}^{L,R})^{2} 
		\left[ \frac{R_{i}}{R_{i}-1} - \frac{R_{i} \log R_{i}}
		{(R_{i}-1)^{2}} \right]
\end{eqnarray}
where $R_{i} \equiv m_{t}^{2}/M_{i}^{2}$.  The correction in the 
SM due to $G^{\pm}$ exchange is denoted by $\delta g^{L,R}_{SM}$.
The non--SM piece of 
$\delta g^{L}$ ($\delta g^{R}$) is positive (negative) definite, 
both of which decrease $R_{b}$.
Therefore, in order for it to be possible to increase
$R_{b}$ through
charged Higgs boson loops, we must have a Higgs sector that contains 
multiplets larger than doublets.

Second, if all the $H_{i}^{+}$ are degenerate with $G^{+}$, we
can sum over the complete sets of states in the second and third 
terms of equation \ref{eq:h+loops2}.  These terms cancel 
and again we are left with,
\begin{eqnarray}
\delta g^{L} &=&  \frac{\lambda_t^2}{16 \pi^2} 
	\frac{e}{2s_Wc_W} \left[ \frac{R}{R-1} - \frac{R \log R}
		{(R-1)^{2}} \right]  \\
\delta g^R &=& - \frac{\lambda_b^2}{16 \pi^2} 
	\frac{e}{2s_Wc_W} \left[ \frac{R}{R-1} - \frac{R \log R}
		{(R-1)^{2}} \right]
\end{eqnarray}
with $R = m_t^2 / M_W^2$.  This formula includes the SM correction
$\delta g^{L,R}_{SM}$.  
As above, the non--SM piece of $\delta g^{L}$ ($\delta g^{R}$) is 
positive (negative) definite, both of which decrease $R_b$. 

In a Higgs sector that contains only
multiplets for which $\rho = 1$ automatically 
(equation \ref{eqn:rhois1}),
the Goldstone boson
does not contribute to the second and third terms of 
equation \ref{eq:h+loops2} because there are no off--diagonal
$Z G^+ H_i^-$ couplings and the $Z G^+ G^-$ coupling is the same 
as in the SM.
Thus in such a model, if all the $H_{i}^{+}$
(excluding $G^{+}$) are degenerate with mass $M$, we can again sum
over the complete sets of states in
the second and third terms of equation \ref{eq:h+loops2}.  These
terms again cancel and we are left with
\begin{eqnarray}
\delta g^L &=& \frac{\lambda_t^2}{16 \pi^2} 
	\left( 1 - \frac{v_2^2}{v_{SM}^2} \right)
	 \frac{e}{2s_Wc_W} \left[ \frac{R}{R-1} - \frac{R \log R}
		{(R-1)^{2}} \right]
	+ \delta g^L_{SM}  \label{eqn:dgLdegenH+}\\
\delta g^R &=& - \frac{\lambda_b^2}{16 \pi^2} 
	\left( 1 - \frac{v_1^2}{v_{SM}^2} \right)
	\frac{e}{2s_Wc_W} \left[ \frac{R}{R-1} - \frac{R \log R}
		{(R-1)^{2}} \right]
	+ \delta g^R_{SM} 
\end{eqnarray}
with $R = m_t^2 / M^2$, for a Type II model.  The correction in a 
Type I model is obtained by replacing
$v_2$ with $v_1$ in equation \ref{eqn:dgLdegenH+}.  As above, the non--SM 
piece of $\delta g^{L}$ ($\delta g^{R}$) is positive (negative) definite,
both of which decrease $R_b$.



\chapter{Neutral Higgs corrections to $Z \rightarrow b \bar{b}$}
\label{sec-H0}

\begin{figure}
\begin{center}
\resizebox{8cm}{!}
        {\includegraphics*[100pt,160pt][370pt,700pt]{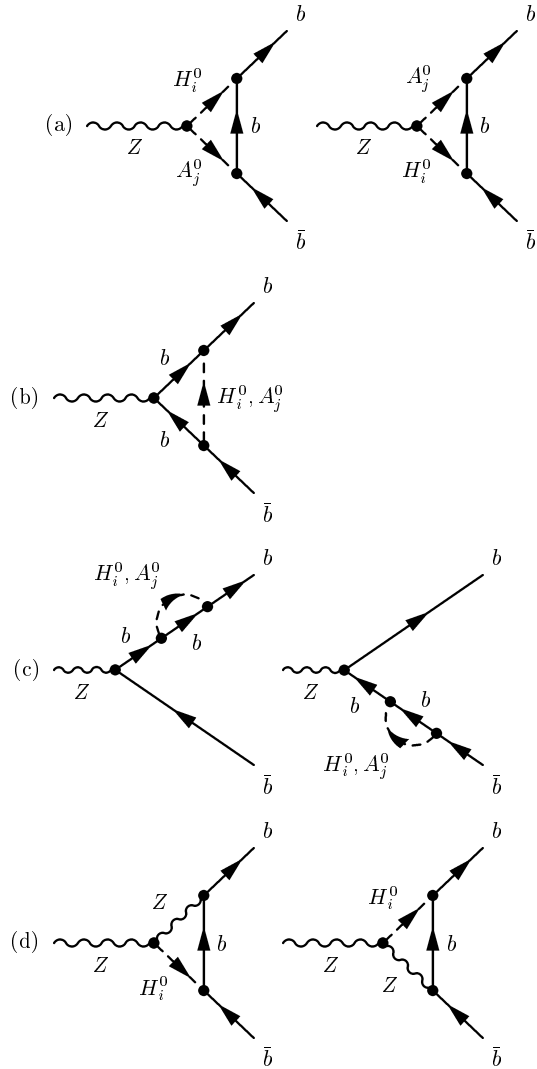}}
\end{center}
\caption{Feynman diagrams for the corrections to $Z \rightarrow b \bar{b}$
involving neutral Higgs bosons in the loop. 
}
\label{fig:h0loops}
\end{figure}

The corrections to $Z \rightarrow b \bar{b}$ from neutral
Higgs boson loops are shown in figure \ref{fig:h0loops}.  
As before, we assume that the Higgs sector is 
CP--conserving.
CP--even
states are denoted by $H_i^0$ and CP--odd states are 
denoted by $A_j^0$. 
These corrections
are proportional to $\lambda_b^2$ because the
neutral Higgs couplings to $b \bar{b}$ are proportional to 
$\lambda_b$.
In a Type I model, $\lambda_b \ll \lambda_t$, so
the corrections from neutral Higgs loops are negligible 
compared to the correction from $H^+$, which is proportional
to $\lambda_t^2$.
In a Type II model, however, $\lambda_b$ increases as $v_1$ decreases.
In the limit of small $v_1$, the corrections involving neutral
Higgs bosons are significant.

In calculating the corrections due to the diagrams in figure 
\ref{fig:h0loops},
we neglect terms proportional to $m_b$ that are not enhanced by small $v_1$.
The diagrams of figure \ref{fig:h0loops}(d)
are suppressed by a factor of $m_b/m_Z$ 
compared to diagrams \ref{fig:h0loops}(a), (b) and (c), 
and so we neglect them as well.
The contributions to $\delta g^{R,L}$ from diagrams 
\ref{fig:h0loops}(a), (b), and (c) are
\begin{eqnarray}
\delta g^{R,L}(a) &=& \pm \frac{1}{16\pi^{2}} \sum_{H^{0}_{i},A^{0}_{j}}
 4 g_{ZH^{0}_{i}A^{0}_{j}} g^{V}_{H^{0}_{i}b\bar{b}} 
	g^{A}_{A^{0}_{j}b\bar{b}}
	C_{24}(m_{b}^2,M_i^2,M_j^2) \nonumber \\
&=& \mp \frac{1}{16\pi^{2}} \lambda_b^2 \frac{e}{s_W c_W}
	\sum_{H^{0}_{i},A^{0}_{j}}
 \langle H^0_i | \phi_1^{0,r} \rangle 
 \langle A^0_j | \phi_1^{0,i} \rangle
  \nonumber \\
 & & \times \sum_{k=1}^{N}
 \langle H^0_i | \phi_k^{0,r} \rangle \langle A^0_j | \phi_k^{0,i} \rangle
 T^3_{\phi_k^0} \times 2 C_{24} (m_b^2, M_i^2, M_j^2)
\label{eq:h0loopsa}
\end{eqnarray}
\pagebreak 
\begin{eqnarray}
&\delta g^{R,L}(b) &
	\nonumber \\
	&=& - \frac{1}{16\pi^{2}} g^{L,R}_{Zb\bar{b}}
	 \left[  
\sum_{H^{0}_{i}} (g^{V}_{H^{0}_{i}b\bar{b}})^{2} 
	\left\{ -2 C_{24} + \frac{1}{2}
	 - M_{Z}^{2} (C_{22}-C_{23}) \right\}  
	 (M_i^2,m_{b}^2, m_{b}^2) \right.
	\nonumber \\
 	 & & \left. - \sum_{A^{0}_{j}} (g^{A}_{A^{0}_{j}b\bar{b}})^{2} 
	\left\{  -2 C_{24} + \frac{1}{2} 
	 - M_{Z}^{2} (C_{22}-C_{23}) \right\} (M_j^2,m_{b}^2, m_{b}^2)
	\right] \nonumber \\
&=& - \frac{1}{16\pi^{2}} g^{L,R}_{Zb\bar{b}} \frac{1}{2} \lambda_b^2 
	\left[
 \sum_{H^0_i} \langle H^0_i | \phi_1^{0,r} \rangle^2
 \left\{ -2 C_{24} + \frac{1}{2} - M_{Z}^{2} (C_{22}-C_{23}) 
 \right\} (M_i^2, m_{b}^2, m_{b}^2) \right. \nonumber \\
 & & + \left. \sum_{A_j^0} \langle A^0_j | \phi_1^{0,i}
 \rangle^2 
 \left\{ -2 C_{24} + \frac{1}{2} - M_{Z}^{2} (C_{22}-C_{23}) 
 \right\} (M_j^2, m_b^2, m_b^2) \right]
\end{eqnarray}
	%
\begin{eqnarray}
\delta g^{R,L}(c) &=& \frac{1}{16\pi^{2}} g^{R,L}_{Zb\bar{b}} \left[ \right. 
	\sum_{H^{0}_{i}} (g^{V}_{H^{0}_{i}b\bar{b}})^{2} 
	B_{1}(m_{b}^{2}; m_{b}^2, M_i^2)	\nonumber \\
 & & - \sum_{A^{0}_{j}} (g^{A}_{A^{0}_{j}b\bar{b}})^{2}
	B_{1}(m_{b}^{2}; m_{b}^2, M_j^2)
\left. \right]	\nonumber \\
&=& \frac{1}{16\pi^{2}} g^{R,L}_{Zb\bar{b}} \frac{1}{2} \lambda_b^2 \left[
 \sum_{H^0_i} \langle H^0_i | \phi_1^{0,r} \rangle^2
 B_1 (m_b^2; m_b^2, M_i^2) \right. \nonumber \\
& & + \left. \sum_{A_j^0} \langle A^0_j | \phi_1^{0,i}
 \rangle^2 B_1 (m_b^2; m_b^2, M_j^2) \right]
\label{eq:h0loopsc}
\end{eqnarray}
The two-- and three--point integrals are specified in appendix
\ref{app:loopints}.  As in section \ref{sec-H+}, we drop the 
first three arguments, $(m_b^2, M_Z^2, m_b^2)$, 
of the three--point integrals for compactness.
Note that $g_{ZH_i^0A_j^0}$ and $g^A_{A_j^0bb}$ are imaginary,
while $g^V_{H_i^0bb}$ is real.
The sum $A^{0}_{j}$ runs over all the CP--odd
neutral Higgs bosons, including $G^0$.  However, the corrections
involving $G^{0}$ can be neglected because the $G^0$ coupling
to $b\bar{b}$ is not enhanced by large $\lambda_b$.  
In particular,
$g^{A}_{G^{0}b\bar{b}} = - m_b / v_{SM}$, independent of
the value of $v_1$.

As in section \ref{sec-H+}, we can use electromagnetic gauge 
invariance to check our calculations.  Electromagnetic gauge
invariance requires that terms proportional to $s^2_W$ sum to
zero in the limit $M_Z \rightarrow 0$.  $\delta g^{R,L}(a)$ is 
independent of $s^2_W$.  In the limit $M_Z \rightarrow 0$,
we find that 
$\delta g^{R,L}(b) + \delta g^{R,L}(c) = 0$, independent of the
Higgs masses.  The terms proportional to $s^2_W$ indeed vanish
in this limit.  Although this is a useful
formal check, the approximation $M_Z \rightarrow 0$ cannot be
applied to the general formulas in equations 
\ref{eq:h0loopsa} -- \ref{eq:h0loopsc} because terms 
proportional to $M_Z^2$ are important here.

We now examine the special case in which all the $H^0_i$ are degenerate
with mass $M_H$, and all the $A^0_j$ are degenerate with
mass $M_A$.  We neglect $G^0$, so that $M_A$ does not have to 
be equal to $M_Z$, which is the $G^0$ mass in the 't Hooft--Feynman
gauge.  In this case, we can sum over complete sets of states
and equations \ref{eq:h0loopsa} -- \ref{eq:h0loopsc} simplify
to,
\begin{equation}
\delta g^{R,L} (a) = \pm \frac{1}{16 \pi^2} \lambda_b^2 
 \left( \frac{e}{s_W c_W} \right) C_{24} (m_b^2, M_H^2, M_A^2)
\end{equation}
\begin{eqnarray}
\delta g^{R,L} (b) &=& - \frac{1}{16 \pi^2} g^{L,R}_{Zb\bar{b}} \frac{1}{2}
 \lambda_b^2 \times \nonumber \\
  & & \left[ \left\{ -2 C_{24} + \frac{1}{2} - M_Z^2 (C_{22} - C_{23})
 \right\} (M_H^2, m_b^2, m_b^2) \right. \nonumber \\
 & & + \left. \left\{ -2 C_{24} + \frac{1}{2} - M_Z^2 (C_{22} - C_{23})
 \right\} (M_A^2, m_b^2, m_b^2) \right]
\end{eqnarray}
\begin{equation}
\delta g^{R,L} (c) = \frac{1}{16 \pi^2} g^{R,L}_{Zb\bar{b}} 
	\frac{1}{2} \lambda_b^2
 	\left[ B_1 (m_b^2; m_b^2, M_H^2) + B_1 (m_b^2; m_b^2, M_A^2) \right].
\end{equation}



\chapter{Corrections to $Z \rightarrow b \bar{b}$ in specific models}
\label{sec:models}

In this chapter we calculate the radiative corrections to
$Z \rightarrow b \bar{b}$ in specific extended Higgs models.
We discuss the form of the corrections in each model.
We also show the constraints on the parameter space of
each model due to the experimental data.

In section \ref{sec:61nonexotic} we calculate the corrections
in models containing only Higgs doublets and/or singlets.
We examine the contributions due to both charged Higgs boson and 
neutral Higgs boson exchange.  We discuss the two Higgs doublet model
in detail, and describe the effects of adding additional Higgs
doublets and singlets.  In section \ref{sec:exotics} we 
calculate the corrections in a number of models containing 
one or more Higgs multiplets larger than doublets. 
We discuss two classes of models which take two different 
approaches to satisfy the experimental 
constraint on the $\rho$ parameter, $\rho \approx 1$.  We first
discuss models which are fine--tuned so that the multiplet larger
than a doublet has a very small vev.  Finally we discuss the
models that preserve ${\rm SU}(2)_c$ symmetry. 

We find that the corrections to $R_b$ are large enough that the measurement
of $R_b$ can be used to constrain the parameter space of specific models.
However, the corrections to $A_b$ are small compared to the uncertainty
in the measurement of $A_b$, and thus cannot be used to further
constrain the models.

\section{Models with Higgs doublets and singlets}
\label{sec:61nonexotic}

\subsection{Charged Higgs boson contributions}

We saw in section \ref{sec-H+} that, in a model containing only Higgs 
doublets and singlets, the radiative corrections due to the charged 
Higgs bosons
are described by equation \ref{eqn:H+doubletsonly}.  We also saw that
these corrections 
have definite signs; in particular, $\delta g^L > 0$ and $\delta g^R < 0$.  
Both of these give $\Delta R_b < 0$, in worse agreement with experiment 
than the SM.

In this section, we calculate the corrections due to charged Higgs bosons
in specific models containing Higgs doublets or singlets or both.  
We can then use the measurement of $R_b$ to constrain the models.
Note that the corrections due to neutral
Higgs boson exchange will also contribute when $\lambda_b$ is enhanced.
They must be taken into account as well in this regime when deriving
constraints from the $R_b$ measurement.
We first consider the corrections in the two Higgs doublet model (2HDM),
then extend the results to multi--doublet models and models with doublets
and singlets.

\subsubsection{Two Higgs doublet model}

The 2HDM contains a single charged Higgs boson, $H^+$.  Its 
contribution to $\delta g^{L,R}$ is found from equation 
\ref{eqn:H+doubletsonly} with only one $H^+$ in the sum.
For the Type II model,
\begin{equation}
\delta g^{L} = \frac{1}{16\pi^{2}} \left(\frac{gm_{t}}{\sqrt{2}M_{W}}
\cot \beta \right)^{2}
\frac{1}{2} \frac{e}{s_{W}c_{W}} \left[ \frac{R}{R-1} - \frac{R \log R}
{(R-1)^{2}}  \right],
\label{eq:dgl2HDM}
\end{equation}
\begin{equation}
\delta g^{R} = -\frac{1}{16\pi^{2}} \left( \frac{gm_{b}}{\sqrt{2}M_{W}}
\tan \beta \right)^{2}
\frac{1}{2} \frac{e}{s_{W}c_{W}} \left[ \frac{R}{R-1} - \frac{R \log R}
{(R-1)^{2}}  \right],
\label{eq:dgr2HDM}
\end{equation}
were $R = m_{t}^{2}/M_{H^{+}}^{2}$.  This correction is in addition
to the correction due to Goldstone boson exchange, which is the same 
as in the SM.
This agrees with the results of references 
\cite{Denner91,Djouadi91,Boulware91,Grant,Kundu96}.  In the Type II
model, $\delta g^L$ is significant at small $\tan\beta$ and is suppressed
at large $\tan\beta$, while $\delta g^R$ is negligible at small $\tan\beta$
but is significant at large $\tan\beta$.

In a Type I model the result is the same except that $\cot^2\beta$ is
replaced with $\tan^2\beta$ in $\delta g^{L}$.  In this case, $\delta g^R$
is negligible compared to $\delta g^L$ at any value of $\tan\beta$.
Both $\delta g^L$ and $\delta g^R$ grow with increasing $\tan\beta$.

This correction decreases for large $M_{H^{+}}$.  It goes to 
zero in the decoupling limit, $M_{H^+} \rightarrow \infty$.  
(For a discussion of the decoupling limit, see reference
\cite{decoupling}.)

\begin{figure}
\resizebox{\textwidth}{!}{\rotatebox{270}{\includegraphics{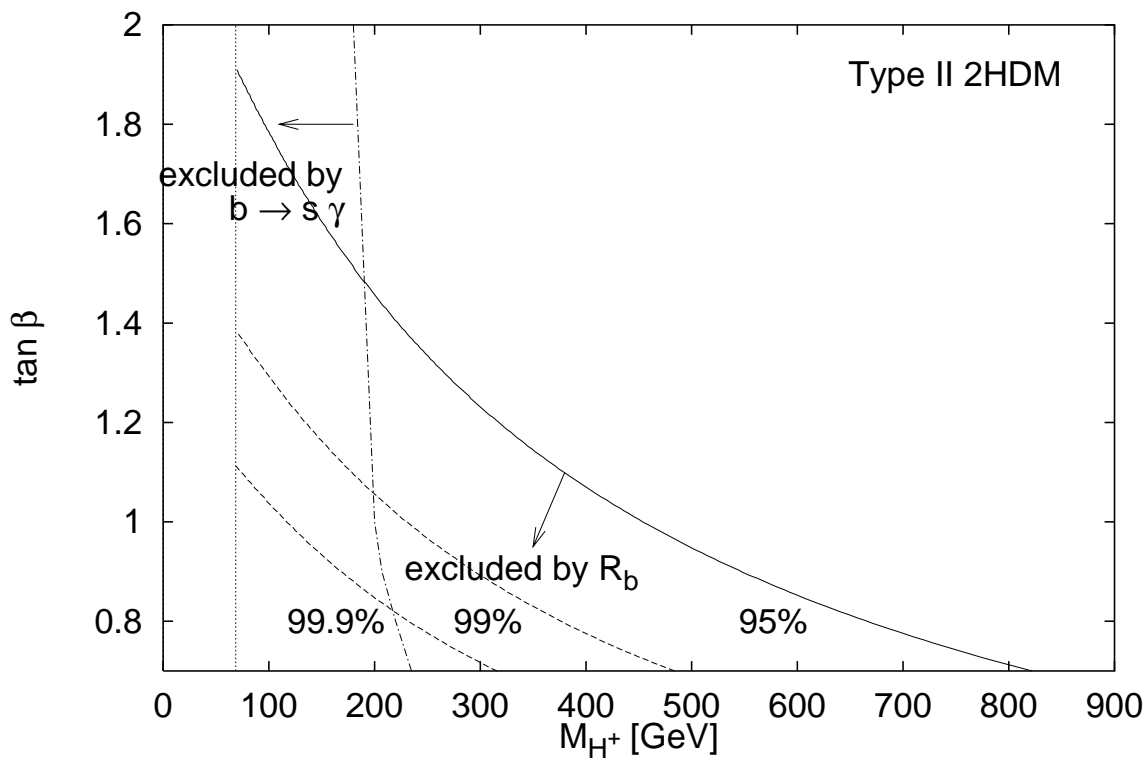}}}
\caption[Constraints from $R_b$ on the charged Higgs mass 
and $\tan\beta$ in the Type II 2HDM]
{Constraints from $R_b$ on the charged Higgs mass and $\tan\beta$ in the
Type II 2HDM.  The area below the solid line is excluded at 95\%
confidence level.  Also shown are
the 99\% and 99.9\% confidence levels (dashed lines).
We also show the 95\% confidence level lower bound on $M_{H^+}$
from the $b \rightarrow s \gamma$ branching ratio as measured by
CLEO \cite{CLEO98,Borzumati98} (dot-dashed).
The vertical dotted line is the direct search bound on the charged
Higgs mass
from the OPAL collaboration, $M_{H^+} > 68.7$ GeV \cite{OPALH+99}, 
from LEP data up to $\sqrt{s} = 189$ GeV.  (For a discussion of the
direct search bound, see appendix \ref{app:dirsearches}.)
A large area below $\tan \beta \approx 1.5$ is excluded by $R_b$.
}
\label{fig:mtwohdm}
\end{figure}

For small $\tan\beta$, the neutral Higgs couplings to $b$ quarks are
small, and contributions to $Z \rightarrow b \bar{b}$ due to 
neutral Higgs boson exchange can be neglected.  In this regime the
corrections due to charged Higgs boson exchange can be used to constrain
the 2HDM.
In figure \ref{fig:mtwohdm} we plot the constraints from $R_b$ on 
$M_{H^+}$ as a function of $\tan\beta$, for a Type II 2HDM.
The input parameters for the calculation are summarized in 
appendix \ref{app:SMinputs}.  We also show the constraints on
the charged Higgs mass from the process $b \to s \gamma$ 
\cite{CLEO98,Borzumati98} and the direct searches at LEP
(for references and a discussion, see appendix \ref{app:dirsearches}).
The constraint on the charged Higgs mass from the D0 experiment
\cite{D0chargedH} are significantly weaker than the constraint 
from $b \to s \gamma$, and are not shown in figure \ref{fig:mtwohdm}.
$R_b$ provides the strongest constraint on $M_{H^+}$ for 
$\tan\beta < 1.5$.  For larger 
$\tan\beta$, the constraint from $b \rightarrow s \gamma$ 
is stronger.  

Radiative corrections to the branching ratio BR($\bar{B} \to X_s \gamma$)
involving charged Higgs boson exchange in 2HDMs have been 
calculated in reference \cite{Borzumati98}.  By comparing 
BR($\bar{B} \to X_s \gamma$) to the value measured at CLEO \cite{CLEO98},
reference \cite{Borzumati98} finds a constraint on the charged Higgs 
boson mass in the Type II 2HDM, as shown in figure \ref{fig:mtwohdm}.  
In the Type I 2HDM, however, the predicted 
range of BR($\bar{B} \to X_s \gamma$) falls within the experimental 
limits.  Thus in a Type I 2HDM, there is no constraint on the 
charged Higgs boson mass from $b \rightarrow s \gamma$.

For large $\tan\beta$, neutral Higgs boson exchange contributes to 
$Z \rightarrow b \bar{b}$ in addition to charged Higgs boson exchange.  
The neutral Higgs boson contributions are discussed in section 
\ref{sec:NeutralHiggs}.

We stress that the bounds in figure \ref{fig:mtwohdm} are valid when the
only non--SM corrections to $Z \rightarrow b \bar{b}$ are due to $H^+$
exchange.  They cannot be applied to models that contain additional 
particles which contribute to the corrections.  In particular, supersymmetric
models give rise to additional corrections to $Z \rightarrow b \bar{b}$
from squark and Higgsino exchange.  However, in the limit that the 
supersymmetric particles are very heavy, their contributions to 
radiative corrections go to zero, and the bounds in figure \ref{fig:mtwohdm}
remain valid.

In the case of a Type I 2HDM, the bound on $M_{H^+}$ from $R_{b}$ 
is the same as in figure \ref{fig:mtwohdm}, but with $\cot\beta$
replacing $\tan\beta$ on the vertical axis.

\subsubsection{Multiple--doublet models and models with singlets}

We now consider charged Higgs boson exchange in a model containing multiple 
Higgs doublets, denoted $\Phi_k$, with hypercharge 1.  
We can add to this model any number of Higgs singlets 
with zero hypercharge.  These contain only neutral degrees of freedom,
and so they have no effect on the charged Higgs sector.

In a Type I model of this type, we let $\Phi_1$ couple to both up--
and down--type quarks, and none of the other doublets couple to quarks.
In a Type II model, we let $\Phi_1$ couple only to down--type quarks, 
and $\Phi_2$ couple to up--type quarks.  Then the Yukawa couplings 
are defined in the same way as in the 2HDM, in equations 
\ref{eqn:lambdatI}--\ref{eqn:lambdabII}.

In a Type II model, the contributions to $Z \rightarrow b \bar{b}$ from
charged Higgs boson exchange are
\begin{eqnarray}
\delta g^L &=& \frac{1}{16 \pi^2} \frac{1}{2} \frac{e}{s_W c_W} 
\left( \frac{gm_t}{\sqrt{2}M_W} \frac{v_{SM}}{v_2} \right)^2
 \sum_{i \neq G^+} \langle H_i^+ | \phi_2^+ \rangle^2
 \left[ \frac{R_i}{R_i-1} - \frac{R_i \log R_i}{(R_i-1)^2} \right]
\label{eqn:dgLMHDM}
\end{eqnarray}
\begin{eqnarray}
\delta g^R &=& -\frac{1}{16 \pi^2} \frac{1}{2} \frac{e}{s_W c_W} 
\left( \frac{gm_b}{\sqrt{2}M_W} \frac{v_{SM}}{v_1} \right)^2
 \sum_{i \neq G^+} \langle H_i^+ | \phi_1^+ \rangle^2
 \left[ \frac{R_i}{R_i-1} - \frac{R_i \log R_i}{(R_i-1)^2} \right]
\end{eqnarray}
where $R_{i} \equiv m_{t}^{2}/M_{H_i^+}^{2}$.  This contribution is in addition
to the contribution due to charged Goldstone boson exchange, which is the
same as in the SM.
In a Type I model, the contribution is the same except that $v_2$ is replaced
with $v_1$ and $\phi_2^+$ is replaced with $\phi_1^+$ in the formula 
for $\delta g^L$.

These corrections to $\delta g^{L,R}$ from charged Higgs boson exchange have
the same dependence on the charged Higgs masses as the corrections in
the 2HDM.  The contribution from each $H_i^+$ is weighted by the
overlap of each $H^+_i$ with the electroweak eigenstate
that couples to the quarks involved.

Note that the Yukawa couplings depend on the ratios $v_{SM}/v_2$ 
and $v_{SM}/v_1$.
This is the same dependence as in the 2HDM.  Recall that in the 2HDM, $v_1$ and
$v_2$ were constrained by the $W$ mass to satisfy the relation, 
$v_1^2 + v_2^2 = v_{SM}^2$, where $v_{SM} = 246$ GeV.  
Thus in the 2HDM, $v_1$ and $v_2$ cannot both
be small at the same time.  However, in a model with more than two
doublets, the $W$ mass constraint involves the vevs of all the doublets,
giving $\sum_k v_k^2 = v_{SM}^2$, where $k$ runs over all the 
Higgs doublets.  In this model, both $v_1$ and $v_2$ can be small at the
same time, leading to singificant contributions to both $\delta g^L$ and
$\delta g^R$.

The corrections to $Z \rightarrow b \bar{b}$ in this model can be understood
by examining their behavior in certain limits.
First, let us examine the limit in which all but one of the $H_i^+$
are very heavy.  The contributions of the heavy $H_i^+$ to 
$\delta g^{L,R}$ go to zero as the masses go to infinity.  The remaining
contribution to $\delta g^{L,R}$ is due to the single light charged Higgs
boson,
and it is of the same form as in the 2HDM.  Comparing with equations
\ref{eq:dgl2HDM}--\ref{eq:dgr2HDM}, we see that in $\delta g^L$,
$\tan\beta$ is replaced by 
$\frac{v_2}{v_{SM}\langle H_i^+ | \phi_2^+ \rangle}$,
and in $\delta g^R$, $\tan\beta$ is replaced by 
$\frac{v_{SM}\langle H_i^+ | \phi_1^+ \rangle}{v_1}$.  The charged Higgs
sector can be constrained by $R_b$ when there are no significant contributions
to $Z \rightarrow b \bar{b}$ coming from neutral Higgs boson exchange.  This is
ensured when $v_1$ is not too small.  In this regime, $\delta g^L$ can
be significant, while $\delta g^R$ is negligible.
The constraint from $R_b$ on the mass of the remaining light 
charged Higgs boson is the same as in figure \ref{fig:mtwohdm}, with
$\tan\beta$ replaced by 
$\frac{v_2}{v_{SM}\langle H_i^+ | \phi_2^+ \rangle}$.  

If $v_2$ and $\langle H_i^+ | \phi_2^+ \rangle$ are held constant while
the masses of the heavy charged Higgs bosons are reduced, the bound shown in
figure \ref{fig:mtwohdm} becomes stronger.  This happens because the
heavy charged Higgs bosons begin to contribute to $\delta g^L$, forcing the
contribution of the light charged Higgs boson to be smaller in order to be
consistent with the measured value of $R_b$.  This is done by raising the 
mass of the light charged Higgs boson.

Finally, if all the charged Higgs bosons are
degenerate, with a common mass $M_H$, then we can sum over a complete set 
of states and the corrections simplify to the following, again for a
Type II model:
\begin{eqnarray}
\delta g^L &=& \frac{1}{16 \pi^2} \frac{1}{2} \frac{e}{s_Wc_W} 
\left( \frac{gm_t}{\sqrt{2}M_W} \right)^2 
\frac{v_{SM}^2 - v_2^2}{v_2^2}
 \left[\frac{R}{R-1} - \frac{R \log R}{(R-1)^2}
 \right] 
\end{eqnarray}
\begin{eqnarray}
\delta g^R &=& \frac{1}{16 \pi^2} \frac{1}{2} \frac{e}{s_Wc_W} 
\left( \frac{gm_b}{\sqrt{2}M_W} \right)^2
\frac{v_{SM}^2 - v_1^2}{v_1^2}
\left[\frac{R}{R-1} - \frac{R \log R}{(R-1)^2}
 \right],
\end{eqnarray}
where $R=m_t^2/M_H^2$.  These corrections are in addition to the corrections
due to charged Goldstone boson exchange in the SM. 
In a Type I model, $v_2$ is replaced by $v_1$ in
$\delta g^L$.

These corrections are the same as the corrections in the 2HDM, with
$\tan\beta$ replaced by $v_2 / \sqrt{v^2_{SM} - v_2^2}$ in $\delta g^L$,
and $\tan\beta$ replaced by $\sqrt{v_{SM}^2 - v_1^2} / v_1$ in 
$\delta g^R$.  As before, the charged Higgs
sector can be constrained by $R_b$ when there are no significant contributions
to $Z \rightarrow b \bar{b}$ coming from neutral Higgs boson exchange.  This is
ensured when $v_1$ is not too small.  In this regime,
the constraint from $R_b$ on the common charged Higgs mass $M_H$ 
is the same as in figure \ref{fig:mtwohdm}, with 
$\tan\beta$ replaced by $v_2 / \sqrt{v^2_{SM} - v_2^2}$.

\subsection{Neutral Higgs boson contributions}
\label{sec:NeutralHiggs}

As we showed in section \ref{sec-H0}, the radiative corrections to 
the process $Z \rightarrow b \bar{b}$ due to neutral Higgs boson exchange 
are proportional to $\lambda_b^2$.  They
are negligible compared to the contributions from charged Higgs boson
exchange which are proportional to $\lambda_t^2$, except when
$\lambda_b$ is enhanced relative to $\lambda_t$.  This happens
in a Type II model when $v_1$ is much smaller than $v_2$.  In what
follows we consider only Type II models.

In this section, we calculate the corrections due to neutral Higgs boson
exchange in specific models containing Higgs doublets and/or singlets.  
We can then use the measurement of $R_b$ to constrain the models.
Note that when $\lambda_b$ in enhanced, the corrections to 
$\delta g^R$ due to charged Higgs boson exchange will also contribute.
We take these into account when deriving constraints from the $R_b$
measurement.

We first consider the corrections in the two Higgs doublet model
(2HDM).  We also examine the corrections in the 2HDM in the 
decoupling limit, in which $h^0$ remains light and its couplings
to SM particles approach those of the SM Higgs boson, while all the other 
Higgs bosons become heavy and decouple from SM particles.  We then 
extend the results to multi--doublet models and models with doublets
and singlets.  Finally we examine the multi--doublet model when 
some of the neutral Higgs bosons are degenerate.

\subsubsection{Two Higgs doublet model}

The corrections due to neutral Higgs boson exchange in the 2HDM depend
on the masses of the three neutral Higgs bosons, $h^0$, $H^0$, and $A^0$,
the mixing angle $\alpha$ of the two CP--even states $h^0$ and 
$H^0$, and of course $\tan\beta$, which determines $\lambda_b$ and 
the mixing between $A^0$ and $G^0$.

The neutral Higgs couplings in the 2HDM are given in appendix
\ref{app:2HDM}.  Inserting these couplings into equations 
\ref{eq:h0loopsa}--\ref{eq:h0loopsc} for the corrections from
neutral Higgs boson exchange, we find,
\begin{eqnarray}
\delta g^{R,L}(a) &=& \pm \frac{1}{16\pi^2} \frac{e}{s_Wc_W}
	\left( \frac{gm_b}{\sqrt{2}M_W} \right)^2 \tan^2\beta
	\nonumber \\
	& & \times \left[ \frac{s_{\alpha}}{s_{\beta}} 
	\cos(\beta - \alpha)
	C_{24}(m_b^2,M_{h^0}^2,M_{A^0}^2)
	\right. \nonumber \\
	& & + \left. \frac{c_{\alpha}}{s_{\beta}} 
	\sin(\beta - \alpha)
	C_{24}(m_b^2,M_{H^0}^2,M_{A^0}^2) \right] 
\end{eqnarray}
\begin{eqnarray}
\delta g^{R,L}(b) &=& -\frac{1}{16\pi^2} g^{L,R}_{Zb\bar{b}}
	\frac{1}{2} \left( \frac{gm_b}{\sqrt{2}M_W} \right)^2
	\tan^2\beta 
	\nonumber \\
	& & \times \left[
	\left(\frac{s_{\alpha}}{s_{\beta}}\right)^2
	\left[-2C_{24} + \frac{1}{2} - M_Z^2(C_{22}-C_{23})\right]
	(M_{h^0}^2,m_b^2,m_b^2)
	\right. \nonumber \\
	& & + \left(\frac{c_{\alpha}}{s_{\beta}}\right)^2
	\left[-2C_{24} + \frac{1}{2} - M_Z^2(C_{22}-C_{23})\right]
	(M_{H^0}^2,m_b^2,m_b^2)
	\nonumber \\
	& & + \left. 
	\left[-2C_{24} + \frac{1}{2} - M_Z^2(C_{22}-C_{23})\right]
	(M_{A^0}^2,m_b^2,m_b^2) \right]  
\end{eqnarray}
\begin{eqnarray}
\delta g^{R,L}(c) &=& \frac{1}{16\pi^2} g^{R,L}_{Zb\bar{b}}
	\frac{1}{2} \left( \frac{gm_b}{\sqrt{2}M_W} \right)^2
	\tan^2\beta 
	\left[
	\left(\frac{s_{\alpha}}{s_{\beta}}\right)^2
	B_1(m_b^2;m_b^2,M_{h^0}^2)
	\right. \nonumber \\
	& & + \left. \left(\frac{c_{\alpha}}{s_{\beta}}\right)^2
	B_1(m_b^2;m_b^2,M_{H^0}^2)
	 + B_1(m_b^2;m_b^2,M_{A^0}^2) \right],
\end{eqnarray}
where $s_{\alpha} = \sin\alpha$, $c_{\alpha}=\cos\alpha$,
$s_{\beta}=\sin\beta$, and $c_{\beta}=\cos\beta$.

The contribution of these corrections to $R_b$ 
can be either positive or negative, depending on the
neutral Higgs masses and the mixing angle $\alpha$.
We plot the corrections for various sets of parameters.

\begin{figure}
\resizebox{\textwidth}{!}{\rotatebox{270}{\includegraphics{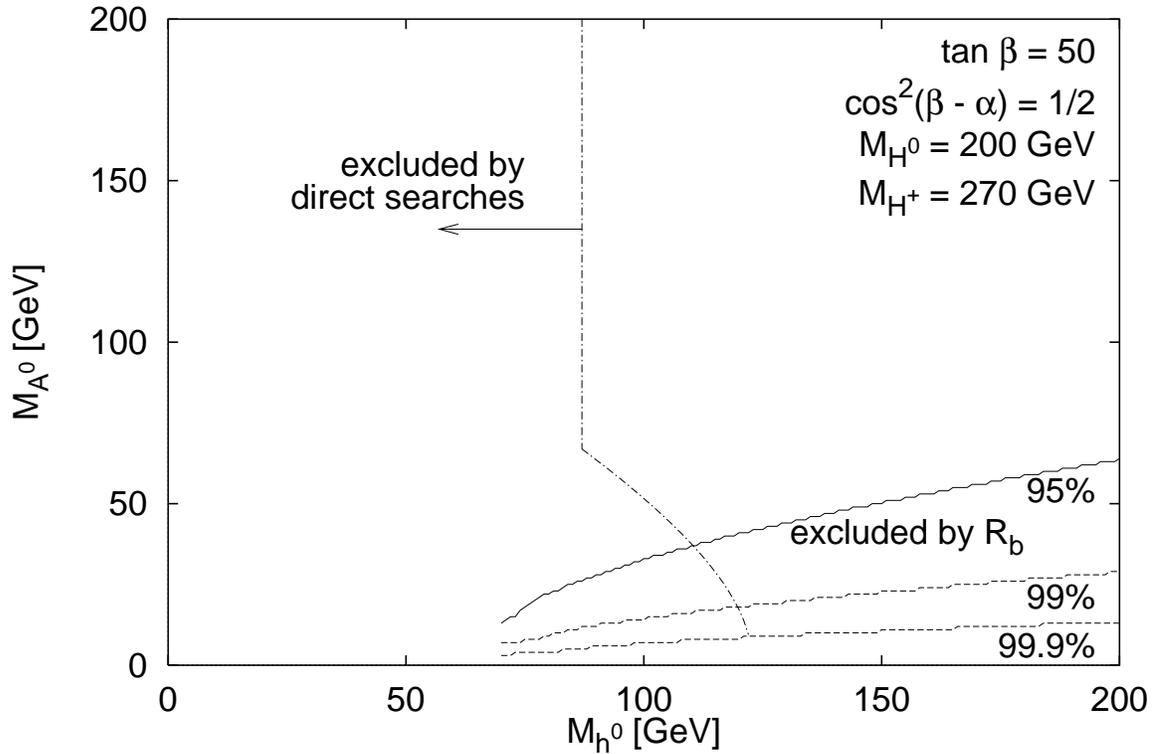}}}
\caption[$R_b$ in the 2HDM with $\tan\beta = 50$,
$\cos^2(\beta - \alpha) = 1/2$, $M_{H^0} = 200$ GeV, and
$M_{H^+} = 270$ GeV]
{$R_b$ in the Type II 2HDM with $\tan\beta = 50$,
$\cos^2(\beta - \alpha) = 1/2$, $M_{H^0} = 200$ GeV, and
$M_{H^+} = 270$ GeV.  The axes are $M_{A^0}$ and $M_{h^0}$.
$\Delta R_b < 0$ for all allowed masses, so this model is 
in worse agreement with experiment than the SM.  The solid
line is the 95\% confidence level lower bound on $M_{A^0}$ from
$R_b$.  We also plot the 99\% and 99.9\% confidence level contours
(dashed lines).
The dot-dashed line is the lower bound on $M_{h^0}$ from direct searches,
as discussed in appendix \ref{app:dirsearches}.
}
\label{fig:Ab0_2_no}
\end{figure}

\begin{figure}
\resizebox{\textwidth}{!}{\rotatebox{270}{\includegraphics{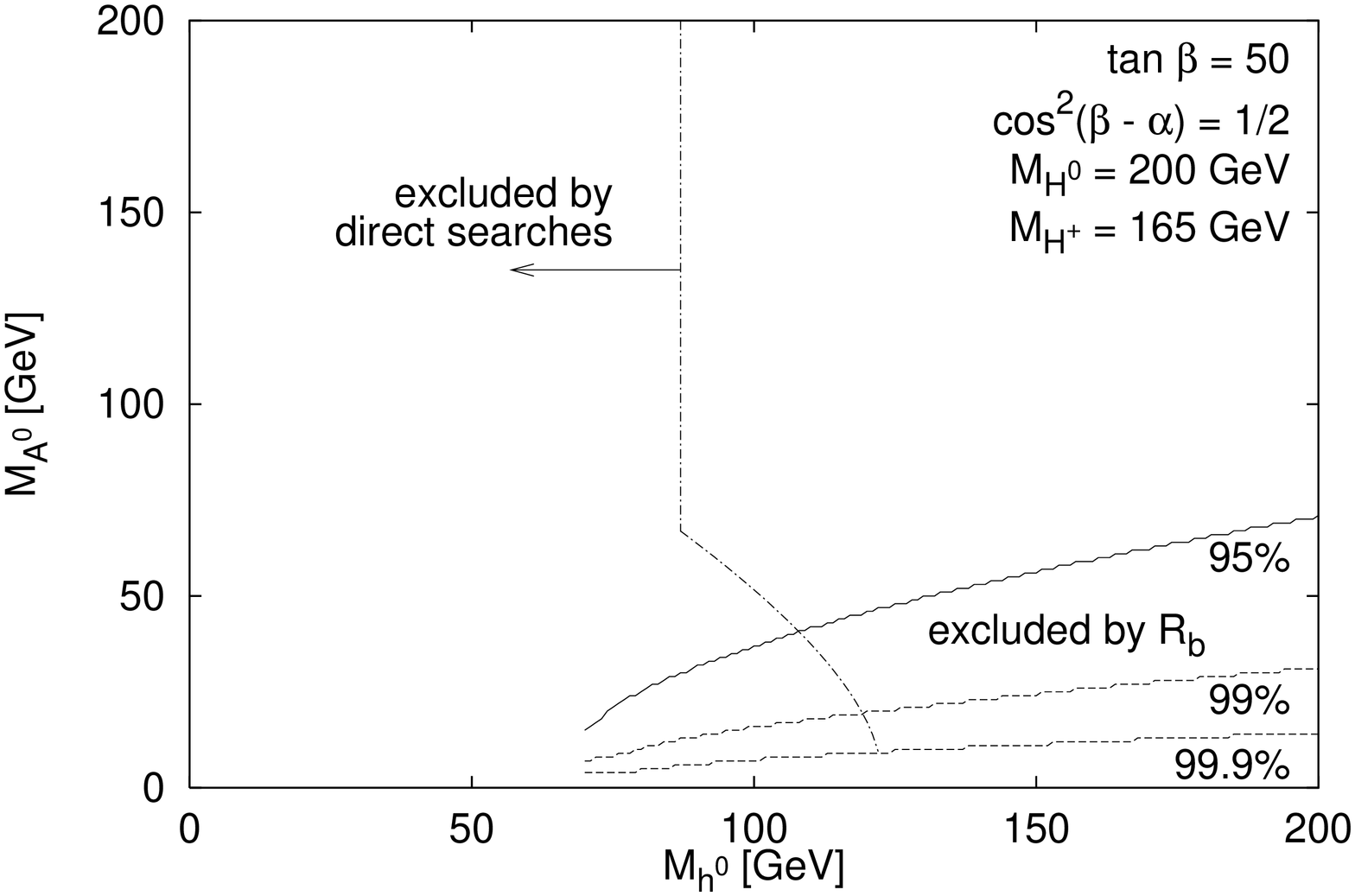}}}
\caption[$R_b$ in the 2HDM with $M_{H^+} = 165$ GeV
and other parameters the same as in figure \ref{fig:Ab0_2_no}]
{$R_b$ in the 2HDM with $M_{H^+} = 165$ GeV (the lower limit
from $b \rightarrow s \gamma$ \cite{Borzumati98}), and other parameters
the same as
in figure \ref{fig:Ab0_2_no}.  Since $\Delta R_b < 0$ from $H^+$,
the 95\% confidence level exclusion curve from $R_b$ (solid line)
moves upward compared to $M_{H^+} \rightarrow \infty$.
Also shown are the 99\% and 99.9\% confidence levels (dashed lines).
The dot-dashed
line is
the lower bound on $M_{h^0}$ from direct searches as in figure
\ref{fig:Ab0_2_no}.
}
\label{fig:Ab0_2_165}
\end{figure}

In figures \ref{fig:Ab0_2_no} and \ref{fig:Ab0_2_165}, we plot 
the constraints on the neutral Higgs sector from $R_b$.  
The parameters in these plots are $\tan\beta = 50$, 
$\cos^2(\beta - \alpha) = 1/2$,
and $M_{H^0} = 200$ GeV.  With $\cos^2(\beta - \alpha) = 1/2$,
the $Zh^0A^0$ and $ZH^0A^0$ couplings are equal, and
$h^0$, $H^0$, and $A^0$ all
contribute to the corrections.  The charged Higgs boson can also contribute.
If the charged Higgs boson were taken to be very heavy, its contribution
to $R_b$ would go to zero and only the effects of the neutral Higgs bosons
would remain.  However, a large mass splitting between the charged Higgs
boson and the neutral Higgs bosons results in large radiative corrections
to the $\rho$ parameter (see, e.g., reference \cite{Haber93rho}
and appendix \ref{app:2HDM}).  
Using the formula for $\Delta \rho_{\rm new}$ from neutral Higgs exchange
in the 2HDM given in appendix \ref{app:2HDM}, and requiring that
$-4.7 \times 10^{-3} < \Delta \rho_{\rm new} < 3.0 \times 10^{-3}$
(see section \ref{sec:finetuned} for further details), we find that
for the
parameters of figures \ref{fig:Ab0_2_no} and \ref{fig:Ab0_2_165}, 
the charged Higgs boson must be lighter than 270 GeV.  
The lower bound on the charged Higgs mass at large $\tan\beta$ is
$M_{H^+} > 165$ GeV, from $b \rightarrow s \gamma$ \cite{CLEO98,Borzumati98}.
In figure \ref{fig:Ab0_2_no} we take $M_{H^+} = 270$ GeV, and 
in figure \ref{fig:Ab0_2_165} we take $M_{H^+} = 165$ GeV, in 
order to show the full allowed range of charged Higgs boson contributions.
For large $\tan\beta$, the charged Higgs boson contributions to 
$\delta g^L$ are negligible.  The charged Higgs boson contributions to
$\delta g^R$ are negative, which reduces $R_b$.

For these parameters, the range of masses of $h^0$ and $A^0$ 
in which $\Delta R_b > 0$ is already excluded by direct searches.  
For all remaining allowed $h^0$ and $A^0$ masses,
$\Delta R_b < 0$, in worse agreement with experiment than the SM.
Since the corrections from both the charged and neutral Higgs bosons
are proportional to $\tan^2\beta$, we can vary $\tan\beta$ within
the large $\tan\beta$ regime and $\Delta R_b$ will still be negative.
Since the corrections grow with $\tan^2\beta$, the region ruled out
by $R_b$ gets larger as $\tan\beta$ increases.

The correction to $A_b$ is very small compared to the experimental
uncertainty in the $A_b$ measurement.  For these parameters, 
$|\Delta A_b| < 0.003$.  Also, $\Delta A_b > 0$ in the regions allowed
by the direct search bounds, in worse agreement
with experiment than the SM.  For $M_{H^+} = 270$ GeV, we find that
$0.936 < A_b < 0.937$ in the allowed region.  For $M_{H^+} = 165$ GeV,
we find that $0.937 < A_b < 0.938$ in the allowed region.

For $\cos^2(\beta - \alpha) = 1/2$, the lower bound on $M_{h^0}$ from 
direct searches is 
87 GeV for arbitrary $M_{A^0}$, as discussed in 
appendix \ref{app:dirsearches}.  Combining the direct search bound and
the constraint from $R_b$, we find that for these parameters with
$M_{H^+} = 270$ GeV, the lower bound on the $A^0$ mass is 
$M_{A^0} > 37$ GeV (shown in figure \ref{fig:Ab0_2_no}).  For an $h^0$ 
mass of 200 GeV or greater, $M_{A^0} > 64$ GeV.
Similarly, if $M_{H^+} = 165$ GeV, we find that the lower bound on the
$A^0$ mass increases to 40 GeV, as shown in figure \ref{fig:Ab0_2_165}.
In this case, for an $h^0$ mass of 200 GeV or greater, 
$M_{A^0} > 71$ GeV.

\begin{figure}
\resizebox{\textwidth}{!}{\rotatebox{270}{\includegraphics{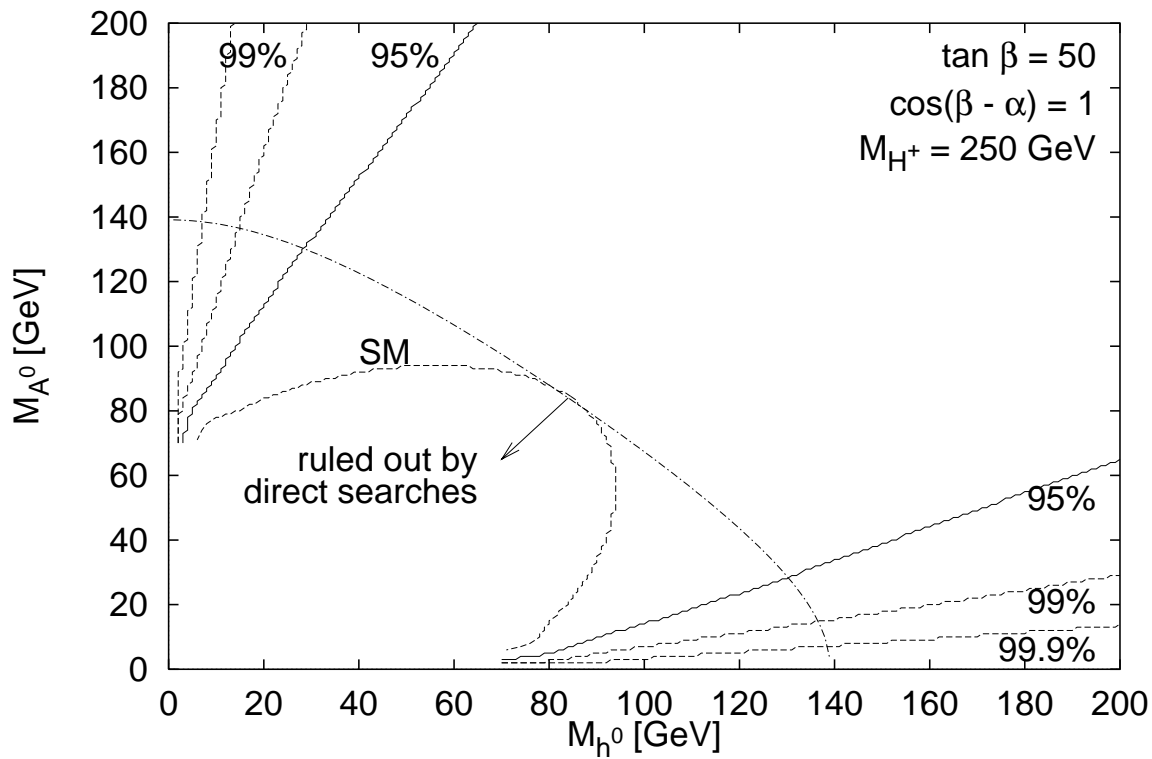}}}
\caption[$R_b$ in the 2HDM with $\tan\beta = 50$, 
$\cos(\beta - \alpha) = 1$,
and $M_{H^+} = 250$ GeV]
{$R_b$ in the Type II 2HDM with $\tan\beta = 50$, 
$\cos(\beta - \alpha) = 1$,
and $M_{H^+} = 250$ GeV.  The axes are $M_{A^0}$ and $M_{h^0}$.  
The solid lines
are the 95\% confidence level lower bounds on $M_{A^0}$ and $M_{h^0}$
from $R_b$.  The dashed lines labelled ``99\%'' and ``99.9\%'' are the 
99\% and 99.9\% confidence level bounds from $R_b$.
The dashed line labelled ``SM'' is where $R_b$ is the same as in the 
SM.  
The region below this line, in which $\Delta R_b > 0$, is almost
entirely excluded by direct searches.
The dot--dashed line is the bound from direct searches,
described in appendix \ref{app:dirsearches}.
}
\label{fig:Rb0_1_no}
\end{figure}

\begin{figure}
\resizebox{\textwidth}{!}{\rotatebox{270}{\includegraphics{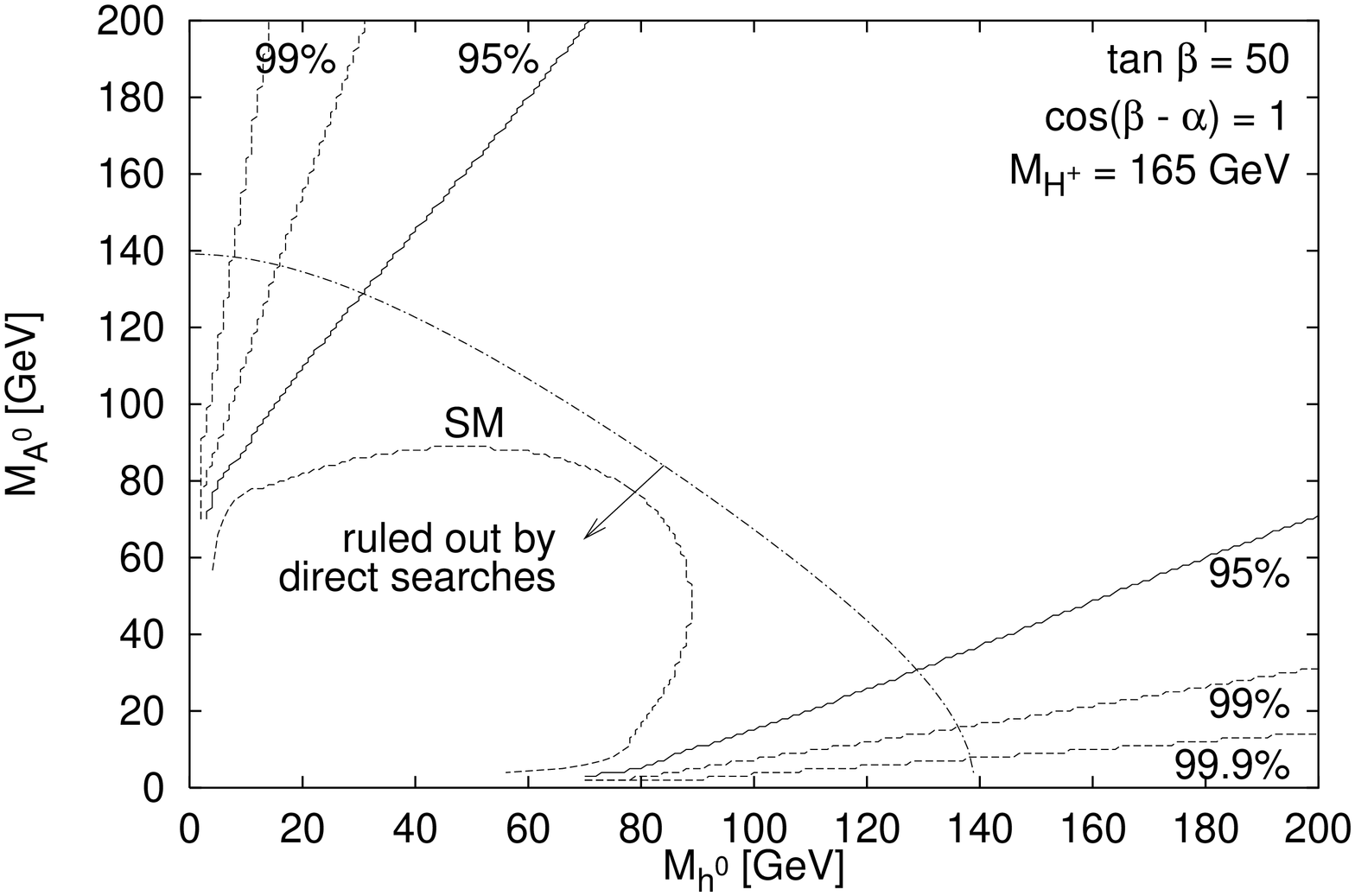}}}
\caption[$R_b$ in the 2HDM with $M_{H^+} = 165$ GeV 
and other parameters the same as in figure \ref{fig:Rb0_1_no}]
{$R_b$ in the 2HDM with $M_{H^+} = 165$ GeV (the lower limit from
$b \rightarrow s \gamma$ \cite{Borzumati98}) and other parameters the same
as in figure
\ref{fig:Rb0_1_no}.  Since $\Delta R_b < 0$ from $H^+$, a larger area
is excluded by $R_b$ at 95\% confidence level (solid lines), and the
region in which $\Delta R_b > 0$ is now entirely excluded by direct searches.
Also shown are the 99\% and 99.9\% confidence level contours (dashed).
The dot--dashed line is the bound from direct searches as in
figure \ref{fig:Rb0_1_no}.
}
\label{fig:Rb0_1_165}
\end{figure}

In figures \ref{fig:Rb0_1_no} and \ref{fig:Rb0_1_165}, we again plot the 
constraints on the neutral Higgs sector from $R_b$.  
This time, the parameters in these plots are
$\tan\beta = 50$ and
$\cos(\beta - \alpha) = 1$.  For $\cos(\beta - \alpha) = 1$, the
$ZH^0A^0$ coupling is zero and the $H^0 b \bar{b}$ coupling is not 
enhanced over the SM $H^0 b \bar{b}$ coupling, so the contribution of
$H^0$ to the corrections is negligible.
As before, the charged Higgs mass is constrained by the $\rho$ parameter.
For the parameters of figures \ref{fig:Rb0_1_no} and \ref{fig:Rb0_1_165},
we find that the charged Higgs boson must be lighter than 250 GeV.
In figure
\ref{fig:Rb0_1_no} we take $M_{H^+} = 250$ GeV, and
in figure \ref{fig:Rb0_1_165}
we take $M_{H^+} = 165$ GeV, which is the 
lower bound from $b \rightarrow s \gamma$ \cite{CLEO98,Borzumati98}.

For these parameters with $M_{H^+} = 250$ GeV, there is
a very small allowed range of $h^0$ and $A^0$ masses 
in which $\Delta R_b > 0$, in better agreement with 
experiment than the SM.  This range is on the verge of being
ruled out experimentally.
This is shown in figure \ref{fig:Rb0_1_no}.  
For $M_{H^+} = 165$ GeV, the corrections from charged Higgs boson exchange
give a negative contribution to $R_b$.
The region where $\Delta R_b > 0$
becomes smaller, and is excluded by the direct search limits.  

If we combine the constraint from $R_b$ with the direct search bounds,
we find absolute lower limits on the $h^0$ and $A^0$ masses for
$\cos(\beta - \alpha) = 1$.  For $M_{H^+} = 250$ GeV, both $h^0$
and $A^0$ must be heavier than 28 GeV.  Similarly, for
$M_{H^+} = 165$ GeV, both $h^0$ and $A^0$ must be heavier than 
30 GeV.  The direct search bounds are discussed in appendix
\ref{app:dirsearches}.
The corrections to $R_b$ are negative for large splittings 
between $M_{h^0}$ and $M_{A^0}$.  Thus areas of low $M_{h^0}$ and 
high $M_{A^0}$, and of low $M_{A^0}$ and 
high $M_{h^0}$, are ruled out by the $R_b$ measurement.

Since both the charged and neutral Higgs boson corrections at large 
$\tan\beta$ are proportional
to $\tan^2\beta$, varying $\tan\beta$ will not change the 
combinations of $M_{h^0}$ and $M_{A^0}$ for which $\Delta R_b = 0$.
Thus the line where $R_b$ is equal to its SM value stays the same
as we vary $\tan\beta$, as long as we remain in the large $\tan\beta$
regime.
Since the corrections grow with $\tan\beta$, the regions ruled out by
$R_b$ in figures \ref{fig:Rb0_1_no} and \ref{fig:Rb0_1_165} get larger 
as $\tan\beta$ increases.

For $\cos(\beta - \alpha) = 1$, the correction to $A_b$ is again very small
compared to the experimental uncertainty in the $A_b$ measurement.
For these parameters, 
$|\Delta A_b| < 0.004$. 
For $M_{H^+} = 250$ GeV,
$0.935 < A_b < 0.937$ in the allowed region.  
For $M_{H^+} = 165$ GeV, $0.936 < A_b < 0.938$ in the allowed region.
This is in slightly
worse agreement with experiment than the SM.

\begin{figure}
\resizebox{\textwidth}{!}{\rotatebox{270}{\includegraphics{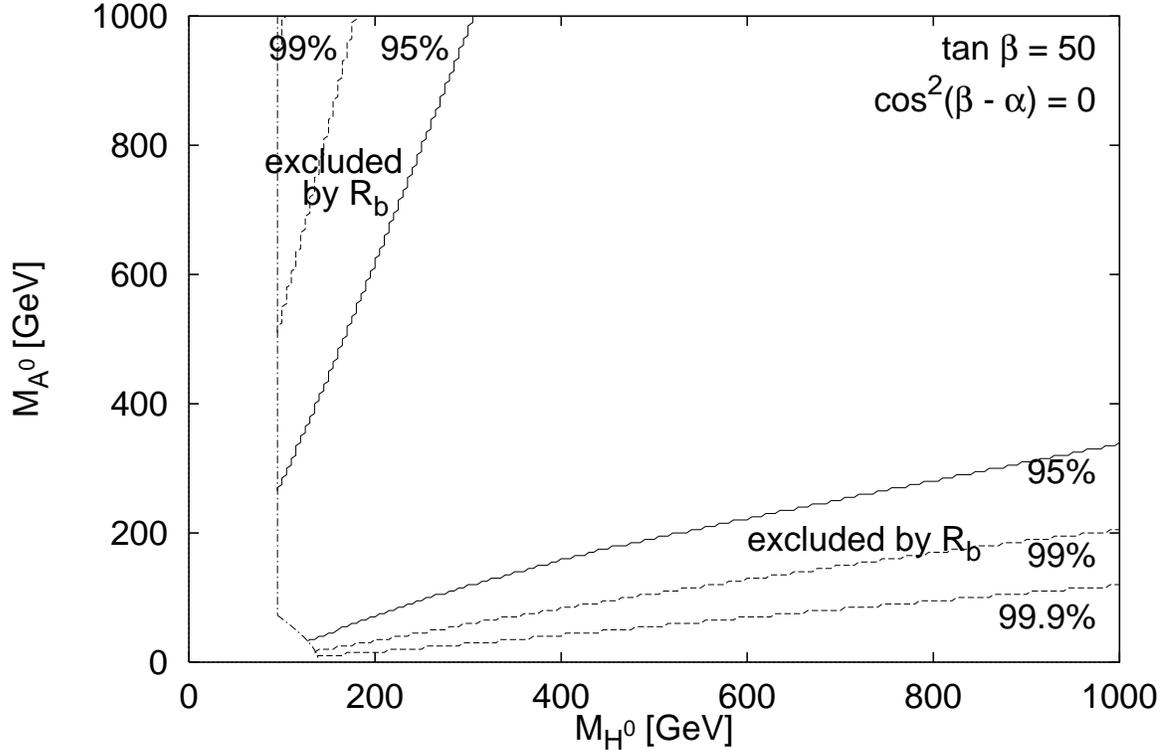}}}
\caption[$R_b$ in the 2HDM with $\tan\beta = 50$ and
$\cos(\beta - \alpha) = 0$]
{$R_b$ in the 2HDM with $\tan\beta = 50$ and 
$\cos(\beta - \alpha) = 0$.
The axes are $M_{H^0}$ and $M_{A^0}$.
For $M_{A^0}>165$ GeV, the charged Higgs mass is set equal to $M_{A^0}$,
while for $M_{A^0} < 165$ GeV, the charged Higgs mass is set equal 
to 165 GeV.
The solid lines are the 95\% confidence level lower bounds on $M_{A^0}$
and $M_{H^0}$ from $R_b$.  The dashed lines labelled 
``99\%'' and ``99.9\%'' are the 
99\% and 99.9\% confidence level bounds from $R_b$.
The dot--dashed line is the bound from direct searches,
described in appendix \ref{app:dirsearches}.
}
\label{fig:Rb0_0_no}
\end{figure}

In the decoupling limit, $\cos(\beta - \alpha) \rightarrow 0$, as
discussed in reference \cite{decoupling}.  Thus the case
$\cos(\beta - \alpha) = 1$ is the opposite of the
decoupling limit, and 
the Higgs sector may develop large, nonperturbative scalar quartic 
couplings as we try to take the charged Higgs mass large.
However, we can take $\cos(\beta - \alpha) = 0$ while interchanging 
$h^0$ and $H^0$ in figures \ref{fig:Rb0_1_no} and \ref{fig:Rb0_1_165},
and the results for $R_b$ and $A_b$ will remain the same.  
This is shown in figure \ref{fig:Rb0_0_no}.
In the limit $\cos(\beta - \alpha) = 0$, the couplings of $h^0$ go
to their SM values.  Therefore, the mass of $h^0$
is constrained by the SM bound,
$M_{h^0} > 95.2$ GeV \cite{Felcini99}.  This is the bound on the
SM Higgs mass from LEP running at $\sqrt{s} = 189$ GeV.
$H^0$ is, by definition, the
heavier CP--even neutral Higgs boson, so $M_{H^0} > M_{h^0} > 95.2$ GeV.
The mass of $H^0$ is also constrained by the search for $H^0 A^0$ 
production, as discussed in appendix \ref{app:dirsearches}.
When $\cos(\beta - \alpha) = 0$, the $Zh^0A^0$ coupling is zero
and the $h^0b\bar{b}$ coupling is not enhanced over the SM coupling,
so $h^0$ does not contribute significantly to the corrections.  We
will neglect it.

As discussed before, a large mass splitting between the charged Higgs boson
and the neutral Higgs bosons results in large radiative corrections
to the $\rho$ parameter.  For this reason, we set $M_{H^+} = M_{A^0}$
in figure \ref{fig:Rb0_0_no}, for $M_{A^0} > 165$ GeV.  For 
$M_{A^0} < 165$ GeV, we set $M_{H^+} = 165$ GeV, which is the lower
bound on the charged Higgs mass from 
$b \rightarrow s \gamma$ \cite{CLEO98,Borzumati98}.

For $\cos(\beta - \alpha) = 0$, the $R_b$ measurement rules out areas
where the mass splitting between $H^0$ and $A^0$ is large.  
For example, 
if the $H^0$ ($A^0$) mass is 1000 GeV, then $A^0$ ($H^0$) must be heavier than 
about 300 GeV.

\subsubsection{Two Higgs doublet model in the decoupling limit}

In the decoupling limit of the 2HDM, $h^0$ remains light
and its couplings to the SM particles approach those of the SM Higgs boson, 
while all the other Higgs bosons become heavy and decouple 
from the SM particles.
This limit is discussed in reference \cite{decoupling}.
In particular, in the decoupling limit,
\begin{eqnarray}
	M_{h^0} &\sim& \mathcal{O}(M_Z), \\
	M_{H^0} &\simeq& M_{A^0}, \\
	| M^2_{H^0} - M^2_{A^0} | &\sim& \mathcal{O}(M_Z^2), \\
	\cos (\beta - \alpha) &\sim& \mathcal{O}
		\left( \frac{M_Z^2}{M_{A^0}^2} \right),
\end{eqnarray}
where $M_{A^0} \gg M_Z$.
We can expand the corrections to $Z \rightarrow b \bar{b}$ from neutral
Higgs boson exchange in the 2HDM in this limit.  We use the expansions
of the three--point integrals given in appendix \ref{app:loopints}.
To leading order in $M_Z^2/M^2_{A^0}$, 
\begin{eqnarray}
\delta g^L &\simeq& \frac{1}{16 \pi^2} \left( \frac{e}{s_Wc_W} \right) 
	\left( \frac{g m_b}{\sqrt{2} M_W} \right)^2 \tan^2 \beta
	\times \frac{M_Z^2}{M_{A^0}^2}  \nonumber \\ 
	& & \times \left[ -\frac{1}{36} + \frac{1}{9} s^2_W 
		\left( \frac{1}{3} 
		+ \log \left( \frac{M^2_{A^0}}{M_Z^2} \right) 
		+ i\pi \right) \right]   \\
\delta g^R &\simeq& \frac{1}{16 \pi^2} \left( \frac{e}{s_Wc_W} \right) 
	\left( \frac{g m_b}{\sqrt{2} M_W} \right)^2 \tan^2 \beta
	\times \frac{M_Z^2}{M_{A^0}^2}  \nonumber \\
	& & \times \left[ -\frac{1}{36} 
		- \frac{1}{6}\left( \log
		\left( \frac{M^2_{A^0}}{M_Z^2} \right) + i\pi \right)
		+ \frac{1}{9} s^2_W 
		\left( \frac{1}{3} 
		+ \log \left( \frac{M^2_{A^0}}{M_Z^2} \right) 
		+ i\pi \right) \right].
\end{eqnarray}
For $\tan\beta = 50$ and $M_{A^0} = 200$ GeV, these corrections give 
$\Delta R_b = -0.00037$, which is half the size of the experimental 
error on the $R_b$ measurement.  As $M_{A^0}$ increases, the correction
decreases; for $M_{A^0} = 500$ GeV, $\Delta R_b = -0.00015$.  

This limit
is approached in figure \ref{fig:Rb0_0_no} 
when $M_{H^0}$ and $M_{A^0}$ are large and similar in size.

\subsubsection{Multiple--doublet models}

We now consider neutral Higgs boson exchange in a model containing
multiple Higgs doublets, denoted $\Phi_k$, with hypercharge $Y=1$.

In a Type I model of this type, we let $\Phi_1$ couple to both up--
and down--type quarks, and none of the other doublets couple to quarks.
In a Type II model, we let $\Phi_1$ couple only to down--type quarks, 
and $\Phi_2$ couple to up--type quarks.  Then the Yukawa couplings 
are defined in the same way as in the 2HDM, in equations 
\ref{eqn:lambdatI}--\ref{eqn:lambdabII}.  As always, the 
contributions to $Z \rightarrow b \bar{b}$ from neutral Higgs boson exchange
are only significant in a Type II model, when $\lambda_b$ is enhanced
by small $v_1$.  We will only consider Type II multi--doublet models
with small $v_1$.

The contributions from neutral Higgs boson exchange in the multi--doublet
model are more complicated than in the 2HDM, simply because there are
more neutral Higgs states.  Only the states which have a nonzero overlap with
$\Phi_1$ can couple to $b$ quarks, so only these states contribute.
The corrections depend on the overlap of each neutral state with $\Phi_1$
and the mass of each state.
As in the 2HDM, the region of parameter space in which the 
correction to $R_b$ is positive is almost entirely ruled out by 
direct searches.

\subsubsection{Multiple--doublet models with Higgs singlets}

We can also consider adding a number of Higgs singlets, with hypercharge
zero, to the multi--doublet model.  The singlets do not couple to 
$Z$ or to quarks.  Their vevs are also unconstrained by the $W$ mass.
In general, the singlets will mix with the neutral components of the 
doublets to form mass eigenstates.  
The couplings
of the physical states to $b\bar{b}$ still depend only on $v_1$, which
fixes $\lambda_b$, and on the overlap of each state with $\Phi_1$.
The couplings of physical states to $Z$ are no longer the same as in
a model containing only doublets.  Instead, they are equal to the 
$Z$ coupling for doublet states weighted by the overlap of each state
with doublets.  Explicitly,
\begin{equation}
g_{ZH^0_iA^0_j} = \frac{-i}{2} \frac{e}{s_Wc_W} \sum_{k}
	\langle H_i^0 | \phi_k^{0,r} \rangle
	\langle A_j^0 | \phi_k^{0,i} \rangle,
\label{eqn:ZHAwithsinglets}
\end{equation}
where $k$ runs only over the Higgs doublets.

In order to understand the effects of singlets on the corrections
to $Z \rightarrow b \bar{b}$, let us imagine replacing each Higgs 
singlet with the neutral component of a doublet, with the appropriate 
CP quantum number, while holding the masses and mixings of the 
physical states constant.  Then, the couplings of each state to 
$b \bar{b}$ remain the same.  However, the couplings of the states
to $Z$ are now equal to,
\begin{equation}
g_{ZH^0_iA^0_j} = \frac{-i}{2} \frac{e}{s_Wc_W},
\end{equation}
which is the coupling in a model containing only Higgs doublets.
Comparing this to equation \ref{eqn:ZHAwithsinglets}, we see that
$\delta g^{R,L}(a)$ in the model with singlets must be smaller in 
magnitude than in the model in which the singlets are replaced by 
doublets.

\subsubsection{Degenerate neutral Higgs bosons in a general extended
Higgs sector}

The corrections to $Z \rightarrow b \bar{b}$ due to neutral Higgs boson
exchange in a general model are quite complicated.  
They depend on the couplings and masses of all the 
neutral Higgs bosons in the model.  However, the corrections can 
be simplified if some of the neutral Higgs bosons are degenerate in 
mass.  We describe these simplifications in this section.

In this section we consider a general extended Higgs sector, 
which can contain Higgs singlets, doublets, and larger multiplets.
We require that the model be Type II, and that $\lambda_b$ 
be enhanced relative to $\lambda_t$.  A Type II model must
contain at least two Higgs doublets, $\Phi_1$ and $\Phi_2$, 
to couple to down-- and up--type quarks, respectively.

Only the neutral Higgs bosons with large couplings to $b\bar{b}$ give
significant contributions to the corrections.  In what follows 
we will only consider these.  States without enhanced $b\bar{b}$ 
couplings, such as $G^0$, do not contribute significantly.  We
will ignore them, and therefore it does not matter what their
masses are.

If all the CP--even neutral Higgs bosons are degenerate with mass $M_H$, 
and all the CP--odd neutral Higgs bosons are degenerate with mass $M_A$,
then we can take the two-- and three--point integrals outside of
the sums in equations \ref{eq:h0loopsa}--\ref{eq:h0loopsc}.
Then we can sum the couplings over complete sets of states.  
Using the couplings in a general model from
equations \ref{eq:gV}, \ref{eq:gA}, and \ref{eq:ZHA}, we find
\begin{eqnarray}
\sum_{H^0_i,A^0_j} g_{ZH^{0}_{i}A^{0}_{j}} g^{V}_{H^{0}_{i}b\bar{b}} 
	g^{A}_{A^{0}_{j}b\bar{b}}
	&=& \frac{1}{2} \frac{e}{s_Wc_W} 
	\left( \frac{gm_b}{\sqrt{2}M_W} \right)^2 
	\left( \frac{v_{SM}}{v_1} \right)^2
	\\
\sum_{H^0_i} (g^{V}_{H^{0}_{i}b\bar{b}})^{2}
	= -\sum_{A^0_j} (g^{A}_{A^{0}_{j}b\bar{b}})^{2}
	&=& \left( \frac{gm_b}{\sqrt{2}M_W} \right)^2
	\left( \frac{v_{SM}}{v_1} \right)^2.
\end{eqnarray}
These sums over the couplings are related to certain couplings in
the 2HDM, as follows.  
On the left--hand side are the couplings in the general model with degenerate 
neutral Higgs bosons, and on the right--hand side are the couplings 
in the 2HDM with $\cos(\beta - \alpha) = 1$.
\begin{eqnarray}
\sum_{H^0_i,A^0_j} g_{ZH^{0}_{i}A^{0}_{j}} g^{V}_{H^{0}_{i}b\bar{b}} 
	g^{A}_{A^{0}_{j}b\bar{b}}
	&=& g_{Zh^0A^0} g^{V}_{h^0b\bar{b}} 
	g^{A}_{A^0b\bar{b}}
	\\
\sum_{H^0_i} (g^{V}_{H^{0}_{i}b\bar{b}})^{2}
	&=& (g^{V}_{h^0b\bar{b}})^{2}
	\\
-\sum_{A^0_j} (g^{A}_{A^{0}_{j}b\bar{b}})^{2}
	&=& -(g^{A}_{A^0b\bar{b}})^{2}.
\end{eqnarray}

Therefore, when all the CP--even neutral Higgs bosons are degenerate 
with mass $M_H$, and all the CP--odd neutral Higgs bosons are degenerate 
with mass $M_A$, the contributions to $Z \rightarrow b\bar{b}$ are
the same as the contributions from the 2HDM with $M_{h^0} = M_H$,
$M_{A^0} = M_A$, and $\cos(\beta - \alpha) = 1$.  The parameter
corresponding to $\tan\beta$ in the extended model is,
\begin{equation}
\sqrt{ \frac{v^2_{SM} - v^2_1}{v^2_1} } = \tan\beta.
\end{equation}
If there are 
no contributions from the charged Higgs bosons, the corrections
to $R_b$ and $A_b$ in this situation are the same as those
for the 2HDM
shown in figure \ref{fig:Rb0_1_no},
for $\tan\beta = 50$.

Similarly, the corrections can be simplified if only the CP--even
states, or only the CP--odd states, are degenerate.  If the 
CP--even states are degenerate, we can sum over the $H^0_i$ 
couplings.  We then get the same
result as if the CP--even neutral Higgs sector consisted 
of a single state $H^0$, which consists entirely of $\phi_1^{0,r}$.
Recall that $\phi_1^{0,r}$ is the CP--even neutral component of
the doublet which couples to down--type quarks.
If, instead, the CP--odd states are degenerate, we can sum over the
$A^0_j$ couplings.  We get the same result as if the CP--odd
neutral Higgs sector consisted of a single state $A^0$, which
consists entirely of $\phi_1^{0,i}$ (up to the small mixing of
$\phi_1^{0,i}$ with $G^0$, which is negligible in the small $v_1$
regime).

\section{Models with Higgs multiplets larger than doublets}
\label{sec:exotics}

We now consider Higgs sectors which contain one or more multiplets 
larger than doublets.
We consider two types of models which use two different approaches
to satisfy $\rho \approx 1$.  We first consider models in which the
vevs of the multiplets larger than doublets are fine--tuned to be very
small, so that their contribution to the $\rho$ parameter is 
negligible.  Second, we consider models which preserve ${\rm SU}(2)_c$
symmetry, ensuring that $\rho = 1$ at tree level.

In section \ref{sec:finetuned} we examine a model containing one
Higgs doublet and one triplet, and in section \ref{sec:2doub1trip}
we examine a model containing two Higgs doublets and one triplet.
In both of these models, the vev of the Higgs triplet must be 
very small in order to be consistent with the measured value of
the $\rho$ parameter.  The model with two doublets and one triplet
is also discussed further in appendix \ref{app:2doub1trip}.

In section \ref{sec:triplets} we examine the model with Higgs triplets
and ${\rm SU}(2)_c$ symmetry introduced by Georgi and Machacek \cite{Georgi1}.
The ${\rm SU}(2)_c$ symmetry ensures that $\rho = 1$ at tree level in this
model, as explained in detail in appendix \ref{app:SU2c}.  We then
extend the results to a class of generalized Georgi--Machacek models
which preserve ${\rm SU}(2)_c$ symmetry.

\subsection{Models with one Higgs doublet and one triplet}
\label{sec:finetuned}

In this section we describe the minimal extension 
of the Higgs sector to include multiplets larger than doublets.  The 
Higgs sector consists of the complex, $Y=1$ doublet of the SM,
denoted by $\Phi$, plus a triplet field.  The vev of the triplet
field must be fine--tuned very small in order to be consistent
with the measured value of the $\rho$ parameter, $\rho \approx 1$.
The triplet field can either be a real triplet with
$Y=0$, or a complex triplet with $Y=2$.  Here we investigate both 
possibilites.

These two models contain only one Higgs doublet, which couples to 
both up-- and down--type quarks, so they are necessarily
Type I models.  Thus $\lambda_b \ll \lambda_t$, and the only 
non--negligible contributions to $Z \rightarrow b \bar{b}$ come
from the contributions to $\delta g^L$ from charged Higgs boson exchange.


We first consider the ``$Y=0$ model'' with one doublet and one real triplet 
field with $Y=0$.  
The triplet field is
$\xi = (\xi^{+}, \xi^{0}, \xi^{-})$.  We define the doublet and 
triplet vevs by $\langle \phi^0 \rangle = v_{\phi}/\sqrt{2}$ and
$\langle \xi^0 \rangle = v_{\xi}$.  The vevs are constrained by the
$W$ mass to satisfy,
\begin{equation}
v_{SM}^2 = v_{\phi}^2 + 4 v_{\xi}^2,
\end{equation}
where $v_{SM} = 246$ GeV.
It is convenient to parameterize
the ratio of the vevs by,
\begin{equation}
\tan \theta_0 = \frac{v_{\phi}}{2 v_{\xi}}.
\end{equation}
In this model, the tree--level $\rho$ parameter is,
\begin{equation}
\rho = \frac{v_{\phi}^{2} + 4 v_{\xi}^{2}}{v_{\phi}^{2}}
	= 1 + \frac{4 v_{\xi}^{2}}{v_{\phi}^{2}} \equiv 1 + \Delta \rho.
\end{equation}
In terms of $\tan\theta_0$, we find
\begin{equation}
\Delta \rho = \frac{1}{\tan^2 \theta_0}.
\end{equation}
We see that in order to have $\rho \approx 1$, the triplet vev
must be very small, giving large $\tan\theta_0$.
The charged states mix to form the charged Goldstone boson and a
single charged physical state,
\begin{eqnarray}
G^{+} &=& \sin\theta_0 \phi^{+} + \cos\theta_0 \xi^{+}  \\
H^{+} &=& \cos\theta_0 \phi^{+} - \sin\theta_0 \xi^{+}.
\end{eqnarray}


We next consider the ``$Y=2$ model'' with one doublet and one complex triplet
field with $Y=2$.  
The triplet field is
$\chi = ( \chi^{++}, \chi^+, \chi^0 )$.  We define the vev of this 
triplet field by $\langle \chi^0 \rangle = v_{\chi}/\sqrt{2}$.
The vevs are constrained by the
$W$ mass to satisfy,
\begin{equation}
v_{SM}^2 = v_{\phi}^2 + 2 v_{\chi}^2.
\end{equation}
It is convenient to parameterize the ratio of the doublet and triplet
vevs by,
\begin{equation}
\tan \theta_2 = \frac{v_{\phi}}{\sqrt{2} v_{\chi}}.
\end{equation}
In this model, the tree--level $\rho$ parameter is,
\begin{equation}
\rho = \frac{ v_{\phi}^2 + 2 v_{\chi}^2 }{ v_{\phi}^2 + 4 v_{\chi}^2 }
	= 1 - \frac{ 2 v_{\chi}^2 }{ v_{\phi}^2 + 4 v_{\chi}^2 }
	\equiv 1 + \Delta \rho.
\end{equation}
In terms of $\tan\theta_2$, we find
\begin{equation}
\Delta \rho = -\frac{1}{\tan^2 \theta_2 + 2}.
\end{equation}
We see that in order to have $\rho \approx 1$, the triplet vev
must be very small, giving large $\tan\theta_2$.
The charged states mix to form the charged Goldstone boson and a
single charged physical state,
\begin{eqnarray}
G^{+} &=& \sin\theta_2 \phi^{+} + \cos\theta_2 \xi^{+}  \\
H^{+} &=& \cos\theta_2 \phi^{+} - \sin\theta_2 \xi^{+}.
\end{eqnarray}

The Higgs couplings to quarks and the $Z$ boson can be parameterized
as follows.  We let $\theta$ denote $\theta_0$ in the $Y=0$ model and
$\theta_2$ in the $Y=2$ model.  We also define a factor $\epsilon$
such that $\epsilon = +1$ in the $Y=0$ model and $\epsilon = -1$ 
in the $Y=2$ model.  The charged Higgs couplings to quarks are,
\begin{eqnarray}
g^{L}_{G^{+}\bar{t}b} &=& \frac{gm_t}{\sqrt{2}M_W}  \\
g^{L}_{H^{+}\bar{t}b} &=& \frac{gm_t}{\sqrt{2}M_W} \cot\theta.
\end{eqnarray}
The $ZH^+_iH^-_j$ couplings are,
\begin{eqnarray}
g_{ZG^{+}G^{-}} &=& -\frac{e}{s_{W}c_{W}} \left(\frac{1}{2} - s^{2}_{W}
	+ \frac{\epsilon}{2}\cos^2 \theta_2 \right)  \\
g_{ZG^{+}H^{-}} &=& \frac{e}{s_{W}c_{W}} \frac{\epsilon}{2} \sin\theta_2 
	\cos\theta_2  \\
g_{ZH^{+}H^{-}} &=& -\frac{e}{s_{W}c_{W}} \left(\frac{1}{2} - s^{2}_{W}
	+ \frac{\epsilon}{2} \sin^2 \theta_2 \right).
\end{eqnarray}
The $ZH^+_iH^-_j$ couplings are different in the two models
because the hypercharge of the triplet is different.

\subsubsection{Contributions to $Z \rightarrow b\bar{b}$}

In both the $Y=0$ and the $Y=2$ models, there is an off--diagonal
$ZG^{+}H^{-}$ coupling, and the diagonal $ZH^{+}H^{-}$ and 
$ZG^{+}G^{-}$ couplings differ from their values in models containing
only Higgs doublets and singlets.  These couplings contribute
to the second and third terms of $\delta g^L$ in equation
\ref{eq:h+loops2}.

The resulting contribution to $\delta g^L$ is,
\begin{eqnarray}
&\delta g^L& = \frac{1}{16 \pi^2} \left( \frac{g m_t}{\sqrt{2} M_W} \right)^2
 \frac{1}{2} \frac{e}{s_Wc_W} \cos^2 \theta 
 \left\{ \frac{1}{\sin^2\theta} 
 \left[\frac{R}{R-1} - \frac{R \log R}{(R-1)^2} \right]
	\right. \nonumber \\
	& & - \left. 2 \epsilon \left[ 
	 C_{24} (m_t^2, M_W^2, M_W^2) 
	+ C_{24} (m_t^2, M_H^2, M_H^2)
	- 2 C_{24} (m_t^2, M_W^2, M_H^2) \right] \right\}, 
\label{eqn:dgL1doub1trip}
\end{eqnarray}
in addition to the SM contribution to $\delta g^L_{SM}$ from $G^+$ 
exchange.  We have defined $R \equiv m_{t}^{2}/M_H^2$, 
$M_H$ is the mass of $H^+$.  
As before, in the $Y=0$ model, $\theta = \theta_0$ and $\epsilon = +1$,
while in the $Y=2$ model, $\theta = \theta_2$ and $\epsilon = -1$.

Note that $\delta g^L$ is proportional to $\cos^2\theta$, which
goes to zero in the large $\tan\theta$ limit.  This is due to the
fact that in the limit of small triplet vev in either of these models, 
the overlap of $H^+$ with the doublet is proportional to $\cos\theta$.
As a result, in the large $\tan\theta$ limit,
$H^+$ is almost entirely triplet and so its couplings to 
quarks are very small.  Also in the large $\tan\theta$ limit,
the off--diagonal $ZG^+H^-$ coupling goes to zero, and the 
$ZG^+G^-$ coupling approaches its SM value.

\subsubsection{Constraints from the $\rho$ parameter}

We must also take into account the constraint on $\tan\theta$ 
from the $\rho$ parameter in each of the models.  Since 
$\Delta \rho$ depends differently on $\tan\theta_0$ than on
$\tan\theta_2$, the constraint on $\tan\theta$ will be different
in the $Y=0$ model than in the $Y=2$ model.

The experimental constraints on $\Delta \rho$ are taken from 
reference \cite{Altarelli98}, in which $\Delta \rho = \epsilon_1$.
Reference \cite{Altarelli98} finds,
\begin{equation}
	\Delta \rho = (3.9 \pm 1.2) \times 10^{-3}.
\end{equation}
However, we cannot take this directly as a constraint on $\tan\theta_0$
and $\tan\theta_2$, because the $\rho$ parameter gets radiative
corrections from SM particles.  We must take into account these 
radiative corrections in order to extract limits on $\Delta \rho$
from new physics.

In the SM, the radiative corrections to $\rho$ depend on the masses 
of the top quark and the SM Higgs boson.  If the top quark and the
SM Higgs boson are taken to lie in certain mass ranges, a range can be
found for the SM prediction for the radiative corrections to $\rho$.
From reference \cite{Altarelli98}, if the mass ranges are taken to
be 170 GeV $< m_t <$ 180 GeV and 70 GeV $< m_H <$ 1000 GeV,
then the SM predictions for $\Delta \rho$ from radiative corrections 
are,
\begin{equation}
3.32 \times 10^{-3} < \Delta \rho_{SM} < 6.19 \times 10^{-3}.
\end{equation}
We must take this into account in order to find the experimental 
limits on $\Delta \rho$ due to new physics.  Varying the SM 
prediction for $\Delta \rho$ within this range, we find the following
limits on $\Delta \rho$ from new physics, at the $2\sigma$ level:
\begin{equation}
-4.7 \times 10^{-3} < \Delta \rho_{\rm new} < 3.0 \times 10^{-3}.
\end{equation}
We now use $\Delta \rho_{\rm new}$ to constrain $\tan\theta_0$ and 
$\tan\theta_2$.  We ignore the radiative corrections from the non--minimal
Higgs sector.
Note that $\Delta \rho_{\rm new}$ can be either positive or negative.
In the $Y=0$ model, $\Delta \rho_{\rm new}$ is positive, while in the
$Y=2$ model, $\Delta \rho_{\rm new}$ is negative.
The resulting $2\sigma$ limits on $\tan\theta_0$ and 
$\tan\theta_2$ are,
\begin{eqnarray}
\tan\theta_0 &>& 18  \\
\tan\theta_2 &>& 15.
\end{eqnarray}

\subsubsection{Results}

The contribution to $\delta g^L$ in both the $Y=0$ model and
the $Y=2$ model is proportional to $\cos^2\theta$ (equation 
\ref{eqn:dgL1doub1trip}).  When the constraints on $\tan\theta$
from the $\rho$ parameter are imposed, the corrections to $R_b$ and $A_b$
are very small.  In the $Y=0$ model, for $\tan\theta_0 = 18$ and
$M_{H^+}$ varying between 10 and 1000 GeV, 
$-6.7 \times 10^{-6} < \Delta R_b < 7.2 \times 10^{-6}$ and
$-2.5 \times 10^{-6} < \Delta A_b < 2.7 \times 10^{-6}$.
In the $Y=2$ model, for $\tan\theta_2 = 15$ and again $M_{H^+}$
varying between 10 and 1000 GeV,
$-1.2 \times 10^{-5} < \Delta R_b < -5.9 \times 10^{-6}$ and
$-4.6 \times 10^{-6} < \Delta A_b < -2.2 \times 10^{-6}$.
These corrections are minuscule compared to the experimental error 
on the $R_b$ and $A_b$ measurements (equations 
\ref{eqn:Rbmeasured} and \ref{eqn:Abmeasured}).

In general,
the contribution to $\delta g^L$ vanishes in the large $\tan\theta$
limit in any model in which the charged
Goldstone boson is made up almost entirely of the doublet
that couples to quarks.
Then the overlap of the other charged
Higgs states with the doublet is very small, so the other charged
Higgs states couple very weakly to quarks.  This 
occurs in any model that contains only one scalar doublet,
plus any number of singlets and multiplets larger than doublets,
as long as the vevs of the multiplets larger than doublets are
forced to be small.

The contributions of multiplets larger than doublets to 
$Z \rightarrow b\bar{b}$ can be large only if the larger 
multiplets mix significantly with doublets, so that the 
resulting Higgs states have non--negligible couplings to quarks.
This can happen in two ways.  First, if the model contains more than 
one doublet, then the charged Goldstone boson will not necessarily 
be made up almost entirely of the doublet which couples to quarks.
In fact, in a Type II model, one doublet couples to up--type quarks
and a different one couples to down--type quarks.  In this case, 
there is no way
that the charged Goldstone boson can consist entirely of both 
the doublets.  The remaining parts of the doublets which couple
to quarks can then be mixed into the physical charged Higgs states,
giving them non--negligible couplings to quarks.  A model 
of this type is discussed in section \ref{sec:2doub1trip}.
Second, if the multiplets larger than doublets have sizeable
vevs, then the charged Goldstone boson must contain some admixture
of the larger multiplets, leaving part of the doublet free to 
mix into the physical charged Higgs states.  However, in order for the
multiplets larger than doublets to have sizeable vevs without
violating the constraint from the $\rho$ parameter, the model 
must preserve ${\rm SU}(2)_c$ symmetry.  Models of this type are 
discussed in section \ref{sec:triplets}.

\subsection{Models with two doublets and one triplet}
\label{sec:2doub1trip}

We now consider a Higgs sector consisting of two doublets and one triplet.  
As in section \ref{sec:finetuned},
the triplet can be real with $Y=0$ or complex with $Y=2$.
The couplings for these models are listed in appendix \ref{app:2doub1trip}.
With two doublets, we can construct either a Type I model or a Type II model.
In this section we consider a Type II model, but we also note the changes in
the formulas that must be made to recover a Type I model.

We will consider both the corrections due to charged Higgs boson exchange and
the corrections due to neutral Higgs boson exchange.  
The corrections from neutral
Higgs boson exchange can be significant in a Type II model with large
$\tan\beta$.
We define $\tan\beta$ in this model exactly as in the 2HDM,
$\tan\beta = v_2/v_1$, where the vevs of the doublets are 
$\langle \phi_1^0 \rangle = v_1/\sqrt{2}$ and 
$\langle \phi_2^0 \rangle = v_2/\sqrt{2}$.

\subsubsection{Charged Higgs boson contributions}

We first consider the corrections due to charged Higgs boson exchange
in the model with two doublets and one real triplet 
field with $Y=0$.  We will refer to this model as the $Y=0$ model.
The triplet field is
$\xi = (\xi^{+}, \xi^{0}, \xi^{-})$.  We define the 
triplet vev by
$\langle \xi^0 \rangle = v_{\xi}$.
In the $Y=0$ model we parameterize the vevs by,
\begin{equation}
\tan \theta_0 = \frac{\sqrt{v_1^2 + v_2^2}}{2 v_\xi},
\end{equation}
in analogy to section \ref{sec:finetuned}.

The charged Higgs states are defined as follows.
The Goldstone boson is,
\begin{equation}
G^+ = \sin\theta_0 (\cos\beta \phi_1^+ + \sin\beta \phi_2^+) 
	+ \cos\theta_0 \xi^+.
\end{equation}
In addition we define two orthogonal states,
\begin{eqnarray}
H_1^{+\prime} &=& \cos\theta_0 (\cos\beta \phi_1^+ + \sin\beta \phi_2^+) 
	- \sin\theta_0 \xi^+ \\
H_2^{+\prime} &=& -\sin\beta \phi_1^+ + \cos\beta \phi_2^+,
\end{eqnarray}
which will mix by an angle $\delta$ to form the mass eigenstates.  
Before mixing
them, however, let us take the limit of large $\tan\theta_0$ 
in order to satisfy the experimental constraint on the $\rho$ 
parameter.  We make the approximation $\sin\theta_0 \approx 1$ and
$\cos\theta_0 \approx 0$.  Then the states are,
\begin{eqnarray}
G^+ &\simeq& \cos\beta \phi_1^+ + \sin\beta \phi_2^+ \\
H_1^{+\prime} &\simeq& -\xi^+ \\
H_2^{+\prime} &=& -\sin\beta \phi_1^+ + \cos\beta \phi_2^+.
\end{eqnarray}
These states mix by an angle $\delta$ to form the mass eigenstates,
which are,
\begin{eqnarray}
H_1^+ &\simeq& \sin\delta (-\sin\beta \phi_1^+ + \cos\beta \phi_2^+)
	- \cos\delta \xi^+ \\
H_2^+ &\simeq& \cos\delta (-\sin\beta \phi_1^+ + \cos\beta \phi_2^+)
	+ \sin\delta \xi^+.
\end{eqnarray}
The exact states are listed in appendix \ref{app:2doub1trip}, where
we do not make the large $\tan\theta_0$ approximation.

We next consider the corrections due to charged Higgs boson exchange
in the model with two doublets and one complex triplet field
with $Y=2$.  We will refer to this model as the $Y=2$ model.
The triplet field is
$\chi = ( \chi^{++}, \chi^+, \chi^0 )$.  We define the triplet
vev by $\langle \chi^0 \rangle = v_{\chi}/\sqrt{2}$.
In the $Y=2$ model we parameterize the vevs by,
\begin{equation}
\tan\theta_2 = \frac{\sqrt{v_1^2 + v_2^2}}{\sqrt{2} v_{\chi}},
\end{equation}
again in analogy to section \ref{sec:finetuned}.

The charged Higgs states in the $Y=2$ model are parameterized in
the same way as the states in the $Y=0$ model.
The Goldstone boson is,
\begin{equation}
G^+ = \sin\theta_2 (\cos\beta \phi_1^+ + \sin\beta \phi_2^+) 
	+ \cos\theta_2 \chi^+.
\end{equation}
In addition we define two orthogonal states,
\begin{eqnarray}
H_1^{+\prime} &=& \cos\theta_2 (\cos\beta \phi_1^+ + \sin\beta \phi_2^+) 
	- \sin\theta_2 \chi^+ \\
H_2^{+\prime} &=& -\sin\beta \phi_1^+ + \cos\beta \phi_2^+
\end{eqnarray}
which will mix by an angle $\delta$ to form the mass eigenstates.  
Before mixing
them, however, let us take the limit of large $\tan\theta_2$ 
in order to satisfy
the experimental constraint on the $\rho$ parameter.  We make the 
approximation $\sin\theta_2 \approx 1$ and $\cos\theta_2 \approx 0$.
Then the states are,
\begin{eqnarray}
G^+ &\simeq& \cos\beta \phi_1^+ + \sin\beta \phi_2^+ \\
H_1^{+\prime} &\simeq& - \chi^+ \\
H_2^{+\prime} &=& -\sin\beta \phi_1^+ + \cos\beta \phi_2^+.
\end{eqnarray}
These states mix by an angle $\delta$ to form the mass eigenstates,
which are,
\begin{eqnarray}
H_1^+ &\simeq& \sin\delta (-\sin\beta \phi_1^+ + \cos\beta \phi_2^+)
	- \cos\delta \chi^+ \\
H_2^+ &\simeq& \cos\delta (-\sin\beta \phi_1^+ + \cos\beta \phi_2^+)
	+ \sin\delta \chi^+.
\end{eqnarray}
The exact states are listed in appendix \ref{app:2doub1trip}, where
we do not make the large $\tan\theta_2$ approximation.

We now calculate the
corrections to $Z \rightarrow b\bar{b}$ from charged
Higgs boson exchange in the Type II $Y=0$ and $Y=2$ models.
As in section \ref{sec:finetuned}, we introduce the parameter 
$\epsilon = +1$ in 
the $Y=0$ model, and $\epsilon = -1$ in the $Y=2$ model.
The contributions to $\delta g^L$ are,
	%
\begin{eqnarray}
\delta g^L &\simeq& 
	\frac{1}{16 \pi^2} \frac{1}{2} \frac{e}{s_Wc_W}
		\left( \frac{g m_t}{\sqrt{2} M_W} \right)^2 
		\cot^2 \beta \nonumber \\
	& & \times \left[ \sin^2 \delta 
		\left[ \frac{R_1}{R_1-1} 
		- \frac{R_1 \log R_1}{(R_1-1)^2} \right]
		+ \cos^2 \delta
		\left[ \frac{R_2}{R_2-1} 
		- \frac{R_2 \log R_2}{(R_2-1)^2} \right] \right]
		\nonumber \\
	& & - \frac{\epsilon}{16 \pi^2} \frac{e}{s_Wc_W}
		\left( \frac{g m_t}{\sqrt{2} M_W} \right)^2 \cot^2 \beta 
		\sin^2 \delta \cos^2 \delta   
		 \left[ C_{24}(m_t^2,M_{H_1^+}^2,M_{H_1^+}^2) \right.
		\nonumber \\
		& &  + \left. 
		C_{24}(m_t^2,M_{H_2^+}^2,M_{H_2^+}^2) 
		- 2 C_{24}(m_t^2,M_{H_1^+}^2,M_{H_2^+}^2)
		\right],
\label{eqn:dgL2doub1trip}
\end{eqnarray}
in addition to the SM correction due to charged Goldstone boson exchange.
We have defined $R_i = m_t^2 / M_{H_i^+}^2$.  In the Type I models,
$\delta g^L$ is the same as above with $\cot^2\beta$ replaced by
$\tan^2\beta$.

The first term of equation \ref{eqn:dgL2doub1trip}
is the same as the correction in a three Higgs doublet
model (3HDM), given in equation \ref{eqn:dgLMHDM}.
It is 
positive, which gives a negative contribution to $R_b$, taking
it farther from the measured value.  The second
term comes from the effects of the triplet.  This second term is 
proportional to $\sin^2\delta \cos^2\delta$, so it is only 
significant for $\delta$ near $\pi/4$, which corresponds to maximal
mixing between the charged doublet and triplet states in $H_1^+$ and
$H_2^+$.  The second term is zero if $H_1^+$ and $H_2^+$ have the 
same mass.  

The sign of the second term depends on the hypercharge
of the Higgs triplet.  In the $Y=0$ model,
the second term is negative.
However, the second term is smaller in magnitude than the first
term, so the overall contribution to $\delta g^L$ is positive in 
the $Y=0$ model.

\begin{figure}
\resizebox{\textwidth}{!}{\rotatebox{270}{\includegraphics{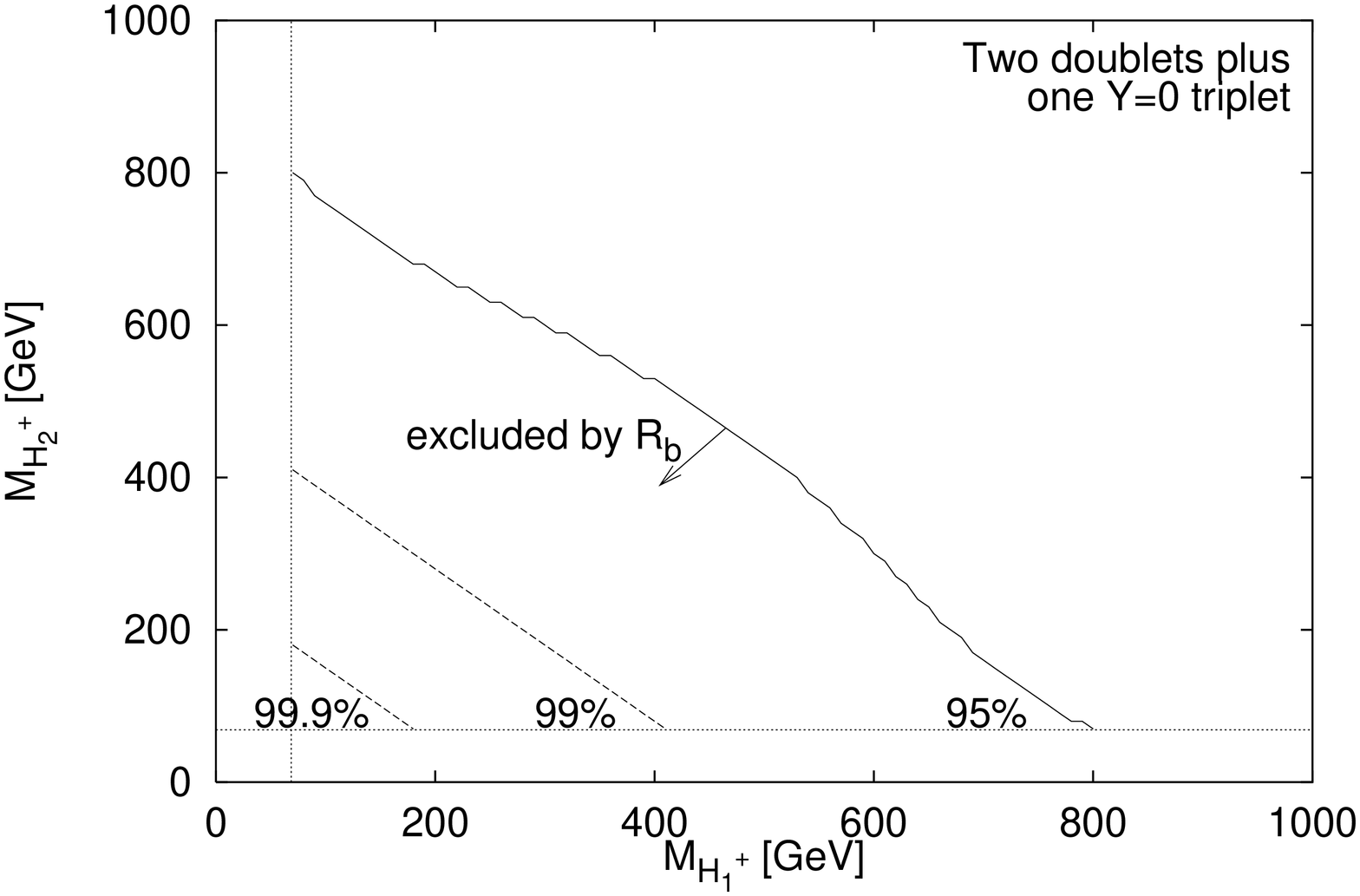}}} 
\caption[Constraints from $R_b$ on the masses of the two charged Higgs
states in the model with two doublets and one real
$Y=0$ triplet]
{Constraints from $R_b$ on the masses of the two charged Higgs
states $H_1^+$ and $H_2^+$ in the model with two doublets and one real
$Y=0$ triplet, with $\tan\beta = 1$ and $\delta = \pi/4$.  The
area below the solid line is excluded at 95\% confidence level.  Also
shown are the 99\% and 99.9\% confidence levels (dashed).  
The dotted lines are the direct search bounds on the $H^+$ mass
from the OPAL collaboration, $M_{H^+} > 68.7$ GeV \cite{OPALH+99}, 
from LEP data up to $\sqrt{s} = 189$ GeV.  (For a discussion of the
direct search bound, see appendix \ref{app:dirsearches}.)
} 
\label{fig:D2T_0+} 
\end{figure}

In figure \ref{fig:D2T_0+}, we plot the constraints on 
$M_{H_1^+}$ and $M_{H_2^+}$ from the $R_b$ measurement in the 
$Y=0$ model, for
maximal doublet--triplet mixing ($\delta = \pi/4$) and $\tan\beta = 1$.
In order for the $Y=0$ model with maximal doublet--triplet mixing
to be consistent with the $R_b$ measurement,
one or both of the charged Higgs bosons must be very heavy.  

In the $Y=2$ model, the second term of equation \ref{eqn:dgL2doub1trip}
is positive, resulting in a 
positive $\delta g^L$ which is larger than in the $Y=0$ model.
As a result, a larger area of parameter space is excluded by the
$R_b$ measurement in the
$Y=2$ model than in the $Y=0$ model.

\begin{figure}
\resizebox{\textwidth}{!}{\rotatebox{270}{\includegraphics{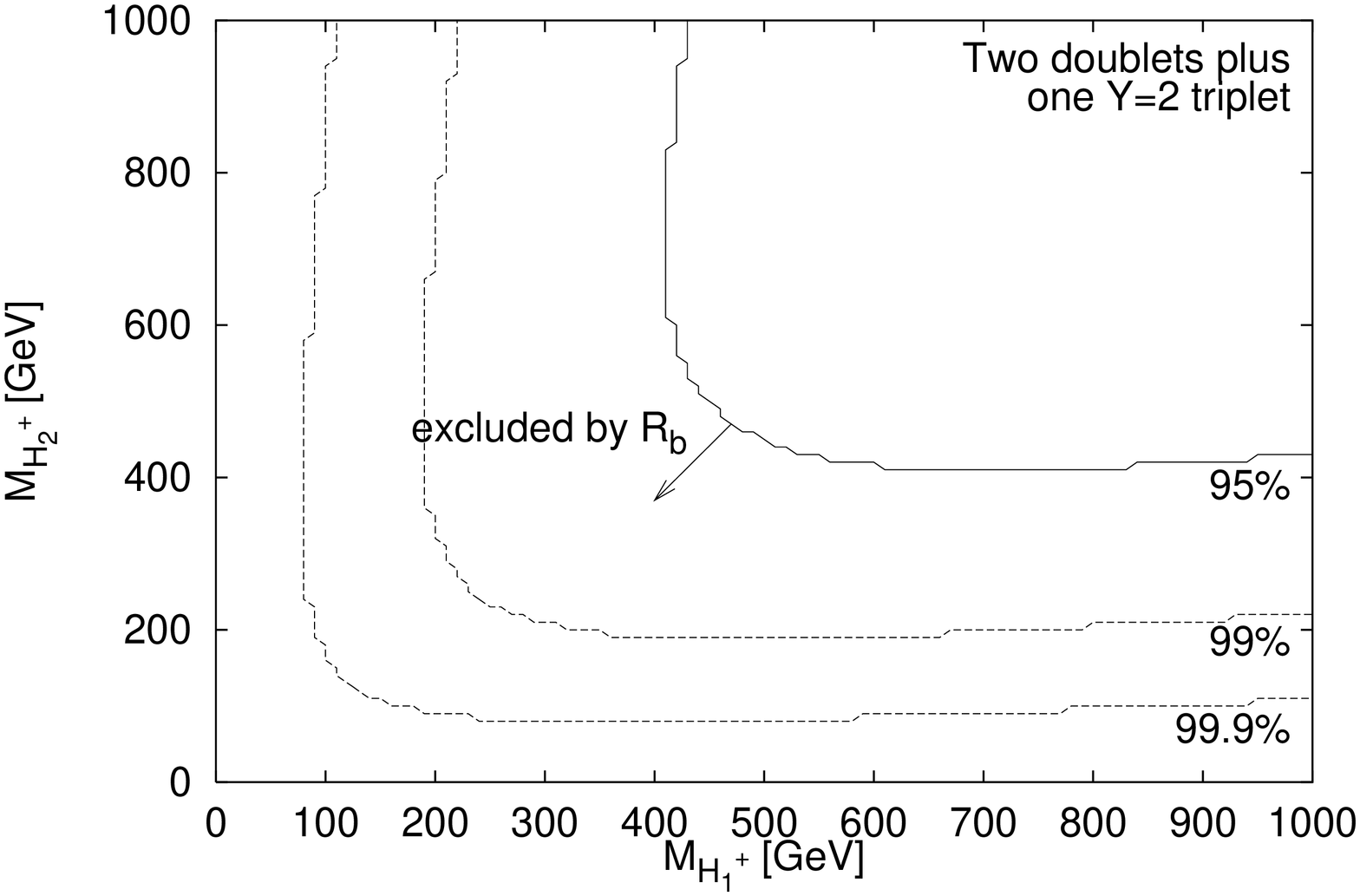}}}
\caption[Constraints from $R_b$ on the masses of the two charged Higgs
states in the model with two doublets and one complex
$Y=2$ triplet]
{Constraints from $R_b$ on the masses of the two charged Higgs
states $H_1^+$ and $H_2^+$ in the model with two doublets and one complex
$Y=2$ triplet, with $\tan\beta = 1$ and $\delta = \pi/4$.  The
area below the solid line is excluded at 95\% confidence level.  For these
values of $\tan\beta$ and $\delta$, $H^+$ masses below 410 GeV are ruled
out. Also
shown are the 99\% and 99.9\% confidence levels (dashed).
}
\label{fig:D2T_2+}
\end{figure}

In figure \ref{fig:D2T_2+}, we
plot the constraints on $M_{H_1^+}$ and $M_{H_2^+}$ from the
$R_b$ measurement on the $Y=2$ model, for maximal doublet--triplet
mixing ($\delta = \pi/4$) and $\tan\beta = 1$.  From the $R_b$ 
constraint with these parameters, we find that both of the 
charged Higgs bosons must be 
heavier than 410 GeV.  If $\delta$ is varied or $\tan\beta$ is 
increased, this bound becomes lower.  Note that we do not plot
a direct search bound on the $H^+$ mass.  In this model, the bound
on the charged Higgs mass quoted by LEP does not apply, as explained 
in appendix \ref{app:dirsearches}.

For completeness, we also write the contributions to $\delta g^R$, which 
are only significant at large $\tan\beta$.
For both 
the Type I and Type II models, they are,
	%
\begin{eqnarray}
\delta g^R &\simeq& - \frac{1}{16 \pi^2} \frac{1}{2} \frac{e}{s_Wc_W}
		\left( \frac{g m_b}{\sqrt{2} M_W} \right)^2 
		\tan^2 \beta \nonumber \\
	& & \times \left[ \sin^2 \delta 
		\left[ \frac{R_1}{R_1-1} 
		- \frac{R_1 \log R_1}{(R_1-1)^2} \right]
		+ \cos^2 \delta
		\left[ \frac{R_2}{R_2-1} 
		- \frac{R_2 \log R_2}{(R_2-1)^2} \right] \right]
		\nonumber \\
	& & - \frac{\epsilon}{16 \pi^2} \frac{e}{s_Wc_W}
		\left( \frac{g m_b}{\sqrt{2} M_W} \right)^2 \tan^2 \beta 
		\sin^2 \delta \cos^2 \delta   
		\left[ C_{24}(m_t^2,M_{H_1^+}^2,M_{H_1^+}^2) \right.
		\nonumber \\
		& & \left.
		+ C_{24}(m_t^2,M_{H_2^+}^2,M_{H_2^+}^2) 
		- 2 C_{24}(m_t^2,M_{H_1^+}^2,M_{H_2^+}^2)
		\right],
\label{eqn:dgR2doub1trip}
\end{eqnarray}
where again $\epsilon = +1$ in the $Y=0$ model and $\epsilon = -1$ 
in the $Y=2$ model.
The first term of equation \ref{eqn:dgR2doub1trip} 
is the same as the correction in a 3HDM.  The second 
term comes from the effects of the triplet.

\subsubsection{Neutral Higgs boson contributions}

Now let us consider the contributions to $Z \rightarrow b \bar{b}$ 
from neutral Higgs boson exchange in these models.  The corrections can
only be significant in the Type II models when $\tan\beta$ is large.
For this reason, we disregard the Type I models here.

The real triplet with $Y=0$ has no CP--odd neutral component, so
there is no $Z \xi^0 A^0$ coupling.  For this reason, $\xi^0$ has the 
same couplings as a Higgs singlet.
(The neutral Higgs couplings for this model are 
listed in appendix \ref{app:2doub1trip}.)  
The corrections from neutral Higgs boson exchange in the
$Y=0$ model thus have the same form as in a model containing two doublets
and a real singlet with $Y=0$.  Models of this type were discussed 
in section \ref{sec:NeutralHiggs}. 

In the $Y=2$ model, the triplet has both a CP--even and a CP--odd
neutral component, and there is a nonzero $Z \chi^{0,r} \chi^{0,i}$ 
coupling.  The neutral Higgs states and couplings in this 
model are listed in appendix \ref{app:2doub1trip}.
We find that
the contributions of the neutral Higgs bosons in this model 
can be split into two pieces.  The first piece is the same as the
contribution in a 3HDM, in which
the neutral Higgs states are given by equations 
\ref{eqn:2d1tG0}, \ref{eqn:2d1tA1}--\ref{eqn:2d1tA2}, and 
\ref{eqn:2d1tH10}--\ref{eqn:2d1tH30},
but with
the triplet states $\chi^{0,r}$ and $\chi^{0,i}$ replaced by the
neutral states of the third doublet.  This piece is denoted
by $\delta g^{R,L}_{\rm 3HDM}$.  The second piece contains
the additional contribution due to the effects of the isospin and 
hypercharge
of the triplet, and is denoted $\delta g^{R,L}_{\rm triplet}$.
That is,
\begin{equation}
\delta g^{R,L} = \delta g^{R,L}_{\rm 3HDM} + \delta g^{R,L}_{\rm triplet}.
\end{equation}
We calculate the contributions to $\delta g^{L,R}$ in the limit of 
large $\tan\theta_2$.  In this limit, the contributions to 
$\delta g^{L,R}$ are as follows.  
The mixing angles $\alpha$, $\gamma$, and $\omega$, which
parameterize the mixing of the neutral states,
are defined in appendix \ref{app:2doub1trip}.  They depend on the 
details of the Higgs potential.
The contribution in the 3HDM is,
\begin{eqnarray}
\delta g^{R,L}_{\rm 3HDM} (a) &\simeq& \mp \frac{1}{16\pi^2}
	\frac{e}{s_Wc_W}
	\left( \frac{gm_b}{\sqrt{2}M_W} \right)^2 \tan^2\beta
	\nonumber \\
	& & \times \left[
	-\cos\gamma \cos\alpha \sin\omega 
		(\cos\gamma \cos\alpha \sin\omega + \sin\gamma \cos\omega)
		C_{24}(m_b^2,M_{H_1^0}^2,M_{A_1^0}^2)
		\right.  \nonumber \\
	& & + \sin\gamma \cos\alpha \sin\omega
		(-\sin\gamma \cos\alpha \sin\omega + \cos\gamma \cos\omega)
		C_{24}(m_b^2,M_{H_2^0}^2,M_{A_1^0}^2)
		\nonumber \\
	& & - \sin^2\alpha \sin^2\omega
		C_{24}(m_b^2,M_{H_3^0}^2,M_{A_1^0}^2)
		\nonumber \\
	& & -\cos\gamma \cos\alpha \cos\omega
		(\cos\gamma \cos\alpha \cos\omega - \sin\gamma \sin\omega)
		C_{24}(m_b^2,M_{H_1^0}^2,M_{A_2^0}^2)
		\nonumber \\
	& & + \sin\gamma \cos\alpha \cos\omega
		(-\sin\gamma \cos\alpha \cos\omega - \cos\gamma \sin\omega)
		C_{24}(m_b^2,M_{H_2^0}^2,M_{A_2^0}^2)
		\nonumber \\
	& & \left. - \sin^2\alpha \cos^2\omega
		C_{24}(m_b^2,M_{H_3^0}^2,M_{A_2^0}^2)  \right]
	\\
\delta g^{R,L}_{\rm 3HDM} (b) &\simeq& -\frac{1}{16\pi^2} g^{L,R}_{Zb\bar{b}}
	\frac{1}{2} \left( \frac{gm_b}{\sqrt{2}M_W} \right)^2 \tan^2\beta
	\nonumber \\
	& & \times \left[
	\cos^2\gamma \cos^2\alpha 
		\left(-2C_{24} + \frac{1}{2} - M_Z^2(C_{22} - C_{23}) \right)
		(M_{H_1^0}^2,m_b^2,m_b^2)
		\right. \nonumber \\
	& & + \sin^2\gamma \cos^2\alpha
		\left(-2C_{24} + \frac{1}{2} - M_Z^2(C_{22} - C_{23}) \right)
		(M_{H_2^0}^2,m_b^2,m_b^2)
		\nonumber \\
	& & + \sin^2\alpha
		\left(-2C_{24} + \frac{1}{2} - M_Z^2(C_{22} - C_{23}) \right)
		(M_{H_3^0}^2,m_b^2,m_b^2)
		\nonumber \\
	& & + \sin^2\omega
		\left(-2C_{24} + \frac{1}{2} - M_Z^2(C_{22} - C_{23}) \right)
		(M_{A_1^0}^2,m_b^2,m_b^2)
		\nonumber \\
	& & \left. + \cos^2\omega
		\left(-2C_{24} + \frac{1}{2} - M_Z^2(C_{22} - C_{23}) \right)
		(M_{A_2^0}^2,m_b^2,m_b^2)
		\right]
		\\
\delta g^{R,L}_{\rm 3HDM} (c) &\simeq& \frac{1}{16\pi^2} g^{R,L}_{Zb\bar{b}}
	\frac{1}{2} \left( \frac{gm_b}{\sqrt{2}M_W} \right)^2 \tan^2\beta
	\nonumber \\
	& & \times \left[
	\cos^2\gamma \cos^2\alpha B_1(m_b^2;m_b^2,M_{H_1^0}^2)
	+ \sin^2\gamma \cos^2\alpha B_1(m_b^2;m_b^2,M_{H_2^0}^2)
	\right. \nonumber \\
	& & + \sin^2\alpha B_1(m_b^2;m_b^2,M_{H_3^0}^2)
	+ \sin^2\omega B_1(m_b^2;m_b^2,M_{A_1^0}^2)
	\nonumber \\
	& & \left.
	+ \cos^2\omega B_1(m_b^2;m_b^2,M_{A_2^0}^2)
	\right].
\end{eqnarray}
The additional contribution due to the effects of the triplet is,
\begin{eqnarray}
\delta g^{R,L}_{\rm triplet} &\simeq& \pm \frac{1}{16 \pi^2}
		\frac{e}{s_Wc_W}
		\left( \frac{g m_b}{\sqrt{2} M_W} \right)^2
		\tan^2 \beta \cos \alpha 
		\cos\gamma \sin\gamma \cos\omega \sin\omega     \nonumber \\
	& & \times \left[ C_{24}(m_b^2,M_{H_1^0}^2,M_{A_1^0}^2)
		+ C_{24}(m_b^2,M_{H_2^0}^2,M_{A_2^0}^2) \right.  \nonumber \\
	& & \left. - C_{24}(m_b^2,M_{H_1^0}^2,M_{A_2^0}^2)
		- C_{24}(m_b^2,M_{H_2^0}^2,M_{A_1^0}^2) \right].
\end{eqnarray}

Note that
$\delta g^{R,L}_{\rm triplet}$ is only significant near maximal 
doublet--triplet mixing in both the CP--odd and CP--even sectors,
which occurs when $\omega$ and $\gamma$ are both near $\pm \pi/4$.
In addition, $\delta g^{R,L}_{\rm triplet}$ is zero if 
$M_{H_1^0} = M_{H_2^0}$ or $M_{A_1^0} = M_{A_2^0}$.  Its sign depends 
on the mixing angles and the Higgs masses.  For all the neutral Higgs
bosons lighter than about 200 GeV and maximal doublet--triplet 
mixing, the contribution to $R_b$ from $\delta g^{R,L}_{\rm triplet}$ is 
smaller than the contribution to $R_b$ from 
$\delta g^{R,L}_{\rm 3HDM}$ over most of the parameter space.
The contribution to $R_b$ from $\delta g^{R,L}_{\rm 3HDM}$ is of
the same order of magnitude as the contribution to $R_b$ from the 
neutral sector of the 2HDM.



\subsection{Georgi--Machacek model with ${\rm SU}(2)_c$ symmetry}
\label{sec:triplets}

In order to obtain $\rho = 1$ at tree level the electroweak symmetry 
breaking must preserve a ``custodial'' SU(2) symmetry, called
${\rm SU}(2)_c$, that ensures
equal masses are given to the $W^{\pm}$ and $W^{3}$.  We refer to 
models with this property as Georgi--Machacek models, after the 
extended model of this type with Higgs triplets created by Georgi and 
Machacek \cite{Georgi1}.  This class of models, and in particular the
triplet Georgi--Machacek model, is described in detail in 
appendix \ref{app:SU2c}, which also contains details on
the ${\rm SU}(2)_L \times 
{\rm SU}(2)_R$ transformations and how ${\rm SU}(2)_c$
is preserved after electroweak symmetry breaking.
In appendix \ref{app:SU2c} we also explain how models that
preserve ${\rm SU}(2)_c$ automatically lead to $\rho = 1$
at tree level, and derive certain Higgs couplings to fermions and 
gauge bosons in a general Georgi--Machacek model.

The triplet Georgi--Machacek model contains a complex
$Y=1$ doublet $\Phi$, a real $Y=0$ triplet $\xi$, and a complex 
$Y=2$ triplet $\chi$.
The Higgs fields take the form
\begin{equation}
\Phi = \left( \begin{array}{cc}
		\phi^{0*} & \phi^{+} \\
		-\phi^{+*}  & \phi^{0}
	      \end{array}	\right)
\end{equation}
\begin{equation}
\chi = \left(  \begin{array}{ccc}
		\chi^{0*}  & \xi^{+} & \chi^{++} \\
		-\chi^{+*}  & \xi^{0} & \chi^{+}  \\
		\chi^{++*} & \xi^{-} & \chi^{0}
	       \end{array}	\right)
\end{equation}
where $\xi^- = -(\xi^+)^*$,
which transform under ${\rm SU}(2)_{L} \times {\rm SU}(2)_{R}$ 
as $(1/2,1/2)$ 
and $(1,1)$ representations, respectively.  
The electroweak symmetry breaking preserves
${\rm SU}(2)_{c}$ when the vevs of the fields are 
diagonal, $\langle \chi \rangle = v_{\chi} \mathbf{I}$ and 
$\langle \phi^0 \rangle = (v_{\phi} / \sqrt{2}) \mathbf{I}$, where
$\mathbf{I}$ is the unit matrix.

Under the electroweak symmetry breaking, the 
${\rm SU}(2)_{L} \times {\rm SU}(2)_{R}$
symmetry is broken down to ${\rm SU}(2)_c$.  A representation $(T,T)$ of 
${\rm SU}(2)_{L} \times {\rm SU}(2)_{R}$ 
decomposes into a set of representations
of ${\rm SU}(2)_c$, in particular,
$2T \oplus 2T-1 \oplus \cdots \oplus 1 \oplus 0$.  In the triplet
Georgi--Machacek model, $\Phi$ breaks down to a 
triplet and a singlet of ${\rm SU}(2)_c$, and $\chi$ breaks down to 
a fiveplet, a triplet, and a singlet of ${\rm SU}(2)_c$.
The $W^{\pm}$ and $Z$ bosons are given mass by absorbing the 
${\rm SU}(2)_{c}$
triplet of Goldstone bosons, $G_{3}^{+,0,-}$.
The remaining physical states are
a fiveplet $H_{5}^{++,+,0,-,--}$, a threeplet 
$H_{3}^{+,0,-}$, and two singlets $H_{1}^{0}$ and $H_{1}^{0\prime}$.
If the Higgs potential is chosen to preserve ${\rm SU}(2)_{c}$, then states 
transforming in different representations of ${\rm SU}(2)_{c}$
cannot mix, and the states in each representation are degenerate.

This model contains only one doublet $\Phi$ which gives mass to both the
top- and bottom-type quarks.  Therefore it is a Type I model and
$\lambda_b \ll \lambda_t$.  
Thus the only sizeable correction to the $Z b \bar{b}$ vertex in this 
model will come from the left-handed charged Higgs boson loops.

The two singly-charged Higgs bosons and $G^{+}$ can be written in terms
of the combinations of triplet fields
\begin{equation}
\psi^{+} = \frac{1}{\sqrt{2}} (\chi^{+} - \xi^{+}),
\end{equation}
which transforms in a triplet of ${\rm SU}(2)_{c}$, and
\begin{equation}
\zeta^{+} = \frac{1}{\sqrt{2}} (\chi^{+} + \xi^{+}),
\end{equation}
which transforms in a fiveplet of ${\rm SU}(2)_{c}$.
We denote the ratio of the vevs of $\chi$ and $\phi$ as
\begin{equation}
\tan\theta_H \equiv \frac{2\sqrt{2} v_{\chi}}{v_{\phi}}.
\end{equation}
Then in terms of the sine and cosine of this angle, denoted 
by $s_H$ and $c_H$, the singly charged Higgs bosons are
\begin{equation}
G^{+}_{3} = c_H \phi^{+} + s_H \psi^{+},
\end{equation}
\begin{equation}
H^{+}_{3} = -s_H \phi^{+} + c_H \psi^{+},
\end{equation}
\begin{equation}
H^{+}_{5} = \zeta^{+}.
\end{equation}

If the Higgs potential is chosen to preserve ${\rm SU}(2)_c$ then
$H^+_3$ and $H^+_5$ are mass eigenstates because they transform
in different representations of ${\rm SU}(2)_c$ \cite{Chanowitz1}. 
Such a potential is desirable because it preserves
${\rm SU}(2)_{c}$ (and $\rho=1$) to all orders in the Higgs
self--couplings.  However, renormalization
of the parameters in the Higgs potential at the one loop level 
introduces quadratically divergent terms that break ${\rm SU}(2)_{c}$
\cite{Gunion2}.  These terms lead to quadratically divergent contributions
to the $\rho$ parameter and to the mixing of some of the Higgs states,
including $H^{+}_{3}$ and $H^{+}_{5}$.  In order to cancel the divergent
corrections, ${\rm SU}(2)_{c}$--breaking 
counterterms must be introduced in the 
bare Lagrangian and fine--tuned to restore $\rho \approx 1$.  These 
${\rm SU}(2)_{c}$--violating 
corrections arise at the two--loop level in $R_{b}$,
so they will be neglected here.

The couplings in this model have been given in \cite{Gunion1,HHG}.  
We have also derived them in appendix \ref{app:SU2c} for a general 
Georgi--Machacek model containing
one multiplet $\Phi = (1/2,1/2)$ and one larger multiplet $X = (T,T)$.
The doublet field $\Phi$ is the only field with
quark Yukawa couplings.  Under ${\rm SU}(2)_{c}$ the doublet decomposes into a
singlet and a triplet.  Thus only ${\rm SU}(2)_{c}$ singlets and triplets can 
contain a doublet admixture and couple to quarks.  This is a general 
feature of any model whose Higgs sector obeys a custodial ${\rm SU}(2)_{c}$
symmetry.  
In the triplet Georgi--Machacek model the charged Higgs couplings to 
quarks are,
\begin{equation}
g^{L}_{G^{+}\bar{t}b} = \frac{gm_{t}}{\sqrt{2}M_{W}}
\end{equation}
\begin{equation}
g^{L}_{H_{3}^{+}\bar{t}b} = \frac{-gm_{t}}{\sqrt{2}M_{W}} \tan\theta_H
\end{equation}
\begin{equation}
g^{L}_{H_{5}^{+}\bar{t}b} = 0.
\end{equation}
These couplings also hold in a general Georgi--Machacek model containing
$\Phi = (1/2,1/2)$ and $X = (T,T)$, if $\tan\theta_H$ is defined as 
\begin{equation}
\tan\theta_H = \frac{v_X \sqrt{\frac{4}{3}T(T+1)(2T+1)}}{v_{\phi}}.
\end{equation}
The loop corrections to $R_{b}$ will only involve the charged Higgs
states that appear in the 
triplet representations of ${\rm SU}(2)_{c}$; namely,
$H^{+}_{3}$ and $G^{+}$.

The relevant $ZH^{+}H^{-}$ couplings for charged Higgs bosons in a 
triplet of ${\rm SU}(2)_c$ are given below, for any model which preserves 
${\rm SU}(2)_c$.
\begin{eqnarray}
g_{ZG^{+}G^{-}} &=& \frac{-e}{s_{W}c_{W}} 
	\left(\frac{1}{2} - s^{2}_{W}\right) \\
g_{ZG^{+}H_{3}^{-}} &=& 0  \\
g_{ZH_{3}^{+}H_{3}^{-}} &=& \frac{-e}{s_{W}c_{W}} 
	\left(\frac{1}{2} - s^{2}_{W}\right),
\end{eqnarray}
as we show in appendix \ref{app:SU2c}.
The loop corrections to $R_{b}$ involving $H^{+}$ are particularly
simple because the $ZG^{+}H_{3}^{-}$ coupling is zero.

In any model which preserves ${\rm SU}(2)_c$ and contains only two multiplets
$\Phi$ and $X$, the correction to $\delta g^L$ is, in addition to
the SM correction due to the charged Goldstone loops,
\begin{equation}
\delta g^{L} = \frac{1}{16\pi^{2}} 
	\left(\frac{gm_{t}}{\sqrt{2}M_{W}} \right)^2 \tan^{2}\theta_H
	\frac{1}{2} \frac{e}{s_{W}c_{W}} 
	\left[ \frac{R}{R-1} - \frac{R \log R}{(R-1)^{2}} \right]
\end{equation}
from loops involving $H_{3}^{+}$, where $R \equiv 
m_{t}^{2}/M_{H_{3}^{+}}^{2}$.  This correction is positive definite and 
has the same form as the correction in the 2HDM (equation
\ref{eq:dgl2HDM}).  

In general for a model with custodial ${\rm SU}(2)_{c}$ 
and more than one exotic
multiplet $X$, the correction becomes
\begin{equation}
\delta g^{L} = \sum_{H_{3i}^{+}} \frac{1}{16\pi^{2}} 
	(g^{L}_{H_{3i}^{+} \bar{t}b})^{2} \frac{1}{2} \frac{e}{s_{W}c_{W}}
	\left[\frac{R_{i}}{R_{i}-1} - \frac{R_{i} \log R_{i}}{(R_{i}-1)^{2}}
	 \right],
\end{equation}
which is positive definite.
Thus when the Higgs potential is invariant under ${\rm SU}(2)_{c}$, 
the corrections always decrease $R_{b}$.

\begin{figure}
\resizebox{\textwidth}{!}{\rotatebox{270}{\includegraphics{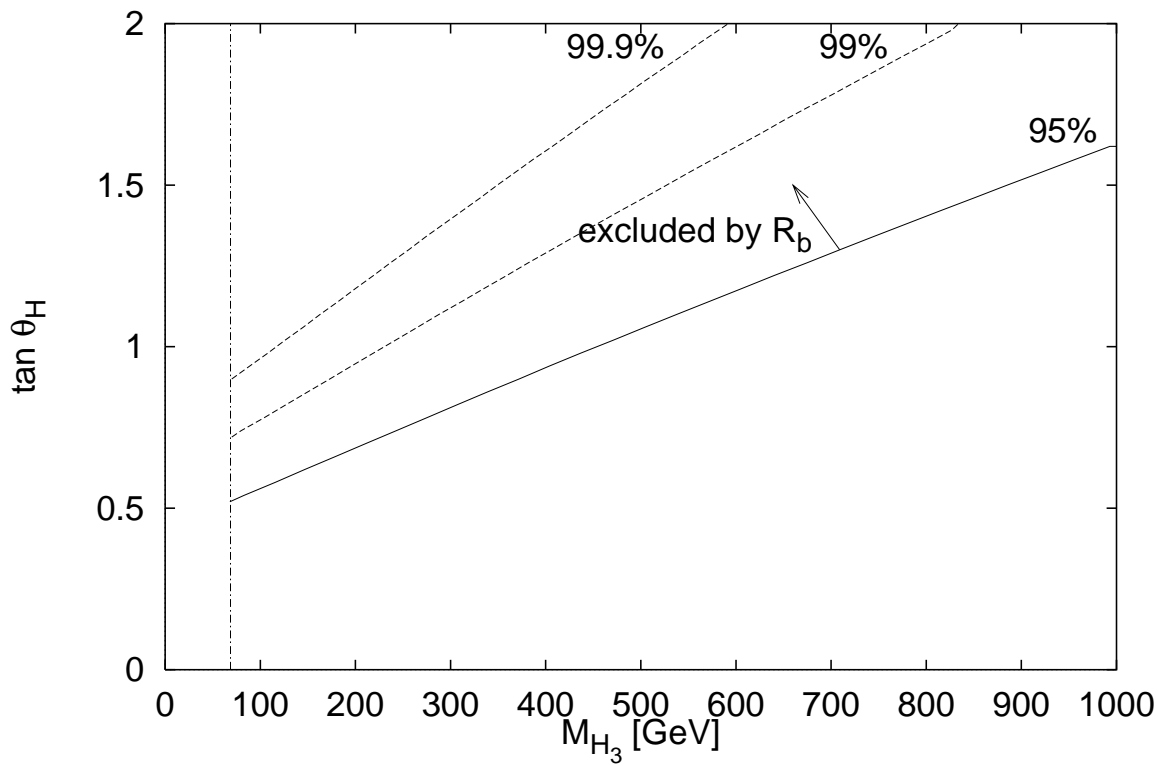}}}
\caption[Bounds from $R_b$ on the Georgi--Machacek model with 
Higgs triplets]
{Bounds from $R_b$ on the Georgi--Machacek model with Higgs triplets
and ${\rm SU}(2)_c$ symmetry.  
On the vertical axis we plot $\tan\theta_H$.  On the
horizontal axis we plot the mass of the ${\rm SU}(2)_c$ triplet $H_3^{\pm}$,
$H_3^0$.  The area above the solid line is ruled out at 95\% confidence
level by $R_b$.  Also shown (top to bottom) are the 99.9\%, 99\%
and 68\%
confidence level contours (dashed).  The dot-dashed line is the direct
search bound on the charged Higgs mass 
from the OPAL collaboration, $M_{H^+} > 68.7$ GeV \cite{OPALH+99}, 
from LEP data up to $\sqrt{s} = 189$ GeV.  (For a discussion of the
direct search bound, see appendix \ref{app:dirsearches}.)
}
\label{fig:tripcustRb}
\end{figure}

As in the 2HDM, the $R_{b}$ measurement can be used to set a lower bound
on the mass of the ${\rm SU}(2)_{c}$ triplet $H_{3}$, which varies with 
$\tan \theta_H$.  This bound in independent of the isospin of the 
exotic ${\rm SU}(2)_L \times {\rm SU}(2)_R$ 
multiplet $X$ (or $\chi$ in the triplet
Georgi--Machacek model).
In figure \ref{fig:tripcustRb} we plot the bound on $M_{H_3}$ as a 
function of $\tan\theta_H$.

For $H_3$ lighter than about 1 TeV, the $R_b$ measurement
puts an upper limit on $\tan\theta_H$, $\tan\theta_H < 2.0$.
In the triplet Georgi--Machacek model, this corresponds to an
upper limit on the triplet vev of $v_{\chi}/v_{\phi} < 0.7$.
As in the Type I 2HDM, the charged Higgs boson contribution to 
$b \rightarrow s \gamma$ is small compared to the 
contribution in the Type II 2HDM \cite{Grinstein90}, and the 
$b \rightarrow s \gamma$ measurement does not provide additional bounds 
on the parameter space. 

In most of the region allowed by $R_b$, $ 0.9345 < A_b < 0.935 $.  There
is a small region of larger $A_b$, up to 0.937, for $\tan\theta_H$
very small.

\subsubsection{Higgs potential without ${\rm SU}(2)_c$ invariance}

If the requirement of ${\rm SU}(2)_c$ symmetry is relaxed, it is no longer
meaningful to write the Higgs fields in 
${\rm SU}(2)_L \times {\rm SU}(2)_R$ matrices.
In the triplet model we must define the vevs of the two 
${\rm SU}(2)_L$ triplets separately,
$\langle \chi^0 \rangle = v_{\chi}$, and $\langle \xi^0 \rangle = v_{\xi}$.
Then ${\rm SU}(2)_c$ symmetry corresponds to $v_{\chi} = v_{\xi}$.
The triplet model
can still satisfy $\rho = 1$ if the Higgs potential is fine-tuned so 
that $v_{\chi} = v_{\xi}$.  In this situation the two physical charged
Higgs bosons $H_{3}^{+}$ and $H_{5}^{+}$ can mix with each other.  If we 
parameterize this mixing with an angle $\alpha$, the new mass eigenstates
are
\begin{equation}
H_{1}^{+} = \sin\alpha H_{3}^{+} + \cos\alpha H_{5}^{+}
\end{equation}
\begin{equation}
H_{2}^{+} = \cos\alpha H_{3}^{+} - \sin\alpha H_{5}^{+}
\end{equation}
and their couplings to the $Z$ and quark pairs are
\begin{equation}
g^{L}_{H_{1}^{+}\bar{t}b}
	= \frac{gm_{t}}{\sqrt{2}M_{W}} \tan\theta_H \sin\alpha
\end{equation}
\begin{equation}
g^{L}_{H_{2}^{+}\bar{t}b} 
	= \frac{gm_{t}}{\sqrt{2}M_{W}} \tan\theta_H \cos\alpha
\end{equation}
\begin{equation}
g_{ZG^{+}H_{1}^{-}} = \frac{-e}{s_{W}c_{W}} \frac{1}{2} s_H \cos\alpha
\end{equation}
\begin{equation}
g_{ZG^{+}H_{2}^{-}} = \frac{e}{s_{W}c_{W}} \frac{1}{2} s_H \sin\alpha
\end{equation}
\begin{equation}
g_{ZH_{1}^{+}H_{1}^{-}} = \frac{-e}{s_{W}c_{W}} (\frac{1}{2} - s^{2}_{W}
		- c_H \sin\alpha \cos\alpha)
\end{equation}
\begin{equation}
g_{ZH_{1}^{+}H_{2}^{-}} = \frac{-e}{s_{W}c_{W}} \frac{1}{2} 
		c_H (\sin^{2}\alpha - \cos^{2}\alpha)
\end{equation}
\begin{equation}
g_{ZH_{2}^{+}H_{2}^{-}} = \frac{-e}{s_{W}c_{W}} (\frac{1}{2} - s^{2}_{W}
		+ c_H \sin\alpha \cos\alpha).
\end{equation}

Now both of the singly charged Higgs bosons 
couple to quarks instead of just one.
There are now off--diagonal $ZH^+_iH^-_j$ couplings with $i\neq j$ 
and non--SM--like terms 
in the diagonal couplings which contribute to $\delta g^L$.

The correction is
\begin{eqnarray}
\delta g^{L}_{H^{+}} &=& \delta g^{L}_{G^{+}} (SM)  \nonumber \\
	&+& \frac{1}{16 \pi^{2}} \frac{1}{2} \frac{e}{s_{W}c_{W}}
		\left( \frac{gm_{t}}{\sqrt{2}M_{W}}
			\right)^{2} \tan^2\theta_H
		\left\{ \sin^{2}\alpha \left[ \frac{R_{1}}{R_{1}-1}
			- \frac{R_{1} \log R_{1}}{(R_{1}-1)^{2}} \right]
			\right. 
				\nonumber \\
	&&		+ \left. \cos^{2}\alpha \left[ \frac{R_{2}}{R_{2}-1}
			- \frac{R_{2} \log R_{2}}{(R_{2}-1)^{2}} \right]
			 \right\}		\nonumber \\
	&+& \frac{1}{16 \pi^{2}} \left(\frac{e}{s_{W}c_{W}} \right)
		\left( \frac{gm_{t}}{\sqrt{2}M_{W}} 
                        \right)^{2} \tan^2\theta_H \times
		2 c_H \sin\alpha \cos\alpha	\nonumber \\
	& & \times \left\{ C_{24}(m_{t}^2,M_{W}^2,M_{2}^2) 
		- C_{24}(m_{t}^2,M_{W}^2,M_{1}^2) 
			\right.	\nonumber \\
	&&	+ \sin^{2}\alpha [C_{24}(m_{t}^2,M_{1}^2,M_{1}^2)
					- C_{24}(m_{t}^2,M_{1}^2,M_{2}^2)]
					\nonumber \\
	& &	+ \left. \cos^{2}\alpha [C_{24}(m_{t}^2,M_{1}^2,M_{2}^2)
				- C_{24}(m_{t}^2,M_{2}^2,M_{2}^2)] \right\},
\label{eq:233mixed}
\end{eqnarray}
where $R_{i} = m_{t}^{2}/M_{i}^{2}$.
The first term is the SM correction due to $G^{+}$.  The second term is 
positive definite and has the same mass dependence as the charged Higgs boson
correction in the 2HDM.  The third term arises from the off-diagonal
$ZH^{+}H^{-}$ couplings and the non--SM parts of the diagonal $ZH^{+}H^{-}$ 
couplings.  This third term can be positive or negative, depending on the
mixing angle $\alpha$.  It is negative for $M_{H_{2}^{+}} > M_{H_{1}^{+}}$ 
when $\sin\alpha \cos\alpha$ is positive, and grows with increasing 
splitting between $M_{H_{1}^{+}}$ and $M_{H_{2}^{+}}$ and between 
$M_{W}$ and the charged Higgs masses.  

This model is fine tuned to $v_{\chi} = v_{\xi}$ to give $\rho =1$; 
when the parameters of the Higgs potential are renormalized
this fine tuning will be lost.  In order to satisfy the 
experimental bounds on $\Delta \rho$ \cite{Altarelli98},
we must have 
\begin{equation}
- 4.7 \times 10^{-3} < \Delta \rho 
	= \frac{4(v_{\xi}^2 - v_{\chi}^2)}{v_{\phi}^2 + 8 v_{\chi}^2}
	< 3.0 \times 10^{-3}
\end{equation}
or $ -(8.4 \mathrm{GeV})^2 < v_{\xi}^2 - v_{\chi}^2 < (6.7 \mathrm{GeV})^2 $.
For the model to
be ``natural'' we require the parameters to be of the same order as their 
fine tuning, or $v_{\chi} \sim v_{\xi} \sim 8$ GeV.  
Then $\tan\theta_H \sim 0.09$
and the correction in equation \ref{eq:233mixed} is suppressed by a factor 
of $\tan^2\theta_H \sim 0.008$.





\chapter{Conclusions}
\label{sec:conclusions}

Radiative corrections to the process $Z \rightarrow b\bar{b}$
arise in extended Higgs sectors due to the exchange of the
additional singly--charged and neutral Higgs bosons in such models.
Because the radiative corrections affect the predictions for $R_b$ and 
$A_b$, the measurements of these quantities can in principle be used
to constrain the parameter space of the models.  
The radiative corrections to $R_b$ from extended Higgs sectors
are typically of the same order of magnitude as the experimental error in the
$R_b$ measurement.  Thus $R_b$ can be used to constrain the models.
However, the radiative corrections to $A_b$ from
extended Higgs sectors are much smaller than the experimental error
in the $A_b$ measurement.  They are also much smaller than the deviation
of the $A_b$ measurement from the SM prediction.
We conclude that if $A_b \neq A_b^{SM}$, the deviation does not arise
from the contributions of an extended Higgs sector.

In this thesis we obtained general formulas for the corrections to the 
$Zb\bar{b}$ vertex, and then used the general formulas
to calculate the contributions to $R_b$ and $A_b$ in specific models.  
Here we summarize our conclusions for the various models.

The contributions from neutral Higgs boson exchange are only significant in a 
Type II model with enhanced $\lambda_b$.  
The regions of parameter space in which 
the contribution to $R_b$ from neutral Higgs boson exchange can be positive
is almost ruled out by direct Higgs boson searches,
as shown in figure \ref{fig:Rb0_1_no}.  
Otherwise, the contribution to $R_b$ is negative, giving a worse agreement
with experiment than the SM.  
A pair of neutral Higgs states, $H^0$ and $A^0$, with a significant $ZH^0A^0$ 
coupling and a large mass splitting, gives a large negative contribution
to $R_b$.  The $R_b$ measurement can then be used to exclude these
regions of parameter space.

The contributions to $R_b$ from charged Higgs boson exchange are negative
in models which contain only doublets and singlets, and in any model 
whose Higgs sector preserves ${\rm SU}(2)_c$ symmetry.
If the contributions from neutral Higgs boson exchange in these models 
are not significant (e.g., if $\lambda_b$ is small), then $R_b$ 
sets a lower bound on the masses of the charged Higgs states.  
The lower bound depends on $\lambda_t$ and the charged Higgs mixing 
angles.

The contribution to $R_b$ from charged Higgs boson exchange can only be
positive if the model contains one of two features.  It must either
contain off--diagonal $ZH^+_iH^-_j$ couplings in which both of the
charged Higgs bosons couple to quarks and have different masses, or it must
contain diagonal $ZH^+_iH^-_i$ couplings which differ from the 
couplings in doublet models, or both.  This can only happen in 
models which contain Higgs multiplets larger than doublets and are
not constrained by ${\rm SU}(2)_c$ symmetry.  In such a model, the vevs 
of the multiplets larger than doublets must be very
small in order for the model to be consistent with the measured 
value of the $\rho$ parameter.
With this constraint, the contribution to $R_b$ can only be positive
when the model contains more than one doublet and there is significant
mixing between the doublets and the larger multiplets.

The precision of the $R_b$ and $A_b$ measurements is not likely to
improve significantly in the near future.
LEP is no longer running at the $Z$ pole and most of LEP's $Z$ pole data
has been analyzed.  SLD will soon stop taking data.
Thus future constraints on extended Higgs sectors must come from other
sources.

The ongoing direct search for Higgs bosons at LEP will 
discover an SM Higgs boson at the 5$\sigma$ level
if its mass is below 104 GeV, or exclude an SM Higgs boson at the 95\%
confidence level up to a mass of
108 GeV \cite{LEPreach98}.
The upcoming search at the Tevatron Run 2 will have a significantly 
greater Higgs mass reach.  
With 10--30 ${\rm fb}^{-1}$ of data per detector,  the 
Tevatron will discover an SM Higgs
boson at the 3--5$\sigma$ level if its mass is below 190 GeV.  If there
is no SM Higgs boson lighter than 190 GeV, the Tevatron will be able
to exclude it at 95\% confidence level with
10 ${\rm fb}^{-1}$ of data \cite{Conway99}.

New virtual constraints on extended Higgs sectors will come from measurements
of $b$ quark decays at BaBar.  For example, the processes $b \to s \gamma$
and $b \to s l^+ l^-$ acquire radiative corrections from charged
Higgs boson exchange.  In addition, the process $b \to s \tau^+ \tau^-$ 
receives a contribution from a neutral Higgs boson coupled to the
$\tau^+ \tau^-$ pair.
The process $b \to c \tau \nu$ receives a contribution from tree--level
charged Higgs boson exchange 
\cite{PKrawczyk88,Kalinowski90,Grossman94}.


\appendix

\chapter{Tree--level $Z q \bar{q}$ couplings in the Standard Model}
\label{sec:Zqqcouplings}

In this section we summarize our conventions for the tree--level 
couplings of $Z$ to fermions.  The tree--level $Zq\bar{q}$ vertex is,
\begin{equation}
-i \gamma_{\mu} (g^L_{Zq\bar{q}} P_L + g^R_{Zq\bar{q}} P_R)
\end{equation}
where $P_{R,L}$ are the right and left handed projection operators,
$P_{R,L} = (1 \pm \gamma_5)/2$, and $g^{R,L}_{Zq\bar{q}}$ are the right and 
left handed $Zq\bar{q}$ couplings.

The tree--level $Zb\bar{b}$ and $Zt\bar{t}$ 
couplings in the SM are,
\begin{equation}
g^{L}_{Zb\bar{b}} = \frac{e}{s_{W}c_{W}} 
	\left(-\nicefrac{1}{2} + \nicefrac{1}{3}s^{2}_{W}\right)
\label{eq:gLZbb}
\end{equation}
\begin{equation}
g^{R}_{Zb\bar{b}} = \frac{e}{s_{W}c_{W}} \left(\nicefrac{1}{3}s^{2}_{W}\right)
\label{eq:gRZbb}
\end{equation}
\begin{equation}
g^{L}_{Zt\bar{t}} = \frac{e}{s_{W}c_{W}} 
	\left(\nicefrac{1}{2} - \nicefrac{2}{3}s^{2}_{W}\right)
\end{equation}
\begin{equation}
g^{R}_{Zt\bar{t}} = \frac{e}{s_{W}c_{W}} 
	\left(- \nicefrac{2}{3}s^{2}_{W}\right).
\end{equation}



\chapter{Higgs--Vector boson couplings}
\label{app:HVcouplings}

In this section we list general formulas for the couplings of Higgs bosons
to vector bosons.  These couplings come from the covariant derivatives in
the kinetic terms for the Higgs bosons in the Lagrangian, 
\begin{equation}
\mathcal{L} = (\mathcal{D}_{\mu} \Phi_k)^{\dagger} 
		(\mathcal{D}^{\mu} \Phi_k)  \\
	+ \frac{1}{2} (\mathcal{D}_{\mu} \eta_i)^T
                (\mathcal{D}^{\mu} \eta_i),
\end{equation}
where the $\Phi_k$ are complex Higgs multiplets with isospin $T_k$ and
hypercharge
$Y_k$, and the $\eta_i$ are real Higgs multiplets with isospin $T_i$ and
hypercharge
zero.  The covariant derivative is
	%
\begin{eqnarray}
\mathcal{D}_{\mu} &=& \partial_{\mu} - igW^a_{\mu}T^a 
		- ig^{\prime} \frac{Y}{2} B_{\mu}  \\
	&=& \partial_{\mu} - i \frac{g}{\sqrt{2}} 
		(W^+_{\mu} T^+ + W^-_{\mu} T^-)  \nonumber \\
	& & - i \frac{g}{\cos\theta_W} Z_{\mu} (T^3 - \sin^2\theta_W Q)
		- ieQA_{\mu}
\end{eqnarray}
where $T^a$ are the SU(2) isospin generators, 
$Q = T^3 + Y/2$ is the electric charge in units of the positron charge,
$T^{\pm} = (T^1 \pm iT^2)$, $\theta_W$ is the weak mixing angle given by
$\tan\theta_W = g^{\prime}/g$, and $e = g g^{\prime} / \sqrt{g^2 +
g^{\prime2}} = g \sin\theta_W$ is the electromagnetic coupling.  The
Higgs--Vector boson couplings are thus entirely determined by the ${\rm SU}(2)
\times {\rm U}(1)$ quantum numbers of the Higgs states and the mixing angles
which determine the mass eigenstates.

We list the Higgs--Vector boson couplings for electroweak
eigenstates in sections \ref{complexreps} and \ref{realreps} below.
Section \ref{complexreps} contains the coupling rules for complex Higgs
representations, and section \ref{realreps} contains the coupling rules
for real Higgs representations (with hypercharge zero).  Finally in
section \ref{mixing} we list the formulas for the couplings of Higgs mass
eigenstates in terms of the couplings of electroweak eigenstates.

The couplings are defined as follows, with all particles incoming to the
vertex:
\vspace{7mm}
\begin{eqnarray}
\begin{fmffile}{fmHHV}
	\parbox{30mm}{\begin{fmfgraph*}(20,15)
	\fmfleft{H2,H1}
	\fmfright{V}
	\fmflabel{$H_1$}{H1}
	\fmflabel{$H_2$}{H2}
	\fmflabel{$V^{\mu}$}{V}
	\fmf{dashes_arrow,label=$p_1$}{H1,v}
	\fmf{dashes_arrow,label=$p_2$}{H2,v}
	\fmf{boson}{v,V}
	\fmfdot{v}
	\end{fmfgraph*}}
\end{fmffile}
&=& i g_{H_1 H_2 V} (p_1 - p_2)^{\mu}  
\label{eqn:HHVdiagram}  \\ [3\baselineskip]
\begin{fmffile}{fmHHVV}
	\parbox{30mm}{\begin{fmfgraph*}(20,15)
	\fmfleft{H2,H1}
	\fmfright{V2,V1}
	\fmflabel{$H_1$}{H1}
	\fmflabel{$H_2$}{H2}
	\fmflabel{$V_1^{\mu}$}{V1}
	\fmflabel{$V_2^{\nu}$}{V2}
	\fmf{dashes}{H1,v,H2}
	\fmf{boson}{V1,v,V2}
	\fmfdot{v}
	\end{fmfgraph*}}
\end{fmffile}
&=& i g_{H_1 H_2 V_1 V_2} g^{\mu\nu}
\label{eqn:HHVVdiagram}  \\ [3\baselineskip]
\begin{fmffile}{fmHVV}
	\parbox{30mm}{\begin{fmfgraph*}(20,15)
	\fmfleft{H}
	\fmfright{V2,V1}
	\fmflabel{$H$}{H}
	\fmflabel{$V_1^{\mu}$}{V1}
	\fmflabel{$V_2^{\nu}$}{V2}
	\fmf{boson}{V1,v}
	\fmf{boson}{V2,v}
	\fmf{dashes}{v,H}
	\fmfdot{v}
	\end{fmfgraph*}}
\end{fmffile}
&=& i g_{HV_1V_2} g^{\mu\nu}  
\label{eqn:HVVdiagram}
\end{eqnarray}
\vspace{7mm}

The Higgs--Higgs--Vector (HHV) couplings come from the terms in the
lagrangian involving two Higgs fields, one vector field and one partial
derivative.  The momentum dependence of the vertex comes from the partial
derivative.  Because of the momentum structure of the vertex, the coupling
$g_{HHV}$ is antisymmetric under interchange of the two Higgs fields.

The Higgs--Higgs--Vector--Vector (HHVV) couplings come from the terms in
the lagrangian involving two Higgs fields and two vector fields. 

The Higgs--Vector--Vector (HVV) couplings come from the terms in the
lagrangian involving one Higgs field, one Higgs vev, and two vector
fields.  Therefore if a multiplet has zero vev, its members will have no
HVV couplings.  The HVV couplings can be immediately obtained from the
HHVV couplings which involve the CP-even neutral Higgs boson 
$\phi^{0,r}$.  This
is done by replacing $\phi^{0,r}$ with $\phi^{0,r} + v$ in the lagrangian
and keeping terms with one Higgs field and one vev.  In order to conserve
electric charge, the HVV couplings can only involve Higgs bosons of charge +2,
+1, 0, -1, or -2.

We note also that in the electroweak basis, the HHV and HHVV couplings
can only involve two Higgs bosons from the same multiplet.

We use the Condon--Shortley phase convention for the SU(2) generators
$T^{\pm}$:
\begin{equation}
T^{\pm} | T, T^3 \rangle = [(T \mp T^3)(T \pm T^3 + 1)]^{1/2} 
	| T, T^3 \pm 1 \rangle
\end{equation}
where $| T, T^3 \rangle$ represents a state with isospin $T$ and third
component of isospin $T^3$.  The electric charge of a state is
related to its hypercharge and third component of isospin by $Q = T^3 +
Y/2$.  Using this, we can write the couplings entirely in terms of the
quantum numbers of the entire Higgs multiplet, $Y$ and
$T$, and the electric charge $Q$ of the particular Higgs state involved.

\section{Higgs--Vector couplings for complex Higgs representations}
\label{complexreps}

In this section we list the Higgs--Vector boson couplings for complex
Higgs representations.  We list separately the couplings involving neutral
Higgs bosons and those involving only charged Higgs bosons 
because the neutral Higgs
states must be separated into CP--even and CP--odd states.  

Our notation is as follows.  We denote the vev of the Higgs multiplet by
$v$, where $\phi^0 = \frac{1}{\sqrt{2}}(\phi^{0,r} + v + i\phi^{0,i})$ and
$v$ is real.  Using this notation, in the Standard Model 
$v = \frac{2 M_W}{g} = 246$ GeV.
We denote the complex conjugate of a Higgs state $\phi^Q$ of charge $Q$ by
$(\phi^Q)^*$.  Note that $(\phi^Q)^*$ is a state of charge $-Q$, but
differs from $\phi^{-Q}$.

\subsection{Higgs--Higgs--Vector couplings}
\label{sec:complexHHV}

In this section we list the HHV couplings, following the notation of
equation \ref{eqn:HHVdiagram}.  We first list the couplings involving only
charged Higgs bosons.
\begin{eqnarray}
g_{\phi^Q (\phi^Q)^{*} \gamma} &=& -eQ  \\
g_{\phi^Q (\phi^Q)^* Z} &=& - \frac{g}{c_W} \left( c^2_W Q - \frac{Y}{2}
	\right)  \\
g_{\phi^Q (\phi^{Q+1})^* W^+} &=& -\frac{g}{\sqrt{2}} \left[\left(
	T - \frac{Y}{2} + 1 + Q \right) \left( T + \frac{Y}{2} - Q
	\right)\right]^{1/2}
\label{eqn:W+HH}
\end{eqnarray}
Note that the $W^-$ coupling can be obtained from equation \ref{eqn:W+HH}
by taking the hermitian conjugate of the Lagrangian.  The couplings
are related by,
\begin{equation}
g_{H_1^* H_2^* W^-} = - (g_{H_1 H_2 W^+})^*.
\label{eqn:HHWhermconj}
\end{equation}
Note that all the couplings are real except for those involving
$\phi^{0,i}$.
In this case,
\begin{equation}
g_{(\phi^Q)^* \phi^{Q+1} W^-} = +\frac{g}{\sqrt{2}} \left[\left(
        T - \frac{Y}{2} + 1 + Q \right) \left( T + \frac{Y}{2} - Q
        \right)\right]^{1/2}.
\end{equation}

We now list the couplings involving one or more neutral Higgs bosons.  The
neutral Higgs couplings to the photon are zero.
Since the $Z$ is CP--odd, it can only couple to one CP--even and one
CP--odd Higgs boson:
\begin{equation}
g_{\phi^{0,r} \phi^{0,i} Z} = -\frac{i g}{c_W} \frac{Y}{2}
\end{equation}
The W--Higgs--Higgs couplings are,
\begin{eqnarray}
g_{\phi^{0,r} (\phi^+)^* W^+} &=& - \frac{g}{2} \left[\left(
	T + \frac{Y}{2} \right) \left( T - \frac{Y}{2} + 1 \right)
	\right]^{1/2}  \\
g_{\phi^{0,i} (\phi^+)^* W^+} &=& -i \frac{g}{2} \left[\left(
        T + \frac{Y}{2} \right) \left( T - \frac{Y}{2} + 1 \right)
        \right]^{1/2}  \\
g_{\phi^- \phi^{0,r} W^+} &=& -\frac{g}{2} \left[\left(
	T - \frac{Y}{2} \right) \left( T + \frac{Y}{2} + 1 \right)
	\right]^{1/2}  \\
g_{\phi^- \phi^{0,i} W^+} &=& i \frac{g}{2} \left[\left(
        T - \frac{Y}{2} \right) \left( T + \frac{Y}{2} + 1 \right)
        \right]^{1/2}  \\
\end{eqnarray}
Again, the $W^-$ couplings are obtained by using equation
\ref{eqn:HHWhermconj}.

\subsection{Higgs--Higgs--Vector--Vector couplings}
\label{sec:HHVVcomplex}

In this section we list the HHVV couplings, following the notation of
equation \ref{eqn:HHVVdiagram}.  We first list the couplings involving
only charged Higgs bosons.
\begin{eqnarray}
g_{\phi^Q (\phi^Q)^* \gamma \gamma} &=& 2 e^2 Q^2  \\
g_{\phi^Q (\phi^Q)^* Z Z} &=& \frac{2 g^2}{c^2_W} 
	\left( c^2_W Q - \frac{Y}{2} \right)^2  \\
g_{\phi^Q (\phi^Q)^* Z \gamma} &=& \frac{2Qeg}{c_W}
	\left( c^2_W Q - \frac{Y}{2} \right)  \\
g_{\phi^Q (\phi^Q)^* W^+ W^-} &=& g^2 \left[
	\left( T + \frac{Y}{2} - Q \right)
	\left( T - \frac{Y}{2} + Q \right) + T \right]  \\
g_{\phi^Q (\phi^{Q+2})^* W^+ W^+} &=& g^2
	\left[\left( T + \frac{Y}{2} - Q \right)
	\left( T - \frac{Y}{2} + Q + 1 \right)\right]^{1/2}
	\nonumber \\
	& & \times \left[\left( T + \frac{Y}{2} - Q - 1 \right)
        \left( T - \frac{Y}{2} + Q + 2 \right)\right]^{1/2}
	\label{eqn:HHW+W+}  \\
g_{\phi^Q (\phi^{Q+1})^* W^+ Z} &=& \frac{g^2}{\sqrt{2} c_W}
	(c^2_W (2Q+1) - Y)
	\nonumber \\
	& & \times \left[\left( T + \frac{Y}{2} - Q \right)
	\left( T - \frac{Y}{2} + 1 + Q \right)\right]^{1/2}  
	\label{eqn:HHW+Z}  \\
g_{\phi^Q (\phi^{Q+1})^* W^+ \gamma} &=& \frac{ge}{\sqrt{2}}
	(2Q+1)
        \left[\left( T + \frac{Y}{2} - Q \right)
        \left( T - \frac{Y}{2} + 1 + Q \right)\right]^{1/2}
	\label{eqn:HHW+gamma}
\end{eqnarray}
The Higgs coupling to $W^-$ pairs can be obtained from equation
\ref{eqn:HHW+W+} using
\begin{equation}
g_{H_1^* H_2 W^- W^-} = (g_{H_1 H_2^* W^+ W^+})^*,
\label{eqn:HHWWhermconj}
\end{equation}
and the Higgs couplings to $W^- V$ where $V = Z, \gamma$ can be obtained
from equations \ref{eqn:HHW+Z} and \ref{eqn:HHW+gamma} using
\begin{equation}
g_{H_1^* H_2 W^- V} = (g_{H_1 H_2^* W^+ V})^*.
\label{eqn:HHWVhermconj}
\end{equation}
We now list the couplings involving one or more neutral Higgs bosons.  The
neutral Higgs couplings to $\gamma \gamma$ and $Z \gamma$ are zero.
\begin{eqnarray}
g_{\phi^{0,r} \phi^{0,r} Z Z} &=&
	g_{\phi^{0,i} \phi^{0,i} Z Z} = \frac{g^2}{c^2_W} \frac{Y^2}{2}  \\
g_{\phi^{0,r} \phi^{0,i} Z Z} &=& 0 \\
g_{\phi^{0,r} \phi^{0,r} W^+ W^-} &=&                                    
        g_{\phi^{0,i} \phi^{0,i} W^+ W^-} = \frac{g^2}{2} 
	\left[ 2T(T+1) - \frac{Y^2}{4} \right]  \\
g_{\phi^{0,r} (\phi^{+2})^* W^+ W^+} 
	&=& -i g_{\phi^{0,i} (\phi^{+2})^* W^+ W^+}  \\
	&=& \frac{g^2}{\sqrt{2}}
	\left[\left( T + \frac{Y}{2} \right)
	\left( T - \frac{Y}{2} + 1 \right)\right]^{1/2}
	\nonumber \\
	& & \times \left[\left( T + \frac{Y}{2} - 1 \right)
	\left( T - \frac{Y}{2} + 2 \right)\right]^{1/2}  
	\label{eqn:h0H*W+W+}  \\
g_{\phi^{0,r} \phi^{-2} W^+ W^+} 
	&=& i g_{\phi^{0,i} \phi^{-2} W^+ W^+}  \\
	&=& \frac{g^2}{\sqrt{2}}
	\left[\left( T - \frac{Y}{2} \right)
        \left( T + \frac{Y}{2} + 1 \right)\right]^{1/2}
	\nonumber \\
        & & \times \left[\left( T - \frac{Y}{2} - 1 \right)
        \left( T + \frac{Y}{2} + 2 \right)\right]^{1/2}
	\label{eqn:h0HW+W+}  \\
g_{\phi^{0,r} (\phi^+)^* W^+ Z} 
	&=& -i g_{\phi^{0,i} (\phi^+)^* W^+ Z}  \\
	&=& \frac{g^2}{2 c_W} (c^2_W - Y) 
	\left[\left( T + \frac{Y}{2} \right)
	\left( T - \frac{Y}{2} + 1 \right)\right]^{1/2}
	\label{eqn:h0H*W+Z}  \\
g_{\phi^{0,r} \phi^- W^+ Z}
	&=& i g_{\phi^{0,i} \phi^- W^+ Z}  \\
	&=& - \frac{g^2}{2 c_W} (c^2_W + Y)
	\left[\left( T - \frac{Y}{2} \right)
        \left( T + \frac{Y}{2} + 1 \right)\right]^{1/2}
	\\
g_{\phi^{0,r} (\phi^+)^* W^+ \gamma}
	&=& -i g_{\phi^{0,i} (\phi^+)^* W^+ \gamma}  \\
	&=& \frac{ge}{2}
	\left[\left( T + \frac{Y}{2} \right)
        \left( T - \frac{Y}{2} + 1 \right)\right]^{1/2}
	\\
g_{\phi^{0,r} \phi^- W^+ \gamma}
        &=& i g_{\phi^{0,i} \phi^- W^+ \gamma}  \\
        &=& -\frac{ge}{2}
        \left[\left( T - \frac{Y}{2} \right)
        \left( T + \frac{Y}{2} + 1 \right)\right]^{1/2}
	\label{eqn:h0HW+gamma}  \\
\end{eqnarray}
The Higgs coupling to $W^-$ pairs can be obtained from equations
\ref{eqn:h0H*W+W+} and \ref{eqn:h0HW+W+} using equation
\ref{eqn:HHWWhermconj} and remembering that $(\phi^{0,r})^* = \phi^{0,r}$
and $(\phi^{0,i})^* = \phi^{0,i}$. 
Similarly the Higgs couplings to $W^- V$ where $V = Z, \gamma$ can
be obtained
from equations \ref{eqn:h0H*W+Z} -- \ref{eqn:h0HW+gamma} using equation
\ref{eqn:HHWVhermconj}.

\subsection{Higgs--Vector--Vector couplings}
\label{sec:HVVcomplex}

In this section we list the HVV couplings, following the notation of
equation \ref{eqn:HVVdiagram}.
The neutral Higgs couplings are,
\begin{eqnarray}
g_{\phi^{0,r} Z Z} &=& \frac{g^2 v}{c^2_W} \frac{Y^2}{2}  \\
g_{\phi^{0,r} W^+ W^-} &=& g^2 v \left( T(T+1) - \frac{Y^2}{4} \right).
\end{eqnarray}
$\phi^{0,i}$ does not couple to gauge boson pairs.

The charged Higgs couplings are,
\begin{eqnarray}
g_{\phi^+ W^- Z} &=& \frac{g^2 v}{2 c_W} (c^2_W - Y) \left[\left( 
	T + \frac{Y}{2} \right) \left( T - \frac{Y}{2} + 1 \right)
	\right]^{1/2}  \\
g_{\phi^+ W^- \gamma} &=& \frac{egv}{2}  \left[\left(    
        T + \frac{Y}{2} \right) \left( T - \frac{Y}{2} + 1 \right)
        \right]^{1/2}  \\
g_{\phi^- W^+ Z} &=& - \frac{g^2 v}{2 c_W} (c^2_W + Y) \left[\left(
        T - \frac{Y}{2} \right) \left( T + \frac{Y}{2} + 1 \right)
        \right]^{1/2}  \\
g_{\phi^- W^+ \gamma} &=& - \frac{egv}{2}  \left[\left(
        T - \frac{Y}{2} \right) \left( T + \frac{Y}{2} + 1 \right)
        \right]^{1/2}  \\
g_{\phi^{+2} W^- W^-} &=& \frac{g^2 v}{\sqrt{2}}
	\left[\left( T - \frac{Y}{2} + 2 \right)
	\left( T + \frac{Y}{2} - 1 \right) \right]^{1/2}
	\nonumber \\
	& & \times \left[\left( T - \frac{Y}{2} + 1 \right)
	\left( T + \frac{Y}{2} \right)\right]^{1/2}  \\
g_{\phi^{-2} W^+ W^+} &=& \frac{g^2 v}{\sqrt{2}}
        \left[\left( T + \frac{Y}{2} + 2 \right)
        \left( T - \frac{Y}{2} - 1 \right) \right]^{1/2}
	\nonumber \\
        & & \times \left[\left( T + \frac{Y}{2} + 1 \right)
        \left( T - \frac{Y}{2} \right)\right]^{1/2}  \\
\end{eqnarray}
The couplings for the conjugate Higgs states $(\phi^Q)^*$ are obtained
using the following equations ($V=Z,\gamma$),
\begin{eqnarray}
g_{(H^{\pm})^* W^{\pm} V} &=& (g_{H^{\pm} W^{\mp} V})^*  \\
g_{(H^{\pm2})^* W^{\pm} W^{\pm}} &=& (g_{H^{\pm2} W^{\mp} W^{\mp}})^*.
\end{eqnarray}

\section{Higgs--Vector couplings for real, $Y=0$ Higgs representations}
\label{realreps}

In this section we list the Higgs--Vector boson couplings for real Higgs
representations.  We consider only real representations with $Y=0$ so that
each electrically charged Higgs boson has an antiparticle of opposite charge. 

Our notation is as follows.  The real Higgs representation is denoted by
$\eta$.  We denote the vev of the Higgs multiplet by $v$, where $\eta^0
\rightarrow \eta^0 + v$, and $v$ is real.

The complex conjugate of a Higgs state $\eta^Q$ is related to the state of
opposite charge by $(\eta^Q)^* = \epsilon_Q \eta^{-Q}$ where 
$\epsilon_Q = (-1)^Q$.  This relation is derived in section
\ref{sec:phase-reln}.
We list the Higgs--Vector boson couplings for two incoming Higgs bosons,
$\eta^Q$ and $\eta^{Q^{\prime}}$.  Note that an incoming $\eta^Q$
corresponds to an outgoing $(\eta^Q)^* = \epsilon_Q \eta^{-Q}$.  This will
be important for finding the correct sign of diagrams involving a Higgs
state that couples to vector bosons at both ends.

The Higgs--Vector couplings for a real Higgs representation will differ
from the couplings for a complex representation because of the phase
relation between $(\eta^Q)^*$ and $\eta^{-Q}$.  The Higgs--Vector--Vector
couplings will also differ because of the different normalization of the
vev of a real representation.

\subsection{Higgs--Higgs--Vector couplings}

In this section we list the HHV couplings, following the notation of
equation \ref{eqn:HHVdiagram}.
\begin{eqnarray}
g_{\eta^Q \eta^{-Q} \gamma} &=& -\epsilon_Q eQ  \\
g_{\eta^Q \eta^{-Q} Z} &=& -\epsilon_Q g c_W Q  \\
g_{\eta^Q \eta^{-Q-1} W^+} &=& \epsilon_Q \frac{g}{\sqrt{2}}
	[(T-Q)(T+Q+1)]^{1/2}  \\
g_{\eta^Q \eta^{-Q+1} W^-} &=& \epsilon_Q \frac{g}{\sqrt{2}}
	[(T+Q)(T-Q+1)]^{1/2}
\end{eqnarray}

These couplings are related to the couplings for complex Higgs
representations given in section \ref{sec:complexHHV} as follows.  Using
$\eta^Q = \epsilon_Q (\eta^{-Q})^*$, $(\epsilon_Q)^2 = 1$, and $\epsilon_Q
\epsilon_{Q+1} = -1$,
\begin{eqnarray}
g_{\eta^Q (\eta^Q)^* \gamma} &=& \epsilon_Q g_{\eta^Q \eta^{-Q} \gamma}
	= -eQ \\
g_{\eta^Q (\eta^Q)^* Z} &=& \epsilon_Q g_{\eta^Q \eta^{-Q} Z}
        = -gc_W Q \\
g_{\eta^Q (\eta^{Q+1})^* W^+} 
	&=& \epsilon_{Q+1} g_{\eta^Q \eta^{-Q-1} W^+}
	\nonumber \\
	&=& - \frac{g}{\sqrt{2}} [(T-Q)(T+Q+1)]^{1/2}  \\
g_{(\eta^Q)^* \eta^{Q+1} W^-} &=& \epsilon_{Q} g_{\eta^{-Q} \eta{Q+1} W^-}
	= \frac{g}{\sqrt{2}} [(T-Q)(T+Q+1)]^{1/2}
\end{eqnarray}
which are the same as the corresponding couplings for a complex Higgs
representation, with $Y=0$.

Note that the $W^+$ and $W^-$ couplings are related by
\begin{eqnarray}
g_{\eta^Q \eta^{-Q-1} W^+} &=& g_{\eta^{-Q} \eta^{Q+1} W^-}  \\
	&=& - (g_{(\eta^Q)^* (\eta^{-Q-1})^* W^-})^*,
\end{eqnarray}
just as for a complex representation.

\subsection{Higgs--Higgs--Vector--Vector couplings}

In this section we list the HHVV couplings, following the notation of
equation \ref{eqn:HHVVdiagram}.
\begin{eqnarray}
g_{\eta^Q \eta^{-Q} \gamma \gamma} &=& 2 \epsilon_Q e^2 Q^2  \\
g_{\eta^Q \eta^{-Q} Z Z} &=& 2 \epsilon_Q g^2 c^2_W Q^2  \\
g_{\eta^Q \eta^{-Q} Z \gamma} &=& 2 \epsilon_Q egc_W Q^2  \\
g_{\eta^Q \eta^{-Q} W^+ W^-} &=& \epsilon_Q g^2 [T(T+1) - Q^2]  \\
g_{\eta^Q \eta^{-Q-2} W^+ W^+} &=& \epsilon_Q g^2 
	[(T-Q)(T+Q+1)]^{1/2}
	\nonumber \\ 
	& & \times [(T-Q-1)(T+Q+2)]^{1/2}  
	\label{eqn:realHHW+W+}  \\
g_{\eta^Q \eta^{-Q-1} W^+ Z} &=& -\epsilon_Q \frac{g^2}{\sqrt{2}} c_W
	(2Q+1) [(T-Q)(T+Q+1)]^{1/2}
	\label{eqn:realHHW+Z}  \\
g_{\eta^Q \eta^{-Q-1} W^+ \gamma} &=& -\epsilon_Q \frac{ge}{\sqrt{2}}
	(2Q+1) [(T-Q)(T+Q+1)]^{1/2}
	\label{eqn:realHHW+gamma}
\end{eqnarray}
The Higgs coupling to $W^-$ pairs can be obtained from equation
\ref{eqn:realHHW+W+} using
\begin{equation}
g_{\eta^{-Q} \eta^{Q+2} W^- W^-} = g_{\eta^Q \eta^{-Q-2} W^+ W^+}
	= (g_{(\eta^{-Q})^* (\eta^{Q+2})^* W^+ W^+})^*
	\label{eqn:realHHWWhermconj}
\end{equation}
and the Higgs couplings to $W^- V$ where $V = Z, \gamma$ can be obtained from 
equations \ref{eqn:realHHW+Z} and \ref{eqn:realHHW+gamma} using 
\begin{equation}
g_{\eta^{-Q} \eta^{Q+1} W^- V} = - g_{\eta^Q \eta^{-Q-1} W^+ V}
	= (g_{(\eta^{-Q})^* (\eta^{Q+1})^* W^+ V})^*,
	\label{eqn:realHHWVhermconj}
\end{equation}
the same as for a complex representation.  Note that the couplings for the
real representation are always real; we write $(g_{HHVV})^*$ in equations
\ref{eqn:realHHWWhermconj} and \ref{eqn:realHHWVhermconj} in order to 
compare with equations
\ref{eqn:HHWWhermconj} and \ref{eqn:HHWVhermconj} for the complex 
representation.

These couplings are related to the couplings for a complex Higgs
representation given in section \ref{sec:HHVVcomplex} as follows.
\begin{eqnarray}
g_{\eta^Q (\eta^Q)^* \gamma \gamma} 
	&=& \epsilon_Q g_{\eta^Q \eta^{-Q} \gamma \gamma}
	= 2 e^2 Q^2  \\
g_{\eta^Q (\eta^Q)^* Z Z}
	&=& \epsilon_Q g_{\eta^Q \eta^{-Q} Z Z}
	= 2 g^2 c^2_W Q^2  \\
g_{\eta^Q (\eta^Q)^* Z \gamma}
	&=& \epsilon_Q g_{\eta^Q \eta^{-Q} Z \gamma}
	= 2 e g c_W Q^2  \\
g_{\eta^Q (\eta^Q)^* W^+ W^-}
	&=& \epsilon_Q g_{\eta^Q \eta^{-Q} W^+ W^-}
	= g^2 [T(T+1) - Q^2]  \\
g_{\eta^Q (\eta^{Q+2})^* W^+ W^+}
	&=& \epsilon_{Q+2} g_{\eta^Q \eta^{-Q-2} W^+ W^+}
	\nonumber \\
	&=& g^2 [(T-Q)(T+Q+1)]^{1/2} 
	\nonumber \\
	& & \times [(T-Q-1)(T+Q+2)]^{1/2}  \\
g_{\eta^Q (\eta^{Q+1})^* W^+ Z}
	&=& \epsilon_{Q+1} g_{\eta^Q \eta^{-Q-1} W^+ Z}
	\nonumber \\
	&=& \frac{g^2}{\sqrt{2}} c_W (2Q+1) [(T-Q)(T+Q+1)]^{1/2}  \\
g_{\eta^Q (\eta^{Q+1})^* W^+ \gamma}
	&=& \epsilon_{Q+1} g_{\eta^Q \eta^{-Q-1} W^+ \gamma}
	\nonumber \\
	&=& \frac{ge}{\sqrt{2}} (2Q+1) [(T-Q)(T+Q+1)]^{1/2}  
\end{eqnarray}
which are the same as the corresponding couplings for a complex Higgs
representation, with $Y=0$.

\subsection{Higgs--Vector--Vector couplings}

In this section we list the HVV couplings, following the notation of
equation \ref{eqn:HVVdiagram}.  The neutral Higgs couplings to $ZZ$,
$\gamma \gamma$, and $Z \gamma$ are zero.  The neutral Higgs coupling to
$W^+ W^-$ is,
\begin{equation}
g_{\eta^0 W^+ W^-} = g^2 v [T(T+1)].
\label{eqn:realHVV1}
\end{equation}

The charged Higgs couplings are,
\begin{eqnarray}
g_{\eta^+ W^- Z} &=& \frac{g^2 c_W v}{\sqrt{2}} [T(T+1)]^{1/2}  \\
g_{\eta^+ W^- \gamma} &=& \frac{gev}{\sqrt{2}} [T(T+1)]^{1/2}  \\
g_{\eta^{+2} W^- W^-} &=& g^2 v [T(T+1)]^{1/2} [(T+2)(T-1)]^{1/2}
\label{eqn:realHVV4}
\end{eqnarray}
The couplings for $\eta^-$ and $\eta^{-2}$ are obtained using the
following equations, with $V=Z,\gamma$,
\begin{eqnarray}
g_{\eta^- W^+ V} &=& - g_{(\eta^+)^* W^+ V} = - (g_{\eta^+ W^- V})^*  \\
g_{\eta^{-2} W^+ W^+} &=& g_{(\eta^{+2})^* W^+ W^+}
    = (g_{\eta^{+2} W^- W^-})^*
\end{eqnarray}
where we have used $\epsilon_1 = -1$ and $\epsilon_2 = +1$.  Note that these
relations are the same as for a complex representation.

Notice that the HVV couplings for a real multiplet are not the same as the
corresponding couplings for a complex multiplet with $Y=0$, given in
section \ref{sec:HVVcomplex}.
The couplings of $\eta^{\pm}$ and $\eta^{\pm2}$ differ
from the corresponding couplings of a complex multiplet by a factor of
$\sqrt{2}$.  This difference comes from the factor of $\sqrt{2}$
difference in the normalization of the vev of the real multiplet compared
to that of the complex multiplet.

\subsection{Derivation of $\epsilon_Q$}
\label{sec:phase-reln}

In this section we derive the phase relation between the states of a real
Higgs representation in the electric charge basis.

The lagrangian for a real Higgs representation,
\begin{equation}
\mathcal{L} = \frac{1}{2} (\mathcal{D}_{\mu} \eta)^T (\mathcal{D}^{\mu}
\eta)
\label{eqn:realLagrangian}
\end{equation}
is written in a Cartesian basis, in which $\eta$ is real and the SU(2)
generators $T^a$ are imaginary and antisymmetric.  However, the Feynman
rules are most useful written in terms of charge eigenstates.  In the
charge basis, equation \ref{eqn:realLagrangian} becomes
\begin{equation}
\mathcal{L} = \frac{1}{2} \sum_{Q=-T}^T ((\mathcal{D}_{\mu} \eta)^Q)^* 
    (\mathcal{D}^{\mu} \eta)^Q
\end{equation}
where $(\mathcal{D}^{\mu} \eta)^Q$ has isospin $T^3 = Q$.  We introduce
this notation in order to treat terms which involve the raising and
lowering operators $T^{\pm}$ correctly.

Let us first examine the kinetic terms for the Higgs states.  The kinetic
part of the Lagrangian is 
\begin{eqnarray}
\mathcal{L} &=& \frac{1}{2} \sum_{Q=-T}^T \partial_{\mu} (\eta^Q)^* 
    \partial^{\mu} \eta^Q  \\
    &=& \frac{1}{2} \partial_{\mu} (\eta^0)^* \partial^{\mu} \eta^0 
    + \sum_{Q=1}^T \partial_{\mu} (\eta^Q)^* \partial^{\mu} \eta^Q  \\
    &=& \frac{1}{2} \epsilon_0 \partial_{\mu} \eta^0 \partial^{\mu} \eta^0
    + \sum_{Q=1}^T \epsilon_Q \partial_{\mu} \eta^{-Q} \partial^{\mu}
\eta^Q  
\end{eqnarray}
We have introduced the notation $(\eta^Q)^* = \epsilon_Q \eta^{-Q}$, where
$\epsilon_Q = \pm 1$ is a phase factor.  Examining the kinetic term for
$\eta^0$, we see immediately that $\epsilon_0 = 1$, or $(\eta^0)^* =
\eta^0$, as is required for a real representation.

We now derive $\epsilon_Q$ for general $Q$.  Consider the following term,
in which $\rho$ and $\eta$ are two real multiplets,
\begin{eqnarray}
\rho^T T^+ \eta &=& \sum_{Q=-T+1}^T (\rho^Q)^* (T^+ \eta)^Q  \\
    &=& \sum_{Q=-T+1}^T \epsilon_Q \rho^{-Q} [(T+Q)(T-Q+1)]^{1/2}
\eta^{Q-1}
    \label{eqn:epsderiv1}
\end{eqnarray}
Taking the complex conjugate of equation \ref{eqn:epsderiv1},
\begin{eqnarray}
(\rho^T T^+ \eta)^* &=& \sum_{Q=-T+1}^T \epsilon_Q (\rho^{-Q})^*
[(T+Q)(T-Q+1)]^{1/2} 
    (\eta^{Q-1})^*  \\
    &=& \sum_{Q=-T+1}^T \epsilon_{Q-1} \rho^Q [(T+Q)(T-Q+1)]^{1/2}
\eta^{-Q+1},
    \label{eqn:epsderiv2}
\end{eqnarray}
where we have used $\epsilon_{-Q} = \epsilon_Q$ and $(\epsilon_Q)^2 = 1$.
The complex conjugate of equation \ref{eqn:epsderiv1} can be derived 
as follows.  Writing the real Higgs multiplets in a Cartesian basis,
the SU(2) generators $T^a$ can be chosen to be imaginary and antisymmetric.
Using 
\begin{equation}
T^{\pm} = T^1 \pm i T^2,
\end{equation}
we find that 
\begin{equation}
(T^+)^* = - T^1 + i T^2 = - T^-.
\end{equation}
The complex conjugate of equation \ref{eqn:epsderiv1} can be written as,
\begin{eqnarray}
(\rho^T T^+ \eta)^* &=& - \rho^T T^- \eta  \\
    &=& - \sum_{Q^{\prime}=-T}^{T-1} (\rho^{Q^{\prime}})^* (T^-
\eta)^{Q^{\prime}}  \\
    &=& - \sum_{Q^{\prime}=-T}^{T-1} (\rho^{Q^{\prime}})^*
[(T-Q^{\prime})(T+Q^{\prime}+1)]^{1/2} \eta^{Q^{\prime}+1}  \\
    &=& - \sum_{Q^{\prime}=-T}^{T-1} \epsilon_{Q^{\prime}}
\rho^{-Q^{\prime}}
    [(T-Q^{\prime})(T+Q^{\prime}+1)]^{1/2} \eta^{Q^{\prime}+1}.
\end{eqnarray}
Changing the summation index to $Q = -Q^{\prime}$ and using the fact that
$\epsilon_{-Q} = \epsilon_Q$,
\begin{equation}
(\rho^T T^+ \eta)^* = - \sum_{Q=-T+1}^T \epsilon_Q \rho^Q
[(T+Q)(T-Q+1)]^{1/2} \eta^{-Q+1}.
\label{eqn:epsderiv3}
\end{equation}
Comparing equations \ref{eqn:epsderiv2} and \ref{eqn:epsderiv3}, we see
that 
$\epsilon_{Q-1} = - \epsilon_Q$.  Together with $\epsilon_0 = 1$, this
gives us a general expression for $\epsilon_Q$,
\begin{equation}
\epsilon_Q = (-1)^Q.
\end{equation}

\section{Higgs--Vector couplings for Higgs mass eigenstates}
\label{mixing}

In computing the amplitudes for real processes we are interested in the
couplings of Higgs mass eigenstates to vector bosons.  In general, the
Higgs mass eigenstates will not correspond to electroweak eigenstates. 
The couplings of Higgs mass eigenstates to vector bosons are obtained from
the couplings given in sections \ref{complexreps} and \ref{realreps} above
using the following formulas.  $H_i$ is a Higgs mass
eigenstate and $\phi_i$ is a Higgs electroweak eigenstate.
\begin{eqnarray}
g_{H V_1 V_2} &=& \sum_{\phi_i} \langle H | \phi_i \rangle 
	g_{\phi_i V_1 V_2}  \\
g_{H_1 H_2 V} &=& \sum_{\phi_i} \sum_{\phi_j} \langle H_1 | \phi_i \rangle
	\langle H_2 | \phi_j \rangle g_{\phi_i \phi_j V}
	\label{eqn:conversionhhv}
\end{eqnarray}
and the $H_1 H_2 V_1 V_2$ coupling is given by equation
\ref{eqn:conversionhhv} with $V$ replaced by $V_1 V_2$.



\chapter{The two Higgs doublet model}
\label{app:2HDM}

In this section we summarize the couplings
and the corrections to the $\rho$ parameter in the two 
Higgs doublet model (2HDM).  
The physical spectrum contains one charged Higgs
state,
\begin{equation}
H^{\pm} = -\sin\beta \phi_1^{\pm} + \cos\beta \phi_2^{\pm},
\end{equation}
one CP--odd neutral state,
\begin{equation}
A^0 = -\sin\beta \phi_1^{0,i} + \cos\beta \phi_2^{0,i},
\end{equation}
and two CP--even neutral states,
\begin{eqnarray}
h^0 &=& -\sin\alpha \phi_1^{0,r} + \cos\alpha \phi_2^{0,r}  \\
H^0 &=& \cos\alpha \phi_1^{0,r} + \sin\alpha \phi_2^{0,r}
\end{eqnarray}
where $\alpha$ is a mixing angle and $M_{h^0} < M_{H^0}$.  There are also
the three Goldstone bosons,
\begin{eqnarray}
G^0 &=& \cos\beta \phi_1^{0,i} + \sin\beta \phi_2^{0,i}  \\
G^{\pm} &=& \cos\beta \phi_1^{\pm} + \sin\beta \phi_2^{\pm}.
\end{eqnarray}

The Higgs Yukawa couplings in the Type II 2HDM are of the 
form
\begin{equation}
i (g^{L}_{H\bar{q}q} P_{L} + g^{R}_{H\bar{q}q} P_{R})
	= i (g^{V}_{H\bar{q}q} + g^{A}_{H\bar{q}q} \gamma_{5}).
\end{equation}
The quark couplings for each Higgs state are,
\begin{equation}
g^{L}_{G^{+}\bar{t}b} = \frac{gm_{t}}{\sqrt{2}M_{W}}
\end{equation}
\begin{equation}
g^{R}_{G^{+}\bar{t}b} = -\frac{gm_{b}}{\sqrt{2}M_{W}}
\end{equation}
\begin{equation}
g^{L}_{H^{+}\bar{t}b} = \frac{gm_{t}}{\sqrt{2}M_{W}} \cot\beta
\label{eq:2HDMgLH+2}
\end{equation}
\begin{equation}
g^{R}_{H^{+}\bar{t}b} = \frac{gm_{b}}{\sqrt{2}M_{W}} \tan\beta
\end{equation}
\begin{equation}
g^{V}_{h^{0}b\bar{b}} = \frac{gm_{b}}{2M_{W}} \frac{\sin\alpha}{\cos\beta}
\end{equation}
\begin{equation}
g^{V}_{H^{0}bb} = -\frac{gm_{b}}{2M_{W}} \frac{\cos\alpha}{\cos\beta}
\end{equation}
\begin{equation}
g^{A}_{G^{0}bb} = -\frac{igm_{b}}{2M_{W}}
\end{equation}
\begin{equation}
g^{A}_{A^{0}bb} = \frac{igm_{b}}{2M_{W}} \tan\beta.
\end{equation}
All the Higgs
couplings in the Type I model are the same as in the Type II model
except for
\begin{equation}
g^{L}_{H^{+}\bar{t}b} = -\frac{gm_{t}}{\sqrt{2}M_{W}} \tan\beta.
\label{eq:2HDMgLH+1}
\end{equation}

The $Z$--Higgs--Higgs couplings take the form,
\begin{equation}
i g_{ZH_{i}H_{j}} (p_{i} - p_{j})_{\mu},
\end{equation}
where $p_i$ ($p_j$) is the incoming momentum of $H_i$ ($H_j$).
The couplings are,
\begin{equation}
g_{ZG^{+}G^{-}} = -\frac{e}{s_{W}c_{W}}(\frac{1}{2} - s^{2}_{W})
\end{equation}
\begin{equation}
g_{ZH^{+}H^{-}} = -\frac{e}{s_{W}c_{W}}(\frac{1}{2} - s^{2}_{W})
\end{equation}
\begin{equation}
g_{ZG^{+}H^{-}} = 0
\end{equation}
\begin{equation}
g_{Zh^{0}G^{0}} = -\frac{i}{2} \frac{e}{s_{W}c_{W}}
	\sin(\beta - \alpha)
\end{equation}
\begin{equation}
g_{ZH^{0}G^{0}} = -\frac{i}{2} \frac{e}{s_{W}c_{W}}
	\cos(\beta - \alpha)
\end{equation}
\begin{equation}
g_{Zh^{0}A^{0}} = -\frac{i}{2} \frac{e}{s_{W}c_{W}}
	\cos(\beta - \alpha)
\end{equation}
\begin{equation}
g_{ZH^{0}A^{0}} = \frac{i}{2} \frac{e}{s_{W}c_{W}}
	\sin(\beta - \alpha).
\end{equation}

We now give the one--loop contribution of the 2HDM Higgs bosons to the $\rho$
parameter, from reference \cite{Haber93rho}.  
This contribution is defined relative to the SM in which the
SM Higgs mass is taken equal to $M_{h^0}$.
We have corrected a typographical error in the formula for $\Delta \rho$
in reference \cite{Haber93rho}.
\def\mw{M_W}
\def\mz{M_Z}
\def\hpm{H^{\pm}}
\def\mhpm{M_{\hpm}}
\def\mhp{M_{\hp}}
\def\hp{H^+}
\def\hm{H^-}
\def\hl{h^0}
\def\hh{H^0}
\def\ha{A^0}
\def\mhl{M_{\hl}}
\def\mhh{M_{\hh}}
\def\mha{M_{\ha}}
	%
\begin{eqnarray}
\Delta \rho &=& \frac{\alpha}{16\pi\mw^2 s_W^2} \Biggl\{
F(\mhpm^2,\mha^2)+\sin^2(\beta-\alpha)\left[F(\mhpm^2,\mhh^2)-
F(\mha^2,\mhh^2)\right]  \nonumber \\ 
& &
+\cos^2(\beta-\alpha)\left[F(\mhpm^2,\mhl^2)
-F(\mha^2,\mhl^2)+F(\mw^2,\mhh^2)\right.  \nonumber \\
& &
-F(\mw^2,\mhl^2)
-F(\mz^2,\mhh^2)+F(\mz^2,\mhl^2) \nonumber  \\
& &
+4\mz^2\left[B_0(0;\mz^2,\mhh^2)-B_0(0;\mz^2,\mhl^2)\right]  \nonumber \\
& & \left.
-4\mw^2\left[B_0(0;\mw^2,\mhh^2)-B_0(0;\mw^2,\mhl^2)\right]\right]\Biggr\}
\end{eqnarray}
where $s_W \equiv \sin\theta_W$, and
\begin{equation}
B_0(0;m_1^2,m_2^2)={A_0(m_1^2)-A_0(m_2^2)\over m_1^2-m_2^2}
\end{equation}
\begin{equation}
A_0(m^2)\equiv m^2(\Delta+1-\log m^2)
\end{equation}
\begin{equation}
F(m_1^2,m_2^2)\equiv\nicefrac{1}{2}(m_1^2+m_2^2)- 
\frac{m_1^2 m_2^2}{m_1^2-m_2^2}\,\log\left(\frac{m_1^2}{m_2^2}\right).
\end{equation}

\chapter{Details of the models with two Higgs doublets and one triplet}
\label{app:2doub1trip}

In this chapter we summarize the details of the models with two
doublets and one triplet, discussed in section \ref{sec:2doub1trip}.
We list the Higgs mass eigenstates, and their couplings which are
relevant in the calculation of the corrections to $Z \rightarrow b\bar{b}$.

The Higgs sector in these models consists of two complex, $Y=1$ 
Higgs doublets, denoted by $\Phi_1$ and $\Phi_2$, plus a triplet
field.  The triplet field can either be a real triplet with
$Y=0$, or a complex triplet with $Y=2$.  We thus have four possible
models to consider:  a Type I model with a $Y=0$ triplet, 
a Type II model with a $Y=0$ triplet, a Type I model with a $Y=2$
triplet, and a Type II model with a $Y=2$ triplet.  We will refer
to the models with a $Y=0$ triplet as the $Y=0$ models, and to the
models with a $Y=2$ triplet as the $Y=2$ models.

We will treat the $Y=0$ and $Y=2$ models separately.  In what follows
we will assume that the models are Type II, but will also give the 
couplings in the Type I models where they differ from those in 
the Type II models.

We define $\tan\beta$ in these models exactly as in the 2HDM:
$\tan\beta = v_2/v_1$, where the vevs of the doublets are 
$\langle \phi_1^0 \rangle = v_1/\sqrt{2}$ and 
$\langle \phi_2^0 \rangle = v_2/\sqrt{2}$.
The vev of the triplet
field must be fine--tuned very small in order to be consistent
with the measured value of the $\rho$ parameter, $\rho \approx 1$.

\section{$Y=0$ model}

We first consider the model with one doublet and one real triplet 
field with $Y=0$.  The triplet field is
$\xi = (\xi^{+}, \xi^{0}, \xi^{-})$.  We define the triplet vev
as $\langle \xi^0 \rangle = v_{\xi}$.
The vevs are constrained by the
$W$ mass to satisfy,
\begin{equation}
v_{SM}^2 = v_1^2 + v_2^2 + 4 v_{\xi}^2,
\end{equation}
where $v_{SM} = 246$ GeV.
It is convenient to parameterize
the ratio of the vevs by,
\begin{equation}
\tan \theta_0 = \frac{\sqrt{v_1^2 + v_2^2}}{2 v_\xi}.
\end{equation}
In this model, the tree--level $\rho$ parameter is,
\begin{equation}
\rho = \frac{v_1^{2} + v_2^2 + 4 v_{\xi}^{2}}{v_1^{2} + v_2^2}
	= 1 + \frac{4 v_{\xi}^{2}}{v_1^{2} + v_2^2} = 1 + \Delta \rho.
\end{equation}
Writing this in terms of $\tan\theta_0$, we find
\begin{equation}
\Delta \rho = \frac{1}{\tan^2 \theta_0}.
\end{equation}
We see that in order to have $\rho \approx 1$, the triplet vev
must be very small, giving large $\tan\theta_0$.

This model contains three charged states, one from each of
the doublets and one from the triplet.
The charged Higgs states mix to form the Goldstone boson,
\begin{equation}
G^+ = \sin\theta_0 (\cos\beta \phi_1^+ + \sin\beta \phi_2^+) 
	+ \cos\theta_0 \xi^+,
\end{equation}
and two orthogonal states which are the physical charged Higgs bosons.
We define two orthogonal states,
\begin{eqnarray}
H_1^{+\prime} &=& \cos\theta_0 (\cos\beta \phi_1^+ + \sin\beta \phi_2^+) 
	- \sin\theta_0 \xi^+ \\
H_2^{+\prime} &=& -\sin\beta \phi_1^+ + \cos\beta \phi_2^+
\end{eqnarray}
which will mix by an angle $\delta$ to form the mass eigenstates.
The mixing angle $\delta$ depends on the details of the Higgs potential.
The mass eigenstates then take the complicated form,
\begin{eqnarray}
H_1^+ &=& (\cos\delta \cos\theta_0 \cos\beta - \sin\delta \sin\beta)
	\phi_1^+  \nonumber \\
	& & + (\cos\delta \cos\theta_0 \sin\beta + \sin\delta \cos\beta)
	\phi_2^+
	- \cos\delta \sin\theta_0 \xi^+
	\\
H_2^+ &=& (-\sin\delta \cos\theta_0 \cos\beta - \cos\delta \sin\beta)
	\phi_1^+  \nonumber \\
	& & + (-\sin\delta \cos\theta_0 \sin\beta + \cos\delta \cos\beta)
	\phi_2^+
	+ \sin\delta \sin\theta_0 \xi^+.
\end{eqnarray}

The Higgs couplings to quarks are of the form,
\begin{equation}
i (g^{L}_{H\bar{q}q} P_{L} + g^{R}_{H\bar{q}q} P_{R})
	= i (g^{V}_{H\bar{q}q} + g^{A}_{H\bar{q}q} \gamma_{5}).
\end{equation}
In the Type II model, the charged Higgs couplings to $\bar{t}b$ are,
\begin{eqnarray}
g^L_{G^+\bar{t}b} &=& \frac{gm_t}{\sqrt{2}M_W}  \\
g^R_{G^+\bar{t}b} &=& -\frac{gm_b}{\sqrt{2}M_W}  \\
g^L_{H_1^+\bar{t}b} &=& \frac{gm_t}{\sqrt{2}M_W} 
	\frac{(\cos\delta \cos\theta_0 \sin\beta + \sin\delta \cos\beta)}
	{\sin\beta \sin\theta_0}  \\
g^R_{H_1^+\bar{t}b} &=& -\frac{gm_b}{\sqrt{2}M_W}
	\frac{(\cos\delta \cos\theta_0 \cos\beta - \sin\delta \sin\beta)}
	{\cos\beta \sin\theta_0}  \\
g^L_{H_2^+\bar{t}b} &=& \frac{gm_t}{\sqrt{2}M_W}
	\frac{(-\sin\delta \cos\theta_0 \sin\beta + \cos\delta \cos\beta)}
	{\sin\beta \sin\theta_0}  \\
g^R_{H_2^+\bar{t}b} &=& -\frac{gm_b}{\sqrt{2}M_W}
	\frac{(-\sin\delta \cos\theta_0 \cos\beta - \cos\delta \sin\beta)}
	{\cos\beta \sin\theta_0}.
\end{eqnarray}
In the Type I model, the couplings are the same except for,
\begin{eqnarray}
g^L_{H_1^+\bar{t}b} &=& \frac{gm_t}{\sqrt{2}M_W} 
	\frac{(\cos\delta \cos\theta_0 \cos\beta - \sin\delta \sin\beta)}
	{\cos\beta \sin\theta_0}  \\
g^L_{H_2^+\bar{t}b} &=& \frac{gm_t}{\sqrt{2}M_W}
	\frac{(-\sin\delta \cos\theta_0 \cos\beta - \cos\delta \sin\beta)}
	{\cos\beta \sin\theta_0}.
\end{eqnarray}

The $Z$--Higgs--Higgs couplings take the form,
\begin{equation}
i g_{ZH_{i}H_{j}} (p_{i} - p_{j})_{\mu},
\end{equation}
where $p_i$ ($p_j$) is the incoming momentum of $H_i$ ($H_j$).
The charged Higgs couplings to $Z$ are,
\begin{eqnarray}
g_{ZG^+(G^+)^*} &=& -\frac{e}{s_Wc_W} \left[ \frac{1}{2} - s^2_W
	+ \frac{1}{2} \cos^2\theta_0 \right]  \\
g_{ZG^+(H_1^+)^*} &=& \frac{e}{s_Wc_W} \frac{1}{2} 
	\sin\theta_0 \cos\theta_0 \cos\delta  \\
g_{ZG^+(H_2^+)^*} &=& -\frac{e}{s_Wc_W} \frac{1}{2} 
	\sin\theta_0 \cos\theta_0 \sin\delta  \\
g_{ZH_1^+(H_1^+)^*} &=& -\frac{e}{s_Wc_W} \left[ \frac{1}{2} - s^2_W
	+ \frac{1}{2} \sin^2\theta_0 \cos^2\delta \right]  \\
g_{ZH_2^+(H_2^+)^*} &=& -\frac{e}{s_Wc_W} \left[ \frac{1}{2} - s^2_W
	+ \frac{1}{2} \sin^2\theta_0 \sin^2\delta \right]  \\
g_{ZH_1^+(H_2^+)^*} &=& \frac{e}{s_Wc_W} \frac{1}{2} 
	\sin\delta \cos\delta \sin^2\theta_0,
\end{eqnarray}
where $s_W$ is the sine of the weak mixing angle.

The states and couplings can be simplified a great deal if we take the limit
of large $\tan\theta_0$, as required by the measured value of the
$\rho$ parameter.  As shown in section \ref{sec:finetuned}, the 
$\rho$ parameter measurement requires $\tan\theta_0 > 18$.  This 
corresponds to $\sin\theta_0 > 0.998$ and $\cos\theta_0 < 0.055$.
We will make the approximation $\sin\theta_0 = 1$ and 
$\cos\theta_0 = 0$.  Then the charged states are,
\begin{eqnarray}
G^+ &\simeq& \cos\beta \phi_1^+ + \sin\beta \phi_2^+ \\
H_1^{+\prime} &\simeq& -\xi^+ \\
H_2^{+\prime} &=& -\sin\beta \phi_1^+ + \cos\beta \phi_2^+.
\end{eqnarray}
Mixing $H_1^{+\prime}$ and $H_2^{+\prime}$ through the mixing angle
$\delta$, the physical mass eigenstates are,
\begin{eqnarray}
H_1^+ &\simeq& \sin\delta (-\sin\beta \phi_1^+ + \cos\beta \phi_2^+)
	- \cos\delta \xi^+ \\
H_2^+ &\simeq& \cos\delta (-\sin\beta \phi_1^+ + \cos\beta \phi_2^+)
	+ \sin\delta \xi^+.
\end{eqnarray}

In the Type II model, the charged Higgs couplings to $\bar{t}b$ 
simplify to,
\begin{eqnarray}
g^L_{H_1^+\bar{t}b} &=& \frac{gm_t}{\sqrt{2}M_W} 
	\cot\beta \sin\delta  \\
g^R_{H_1^+\bar{t}b} &=& +\frac{gm_b}{\sqrt{2}M_W}
	\tan\beta \sin\delta  \\
g^L_{H_2^+\bar{t}b} &=& \frac{gm_t}{\sqrt{2}M_W}
	\cot\beta \cos\delta  \\
g^R_{H_2^+\bar{t}b} &=& +\frac{gm_b}{\sqrt{2}M_W}
	\tan\beta \cos\delta.
\end{eqnarray}
In the Type I model, $\cot\beta$ is replaced by $\tan\beta$ in
the left--handed couplings.

The charged Higgs couplings to $Z$ simplify to,
\begin{eqnarray}
g_{ZG^+(G^+)^*} &=& -\frac{e}{s_Wc_W} \left[ \frac{1}{2} - s^2_W
	\right]  \\
g_{ZG^+(H_1^+)^*} &=& 0 \\
g_{ZG^+(H_2^+)^*} &=& 0  \\
g_{ZH_1^+(H_1^+)^*} &=& -\frac{e}{s_Wc_W} \left[ \frac{1}{2} - s^2_W
	+ \frac{1}{2} \cos^2\delta \right]  \\
g_{ZH_2^+(H_2^+)^*} &=& -\frac{e}{s_Wc_W} \left[ \frac{1}{2} - s^2_W
	+ \frac{1}{2} \sin^2\delta \right]  \\
g_{ZH_1^+(H_2^+)^*} &=& \frac{e}{s_Wc_W} \frac{1}{2} 
	\sin\delta \cos\delta.
\end{eqnarray}

\subsubsection{Neutral Higgs bosons in the $Y=0$ model}

In the $Y=0$ model there are three CP--even neutral degrees of
freedom and only two CP--odd neutral degrees of freedom, because
the real triplet contributes only a CP--even degree of freedom.  
These states mix to
form three CP--even states, one CP--odd state, and the CP--odd
neutral Goldstone boson.

The neutral Goldstone boson is,
\begin{equation}
G^0 = \cos\beta \phi_1^{0,i} + \sin\beta \phi_2^{0,i},
\end{equation}
and the physical CP--odd neutral Higgs boson is the orthogonal state,
\begin{equation}
A^0 = -\sin\beta \phi_1^{0,i} + \cos\beta \phi_2^{0,i}.
\end{equation}
Note that $G^0$ and $A^0$ are the same as in the 2HDM.

We next define three orthogonal CP--even Higgs states.
We allow the CP--even components of the two doublets to mix by 
an angle $\alpha$, in analogy with the 2HDM.  This gives two 
orthogonal states,
\begin{eqnarray}
H^{0\prime} &=& \cos\alpha \phi_1^{0,r} + \sin\alpha \phi_2^{0,r}  \\
h^{0\prime} &=& -\sin\alpha \phi_1^{0,r} + \cos\alpha \phi_2^{0,r}.
\end{eqnarray}
In general, both $H^{0\prime}$ and $h^{0\prime}$ mix with 
$\xi^0$ by angles determined by the details of the Higgs
potential.  However, for simplicity we will consider the case 
in which only $H^{0\prime}$ mixes with the triplet.  We will 
parameterize this doublet--triplet mixing with the mixing angle
$\gamma$.  The resulting mass eigenstates are,
\begin{eqnarray}
H_1^0 &=& \cos\gamma (\cos\alpha \phi_1^{0,r} 
	+ \sin\alpha \phi_2^{0,r}) + \sin\gamma \xi^0  
	\\
H_2^0 &=& -\sin\gamma (\cos\alpha \phi_1^{0,r} 
	+ \sin\alpha \phi_2^{0,r}) + \cos\gamma \xi^0  
	\\
H_3^0 &=& -\sin\alpha \phi_1^{0,r} + \cos\alpha \phi_2^{0,r}.
\end{eqnarray}

The neutral Higgs couplings to quarks are,
\begin{eqnarray}
g^V_{H_1^0b\bar{b}} &=& -\frac{1}{\sqrt{2}}
	\left( \frac{gm_b}{\sqrt{2}M_W} \right)
	\frac{\cos\gamma \cos\alpha}{\cos\beta \sin\theta_0}  \\
g^V_{H_2^0b\bar{b}} &=& -\frac{1}{\sqrt{2}}
	\left( \frac{gm_b}{\sqrt{2}M_W} \right)
	\frac{-\sin\gamma \cos\alpha}{\cos\beta \sin\theta_0}  \\
g^V_{H_3^0b\bar{b}} &=& -\frac{1}{\sqrt{2}}
	\left( \frac{gm_b}{\sqrt{2}M_W} \right)
	\frac{-\sin\alpha}{\cos\beta \sin\theta_0}  \\
g^A_{G^0b\bar{b}} &=& -\frac{i}{\sqrt{2}}
	\left( \frac{gm_b}{\sqrt{2}M_W} \right)
	\frac{1}{\sin\theta_0}  \\
g^A_{A^0b\bar{b}} &=& -\frac{i}{\sqrt{2}}
	\left( \frac{gm_b}{\sqrt{2}M_W} \right)
	\frac{-\tan\beta}{\sin\theta_0}.
\end{eqnarray} 

The neutral Higgs couplings to $Z$ are,
\begin{eqnarray}
g_{ZH_1^0G^0} &=& \frac{ie}{s_Wc_W} \left[ -\frac{1}{2} \cos\gamma
	\cos(\beta - \alpha) \right]  \\
g_{ZH_1^0A^0} &=& \frac{ie}{s_Wc_W} \left[ \frac{1}{2} \cos\gamma
	\sin(\beta - \alpha) \right] \\
g_{ZH_2^0G^0} &=& \frac{ie}{s_Wc_W} \left[ \frac{1}{2} \sin\gamma
	\cos(\beta - \alpha) \right]  \\
g_{ZH_2^0A^0} &=& \frac{ie}{s_Wc_W} \left[ -\frac{1}{2} \sin\gamma
	\sin(\beta - \alpha) \right] \\
g_{ZH_3^0G^0} &=& \frac{ie}{s_Wc_W} \left[ -\frac{1}{2} 
	\sin(\beta - \alpha) \right] \\
g_{ZH_3^0A^0} &=& \frac{ie}{s_Wc_W} \left[ -\frac{1}{2} 
	\cos(\beta - \alpha) \right].
\end{eqnarray}

The Higgs couplings to $b$ quarks can be simplified if we take the
limit of large $\tan\theta_0$, as required by the measured value 
of the $\rho$ parameter.  When we make the approximation 
$\sin\theta_0 \approx 1$, the neutral Higgs couplings to $b$ quarks
become,
\begin{eqnarray}
g^V_{H_1^0b\bar{b}} &=& -\frac{1}{\sqrt{2}}
	\left( \frac{gm_b}{\sqrt{2}M_W} \right)
	\frac{\cos\gamma \cos\alpha}{\cos\beta}  \\
g^V_{H_2^0b\bar{b}} &=& -\frac{1}{\sqrt{2}}
	\left( \frac{gm_b}{\sqrt{2}M_W} \right)
	\frac{-\sin\gamma \cos\alpha}{\cos\beta}  \\
g^V_{H_3^0b\bar{b}} &=& -\frac{1}{\sqrt{2}}
	\left( \frac{gm_b}{\sqrt{2}M_W} \right)
	\frac{-\sin\alpha}{\cos\beta}  \\
g^A_{G^0b\bar{b}} &=& -\frac{i}{\sqrt{2}}
	\left( \frac{gm_b}{\sqrt{2}M_W} \right)
	\\
g^A_{A^0b\bar{b}} &=& -\frac{i}{\sqrt{2}}
	\left( \frac{gm_b}{\sqrt{2}M_W} \right)
	(-\tan\beta).
\end{eqnarray}

\section{$Y=2$ model}

We now consider the model with one doublet and one complex triplet 
field with $Y=2$.  The triplet field is
$\chi = ( \chi^{++}, \chi^+, \chi^0 )$.
We define the triplet vev
as $\langle \chi^0 \rangle = v_{\chi}/\sqrt{2}$.
The vevs are constrained by the
$W$ mass to satisfy,
\begin{equation}
v_{SM}^2 = v_1^2 + v_2^2 + 2 v_{\chi}^2,
\end{equation}
where $v_{SM} = 246$ GeV.
It is convenient to parameterize
the ratio of the vevs by,
\begin{equation}
\tan \theta_2 = \frac{\sqrt{v_1^2 + v_2^2}}{\sqrt{2} v_{\chi}}.
\end{equation}
In this model, the tree--level $\rho$ parameter is,
\begin{equation}
\rho = \frac{ v_1^2 + v_2^2 + 2 v_{\chi}^2 }{ v_1^2 + v_2^2 + 4 v_{\chi}^2 }
	= 1 - \frac{ 2 v_{\chi}^2 }{ v_1^2 + v_2^2 + 4 v_{\chi}^2 }
	= 1 + \Delta \rho.
\end{equation}
Writing this in terms of $\tan\theta_2$, we find
\begin{equation}
\Delta \rho = -\frac{1}{\tan^2 \theta_2 + 2}.
\end{equation}
We see that in order to have $\rho \approx 1$, the triplet vev
must be very small, giving large $\tan\theta_2$.

This model contains three singly--charged states, one from each of
the doublets and one from the triplet.
The charged Higgs states mix to form the Goldstone boson,
\begin{equation}
G^+ = \sin\theta_2 (\cos\beta \phi_1^+ + \sin\beta \phi_2^+) 
	+ \cos\theta_2 \chi^+,
\end{equation}
and two orthogonal states which are the physical charged Higgs bosons.
We define two orthogonal states,
\begin{eqnarray}
H_1^{+\prime} &=& \cos\theta_2 (\cos\beta \phi_1^+ + \sin\beta \phi_2^+) 
	- \sin\theta_2 \chi^+ \\
H_2^{+\prime} &=& -\sin\beta \phi_1^+ + \cos\beta \phi_2^+,
\end{eqnarray}
which will mix by an angle $\delta$ to form the mass eigenstates.
The mixing angle $\delta$ depends on the details of the Higgs potential.
The mass eigenstates then take the complicated form,
\begin{eqnarray}
H_1^+ &=& (\cos\delta \cos\theta_2 \cos\beta - \sin\delta \sin\beta)
	\phi_1^+  \nonumber \\
	& & + (\cos\delta \cos\theta_2 \sin\beta + \sin\delta \cos\beta)
	\phi_2^+
	- \cos\delta \sin\theta_2 \xi^+
	\\
H_2^+ &=& (-\sin\delta \cos\theta_2 \cos\beta - \cos\delta \sin\beta)
	\phi_1^+  \nonumber \\
	& & + (-\sin\delta \cos\theta_2 \sin\beta + \cos\delta \cos\beta)
	\phi_2^+
	+ \sin\delta \sin\theta_2 \xi^+.
\end{eqnarray}

The Higgs couplings to quarks are of the form,
\begin{equation}
i (g^{L}_{H\bar{q}q} P_{L} + g^{R}_{H\bar{q}q} P_{R})
	= i (g^{V}_{H\bar{q}q} + g^{A}_{H\bar{q}q} \gamma_{5}).
\end{equation}
In the Type II model, the charged Higgs couplings to $\bar{t}b$ are,
\begin{eqnarray}
g^L_{G^+\bar{t}b} &=& \frac{gm_t}{\sqrt{2}M_W}  \\
g^R_{G^+\bar{t}b} &=& -\frac{gm_b}{\sqrt{2}M_W}  \\
g^L_{H_1^+\bar{t}b} &=& \frac{gm_t}{\sqrt{2}M_W} 
	\frac{(\cos\delta \cos\theta_2 \sin\beta + \sin\delta \cos\beta)}
	{\sin\beta \sin\theta_2}  \\
g^R_{H_1^+\bar{t}b} &=& -\frac{gm_b}{\sqrt{2}M_W}
	\frac{(\cos\delta \cos\theta_2 \cos\beta - \sin\delta \sin\beta)}
	{\cos\beta \sin\theta_2}  \\
g^L_{H_2^+\bar{t}b} &=& \frac{gm_t}{\sqrt{2}M_W}
	\frac{(-\sin\delta \cos\theta_2 \sin\beta + \cos\delta \cos\beta)}
	{\sin\beta \sin\theta_2}  \\
g^R_{H_2^+\bar{t}b} &=& -\frac{gm_b}{\sqrt{2}M_W}
	\frac{(-\sin\delta \cos\theta_2 \cos\beta - \cos\delta \sin\beta)}
	{\cos\beta \sin\theta_2}.
\end{eqnarray}
In the Type I model, the couplings are the same except for,
\begin{eqnarray}
g^L_{H_1^+\bar{t}b} &=& \frac{gm_t}{\sqrt{2}M_W} 
	\frac{(\cos\delta \cos\theta_2 \cos\beta - \sin\delta \sin\beta)}
	{\cos\beta \sin\theta_2}  \\
g^L_{H_2^+\bar{t}b} &=& \frac{gm_t}{\sqrt{2}M_W}
	\frac{(-\sin\delta \cos\theta_2 \cos\beta - \cos\delta \sin\beta)}
	{\cos\beta \sin\theta_2}.
\end{eqnarray}

The $Z$--Higgs--Higgs couplings take the form,
\begin{equation}
i g_{ZH_{i}H_{j}} (p_{i} - p_{j})_{\mu},
\end{equation}
where $p_i$ ($p_j$) is the incoming momentum of $H_i$ ($H_j$).
The charged Higgs couplings to $Z$ are,
\begin{eqnarray}
g_{ZG^+(G^+)^*} &=& -\frac{e}{s_Wc_W} \left[ \frac{1}{2} - s^2_W
	- \frac{1}{2} \cos^2\theta_2 \right]  \\
g_{ZG^+(H_1^+)^*} &=& -\frac{e}{s_Wc_W} \frac{1}{2} 
	\sin\theta_2 \cos\theta_2 \cos\delta  \\
g_{ZG^+(H_2^+)^*} &=& \frac{e}{s_Wc_W} \frac{1}{2} 
	\sin\theta_2 \cos\theta_2 \sin\delta  \\
g_{ZH_1^+(H_1^+)^*} &=& -\frac{e}{s_Wc_W} \left[ \frac{1}{2} - s^2_W
	- \frac{1}{2} \sin^2\theta_2 \cos^2\delta \right]  \\
g_{ZH_2^+(H_2^+)^*} &=& -\frac{e}{s_Wc_W} \left[ \frac{1}{2} - s^2_W
	- \frac{1}{2} \sin^2\theta_2 \sin^2\delta \right]  \\
g_{ZH_1^+(H_2^+)^*} &=& -\frac{e}{s_Wc_W} \frac{1}{2} 
	\sin\delta \cos\delta \sin^2\theta_2,
\end{eqnarray}
where $s_W$ is the sine of the weak mixing angle.

The states and couplings can be simplified a great deal if we take the 
limit
of large $\tan\theta_2$, as required by the measured value of the
$\rho$ parameter.  As shown in section \ref{sec:finetuned}, the 
$\rho$ parameter measurement requires $\tan\theta_2 > 15$.  This 
corresponds to $\sin\theta_2 > 0.998$ and $\cos\theta_2 < 0.067$.
We will make the approximation $\sin\theta_2 = 1$ and 
$\cos\theta_2 = 0$.  Then the charged states are,
\begin{eqnarray}
G^+ &\simeq& \cos\beta \phi_1^+ + \sin\beta \phi_2^+ \\
H_1^{+\prime} &\simeq& - \chi^+ \\
H_2^{+\prime} &=& -\sin\beta \phi_1^+ + \cos\beta \phi_2^+.
\end{eqnarray}
Mixing $H_1^{+\prime}$ and $H_2^{+\prime}$ through the mixing angle
$\delta$, the physical mass eigenstates are,
\begin{eqnarray}
H_1^+ &\simeq& \sin\delta (-\sin\beta \phi_1^+ + \cos\beta \phi_2^+)
	- \cos\delta \chi^+ \\
H_2^+ &\simeq& \cos\delta (-\sin\beta \phi_1^+ + \cos\beta \phi_2^+)
	+ \sin\delta \chi^+.
\end{eqnarray}

In the Type II model, the charged Higgs couplings to $\bar{t}b$ 
simplify to,
\begin{eqnarray}
g^L_{H_1^+\bar{t}b} &=& \frac{gm_t}{\sqrt{2}M_W} 
	\cot\beta \sin\delta  \\
g^R_{H_1^+\bar{t}b} &=& +\frac{gm_b}{\sqrt{2}M_W}
	\tan\beta \sin\delta  \\
g^L_{H_2^+\bar{t}b} &=& \frac{gm_t}{\sqrt{2}M_W}
	\cot\beta \cos\delta  \\
g^R_{H_2^+\bar{t}b} &=& +\frac{gm_b}{\sqrt{2}M_W}
	\tan\beta \cos\delta.
\end{eqnarray}
In the Type I model, $\cot\beta$ is replaced by $\tan\beta$ in
the left--handed couplings.

The charged Higgs couplings to $Z$ simplify to,
\begin{eqnarray}
g_{ZG^+(G^+)^*} &=& -\frac{e}{s_Wc_W} \left[ \frac{1}{2} - s^2_W
	\right]  \\
g_{ZG^+(H_1^+)^*} &=& 0 \\
g_{ZG^+(H_2^+)^*} &=& 0  \\
g_{ZH_1^+(H_1^+)^*} &=& -\frac{e}{s_Wc_W} \left[ \frac{1}{2} - s^2_W
	- \frac{1}{2} \cos^2\delta \right]  \\
g_{ZH_2^+(H_2^+)^*} &=& -\frac{e}{s_Wc_W} \left[ \frac{1}{2} - s^2_W
	- \frac{1}{2} \sin^2\delta \right]  \\
g_{ZH_1^+(H_2^+)^*} &=& -\frac{e}{s_Wc_W} \frac{1}{2} 
	\sin\delta \cos\delta.
\end{eqnarray}

\subsubsection{Neutral Higgs bosons in the $Y=2$ model}

In the $Y=2$ model there are three CP--even neutral degrees of
freedom and three CP--odd neutral degrees of freedom.  These 
mix to form three CP--even states, two CP--odd states, and
the CP--odd neutral Goldstone boson.

The neutral Goldstone boson is,
\begin{equation}
G^0 = \sin\theta_2 (\cos\beta \phi_1^{0,i} + \sin\beta \phi_2^{0,i}) 
	+ \cos\theta_2 \chi^{0,i},
\label{eqn:2d1tG0}
\end{equation}
and there are two orthogonal states which 
are the physical CP--odd neutral Higgs bosons.
We define two orthogonal states,
\begin{eqnarray}
A_1^{0\prime} &=& \cos\theta_2 (\cos\beta \phi_1^{0,i} 
	+ \sin\beta \phi_2^{0,i}) - \sin\theta_2 \chi^{0,i}  \\
A_2^{0\prime} &=& -\sin\beta \phi_1^{0,i} + \cos\beta \phi_2^{0,i}.
\end{eqnarray}
Note that $A_2^{0\prime}$ corresponds to $A^0$ in the 2HDM.
These states will mix by an angle $\omega$ to form the mass eigenstates.
The mixing angle $\omega$ depends on the details of the Higgs potential.
The mass eigenstates then take the complicated form,
\begin{eqnarray}
A_1^0 &=& (\cos\omega \cos\theta_2 \cos\beta - \sin\omega \sin\beta)
	\phi_1^{0,i} 
	\nonumber \\
	& & + (\cos\omega \cos\theta_2 \sin\beta + \sin\omega \cos\beta)
	\phi_2^{0,i} - \cos\omega \sin\theta_2 \chi^{0,i}  
	\label{eqn:2d1tA1}\\
A_2^0 &=& (-\sin\omega \cos\theta_2 \cos\beta - \cos\omega \sin\beta)
	\phi_1^{0,i}
	\nonumber \\
	& & + (-\sin\omega \cos\theta_2 \sin\beta + \cos\omega \cos\beta)
	\phi_2^{0,i} + \sin\omega \sin\theta_2 \chi^{0,i}.
	\label{eqn:2d1tA2}
\end{eqnarray}

Similarly, we can define three orthogonal CP--even Higgs states.
We allow the CP--even components of the two doublets to mix by 
an angle $\alpha$, in analogy with the 2HDM.  This gives two 
orthogonal states,
\begin{eqnarray}
H^{0\prime} &=& \cos\alpha \phi_1^{0,r} + \sin\alpha \phi_2^{0,r}  \\
h^{0\prime} &=& -\sin\alpha \phi_1^{0,r} + \cos\alpha \phi_2^{0,r}.
\end{eqnarray}
In general, both $H^{0\prime}$ and $h^{0\prime}$ mix with 
$\chi^{0,r}$ by angles determined by the details of the Higgs
potential.  However, for simplicity we will consider the case 
in which only $H^{0\prime}$ mixes with the triplet.  We will 
parameterize this doublet--triplet mixing with the mixing angle
$\gamma$.  The resulting mass eigenstates are,
\begin{eqnarray}
H_1^0 &=& \cos\gamma (\cos\alpha \phi_1^{0,r} 
	+ \sin\alpha \phi_2^{0,r}) + \sin\gamma \chi^{0,r}  
	\label{eqn:2d1tH10}\\
H_2^0 &=& -\sin\gamma (\cos\alpha \phi_1^{0,r} 
	+ \sin\alpha \phi_2^{0,r}) + \cos\gamma \chi^{0,r}  
	\label{eqn:2d1tH20} \\
H_3^0 &=& -\sin\alpha \phi_1^{0,r} + \cos\alpha \phi_2^{0,r}.
	\label{eqn:2d1tH30}
\end{eqnarray}

The neutral Higgs couplings to quarks are,
\begin{eqnarray}
g^V_{H_1^0b\bar{b}} &=& -\frac{1}{\sqrt{2}}
	\left( \frac{gm_b}{\sqrt{2}M_W} \right)
	\frac{\cos\gamma \cos\alpha}{\cos\beta \sin\theta_2}  \\
g^V_{H_2^0b\bar{b}} &=& -\frac{1}{\sqrt{2}}
	\left( \frac{gm_b}{\sqrt{2}M_W} \right)
	\frac{-\sin\gamma \cos\alpha}{\cos\beta \sin\theta_2}  \\
g^V_{H_3^0b\bar{b}} &=& -\frac{1}{\sqrt{2}}
	\left( \frac{gm_b}{\sqrt{2}M_W} \right)
	\frac{-\sin\alpha}{\cos\beta \sin\theta_2}  \\
g^A_{G^0b\bar{b}} &=& -\frac{i}{\sqrt{2}}
	\left( \frac{gm_b}{\sqrt{2}M_W} \right)  \\
g^A_{A_1^0b\bar{b}} &=& -\frac{i}{\sqrt{2}}
	\left( \frac{gm_b}{\sqrt{2}M_W} \right)
	\frac{(\cos\omega \cos\theta_2 \cos\beta - \sin\omega \sin\beta)}
	{\cos\beta \sin\theta_2}  \\
g^A_{A_2^0b\bar{b}} &=& -\frac{i}{\sqrt{2}}
	\left( \frac{gm_b}{\sqrt{2}M_W} \right)
	\frac{(-\sin\omega \cos\theta_2 \cos\beta - \cos\omega \sin\beta)}
	{\cos\beta \sin\theta_2}.
\end{eqnarray} 

The neutral Higgs couplings to $Z$ are,
\begin{eqnarray}
g_{ZH_1^0G^0} &=& \frac{ie}{s_Wc_W} \left[
	-\frac{1}{2} 
		\cos\gamma \sin\theta_2
		\cos(\beta - \alpha)
	- \sin\gamma \cos\theta_2 \right]  \\
g_{ZH_1^0A_1^0} &=& \frac{ie}{s_Wc_W} \left[
	-\frac{1}{2} 
		(\cos\gamma \cos\omega \cos\theta_2 \cos(\beta - \alpha)
		- \cos\gamma \sin\omega \sin(\beta - \alpha))
	\right. \nonumber \\
	& & \left.
	+ \sin\gamma \cos\omega \sin\theta_2 \right]  \\
g_{ZH_1^0A_2^0} &=& \frac{ie}{s_Wc_W} \left[
	-\frac{1}{2} 
		(-\cos\gamma \sin\omega \cos\theta_2 \cos(\beta - \alpha)
		- \cos\gamma \cos\omega \sin(\beta - \alpha))
	\right. \nonumber \\
	& & \left.
	- \sin\gamma \sin\omega \sin\theta_2 \right]  \\
g_{ZH_2^0G^0} &=& \frac{ie}{s_Wc_W} \left[
	\frac{1}{2} \sin\gamma \sin\theta_2 \cos(\beta - \alpha)
	- \cos\gamma \cos\theta_2 \right]  \\
g_{ZH_2^0A_1^0} &=& \frac{ie}{s_Wc_W} \left[
	-\frac{1}{2} 
		(-\sin\gamma \cos\omega \cos\theta_2 \cos(\beta - \alpha)
		+ \sin\gamma \sin\omega \sin(\beta - \alpha))
	\right. \nonumber \\
	& & \left.
	+ \cos\gamma \cos\omega \sin\theta_2 \right]  \\
g_{ZH_2^0A_2^0} &=& \frac{ie}{s_Wc_W} \left[
	-\frac{1}{2} 
		(\sin\gamma \sin\omega \cos\theta_2 \cos(\beta - \alpha)
		+ \sin\gamma \cos\omega \sin(\beta - \alpha))
	\right. \nonumber \\
	& & \left.
	- \cos\gamma \sin\omega \sin\theta_2 \right]  \\
g_{ZH_3^0G^0} &=& \frac{ie}{s_Wc_W} \left[
	-\frac{1}{2} \sin\theta_2 \sin(\beta - \alpha) \right]  \\
g_{ZH_3^0A_1^0} &=& \frac{ie}{s_Wc_W} \left[
	-\frac{1}{2} 
		(\cos\omega \cos\theta_2 \sin(\beta - \alpha)
		+ \sin\omega \cos(\beta - \alpha)) \right]  \\
g_{ZH_3^0A_2^0} &=& \frac{ie}{s_Wc_W} \left[
	-\frac{1}{2}
		(-\sin\omega \cos\theta_2 \sin(\beta - \alpha)
		+ \cos\omega \cos(\beta - \alpha)) \right].
\end{eqnarray}

The states and couplings can be simplified a great deal if we take the 
limit
of large $\tan\theta_2$, as required by the measured value of the
$\rho$ parameter.  We make the approximation, $\cos\theta_2 \approx 0$
and $\sin\theta_2 \approx 1$.  Then the CP--odd neutral states 
simplify to,
\begin{eqnarray}
G^0 &\simeq& \cos\beta \phi_1^{0,i} + \sin\beta \phi_2^{0,i}  \\
A_1^0 &\simeq& - \sin\omega \sin\beta \phi_1^{0,i} 
	 + \sin\omega \cos\beta
	\phi_2^{0,i} - \cos\omega \chi^{0,i}  \\
A_2^0 &\simeq& - \cos\omega \sin\beta
	\phi_1^{0,i}
	+ \cos\omega \cos\beta
	\phi_2^{0,i} + \sin\omega \chi^{0,i}.
\end{eqnarray}
The couplings of $A_1^0$ and $A_2^0$ to quarks simplify to,
\begin{eqnarray}
g^A_{A_1^0b\bar{b}} &\simeq& +\frac{i}{\sqrt{2}}
	\left( \frac{gm_b}{\sqrt{2}M_W} \right)
	\tan\beta \sin\omega  \\
g^A_{A_2^0b\bar{b}} &\simeq& +\frac{i}{\sqrt{2}}
	\left( \frac{gm_b}{\sqrt{2}M_W} \right)
	\tan\beta \cos\omega.
\end{eqnarray} 
Likewise, the neutral Higgs couplings to $Z$ simplify to,
\begin{eqnarray}
g_{ZH_1^0G^0} &\simeq& \frac{ie}{s_Wc_W} \left[
	-\frac{1}{2} 
		\cos\gamma
		\cos(\beta - \alpha)
	\right]  \\
g_{ZH_1^0A_1^0} &\simeq& \frac{ie}{s_Wc_W} \left[
	\frac{1}{2} 
		 \cos\gamma \sin\omega \sin(\beta - \alpha)
	+ \sin\gamma \cos\omega \right]  \\
g_{ZH_1^0A_2^0} &\simeq& \frac{ie}{s_Wc_W} \left[
	\frac{1}{2} 
		 \cos\gamma \cos\omega \sin(\beta - \alpha)
	- \sin\gamma \sin\omega \right]  \\
g_{ZH_2^0G^0} &\simeq& \frac{ie}{s_Wc_W} \left[
	\frac{1}{2} \sin\gamma \cos(\beta - \alpha)
	\right]  \\
g_{ZH_2^0A_1^0} &\simeq& \frac{ie}{s_Wc_W} \left[
	-\frac{1}{2} 
		\sin\gamma \sin\omega \sin(\beta - \alpha)
	+ \cos\gamma \cos\omega \right]  \\
g_{ZH_2^0A_2^0} &\simeq& \frac{ie}{s_Wc_W} \left[
	-\frac{1}{2} 
		 \sin\gamma \cos\omega \sin(\beta - \alpha)
	- \cos\gamma \sin\omega \right]  \\
g_{ZH_3^0G^0} &\simeq& \frac{ie}{s_Wc_W} \left[
	-\frac{1}{2} \sin(\beta - \alpha) \right]  \\
g_{ZH_3^0A_1^0} &\simeq& \frac{ie}{s_Wc_W} \left[
	-\frac{1}{2} 
		 \sin\omega \cos(\beta - \alpha) \right]  \\
g_{ZH_3^0A_2^0} &\simeq& \frac{ie}{s_Wc_W} \left[
	-\frac{1}{2}
		 \cos\omega \cos(\beta - \alpha)  \right].
\end{eqnarray}


\chapter{Custodial SU(2) symmetry and the Georgi--Machacek 
class of models}
\label{app:SU2c}

The Standard Model (SM) makes a number of successful predictions
which arise naturally from the simple form of the SM; for example,
the absence of tree--level flavor--changing neutral currents and
the equality of the $\rho$ parameter to one at tree--level.  
Many of these 
built--in successes are lost when the SM is extended, often
requiring parameters of the extended models to be fine--tuned 
in order to agree with experimental results.

In particular, extended Higgs sectors which contain isospin multiplets
larger than doublets will in general yield a value of the 
$\rho$ parameter different from one.  The $\rho$ parameter 
parameterizes the relation between the $W^{\pm}$ and $Z$ boson
masses and the electroweak mixing angle, as will be further 
described in section \ref{sec:GMmotivation}.  In such a model,
certain parameters must be fine--tuned in
order to agree with the measured value of $\rho$.

Fortunately, there is a class of extended Higgs sectors that 
automatically yield $\rho = 1$ at tree level, without fine--tuning
of parameters.  The Higgs doublet of the SM is a member of 
this class.  These models preserve a symmetry, known as
``custodial'' SU(2) symmetry, which ensures that $\rho = 1$ at 
tree level.  The first extended model of this type was introduced
by Georgi and Machacek \cite{Georgi1}, and contains isospin 
triplets in addition to the standard Higgs doublet.  We will
refer to this model as the Georgi--Machacek model.  We have 
generalized the Georgi--Machacek model to include multiplets
of arbitrary isospin, while still preserving the custodial 
symmetry that leads to $\rho = 1$ \cite{GHL99}.

In section \ref{sec:GMnotation} we introduce the ${\rm SU}(2)_L \times 
{\rm SU}(2)_R$ transformations and describe how the custodial symmetry
is preserved after electroweak symmetry breaking.  We also describe
the Higgs doublet of the SM in this framework.
In section \ref{sec:GMmotivation} we explain how models that
preserve the custodial symmetry automatically lead to $\rho = 1$
at tree level.  This is the motivation for 
considering models that preserve the custodial symmetry.
In section \ref{sec:GMtripletmodel} we describe the
Georgi--Machacek model with 
Higgs triplets and custodial symmetry, first described in reference
\cite{Georgi1}.  In section \ref{sec:GMgeneralized} we 
extend the Georgi--Machacek model to include Higgs multiplets of 
arbitrary size while still preserving the custodial symmetry and 
$\rho = 1$ at tree level.  For the generalized models we
 derive certain Higgs couplings to 
fermions in section \ref{sec:GMgeneral_ffcouplings}, and 
to gauge bosons in section \ref{sec:GMgaugecouplings}.
Finally in section \ref{sec:GMnoAxion} we show that the 
generalized Georgi--Machacek model is not invariant under any
accidental global U(1) symmetries, and thus does not contain
a massless axion, as some extended Higgs sectors do.

\section{Symmetries, notation and conventions}
\label{sec:GMnotation}

In the SM, the standard Higgs doublet can be 
written as a matrix,
\begin{equation}
\Phi = \left( \begin{array}{cc}
	\phi^{0*} & \phi^+ \\
	-\phi^{+*} & \phi^0
	\end{array} \right)
\end{equation}
which transforms under ${\rm SU}(2)_L \times {\rm SU}(2)_R$ as 
$\Phi \rightarrow U_L^{\dagger} \Phi U_R$.
(The neutral states decompose into CP--even and CP--odd parts
as $\phi^0 = (\phi^{0,r} + i\phi^{0,i})/\sqrt{2}$.)
The transformation
matrices are
\begin{equation}
U_{L,R} = \mathrm{exp}[-i \theta_{L,R} \hat{n}_{L,R} 
	\cdot \vec{T}_{L,R}]
\end{equation}
where $\vec{T} = (T^1, T^2, T^3)$ are the generators of SU(2).
$\Phi$ transforms as a $T_L=1/2$ representation of ${\rm SU}(2)_L$ and a 
$T_R = 1/2$ representation of 
${\rm SU}(2)_R$.  We denote this by introducing the notation 
$(T_L,T_R)$ for a multiplet tranforming under 
${\rm SU}(2)_L \times {\rm SU}(2)_R$;
in this notation, $\Phi = (1/2,1/2)$.

The ${\rm SU}(2)_L \times {\rm SU}(2)_R$ doublet $\Phi$ does not transform
as a multiplet of
the usual electroweak gauge symmetries 
${\rm SU}(2)_L \times {\rm U}(1)_Y$, since the two columns of $\Phi$ have 
different values of hypercharge.  In fact, hypercharge corresponds
to the third component of ${\rm SU}(2)_R$, $Y = -2T^3_R$.  Because of 
this, a Higgs potential for $\Phi$ which is invariant under the 
full ${\rm SU}(2)_L \times {\rm SU}(2)_R$ is also invariant under 
${\rm SU}(2)_L \times {\rm U}(1)_Y$.  The SM Higgs potential can be written 
in a ${\rm SU}(2)_L \times {\rm SU}(2)_R$ symmetric form without loss of 
generality.

However, because the hypercharge operator corresponds to the
$T_3$ of ${\rm SU}(2)_R$, radiative corrections involving 
the interaction of the ${\rm U}(1)_Y$ gauge boson $B_{\mu}$ with the
Higgs sector break the ${\rm SU}(2)_c$ symmetry.  This violation of
${\rm SU}(2)_c$ leads to one--loop corrections to $\rho$ which are
quadratically divergent.  This problem was first noted in
reference \cite{Chanowitz1}, and further elaborated in reference
\cite{Gunion2}.  As pointed out in \cite{Gunion2}, fine--tuning
is required to keep $\rho$ near one in a grand--unification 
scheme in which the quadratic divergence is cut off by the 
grand--unification scale.  However, this fine--tuning is no worse
than the fine--tuning required in the SM to control the quadratically
divergent one--loop corrections to the SM Higgs mass.  In what 
follows we will ignore the breaking of ${\rm SU}(2)_c$ and the 
fine--tuning problem; these issues are addressed at length 
in reference \cite{Gunion2}.

Under the standard electroweak symmetry breaking, $\phi^0$ acquires
a vacuum expectation value (vev) 
$\langle \phi^0 \rangle = v_{SM}/\sqrt{2}$, breaking 
${\rm SU}(2)_L \times {\rm U}(1)_Y$ down to ${\rm U}(1)_{EM}$.  
In the matrix notation
the vev of $\Phi$ is 
\begin{equation}
\langle \Phi \rangle = \frac{v_{SM}}{\sqrt{2}} \left(
	\begin{array}{cc}
	1 & 0  \\
	0 & 1
	\end{array} \right).
\end{equation}
${\rm SU}(2)_L \times {\rm SU}(2)_R$ is broken down to a ``custodial'' symmetry
${\rm SU}(2)_c$, under which $\Phi$ transforms as 
$\Phi \rightarrow U_c^{\dagger} \Phi U_c$, with the transformation
matrices
\begin{equation}
U_c = \mathrm{exp}[-i \theta_c \hat{n} \cdot \vec{T}].
\end{equation}
This is the same as an ${\rm SU}(2)_L \times {\rm SU}(2)_R$ transformation 
with $\theta_L \hat{n}_L = \theta_R \hat{n}_R$.
After electroweak symmetry breaking, $\Phi$ decomposes into a 
singlet and a triplet under ${\rm SU}(2)_c$.  The ${\rm SU}(2)_c$ singlet is 
\begin{equation}
H_1^0 = \phi^{0,r}
\end{equation}
and the ${\rm SU}(2)_c$ triplet is 
\begin{equation}
\Phi_3 = ( \Phi_3^+, \Phi_3^0, \Phi_3^- ) = (\phi^+, \phi^{0,i}, -\phi^{+*}).
\end{equation}
In the SM, the ${\rm SU}(2)_c$ singlet is the physical Higgs boson and the 
states of the ${\rm SU}(2)_c$ triplet are the Goldstone bosons.  
Electric charge corresponds to the third component of ${\rm SU}(2)_c$,
$Q = T^3_c$.

Let us now extend this notation to an arbitrary multiplet $X$ that 
transforms as $(T,T)$ under ${\rm SU}(2)_L \times {\rm SU}(2)_R$.  $X$ can be
written as a $(2T+1)$ by $(2T+1)$ matrix.  We will denote the vev
of $X$ by
\begin{equation}
\langle X \rangle = v_X \mathbf{I},
\end{equation}
where $\mathbf{I}$ is the identity matrix.  In this notation, the vev
of the SM doublet is $v_X = \sqrt{2} v_{SM}$.

When ${\rm SU}(2)_L \times {\rm SU}(2)_R$ is broken down to 
${\rm SU}(2)_c$, $X$ 
decomposes as
\begin{equation}
(T,T) \rightarrow 2T \oplus 2T-1 \oplus \cdots \oplus 1 \oplus 0.
\end{equation}
In this notation, the SM doublet decomposes as 
$(1/2,1/2) \rightarrow 1 \oplus 0$, as described before.

Note that when the Higgs potential is invariant under ${\rm SU}(2)_c$,
all the states which transform in a single representation of 
${\rm SU}(2)_c$ must be degenerate in mass.

\section{Motivation: the $\rho$ parameter under custodial SU(2) symmetry}
\label{sec:GMmotivation}

Perhaps the most important clue we have to the form of the Higgs
sector is the relation between the $W^{\pm}$ and $Z$ masses and the
electroweak mixing angle.  This relation is parameterized by the 
$\rho$ parameter,
\begin{equation}
\rho \equiv \frac{M_W^2}{M_Z^2 c^2_W}
\end{equation}
where $M_W$ and $M_Z$ are the $W$ and $Z$ masses and 
$c^2_W = \cos^2\theta_W$ is the cosine of the electroweak mixing angle.
Experimentally, $\rho = 1$ to within one percent (\cite{Altarelli98}, 
in which $\epsilon_1 = \rho - 1$).  This follows if the electroweak 
symmetry breaking gives equal masses to $W^{\pm}$ and $W^3$.  
This is ensured in a model in which the symmetry breaking preserves
${\rm SU}(2)_c$, giving $\langle X \rangle = v_X \mathbf{I}$, 
as we will now prove.

For simplicity, we will consider only one Higgs multiplet, $X = (T,T)$
under ${\rm SU}(2)_L \times {\rm SU}(2)_R$.  Once we show that $X$
gives equal masses to $W^{\pm}$ and $W^3$, the result is easily extended
to a model consisting of an arbitrary set of multiplets $X$ by noting
that the resulting gauge boson mass--squared matrix is the sum of the
mass--squared matrices coming from each $X$.

The electroweak gauge boson mass terms come from the covariant 
derivatives in the Lagrangian with the vev of $X$ inserted.  The 
covariant derivatives of $X$ in the Lagrangian are
\begin{equation}
\mathcal{L} = \frac{1}{2} \mathrm{Tr} [(\mathcal{D}_{\mu} X)^{\dagger}
	(\mathcal{D}^{\mu} X)] 
\end{equation}
where the covariant derivative is 
\begin{equation}
\mathcal{D}_{\mu} = \partial_{\mu} - igW_{\mu}^a T^a 
	- ig^{\prime}\frac{Y}{2} B_{\mu}.
\end{equation}
$T^a$ are the generators of ${\rm SU}(2)_L$ and $Y$ is the hypercharge.
The hypercharge is related to the third component of the ${\rm SU}(2)_R$
by,
\begin{equation}
\frac{Y}{2} X = - X T^3.
\end{equation}
The ${\rm SU}(2)_c$--symmetric vev of $X$ is 
$\langle X \rangle = v_X \mathbf{I}$.
Putting all this together, the relevant part of the Lagrangian becomes
\begin{eqnarray}
\mathcal{L} &=& \frac{1}{2} \mathrm{Tr} 
	\left[ 
	\left( igW_{\mu}^a v_X T^a - ig^{\prime}B_{\mu} T^3 v_X \right)
	\left( -igW^{b\mu} T^b v_X + ig^{\prime}B^{\mu} v_X T^3 \right)
	\right]  \\
	&=& \frac{g^2}{2} v_X^2 W_{\mu}^a W^{b\mu} \mathrm{Tr}[T^aT^b]
	- 2 \frac{gg^{\prime}}{2} v_X^2 W_{\mu}^a B^{\mu} \mathrm{Tr}[T^aT^3]
	\nonumber \\
	& & + \frac{g^{\prime 2}}{2} v_X^2 B_{\mu}B^{\mu} \mathrm{Tr}[T^3T^3].
\end{eqnarray}
The traces over the SU(2) generators of a representation $r$ with
isospin $T$ are given by
$\mathrm{Tr}[T^a_r T^b_r] = C(r) \delta^{ab}$, where 
$C(r) = T(T+1)(2T+1)/3$ is a constant for each representation $r$.
We find,
\begin{equation}
\mathcal{L} = \frac{T(T+1)(2T+1)}{3} \frac{v_X^2}{2}
	[g^2 W_{\mu}^a W^{a\mu} - 2gg^{\prime}W_{\mu}^3 B^{\mu}
	+ g^{\prime 2} B_{\mu} B^{\mu}].
\end{equation}
The resulting mass--squared matrix for the gauge bosons is,
\begin{equation}
M^2 = \frac{T(T+1)(2T+1)}{3} v_X^2
	\left(  \begin{array}{cccc}  
	g^2 & 0 & 0 & 0  \\
	0 & g^2 & 0 & 0  \\
	0 & 0 & g^2 & -g g^{\prime}  \\
	0 & 0 & - g g^{\prime} & g^{\prime 2}
	\end{array} \right)
\end{equation}
where the rows and columns correspond to $W^1_{\mu}$, $W^2_{\mu}$, 
$W^3_{\mu}$, and $B_{\mu}$.  Note that equal masses are given to 
all three $W^a$.

The gauge boson masses are as follows.  $W^1_{\mu}$ and $W^2_{\mu}$ 
mix to form the charge eigenstates 
\begin{equation}
W_{\mu}^{\pm} = \frac{1}{\sqrt{2}} (W^1_{\mu} \mp W^2_{\mu}),
\end{equation}
with mass $M_W^2 = g^2 v_X^2 T(T+1)(2T+1)/3$.  $W^3_{\mu}$ and $B_{\mu}$
mix to form the $Z_{\mu}$ and the (massless) photon $A_{\mu}$, 
\begin{eqnarray}
Z_{\mu} &=& \frac{1}{\sqrt{g^2 + g^{\prime 2}}} 
	(g W^3_{\mu} - g^{\prime} B_{\mu})  \\
	&=& c_W W^3_{\mu} - s_W B_{\mu}  \\
A_{\mu} &=& \frac{1}{\sqrt{g^2 + g^{\prime 2}}}
	(g^{\prime} W^3_{\mu} + g B_{\mu}) \\
	&=& s_W W^3_{\mu} + c_W B_{\mu}
\end{eqnarray}
where $c_W= \cos\theta_W$, $s_W= \sin\theta_W$, and $c_W$ is,
\begin{equation}
c_W = \frac{g}{\sqrt{g^2 + g^{\prime 2}}}.
\end{equation}
Finally the $Z$ mass is $M_Z^2 = (g^2 + g^{\prime 2}) v_X^2 T(T+1)(2T+1)/3$.

Combining $M_W$, $M_Z$, and $c_W$, we find that the $\rho$ parameter is
\begin{equation}
\rho = \frac{M_W^2}{M_Z^2 c^2_W} = 1
\end{equation}
in this model, for any values of $v_X$ and $T$.  This result remains true
for a Higgs sector consisting of any combination of multiplets $X$ 
as long as $\langle X \rangle$ is invariant under ${\rm SU}(2)_c$ for each
$X$.

In the SM, $T=1/2$ giving $C(r) = 1/2$.  The resulting $W$ and $Z$ 
masses are, using $v_{SM} = \sqrt{2} v_X = 246$ GeV,
\begin{eqnarray}
M_W^2 &=& \frac{g^2}{4} v_{SM}^2  \\
M_Z^2 &=& \frac{g^2 + g^{\prime 2}}{4} v_{SM}^2.
\end{eqnarray}

\section{The Georgi--Machacek model with Higgs triplets}
\label{sec:GMtripletmodel}

In this section we review in detail a model with custodial SU(2) 
symmetry and Higgs triplets.
This model was first constructed by Georgi and Machacek \cite{Georgi1}.
It was considered in greater depth by Chanowitz and Golden 
\cite{Chanowitz1}, who showed that a Higgs potential for the model could be
constructed that was invariant under the full 
${\rm SU}(2)_L \times {\rm SU}(2)_R$.  
This ensured that radiative corrections from Higgs self--interactions 
preserved ${\rm SU}(2)_c$.  A more detailed study of the phenomenology of the
model \cite{Gunion1} and naturalness problems from one--loop effects 
\cite{Gunion2} was made by Gunion, Vega, and Wudka.  This model is also
reviewed in \cite{HHG}.

\subsubsection{Notation and conventions}

This model contains a complex
$Y=1$ doublet $\Phi = (\phi^+, \phi^0)$, a real $Y=0$ triplet 
$\xi = (\xi^+, \xi^0, \xi^-)$ (with $\xi^{0*} = \xi^0$ and 
$\xi^- = -\xi^{+*}$), and a complex 
$Y=2$ triplet $\chi = (\chi^{++}, \chi^+, \chi^0)$.
In the ${\rm SU}(2)_L \times {\rm SU}(2)_R$ 
notation, the Higgs fields take the form
\begin{equation}
\Phi = \left( \begin{array}{cc}
		\phi^{0*} & \phi^{+} \\
		-\phi^{+*}  & \phi^{0}
	      \end{array}	\right)
\end{equation}
\begin{equation}
\chi = \left(  \begin{array}{ccc}
		\chi^{0*}  & \xi^{+} & \chi^{++} \\
		-\chi^{+*}  & \xi^{0} & \chi^{+}  \\
		\chi^{++*} & \xi^{-} & \chi^{0}
	       \end{array}	\right)
\end{equation}
which transform respectively as a $(1/2,1/2)$ and $(1,1)$ of 
${\rm SU}(2)_{L} \times {\rm SU}(2)_{R}$.  
This definition differs slightly from
that of references \cite{Georgi1,Gunion1,Gunion2,HHG}, which use
$\chi = (\chi^{++}, \chi^+, \chi^{0*})$, so that $\chi^{0*}$ appears
in place of $\chi^0$.  Otherwise the phase conventions here are the same.
(References \cite{Georgi1,Gunion1,Gunion2,HHG} define the negative--charged
states $\phi^- = -(\phi^+)^*$, $\chi^{--} = (\chi^{++})^*$, 
and $\chi^- = -(\chi^+)^*$.  We avoid these definitions in order to 
avoid confusion when the Georgi--Machacek model is extended to larger
representations of ${\rm SU}(2)_L \times {\rm SU}(2)_R$.)

The electroweak symmetry breaking preserves
tree level custodial ${\rm SU}(2)_{c}$ when the vevs of the fields are 
diagonal, 
\begin{equation}
\langle \Phi \rangle = \left( \begin{array}{cc}
		v_{\phi}/\sqrt{2} & 0 \\
		0  & v_{\phi}/\sqrt{2}
	      \end{array}	\right)
\end{equation}
\begin{equation}
\langle \chi \rangle = \left(  \begin{array}{ccc}
		v_{\chi}  & 0 & 0 \\
		0  & v_{\chi} & 0  \\
		0 & 0 & v_{\chi}
	       \end{array}	\right)
\end{equation}
For the triplets, this means 
$\langle \chi^0 \rangle = \langle \xi^0 \rangle$.
(In the notation of \cite{Georgi1,Gunion1,Gunion2,HHG} the vevs are
$v_{\phi} = a$, $v_{\chi} = b$.)

The $W^{\pm}$ and $Z$ masses fix a combination of the vevs.  At 
tree level,
\begin{eqnarray}
M_W^2 &=& M_Z^2 \cos^2\theta_W = \frac{g^2}{4} v^2_{SM} \\
	&=& \frac{g^2}{4} (v_{\phi}^2 + 8 v_{\chi}^2).
\end{eqnarray}

As in \cite{Georgi1,Gunion1,Gunion2,HHG} we define the doublet--triplet
mixing angle $\theta_H$ by
\begin{equation}
\tan\theta_H = \frac{2\sqrt{2} v_{\chi}}{v_{\phi}},
\end{equation}
so that
\begin{eqnarray}
c_H &\equiv& \cos\theta_H = \frac{v_{\phi}}
	{\sqrt{v_{\phi}^2 + 8 v_{\chi}^2}}  \\
s_H &\equiv& \sin\theta_H = \sqrt{\frac{8v_{\chi}}
	{v_{\phi}^2 + 8 v_{\chi}^2}}.
\end{eqnarray}
We also define the charged states of $\chi$ transforming in the 
triplet and fiveplet representations of ${\rm SU}(2)_c$,
\begin{eqnarray}
\chi_3^+ &=& \frac{\chi^+ + \xi^+}{\sqrt{2}}  \\
\chi_5^+ &=& \frac{\chi^+ - \xi^+}{\sqrt{2}},
\end{eqnarray}
which obey the phase convention $\chi_3^- = -(\chi_3^+)^*$,
$\chi_5^- = -(\chi_5^+)^*$.
We use a slightly different notation
from references \cite{Georgi1,Gunion1,Gunion2,HHG}.  
This notation will make the 
analysis of the generalized Georgi--Machacek model in section 
\ref{sec:GMgeneralized} more straightforward.
In the notation of \cite{Georgi1,Gunion1,Gunion2,HHG},
$\chi_3^{\pm} = \psi^{\pm}$ and $\chi_5^{\pm} = \zeta^{\pm}$.

The $W$ and $Z$ are given mass by absorbing the Goldstone bosons,
\begin{eqnarray}
G_3^0 &=& c_H \phi^{0,i} + s_H \chi^{0,i}  \\
G_3^+ &=& c_H \phi^+ + s_H \chi_3^+
\end{eqnarray}
which together form an ${\rm SU}(2)_c$ triplet, with $G_3^- = -(G_3^+)^*$.
The remaining physical states are classified according to their 
transformation properties under ${\rm SU}(2)_c$.  The model contains a 
five-plet $H_5^{++,+,0,-,--}$, a triplet $H_3^{+,0,-}$, and two
singlets, $H_1^0$ and $H_1^{0\prime}$.  In terms of the fields of
$\Phi$ and $\chi$, these states are,
\begin{eqnarray}
H_5^{++} &=& \chi^{++}  \\
H_5^+ &=& \chi_5^+  \\
H_5^0 &=& \frac{\sqrt{2}\xi^0 - \chi^{0,r}}{\sqrt{3}}  \\
H_3^+ &=& - s_H \phi^+ + c_H \chi_3^+  \\
H_3^0 &=& - s_H \phi^{0,i} + c_H \chi^{0,i}  \\
H_1^0 &=& \phi^{0,r}  \\
H_1^{0\prime} &=& \frac{\sqrt{2}\chi^{0,r} + \xi^0}{\sqrt{3}}.
\label{eqn:GM3physstates}
\end{eqnarray}
According to our phase conventions, the charged states obey
$H_5^{--} = (H_5^{++})^*$, $H_5^- = -(H_5^+)^*$, and $H_3^- = -(H_3^+)^*$.
$H_3^0$ is CP--odd while all the other neutral states are 
CP--even.

Not all of these states need be mass eigenstates; in general they 
will mix with each other in a way determined by the details of the
Higgs potential.  $H_5^{++}$ is always a mass eigenstate 
since there is no other doubly--charged Higgs state for it to mix 
with.
If the Higgs potential is CP--invariant, then $H_3^0$ will be a 
mass eigenstate, since
$H_3^0$ is CP--odd while all the other neutral Higgs 
states are CP--even.
Finally, if the Higgs potential is invariant under the full
${\rm SU}(2)_L \times {\rm SU}(2)_R$ 
symmetry, the mixing is further constrained.
After electroweak symmetry
breaking, such a  Higgs potential would remain invariant under 
${\rm SU}(2)_c$.
Then states which transform under different
representations of ${\rm SU}(2)_c$ cannot mix with each other.  Thus in this
case only $H_1^0$ and $H_1^{0\prime}$ can mix; all the other states 
in equation \ref{eqn:GM3physstates} are mass eigenstates.

Chanowitz and Golden \cite{Chanowitz1} showed that a Higgs potential 
could be constructed which was invariant under ${\rm SU}(2)_R \times 
{\rm SU}(2)_L$, and therefore preserved ${\rm SU}(2)_c$ 
after electroweak symmetry
breaking.  This potential has two benefits.  First, the 
${\rm SU}(2)_c$--preserving vevs of the Higgs multiplets are required
by the symmetry of the potential.  Second,
when radiative corrections are included
the ${\rm SU}(2)_c$ symmetry is preserved to all orders in the Higgs 
self couplings.  Then the radiative corrections to $\rho$ from Higgs
self--interactions are zero.
In addition, if the Higgs potential is invariant under ${\rm SU}(2)_c$,
then Higgs states which transform in a single representation of 
${\rm SU}(2)_c$ will be degenerate at tree level.  This degeneracy will
be preserved to all orders in the Higgs self couplings.  In particular,
$H_3^0$ and $H_3^{\pm}$ will have the same mass, denoted $M_{H_3}$, 
and $H_5^0$, $H_5^{\pm}$, and $H_5^{\pm\pm}$ will have the same mass,
denoted $M_{H_5}$.

In the following sections we list the couplings for the Higgs states given 
in equation \ref{eqn:GM3physstates}.

\subsubsection{Couplings to fermions}

The fermion couplings in this model have been studied in 
\cite{Gunion1,HHG}.  Only ${\rm SU}(2)_L$ doublets with $Y=\pm1$ can 
have ${\rm SU}(2)_L \times {\rm U}(1)_Y$ invariant couplings to fermions 
without violating lepton number.  We will only consider couplings
of this type.  In principle the $Y=2$ triplet Higgs field can 
couple to the lepton--lepton channel, violating lepton number, 
but we will assume that this does not happen.

Since only the doublet field couples to fermions, the fermion 
couplings of Higgs mass eigenstates will arise from the overlap of
the mass eigenstates with $\Phi$.  Under ${\rm SU}(2)_c$ $\Phi$ contains 
a singlet ($H_1^0$) and a triplet.  Therefore if the Higgs potential
preserves the ${\rm SU}(2)_c$ symmetry, only ${\rm SU}(2)_c$ 
singlets and triplets
can have an overlap with $\Phi$.  This is evident in the states of
equation \ref{eqn:GM3physstates}.  Only $H_3^{\pm}$, $H_3^0$, and 
$H_1^0$ will have nonzero fermion--antifermion couplings.
The couplings of the Higgs mass eigenstates to fermions in this model
are of the form
\begin{equation}
i (g^{L}_{H\bar{f}f^{\prime}} P_{L} + g^{R}_{H\bar{f}f^{\prime}} P_{R})
	= i (g^{V}_{H\bar{f}f^{\prime}} + g^{A}_{H\bar{f}f^{\prime}} 
		\gamma_{5})
\end{equation}
where $P_{R,L} = (1 \pm \gamma_5)/2$.  Using third--generation
notation,
\begin{eqnarray}
g^V_{H_1^0 f\bar{f}} &=& -\frac{g m_f}{2 M_W c_H}  \\
g^A_{H_3^0 t\bar{t}} &=& -\frac{i g m_t s_H}{2 M_W c_H}  \\
g^A_{H_3^0 b\bar{b}} &=& \frac{i g m_b s_H}{2 M_W c_H}  \\
g^L_{H_3^+ \bar{t}b} &=& \frac{- g m_t s_H}{\sqrt{2} M_W c_H}  \\
g^R_{H_3^+ \bar{t}b} &=& \frac{g m_b s_H}{\sqrt{2} M_W c_H}.
\end{eqnarray}
Analogous expressions hold for the couplings to leptons.
$H_1^0$ is CP--even so it has only vector couplings to fermions.
$G^0$ and $H_3^0$ are CP--odd so they have only axial--vector
couplings to fermions.
For completeness we also list the couplings of the Goldstone bosons 
to fermions,
\begin{eqnarray}
g^A_{G^0 t\bar{t}} &=& \frac{i g m_t}{2 M_W}  \\
g^A_{G^0 b\bar{b}} &=& -\frac{i g m_b}{2 M_W}  \\
g^L_{G^+ \bar{t}b} &=& \frac{g m_t}{\sqrt{2} M_W}  \\
g^R_{G^+ \bar{t}b} &=& \frac{- g m_b}{\sqrt{2} M_W}.
\end{eqnarray}

These couplings agree with those in reference \cite{HHG} 
except for the $H_3^0$
couplings, which differ by a minus sign.  This sign comes from the 
sign of $\langle H_3^0 | \phi^{0,i} \rangle = -s_H$, whereas in reference 
\cite{HHG}, $\langle H_3^0 | \phi^{0,i} \rangle = +s_H$.  This difference
is due ultimately to reference \cite{HHG} defining the neutral state 
of the $Y=2$ ${\rm SU}(2)_L$ triplet to be $\chi^{0*}$ instead of $\chi^0$.

\subsubsection{Couplings to gauge bosons}

In this section we list the Higgs couplings to vector bosons.
The Higgs--Vector--Vector (HVV) couplings in this model were first
studied in \cite{Georgi1}, while the Higgs--Higgs--Vector (HHV) couplings
were first studied in \cite{Gunion1}.
For completeness we list them here, using our phase conventions.

We list first the HVV couplings.  The $HV_1V_2$ vertex is 
$ig_{HV_1V_2}g^{\mu\nu}$, where $V_1$ and $V_2$ are any two vector 
bosons.  The couplings are,
\begin{eqnarray}
g_{H_5^{++}W^-W^-} &=& \sqrt{2} g M_W s_H  \\
g_{H_5^+ W^-Z} &=& -\frac{g M_W s_H}{c_W}  \\
g_{H_5^+ W^-\gamma} &=& 0 \\
g_{H_5^0 W^-W^+} &=& \frac{g M_W s_H}{\sqrt{3}}  \\
g_{H_5^0 ZZ} &=& -\frac{2 g M_W s_H}{\sqrt{3} c^2_W}  \\
g_{H_1^0 W^-W^+} &=& g M_W c_H  \\
g_{H_1^0 ZZ} &=& \frac{g M_W c_H}{c^2_W}  \\
g_{H_1^{0\prime} W^- W^+} &=& \frac{2\sqrt{2}}{\sqrt{3}} g M_W s_H  \\
g_{H_1^{0\prime} ZZ} &=& \frac{2\sqrt{2}}{\sqrt{3}} 
	\frac{g M_W s_H}{c^2_W}.
\end{eqnarray}
The $H_3^{\pm}$, $H_3^0$ states do not couple to vector boson pairs.

In the limit of $s_H \rightarrow 0$, $H_1^0$ plays the role of the SM
Higgs boson with SM couplings, while the HVV and Higgs--fermion
couplings of the rest of the states go to zero.  However we
reemphasize that in this model with ${\rm SU}(2)_c$, $s_H$ is not required
to be small.

There is a nonzero $H_5^+ W^- Z$ coupling in this model, which is 
absent in any model containing only Higgs doublets and singlets.
This coupling raises the possibility of charged Higgs production
via $Z \rightarrow H_5^+ W^-$ or $W^+ \rightarrow H_5^+ Z$, which
are unmistakeable indications of a Higgs sector containing 
multiplets larger than doublets.

Finally we note that the couplings of $H_3$ and $H_5$ are strikingly 
different.  $H_3$ couples to fermion--antifermion pairs and not to 
gauge boson pairs, while $H_5$ couples to gauge boson pairs and not 
to fermion--antifermion pairs.  Thus, disregarding $HV$ and $HH$ 
channels, the $H_3$ can only couple and decay to fermion--antifermion
pairs while the $H_5$ can only couple and decay to gauge boson pairs.

We now list the HHV couplings.  The $H_1H_2V$ vertex is 
$ig_{H_1H_2V}(p_1 - p_2)^{\mu}$, where $p_1$ ($p_2$) is the 
incoming momentum of $H_1$ ($H_2$).  The couplings are,
\begin{eqnarray}
g_{H_1^0 (H_3^+)^* W^+} &=& + \frac{g s_H}{2} \\
g_{H_1^0 (H_5^+)^* W^+} &=& 0  \\
g_{H_1^{0\prime} (H_3^+)^* W^+} &=& -\sqrt{\frac{2}{3}} g c_H  \\
g_{H_1^{0\prime} (H_5^+)^* W^+} &=& 0  \\
g_{H_5^0 (H_3^+)^* W^+} &=& \frac{- g c_H}{2\sqrt{3}}  \\
g_{H_5^0 (H_5^+)^* W^+} &=& \frac{\sqrt{3}}{2} g  \\
g_{H_3^0 (H_5^+)^* W^+} &=& -i \frac{g}{2} c_H  \\
g_{H_3^0 (H_3^+)^* W^+} &=& -i \frac{g}{2}  \\
g_{H_5^+ (H_5^{++})^* W^+} &=& \frac{-g}{\sqrt{2}}  \\
g_{H_3^+ (H_5^{++})^* W^+} &=& \frac{-g c_H}{\sqrt{2}}  \\
g_{H_3^0 H_1^0 Z} &=& \frac{-i g s_H}{2 c_W}  \\
g_{H_3^0 H_1^{0\prime} Z} &=& \frac{\sqrt{2} i g c_H}{\sqrt{3} c_W}  \\
g_{H_3^0 H_5^0 Z} &=& \frac{-i g c_H}{\sqrt{3} c_W}  \\
g_{(H_5^+)^* H_3^+ Z} &=& \frac{-g c_H}{2 c_W} \\
g_{(H_3^+)^* H_3^+ Z} &=& \frac{g}{c_W} \left(\frac{1}{2} - s^2_W \right) \\
g_{(H_5^+)^* H_5^+ Z} &=& \frac{g}{c_W} \left(\frac{1}{2} - s^2_W \right) \\
g_{(H_5^{++})^* H_5^{++} Z} &=& \frac{g}{c_W} (1-2s^2_W).
\end{eqnarray}
Couplings of a pair of charged Higgs bosons to the photon are diagonal
and are determined solely by the charge,
\begin{eqnarray}
g_{H_3^+ (H_3^+)^* \gamma} = g_{H_5^+ (H_5^+)^* \gamma} &=& -e  \\
g_{H_5^{++} (H_5^{++})^* \gamma} &=& -2e.
\end{eqnarray}

\section{The generalized Georgi--Machacek models}
\label{sec:GMgeneralized}

In this section we describe our generalization of the 
Georgi--Machacek model to include multiplets of arbitrary 
isospin, while still preserving ${\rm SU}(2)_c$ \cite{GHL99}.

We construct a generalized Georgi--Machacek model out of 
a set of multiplets $X = (T,T)$ of ${\rm SU}(2)_L \times {\rm SU}(2)_R$.
The ${\rm SU}(2)_L \times {\rm U}(1)_Y$ content of some of these are listed
in table \ref{table:GMgeneralized}.
\begin{table}
\begin{center}
\begin{tabular}{|c|c|c|}
  \hline 
T & Y & $(T_L, T_R)$\\
\hline
1/2 & 1 & (1/2,1/2) \\ \hline
1 & 0 & \\
1 & 2 & (1,1) \\ \hline
3/2 & 1 & \\
3/2 & 3 & (3/2,3/2) \\ \hline
2 & 0 & \\
2 & 2 & (2,2) \\
2 & 4 & \\ \hline
5/2 & 1 & \\
5/2 & 3 & (5/2,5/2) \\
5/2 & 5 & \\ \hline
3 & 0 & \\
3 & 2 & \\
3 & 4 & (3,3) \\
3 & 6 & \\ \hline
\vdots & \vdots & \vdots \\ \hline
\end{tabular}
\end{center}
\caption[Higgs multiplets in the generalized Georgi--Machacek model]
{Sets of multiplets of ${\rm SU}(2)_L \times {\rm U}(1)_Y$ which combine into a
single multiplet of ${\rm SU}(2)_L \times {\rm SU}(2)_R$ in the generalized
Georgi--Machacek models.  Multiplets with $Y=0$ are real; the rest
are complex.
}
\label{table:GMgeneralized}
\end{table}
The multiplets $X$ acquire vevs proportional to the unit matrix,
$\langle X \rangle = v_X \mathbf{I}$, in order to preserve 
${\rm SU}(2)_c$.  We assume that the Higgs potential is symmetric under
the full ${\rm SU}(2)_L \times {\rm SU}(2)_R$, so that after electroweak 
symmetry breaking the Higgs potential preserves ${\rm SU}(2)_c$.
This has three significant consequences.  First,
the vevs of the multiplets $X$ automatically preserve ${\rm SU}(2)_c$.
Second, ${\rm SU}(2)_c$ is not broken by radiative corrections in the
Higgs sector, so $\rho = 1$ to all orders in the Higgs self 
couplings.  Finally, the physical Higgs states can be classified
by their transformation properties under ${\rm SU}(2)_c$, and states in
different representations of ${\rm SU}(2)_c$ do not mix with each other.

Let us define the following notation for the states that transform
under ${\rm SU}(2)_c$.  A Higgs doublet $\Phi = (1/2,1/2)$ decomposes after
electroweak symmetry breaking into a singlet and a triplet under 
${\rm SU}(2)_c$.  We will denote the singlet by $\Phi_1 = H_1^0$ (in 
analogy to the triplet Georgi--Machacek model), and the triplet
by $\Phi_3 = (\Phi_3^+, \Phi_3^0, \Phi_3^-)$.  Similarly, 
a general multiplet $X = (T,T)$ decomposes into the ${\rm SU}(2)_c$ 
multiplets $2T \oplus 2T-1 \oplus \cdots \oplus 1 \oplus 0$.
We will denote these as $X_{4T+1}, X_{4T-1}, \ldots, X_3, X_1$.
We will also define $X_1 = H_1^{0\prime}$, in analogy to the 
triplet Georgi--Machacek model.  

For simplicity we consider
a model which contains only two multiplets of fields, one 
doublet, $\Phi = (1/2,1/2)$, and one larger multiplet, $X = (T,T)$.
In general, the two ${\rm SU}(2)_c$ singlets, $\Phi_1$ and $X_1$, 
will mix with each other to form mass eigenstates.  The 
two ${\rm SU}(2)_c$ triplets, $\Phi_3$ and $X_3$, will also mix
with each other to form the triplet of Goldstone bosons and 
a physical triplet of Higgs states.  The remaining ${\rm SU}(2)_c$
multiplets of $X$ will not mix and will be mass eigenstates.

In section \ref{sec:GMgeneral_ffcouplings} we derive the 
fermion--antifermion couplings of all the Higgs states in a 
generalized Georgi--Machacek model.
In section \ref{sec:GMgaugecouplings} we derive 
some of the Higgs couplings to vector bosons.  In particular,
we find the $Z H_3^+ H_3^-$ coupling, which appears in the loop
corrections to the $Z b \bar{b}$ vertex, and the $H_1^0$, 
$H_1^{0\prime}$ couplings to vector boson pairs, which are relevant 
for Higgs boson production at LEP and the Tevatron.

\subsection{Couplings to fermions}
\label{sec:GMgeneral_ffcouplings}

In this section we describe how the Higgs couplings to fermions
arise in the generalized Georgi--Machacek models.  For simplicity
we consider a model containing two multiplets of fields, 
$\Phi = (1/2,1/2)$ and $X = (T,T)$.  $\Phi$ is the usual ${\rm SU}(2)_L$ 
doublet with hypercharge 1.  $X$ is an exotic multiplet with
isospin $T$ under both ${\rm SU}(2)_L$ and ${\rm SU}(2)_R$.  

$\Phi$ couples to fermions in the normal way, while $X$ does not 
couple to fermions.  The Yukawa coupling to a fermion $f$ is
\begin{equation}
\lambda_f = \frac{\sqrt{2} m_f}{v_{\phi}} 
\end{equation}
where $\langle \phi^0 \rangle = v_{\phi}/\sqrt{2}$.  This is a 
Type I model since a single Higgs doublet gives mass to all fermions.

The vevs are constrained by the $W$ mass to obey the relation
\begin{equation}
\left(\frac{2M_W}{g}\right)^2 = v_{SM}^2 
	= \sum_k 2 v_k^2 (T_k(T_k+1) - Y_k^2/4) 
	+ \sum_i 2v_i^2 T_i (T_i+1)
\end{equation}
where $k$ runs over the complex representations of ${\rm SU}(2)_L$, $i$
runs over the real representations of ${\rm SU}(2)_L$ with $Y_i=0$, and the 
vevs are normalized according to 
$\langle \phi_k^0 \rangle = v_k/\sqrt{2}$ and 
$\langle \eta_i^0 \rangle = v_i$.
Summing over $\Phi$ and the multiplets of $X$, and using 
$v_k = \sqrt{2} v_X$ for the complex representations in $X$, we obtain
\begin{equation}
\left(\frac{2M_W}{g}\right)^2 = v_{SM}^2
	= v_{\phi}^2 + \frac{4}{3} T (T+1) (2T+1) v_X^2
\end{equation}
for any $T$.  Using this, we can write $v_{\phi}$ as
\begin{equation}
v_{\phi} = v_{SM} \cos\theta_H
\end{equation}
where we have defined a mixing angle $\theta_H$ in analogy with
the triplet Georgi--Machacek model as
\begin{equation}
\tan\theta_H = \frac{\sqrt{\frac{4}{3}T(T+1)(2T+1)} v_X}{v_{\phi}}.
\label{eqn:GMthetaH}
\end{equation}
This mixing angle also appears in the Goldstone bosons, as we will show
below.  In terms of $\theta_H$ the Yukawa coupling for fermion $f$ is
\begin{equation}
\lambda_f = \frac{g m_f}{\sqrt{2} M_W \cos\theta_H}.
\end{equation}
Note that the Yukawa couplings increase as $v_{\phi}$ decreases.

The couplings of the Higgs mass eigenstates to fermions in this model
are of the form
\begin{equation}
i (g^{L}_{H\bar{f}f^{\prime}} P_{L} + g^{R}_{H\bar{f}f^{\prime}} P_{R})
	= i (g^{V}_{H\bar{f}f^{\prime}} + g^{A}_{H\bar{f}f^{\prime}} 
		\gamma_{5})
\end{equation}
where $P_{R,L} = (1 \pm \gamma_5)/2$ and using third--generation
notation,
\begin{eqnarray}
g_{H_{i}^{0}f\bar{f}}^{V} &=& - \frac{1}{\sqrt{2}} \lambda_f
	\langle H_{i}^{0} | \phi^{0,r} \rangle  \\
g_{A_{i}^{0}t\bar{t}}^{A} &=& + \frac{i}{\sqrt{2}} \lambda_t
        \langle A_{i}^{0} | \phi^{0,i} \rangle   \\
g_{A_{i}^{0}b\bar{b}}^{A} &=& - \frac{i}{\sqrt{2}} \lambda_b
        \langle A_{i}^{0} | \phi^{0,i} \rangle   \\
g_{H_{i}^{+}\bar{t}b}^{R} &=& - \lambda_b 
	\langle H_{i}^{+} | \phi^{+} \rangle  \\
g_{H_{i}^{+}\bar{t}b}^{L} &=& + \lambda_t
        \langle H_{i}^{+} | \phi^{+} \rangle.
\end{eqnarray}
Thus the Higgs--fermion couplings depend on the overlap of Higgs
mass eigenstates with $\Phi$.
Since $\Phi$ contains only a singlet and
a triplet of ${\rm SU}(2)_c$, only singlets and triplets of 
${\rm SU}(2)_c$ can 
couple to fermions.

\subsubsection{${\rm SU}(2)_c$ singlets}

The fermion couplings of the ${\rm SU}(2)_c$ singlets are straightforward to
find.  The two singlets are,
\begin{eqnarray}
H_1^0 &=& \Phi_1 = \phi^{0,r}  \\
H_1^{0\prime} &=& X_1.
\end{eqnarray}
An expression for $X_1$ is derived in section \ref{sec:GMgaugecouplings}.
$H_1^0$ and $H_1^{0\prime}$ mix by an arbitrary angle to form the 
mass eigenstates.  Their couplings to fermions are
\begin{eqnarray}
g^V_{H_1^0 f\bar{f}} &=& -\frac{g m_f}{2 M_W \cos\theta_H}  \\
g^V_{H_1^{0\prime} f\bar{f}} &=& 0.
\end{eqnarray}
Since $H_1^0$ and $H_1^{0\prime}$ are CP--even they have only vector
couplings to fermions.  Their axial--vector couplings $g^A$ are zero.
Note that in terms of $\theta_H$, these couplings are the same as 
the corresponding couplings in the triplet Georgi--Machacek model.
If $H_1^0$ and $H_1^{0\prime}$ mix by an angle $\alpha$ to form the mass
eigenstates
\begin{eqnarray}
H^0 &=& \cos\alpha H_1^0 + \sin\alpha H_1^{0\prime}
	= \cos\alpha \phi^{0,r} + \sin\alpha X_1  \\
h^0 &=& -\sin\alpha H_1^0 + \cos\alpha H_1^{0\prime}
	= -\sin\alpha \phi^{0,r} + \cos\alpha X_1,
\end{eqnarray}
then the couplings to fermions will be
\begin{eqnarray}
g^V_{H^0 f\bar{f}} &=& -\frac{g m_f \cos\alpha}{2 M_W \cos\theta_H}  \\
g^V_{h^0 f\bar{f}} &=& +\frac{g m_f \sin\alpha}{2 M_W \cos\theta_H}.
\end{eqnarray}

\subsubsection{${\rm SU}(2)_c$ triplets}

In order to find the fermion couplings of the ${\rm SU}(2)_c$ triplets, we must
first find the mass eigenstates.  The ${\rm SU}(2)_c$ triplets $\Phi_3$ and 
$X_3$ mix to form a triplet of Goldstone bosons and a triplet 
of physical states.  We will denote the triplet of Goldstone bosons by
$G_3 = (G^+,G^0,G^-)$ and the triplet of physical states by 
$H_3 = (H_3^+,H_3^0,H_3^-)$.

The neutral Goldstone boson is given by
\begin{equation}
G^0 = \frac{ \sum_k Y_k v_k \phi_k^{0,i} }
	{ \sqrt{\sum_k Y_k^2 v_k^2} }
\label{eqn:GM_Goldstone0}
\end{equation}
where again $k$ runs over the complex representations of ${\rm SU}(2)_L$ 
and the vevs are normalized according to 
$\langle \phi_k^0 \rangle = v_k/\sqrt{2}$.
$X_3^0$ is the combination of the multiplets of $X$ that appears 
in $G^0$, which can be obtained by setting $v_{\phi} = 0$,
\begin{equation}
X_3^0 = \frac{ \sum_{k \in X} Y_k v_k \chi_k^{0,i} }
	{ \sqrt{\sum_{k \in X} Y_k^2 v_k^2} }.
\end{equation}
Inserting this and the contribution from $\Phi$ into equation
\ref{eqn:GM_Goldstone0} we obtain
\begin{equation}
G^0 = \frac{ \sqrt{\sum_{k \in X}(Y_k^2 v_k^2}) X_3^0 + v_{\phi} \Phi_3^0 }
	{\sqrt{ \sum_{k \in X}(Y_k^2 v_k^2) + v_{\phi}^2 }}
\end{equation}
where $\Phi_3^0 = \phi^{0,i}$.
Summing over the multiplets of $X$, and using 
$v_k = \sqrt{2} v_X$ for the complex representations in $X$, we obtain
\begin{eqnarray}
G^0 &=& \frac{ v_X \sqrt{\frac{4}{3}T(T+1)(2T+1)} X_3^0 + v_{\phi} \Phi_3^0 }
	{\sqrt{ v_X^2 \frac{4}{3}T(T+1)(2T+1) + v_{\phi}^2 }}  \\
	&=& \sin\theta_H X_3^0 + \cos\theta_H \Phi_3^0
\end{eqnarray}
where $\theta_H$ is the mixing angle defined in equation \ref{eqn:GMthetaH}.
The orthogonal state is the neutral member of the physical ${\rm SU}(2)_c$ 
triplet,
\begin{equation}
H_3^0 = \cos\theta_H X_3^0 - \sin\theta_H \Phi_3^0.
\end{equation}
We can now find the fermion couplings of the neutral members of the 
${\rm SU}(2)_c$ triplets.  They are, using third--generation notation,
\begin{eqnarray}
g^A_{G^0 t\bar{t}} &=& \frac{i g m_t}{2 M_W}  \\
g^A_{G^0 b\bar{b}} &=& -\frac{i g m_b}{2 M_W}  \\
g^A_{H_3^0 t\bar{t}} &=& -\frac{i g m_t \tan\theta_H}{2 M_W}  \\
g^A_{H_3^0 b\bar{b}} &=& \frac{i g m_b \tan\theta_H}{2 M_W}.
\end{eqnarray}
Since $G^0$ and $H_3^0$ are CP--odd they have only axial--vector
couplings to fermions.  Their vector couplings $g^V$ are zero.
Note that in terms of $\theta_H$, these couplings are the same as 
the corresponding couplings in the triplet Georgi--Machacek model.

We now find the couplings for the charged states $G^{\pm}$ and 
$H_3^{\pm}$.  The charged Goldstone boson is given by
\begin{eqnarray}
G^+ &=& \left\{ \sum_k \left[ 
	\left[ T_k(T_k+1) - Y_k(Y_k-2)/4 \right]^{1/2}
	v_k \phi_k^+  \right.\right.
	\nonumber \\
	& & \left.\left. - 
	\left[ T_k(T_k+1) - Y_k(Y_k+2)/4 \right]^{1/2}
	 v_k (\phi_k^-)^* 
	\right]
	+ \sum_i 
	\left[ 2T_i(T_i+1) \right]^{1/2} v_i \eta_i^+ \right\}  
	\nonumber \\
	& & \times \left\{
	\sum_k 2 v_k^2 [T_k(T_k+1) - Y_k^2/4] 
	+ \sum_i 2 v_i^2 T_i(T_i+1) \right\}^{-1/2}
\label{eqn:GM_Goldstone+}
\end{eqnarray}
where again $k$ runs over the complex representations of ${\rm SU}(2)_L$ 
and the vevs are normalized according to 
$\langle \phi_k^0 \rangle = v_k/\sqrt{2}$.
$X_3^+$ is the combination of the multiplets of $X$ that appears 
in $G^+$, which can be obtained by setting $v_{\phi} = 0$ in equation
\ref{eqn:GM_Goldstone+},
\begin{eqnarray}
X_3^+ &=& \left\{ \sum_{k \in X} \left[ 
	\left[ T(T+1) - Y_k(Y_k-2)/4 \right]^{1/2}
	 v_k \chi_k^+  \right.\right.
	\nonumber \\
	& & - \left.\left. 
	\left[ T(T+1) - Y_k(Y_k+2)/4 \right]^{1/2}
	v_k (\chi_k^-)^* 
	\right]
	+ \sum_{i \in X} 
	\left[ 2T(T+1) \right]^{1/2} v_i \xi_i^+ \right\}  
	\nonumber \\
	& & \times \left\{
	\sum_{k \in X} 2 v_k^2 [T(T+1) - Y_k^2/4] 
	+ \sum_{i \in X} 2 v_i^2 T(T+1) \right\}^{-1/2}.
\label{eqn:GMwhatisX3+}
\end{eqnarray}
Inserting this and the contribution from $\Phi$ into equation
\ref{eqn:GM_Goldstone+}, we obtain
\begin{equation}
G^+ = \frac{ 
	\sqrt{\sum_{k \in X} 2 v_k^2 (T(T+1) - Y_k^2/4) 
	+ \sum_{i \in X} 2 v_i^2 T(T+1)} X_3^+ 
	+ v_{\phi} \Phi_3^+ }
	{ \sqrt{\sum_{k \in X}( 2 v_k^2 (T(T+1) - Y_k^2/4)) 
	+ \sum_{i \in X}( 2 v_i^2 T(T+1)) + v_{\phi}^2} }
\end{equation}
where $\Phi_3^+ = \phi^+$.
Summing over the multiplets of $X$, using 
$v_k = \sqrt{2} v_X$ for the complex representations in $X$, and 
remembering that $X$ only contains a real representation if $T$ is an
integer (in which case $v_i = v_X$), we obtain for any $T$,
\begin{eqnarray}
G^+ &=&  \frac{ v_X \sqrt{\frac{4}{3}T(T+1)(2T+1)} X_3^+ + v_{\phi} \Phi_3^+ }
	{\sqrt{ v_X^2 \frac{4}{3}T(T+1)(2T+1) + v_{\phi}^2 }}  \\
	&=& \sin\theta_H X_3^+ + \cos\theta_H \Phi_3^+
\end{eqnarray}
where again $\theta_H$ is the mixing angle defined in equation 
\ref{eqn:GMthetaH}.
The orthogonal state is the charged member of the physical ${\rm SU}(2)_c$ 
triplet,
\begin{equation}
H_3^+ = \cos\theta_H X_3^+ - \sin\theta_H \Phi_3^+.
\end{equation}
We can now find the fermion couplings of the charged members of the 
${\rm SU}(2)_c$ triplets.  They are, using third--generation notation,
\begin{eqnarray}
g^L_{G^+ \bar{t}b} &=& \frac{g m_t}{\sqrt{2} M_W}  \\
g^R_{G^+ \bar{t}b} &=& \frac{- g m_b}{\sqrt{2} M_W}  \\
g^L_{H_3^+ \bar{t}b} &=& \frac{- g m_t \tan\theta_H}{\sqrt{2} M_W}  \\
g^R_{H_3^+ \bar{t}b} &=& \frac{g m_b \tan\theta_H}{\sqrt{2} M_W}
\end{eqnarray}
Note that in terms of $\theta_H$, these couplings are the same as 
the corresponding couplings in the triplet Georgi--Machacek model.

\subsection{Couplings to gauge bosons}
\label{sec:GMgaugecouplings}

In this section we derive some of the Higgs couplings to vector 
bosons in the generalized Georgi--Machacek models.  In particular,
we find the $Z H_3^+ H_3^-$ coupling, which appears in the loop
corrections to the $Z b \bar{b}$ vertex, and the $H_1^0$, 
$H_1^{0\prime}$ couplings to vector boson pairs, which are relevant 
for Higgs boson production at LEP and the Tevatron.

\subsubsection{$ZH_3^+H_3^-$ couplings}

In general, all the charged Higgs bosons will couple to $Z$.  However, since
only singlets and triplets of ${\rm SU}(2)_c$ can couple to fermions, only the
$Z H_3^+ H_3^-$, $Z G^+ G^-$, and $Z H_3^+ G^-$ couplings will affect the 
$Z b \bar{b}$ vertex.

The charged Higgs couplings to $Z$ come from the covariant derivatives
in the Lagrangian.  The $\phi^+ (\phi^+)^* Z$ vertex is 
$i g_{\phi^+ (\phi^+)^* Z} (p_1 - p_2)^{\mu}$, where $p_1$ is the incoming
momentum of $\phi^+$ and $p_2$ is the incoming momentum of $(\phi^+)^*$.
For a complex representation with hypercharge $Y$ or a real representation
with hypercharge $Y=0$ the coupling is
\begin{equation}
g_{\phi^+ (\phi^+)^* Z} = -\frac{e}{s_W c_W} (c^2_W - Y/2).
\end{equation}
The electroweak eigenstates mix to form the mass eigenstates $H_3^+$ and
$G^+$.  The couplings of these mass eigenstates are given by
\begin{equation}
g_{H_1 H_2 Z} = \sum_{\phi_i} \sum_{\phi_j} 
	\langle H_1 | \phi_i \rangle \langle H_2 | \phi_j \rangle
	g_{\phi_i \phi_j Z},
\label{eqn:GMmixHHZ}
\end{equation}
or, for $H_3^+$ and $G^+$,
\begin{eqnarray}
g_{H_3^+ (H_3^+)^* Z} &=& \sum_{\phi_k}
	|\langle H_3^+ | \phi_k^+ \rangle|^2 g_{\phi_k^+ (\phi_k^+)^* Z}  \\
g_{G^+ (G^+)^* Z} &=& \sum_{\phi_k}
	|\langle G^+ | \phi_k^+ \rangle|^2 g_{\phi_k^+ (\phi_k^+)^* Z}  \\
g_{H_3^+ (G^+)^* Z} &=& \sum_{\phi_k}
	\langle H_3^+ | \phi_k^+ \rangle \langle \phi_k^+ | G^+ \rangle
	g_{\phi_k^+ (\phi_k^+)^* Z}.
\end{eqnarray}
We derive the couplings in two stages.  We first derive the 
$\Phi_3^+ (\Phi_3^+)^* Z$ and $X_3^+ (X_3^+)^* Z$ couplings, then
we take into account the mixing between $\Phi$ and $X$.

The $\Phi_3$ coupling is simple to derive since $\Phi$ involves only
one ${\rm SU}(2)_L$ multiplet, the $Y=1$ doublet.  The $\Phi_3^+$ coupling 
is, as in the SM,
\begin{equation}
g_{\Phi_3^+ (\Phi_3^+)^* Z} = g_{\phi^+ (\phi^+)^* Z} 
	= - \frac{e}{s_W c_W} \left( \frac{1}{2} - s^2_W \right).
\end{equation}

In order to find the $X_3^+$ coupling, we must use equation 
\ref{eqn:GMmixHHZ} to sum up the contributions of each of the 
${\rm SU}(2)_L$ multiplets in $X$.
From equation \ref{eqn:GMwhatisX3+} we can find the overlap of $X_3^+$ with 
each of the ${\rm SU}(2)_L$ multiplets of $X$.  For the complex multiplets,
using $v_k = \sqrt{2} v_X$ and evaluating the sums in the denominator,
\begin{eqnarray}
\langle X_3^+ | \chi_k^+ \rangle^2 
	&=& \frac{2T(T+1) - Y_k(Y_k-2)/2}{\frac{4}{3}T(T+1)(2T+1)}  \\
\langle X_3^+ | (\chi_k^-)^* \rangle^2
	&=& \frac{2T(T+1) - Y_k(Y_k+2)/2}{\frac{4}{3}T(T+1)(2T+1)}.
\end{eqnarray}
If $T$ is an integer there is also one real multiplet, with $Y=0$ and 
$v_i = v_X$, which gives
\begin{equation}
\langle X_3^+ | \xi^+ \rangle^2
	= \frac{2T(T+1)}{\frac{4}{3}T(T+1)(2T+1)}.
\end{equation}
The coupling of each of the ${\rm SU}(2)_L$ multiplets is
\begin{eqnarray}
g_{\chi_k^+ (\chi_k^+)^* Z} &=& -\frac{e}{s_W c_W}(c^2_W - Y_k/2)  \\
g_{(\chi_k^-)^* \chi_k^- Z} &=& +\frac{e}{s_W c_W}(-c^2_W - Y_k/2)  \\
g_{\xi^+ (\xi^+)^* Z} &=& -\frac{e}{s_W c_W}(c^2_W).
\end{eqnarray}
Inserting these into equation \ref{eqn:GMmixHHZ} and summing over the
multiplets of $X$, we find, for any $T$,
\begin{equation}
g_{X_3^+ (X_3^+)^* Z} = -\frac{e}{s_W c_W} \left( \frac{1}{2} - s^2_W \right),
\end{equation}
the same as for an ${\rm SU}(2)_L$ doublet.

When $\Phi_3^+$ and $X_3^+$ mix to form mass eigenstates, the couplings
remain the same.  They are,
\begin{eqnarray}
g_{H_3^+ (H_3^+)^* Z} = g_{G^+ (G^+)^* Z}
&=& -\frac{e}{s_W c_W} \left( \frac{1}{2} - s^2_W \right)  \\
g_{H_3^+ (G^+)^* Z} &=& 0.
\end{eqnarray}
Note that the $H_3^+ (H_3^+)^* Z$ coupling is the same as the 
$H^+ H^- Z$ coupling in the two Higgs doublet model, and there 
is no $H_3^+ G^- Z$ coupling.

In general, for a model with ${\rm SU}(2)_c$ containing any number of 
multiplets $X$, the couplings of the ${\rm SU}(2)_c$ triplet mass 
eigenstates are
\begin{eqnarray}
g_{H_{3}^{+}(H_{3}^{+\prime})^* Z} 
	&=& \sum_{X} g_{X_{3}^{+}(X_{3}^{+})^* Z}
	\langle H_{3}^{+} | X_{3}^{+} \rangle 
	\langle X_{3}^{+} | H_{3}^{+\prime} \rangle \nonumber \\
&=& -\frac{e}{s_{W}c_{W}} (\frac{1}{2}-s^{2}_{W}) 
	\langle H_{3}^{+} | H_{3}^{+\prime} \rangle.
\end{eqnarray}

\subsubsection{$H_1^0$, $H_1^{0\prime}$ couplings to 
vector boson pairs}

We now consider the couplings of the ${\rm SU}(2)_c$ singlets to vector 
boson pairs.  The neutral Higgs couplings to $W^+W^-$ and $ZZ$ 
come from the covariant derivatives in the Lagrangian.  The 
$\phi^{0,r} V_1 V_2$ vertex is $i g_{\phi^{0,r} V_1 V_2} g^{\mu\nu}$,
where $V_1$ and $V_2$ are any two vector bosons.  Note that only 
CP--even states couple in this way to vector boson pairs.
For a complex representation with hypercharge $Y$ or a real representation
with hypercharge $Y=0$ the couplings are
\begin{eqnarray}
g_{\phi^{0,r} W^+W^-} &=& g^2 v 
	\left( T(T+1) - Y^2/4 \right) \\
g_{\phi^{0,r} ZZ} &=& \frac{g^2 v}{c^2_W} \frac{Y^2}{2}
\end{eqnarray}
where $\langle \phi^0 \rangle = v/\sqrt{2}$ for a complex representation
and $\langle \eta^0 \rangle = v$ for a real representation.

The couplings of $H_1^0$ are easy to derive, because 
$H_1^0 = \phi^{0,r}$ involves only one ${\rm SU}(2)_L$ multiplet, the 
$Y=1$ doublet.  The $H_1^0$ couplings are,
\begin{eqnarray}
g_{H_1^0 W^+ W^-} &=& \frac{g^2 v_{\phi}}{2}
	= g M_W \cos\theta_H  \\
g_{H_1^0 ZZ} &=& \frac{g^2 v_{\phi}}{2 c^2_W}
	= \frac{g M_W \cos\theta_H}{c^2_W}
\end{eqnarray}
where we have used $v_{\phi}/v_{SM} = \cos\theta_H$ and 
$v_{SM} = 2M_W/g$.

In order to find the couplings of $H_1^{0\prime}$, we must sum up the 
contribution of each electroweak eigenstate in $H_1^{0\prime}$ using
\begin{equation}
g_{HV_1V_2} = \sum_{\phi_i}
	\langle H | \phi_i \rangle g_{\phi_i V_1V_2}.
\label{eqn:GMmixHVV}
\end{equation}
We now need an expression
for the composition of $H_1^{0\prime}$ in terms of electroweak 
eigenstates.  
$H_1^{0\prime}$ is the ${\rm SU}(2)_c$ singlet of $X$; that is,
$H_1^{0\prime}$ is the state that remains invariant under simultaneous
${\rm SU}(2)_L$ and ${\rm SU}(2)_R$ 
rotations of $X$ by the same angle.  Clearly,
$H_1^{0\prime}$ must be proportional to the unit matrix, just as 
$\langle X \rangle$ is.  In terms of the electroweak eigenstates,
for integer $T$,
\begin{equation}
H_1^{0\prime} = \frac{\sum_{complex} (\chi_k^0 + (\chi_k^0)^*) + \xi^0}
	{\sqrt{2T+1}}
	= \frac{\sum_{k=1}^T \sqrt{2} \chi_k^{0,r} + \xi^0}
	{\sqrt{2T+1}},
\end{equation}
and for half--odd--integer $T$,
\begin{equation}
H_1^{0\prime} = \frac{\sum_{complex} (\chi_k^0 + (\chi_k^0)^*)}
	{\sqrt{2T+1}}
	= \frac{\sum_{k=1}^{T+1/2} \sqrt{2} \chi_k^{0,r}}
	{\sqrt{2T+1}}.
\end{equation}
We have denoted the ${\rm SU}(2)_L$ components of $X$ as follows.
We denote the complex multiplets of ${\rm SU}(2)_L$ as $\chi_k$ 
and, for integer $T$, the real multiplet as $\xi$.
The couplings of each of the ${\rm SU}(2)_L$ multiplets are
\begin{eqnarray}
g_{\chi_k^{0,r} W^+W^-} &=& \sqrt{2} g^2 v_X (T(T+1) - Y_k^2/4)  \\
g_{\xi^0 W^+ W^-} &=& g^2 v_X T(T+1)  \\
g_{\chi_k^{0,r} ZZ} &=& \frac{\sqrt{2} g^2 v_X}{c^2_W} \frac{Y^2_k}{2}  \\
g_{\xi^0 ZZ} &=& 0
\end{eqnarray}
where we have used $\langle \chi^0_k \rangle = v_X$
and $\langle \xi^0 \rangle = v_X$.
Inserting these into equation \ref{eqn:GMmixHVV} and summing over 
the multiplets of $X$, we find,
\begin{eqnarray}
g_{H_1^{0\prime} W^+W^-} &=& g M_W \sqrt{\frac{4}{3}T(T+1)} \sin\theta_H  \\
g_{H_1^{0\prime} ZZ} &=& \frac{g M_W}{c^2_W} 
	\sqrt{\frac{4}{3}T(T+1)} \sin\theta_H
\end{eqnarray}
where we have used 
$\sin\theta_H = \sqrt{\frac{4}{3}T(T+1)(2T+1)}v_X/v_{SM}$ and 
$v_{SM} = 2M_W/g$.

\section{Absence of a massless Goldstone boson in the generalized 
Georgi--Machacek models}
\label{sec:GMnoAxion}

In a general extended Higgs sector which does not preserve 
${\rm SU}(2)_c$, the Higgs potential will sometimes have an accidental
continuous global symmetry.  For example, the Higgs potential may
be invariant under a separate ${\rm U}(1)$ rotation of each of the
multiplets.  This is the case, for example, in a model containing
one doublet and one larger multiplet with $(T,Y)=(3,4)$, which 
preserves $\rho=1$.  The Higgs potential for this model has the
symmetries ${\rm SU}(2)_L \times {\rm U}(1)_Y \times {\rm U}(1)$, 
where the second
U(1) is a rotation of the larger multiplet, leaving the
doublet invariant.  This second U(1) is a global symmetry.

In general,
the global symmetry is spontaneously broken when the multiplets
larger than doublets
get vevs, resulting in massless Goldstone bosons in the physical 
spectrum.  As described in reference \cite{Nir89}, the physical
spectrum must also contain a light CP--even Higgs boson $H^0$ with mass on
the order of the vev of the larger multiplet.  This is required 
because the mass splitting between the massless Goldstone boson
and $H^0$ is on the order
of the vev that breaks the accidental global symmetry.
Such a massless Goldstone boson 
is then ruled out by the experimental limits on 
$Z \rightarrow a^0 H^{0}$, where $a^0$ is the massless Goldstone boson
\cite{Nir89}.  

In this section we will show that the generalized Georgi--Machacek 
models do not suffer from this massless Goldstone boson problem.  
The key feature of 
the Georgi--Machacek models that we will use is the fact that a
multiplet $X$ of ${\rm SU}(2)_L \times {\rm SU}(2)_R$ can be thought of as a 
real multiplet.  $X$ can be transformed into a Cartesian basis,
in which all the components of $X$ are real, through independent
left-- and right--handed unitarty rotations.  This is possible 
because an $n \times n$ matrix $X$ contains $n^2$ degrees of freedom,
so it can be written as an $n \times n$ real matrix.  In contrast,
a complex representation of ${\rm SU}(2)_L \times {\rm U}(1)_Y$ which is an
$n$ component vector has $2n$ degrees of freedom, and so cannot 
be rotated into a Cartesian basis.

In the Cartesian basis, the ${\rm SU}(2)_L$ and ${\rm SU}(2)_R$ generators 
$\vec{\tau}$ can be chosen to be imaginary and antisymmetric.
Then the ${\rm SU}(2)_L \times {\rm SU}(2)_R$ transformation matrices have
the property that,
\begin{eqnarray}
U_{R,L}^{T} &=& {\rm exp}[-i\theta_{R,L}\hat{n}_{R,L}\cdot \vec{T}_{R,L}^{T}]
	= {\rm exp}[+i \theta_{R,L}\hat{n}_{R,L}\cdot \vec{T}_{R,L}]
	= U_{R,L}^{\dagger}  \\
U_{R,L}^* &=& {\rm exp}[+i\theta_{R,L}\hat{n}_{R,L}\cdot \vec{T}_{R,L}^{*}]
	= {\rm exp}[-i \theta_{R,L}\hat{n}_{R,L}\cdot \vec{T}_{R,L}]
	= U_{R,L}.
\end{eqnarray}
Using these, we find that in the Cartesian basis, the term 
$\mathrm{Tr} (X^T X)$ transforms as follows under ${\rm SU}(2)_L \times
{\rm SU}(2)_R$,
\begin{eqnarray}
\mathrm{Tr}(X^T X) &\to& \mathrm{Tr}(U_R^T X^T U_L^* U_L^{\dagger} X U_R)
	\nonumber \\
	&=& \mathrm{Tr} (U_R^{\dagger} X^T U_L U_L^{\dagger} X U_R)
	\nonumber \\
	&=& \mathrm{Tr} (X^T X).
\end{eqnarray}
Thus in the Cartesian basis, the term $\mathrm{Tr} (X^T X)$ is invariant
under ${\rm SU}(2)_L \times {\rm SU}(2)_R$, 
and can appear in the Higgs potential.
This term is not invariant under a global U(1) transformation of $X$,
under which $X \to e^{i\theta}X$.  Instead, the term transforms as,
\begin{eqnarray}
\mathrm{Tr}(X^T X) &\to& \mathrm{Tr}(e^{i\theta} X^T e^{i\theta} X)
	\nonumber \\
	&=& e^{2i\theta} \mathrm{Tr}(X^T X)
	\neq \mathrm{Tr} (X^T X).
\end{eqnarray}
Thus we conclude that the Higgs potential of a generalized 
Georgi--Machacek model is not invariant under any additional U(1) 
rotations.



\chapter{Calculation of one--loop integrals}
\label{app:loopints}

In this section I describe the techniques used to calculate the integrals
that arise in one--loop
radiative corrections.  
The methods used here are based on the work of 't Hooft, Veltman and 
Passarino \cite{tHooftVeltman,PassarinoVeltman}.

In 
section \ref{1loop-defs} I introduce the one--loop 
integrals, following Hollik \cite{Hollik1}, and in section 
\ref{1loop-calc} I provide details of their calculation.  In section
\ref{1loop-deriv} I list the derivatives of the one--loop integrals
which appear in the calculation of wave--function renormalization.
Finally, in 
section \ref{1loop-sym} I list the symmetries of the one--loop integrals
and in section \ref{1loop-approx} I give expansions of some of the 
integrals in certain limits, following \cite{PHI}.

\section{Notation for one--loop integrals}
\label{1loop-defs}

In this section I list the definitions of the one--loop integrals used.
I use the metric $g_{\mu\nu} = {\rm diag}(1, -1, -1, -1)$.
This section follows the discussion of Hollik \cite{Hollik1}.

\subsubsection{One--point integral:}
This integral occurs in the calculation of tadpole diagrams, where the 
particle in the loop has mass $m$ and $k$ is the loop momentum.
\begin{equation}
\mu^{4-D} \int \frac{d^D k}{(2\pi)^D} \frac{1}{k^2 - m^2} 
   = \frac{i}{16\pi^2} A(m^2).
\end{equation}
The integral will be evaluated in dimensional regularization 
(see section \ref{1loop-calc}) in $D$ dimensions.  The parameter 
$\mu$ is the renormalization scale with dimensions of mass.

\subsubsection{Two--point integrals:}
These integrals occur in the calculation of corrections to a particle
propagator, with loop momentum $k$.  The two particles in the loop have
masses $m_1$ and $m_2$.  The external momentum is $q$.
\begin{equation}
\mu^{4-D} \int \frac{d^D k}{(2\pi)^D} \frac{1; k_{\mu}; k_{\mu}k_{\nu}}
{[k^2 - m_1^2][(k+q)^2 - m_2^2]}
   = \frac{i}{16\pi^2} B_{0; \mu; \mu\nu}(q^2; m_1^2, m_2^2). 
\end{equation}
The vector and tensor integrals $B_{\mu}$ and $B_{\mu\nu}$ can be expanded
into scalar coefficients and Lorentz covariants as follows:
\begin{eqnarray}
B_{\mu} &=& q_{\mu} B_1 \\
B_{\mu\nu} &=& g_{\mu\nu} B_{22} 
   + q_{\mu} q_{\nu} B_{21}.
\end{eqnarray}
Note that the scalar coefficients depend only on the square of the external
four--momentum $q^2$.

\subsubsection{Three--point integrals:}
These integrals occur in the calculation of corrections to a three--particle
vertex.  The loop momentum is $k$ and the three particles in the loop 
have masses $m_1$, $m_2$ and $m_3$.  The external momenta are $p_1$, 
$p_2$, and $p = -(p_1 + p_2)$, where all momenta are defined to flow into
the vertex.
\begin{eqnarray}
&\mu^{4-D}& \int \frac{d^D k}{(2\pi)^D} \frac{1; k_{\mu}; k_{\mu}k_{\nu}}
{[k^2 - m_1^2][(k + p_1)^2 - m_2^2][(k + p_1 + p_2)^2 - m_3^2]} 
  \nonumber \\
& & = \frac{i}{16\pi^2} C_{0; \mu; \mu\nu}
   (p_1^2, p_2^2, p^2; m_1^2, m_2^2, m_3^2).
	\label{eqn:3pointloopintdef}
\end{eqnarray}
The vector and tensor integrals $C_{\mu}$ and $C_{\mu\nu}$ can again be
expanded into scalar coefficients and Lorentz covariants:
\begin{eqnarray}
C^{\mu} &=& p_1^{\mu} C_{11} + p_2^{\mu} C_{12} \\
C^{\mu\nu} &=& g^{\mu\nu} C_{24} + p_1^{\mu} p_1^{\nu} C_{21} 
   + p_2^{\mu} p_2^{\nu} C_{22} \\
   & & + (p_1^{\mu} p_2^{\nu} + p_2^{\mu} p_1^{\nu}) C_{23}.
\end{eqnarray}
Note that the scalar coefficients again depend only on the squares of 
the external four--momenta.

In reference \cite{Hollik1} the three--point integrals
are defined exactly as in equation \ref{eqn:3pointloopintdef}.
Note, however, that reference \cite{Hollik1} uses the notation
$C_{0;\mu;\mu\nu}(p_1,p_2;m_1,m_2,m_3)$ for the three--point 
integrals.

\section{Calculation of one--loop integrals}
\label{1loop-calc}

The coefficients of the vector and tensor one--loop integrals can all
be expressed algebraically in terms of the scalar integrals 
$A$, $B_0$ and $C_0$.  (For details of the calculation see 
\cite{Hollik1}.)  We have, for the two--point integrals,

\begin{eqnarray}
B_1(q^2; m_1^2, m_2^2) &=& \frac{1}{2 q^2} \left[ A(m_1^2) - A(m_2^2)
	\right. \nonumber \\
	& & \left. + (m_2^2 - m_1^2 - q^2) B_0(q^2; m_1^2, m_2^2) \right] 
	\label{eqn-B1exp} \\
B_{22}(q^2; m_1^2, m_2^2) &=& \frac{1}{6} \left[ A(m_2^2) 
	+ 2 m_1^2 B_0(q^2; m_1^2, m_2^2) \right. \nonumber \\
	& & \left. + (q^2 + m_1^2 - m_2^2) B_1(q^2; m_1^2, m_2^2)
	+ m_1^2 + m_2^2 - \frac{q^2}{3} \right] \\
B_{21}(q^2; m_1^2, m_2^2) &=& \frac{1}{3q^2} \left[ A(m_2^2) 
	- m_1^2 B_0(q^2; m_1^2, m_2^2) \right. \nonumber \\
	& & \left. -2(q^2 + m_1^2 - m_2^2) B_1(q^2; m_1^2, m_2^2)
	- \frac{m_1^2 + m_2^2}{2} + \frac{q^2}{6} \right].
\end{eqnarray}
Note that equation \ref{eqn-B1exp} for $B_1$ differs from the expression in 
\cite{Hollik1} in that $m_1^2$ and $m_2^2$ are interchanged in the
coefficient of $B_0$.  This is necessary in order for $B_1$ to have 
the correct divergent piece, as discussed below.

The three--point integrals are listed below.  The arguments of all the
three--point integrals are assumed to be 
$(p_1^2, p_2^2, p^2; m_1^2, m_2^2, m_3^2)$, as in section \ref{1loop-defs}.

\begin{eqnarray}
C_{11} &=& \frac{1}{\kappa}[p_2^2 R_1 - p_1 \cdot p_2 R_2] \\
C_{12} &=& \frac{1}{\kappa}[-p_1 \cdot p_2 R_1 + p_1^2 R_2] \\
C_{24} &=& \frac{1}{4}[B_0(p_2^2; m_2^2, m_3^2) + r_1 C_{11} + r_2 C_{12}
	+ 2 m_1^2 C_0 + 1]  \\
C_{21} &=& \frac{1}{\kappa}[p_2^2 R_3 - p_1 \cdot p_2 R_5]  \\
C_{23} &=& \frac{1}{\kappa}[-p_1 \cdot p_2 R_3 + p_1^2 R_5]  
		= \frac{1}{\kappa}[p_2^2 R_4 - p_1 \cdot p_2 R_6] \\
C_{22} &=& \frac{1}{\kappa}[-p_1 \cdot p_2 R_4 + p_1^2 R_6]
\end{eqnarray}
where 
\begin{eqnarray}
\kappa &=& p_1^2 p_2^2 - (p_1 \cdot p_2)^2 \\
r_1 &=& p_1^2 + m_1^2 - m_2^2 \\
r_2 &=& p^2 - p_1^2 + m_2^2 - m_3^2  \\
R_1 &=& \frac{1}{2}[B_0(p^2; m_1^2, m_3^2) - B_0(p_2^2; m_2^2, m_3^2)
	- (p_1^2 + m_1^2 - m_2^2) C_0] \\
R_2 &=& \frac{1}{2}[B_0(p_1^2; m_1^2, m_2^2) - B_0(p^2; m_1^2, m_3^2)
	+ (p_1^2 - p^2 - m_2^2 + m_3^2) C_0] \\
R_3 &=& -C_{24} - \frac{1}{2}[r_1 C_{11} - B_1(p^2; m_1^2, m_3^2)
	- B_0(p_2^2; m_2^2, m_3^2)]  \\
R_4 &=& -\frac{1}{2}[r_1 C_{12} - B_1(p^2; m_1^2, m_3^2)
	+ B_1(p_2^2; m_2^2, m_3^2)]  \\
R_5 &=& -\frac{1}{2}[r_2 C_{11} - B_1(p_1^2; m_1^2, m_2^2) 
	+ B_1(p^2; m_1^2, m_3^2)]  \\
R_6 &=& -C_{24} - \frac{1}{2}[r_2 C_{12} + B_1(p^2; m_1^2, m_3^2)].
\end{eqnarray}

Finally the integrals $A$, $B_0$ and $C_0$ are calculated using 
dimensional regularization \cite{Bollini72,Ashmore72,tHooft72}
in $D$ dimensions with $D = 4 - \epsilon$ and $\epsilon \rightarrow 0$ 
(for details see \cite{Hollik1}).
The one--point integral is
\begin{equation}
A(m^2) = m^2 \left( \Delta - \log \frac{m^2}{\mu^2} + 1 \right)
	+ \mathcal{O}(\epsilon).
\end{equation}
The divergent part is
	\begin{equation}
	\Delta = \frac{2}{\epsilon} - \gamma + \log 4\pi,
	\end{equation}
where $\gamma = 0.577 \ldots$ is Euler's constant, and $\mu$ is the
renormalization scale with dimensions of mass.  Note that in computations
of physical processes both $\Delta$ and $\mu$ must cancel out of the 
final result.  This provides a useful check on computations.  The 
divergent parts of all the one--loop integrals are listed at the end of 
this section.  Note also that the divergent term can pick out the 
$\epsilon$ in $D = 4-\epsilon$ yielding a finite additional term,
\begin{equation}
D \cdot \Delta = 4 \Delta - 2 + \mathcal{O}(\epsilon).
\end{equation}

The two--point integral $B_0$ is
\begin{equation}
B_0(q^2; m_1^2, m_2^2) = \Delta - \int_0^1 dx 
	\log \left( 
	\frac{x^2 q^2 - x(q^2 + m_1^2 - m_2^2) + m_1^2 - i\varepsilon}
	{\mu^2} \right).
\end{equation}

The three--point integral $C_0(p_1^2, p_2^2, p^2; m_1^2, m_2^2, m_3^2)$ is 
\begin{equation}
C_0 = - \int_0^1 dx \int_0^x dy \frac{1}
	{ax^2 + by^2 + cxy + dx + ey + f}
\end{equation}
where
\begin{eqnarray}
a &=& p^2  \\
b &=& p_2^2  \\
c &=& p_1^2 - p_2^2 - p^2  \\
d &=& m_3^2 - m_1^2 - p^2  \\
e &=& m_2^2 - m_3^2 + p^2 - p_1^2  \\
f &=& m_1^2 - i \varepsilon.
\end{eqnarray}
Note that there is a mistake in the expression for $e$
in reference \cite{Hollik1}, in which $m_3^2$ is replaced with $m_1^2$.

It is easy to find the divergent terms in the one--loop integrals.  They 
are,
\begin{eqnarray}
A(m^2) &=& m^2 \Delta + {\rm finite}  \\
B_0 &=& \Delta + {\rm finite}  \\
B_1 &=& -\frac{1}{2} \Delta + {\rm finite}  \\
B_{22}(q^2;m_1^2,m_2^2) &=& \frac{1}{4}\left[m_1^2 + m_2^2 - \frac{q^2}{3}
	\right] \Delta + {\rm finite}  \\
B_{21} &=& \frac{1}{3} \Delta + {\rm finite}  \\
C_{24} &=& \frac{1}{4} \Delta + {\rm finite}.
\end{eqnarray}
The integrals $C_0$, $C_{11}$, $C_{12}$, $C_{21}$, $C_{22}$, and 
$C_{23}$ are all finite, as are the combinations $R_1, \ldots, R_6$.
This list is useful because it allows one to easily check that the 
divergences cancel in the calculation of a physical process.  The 
divergent terms are also relevant in the calculation of renormalization
group equations.

\section{Derivatives of two--point integrals}
\label{1loop-deriv}

In this section I list the derivatives of the one--loop integrals
$B_0$, $B_1$, $B_{22}$, and $B_{21}$ with respect to the external 
momentum.  These derivatives appear in calculations of corrections
to a process from loop corrections to the external legs.

Let us define the shorthand notation
\begin{equation}
B^{\prime}(M^2;m_1^2,m_2^2) \equiv \left[ \frac{\partial}{\partial q^2} B
		(q^2;m_1^2,m_2^2) \right]_{q^2 = M^2}
\end{equation}
for the type of derivative which typically appears in the wave--function
renormalization.  All of the derivatives of the two--point integrals
can be expressed in terms of other two--point integrals and the
derivative of $B_0$.
\begin{eqnarray}
B^{\prime}_1 (q^2;m_1^2,m_2^2) &=& -\frac{1}{q^2} \left[ B_1 
	+ \frac{1}{2} B_0 + \frac{1}{2} (q^2 + m_1^2 - m_2^2) B^{\prime}_0
	\right] \\
B^{\prime}_{22} (q^2;m_1^2,m_2^2) &=& \frac{1}{6} \left[ B_1
	- \frac{1}{3} + 2 m_1^2 B^{\prime}_0
	+ (q^2 + m_1^2 - m_2^2) B^{\prime}_1 \right]  \\
B^{\prime}_{21} (q^2;m_1^2,m_2^2) &=& -\frac{1}{3q^2} \left[
	3 B_{21} + 2 B_1 - \frac{1}{6} + m_1^2 B^{\prime}_0 
	\right. \nonumber \\
	& & \left. + 2(q^2 + m_1^2 - m_2^2) B^{\prime}_1 \right].
\end{eqnarray}

Finally, $B^{\prime}_0$ is given by
\begin{equation}
B^{\prime}_0 (q^2;m_1^2,m_2^2) = \int_0^1 dx \frac{x(1-x)}
	{x^2 q^2 - x(q^2 + m_1^2 - m_2^2) + m_1^2}.
\end{equation}

The integrals $B^{\prime}_0$, $B^{\prime}_1$, and $B^{\prime}_{21}$
are finite.  The integral $B^{\prime}_{22}$ has a divergent term,
\begin{equation}
B^{\prime}_{22} = -\frac{1}{12} \Delta + {\rm finite}.
\end{equation}

\section{Symmetries of the one--loop integrals under permutaion of 
their arguments}
\label{1loop-sym}

In this section I list the symmetry properties of the one--loop integrals
under interchange of their arguments.

The two--point integrals obey symmetry relations under the interchange
of $m_1$ and $m_2$.  $B_0$ is symmetric under the exchange,
	\begin{equation}
	B_0(q^2; m_1^2, m_2^2) = B_0(q^2; m_2^2, m_1^2).
	\label{eqn-B0sym}
	\end{equation}
The symmetry properties of the remaining two--point integrals are more
complicated because they involve powers of the loop momentum in the 
numerator.  $B_1$ obeys the following relation,
\begin{equation}
B_1(q^2; m_1^2, m_2^2) = -B_1(q^2; m_2^2, m_1^2) - B_0(q^2; m_2^2, m_1^2)
\label{eqn-B1sym}
\end{equation}
	or, using equation \ref{eqn-B0sym} to symmetrize this,
\begin{equation}
\left[B_1 + \frac{1}{2} B_0\right](q^2;m_1^2,m_2^2)
	= -\left[B_1 + \frac{1}{2} B_0\right](q^2;m_2^2,m_1^2).
\end{equation}
$B_{22}$ and $B_{21}$ obey the following relations,
\begin{equation}
B_{22}(q^2;m_1^2,m_2^2) = B_{22}(q^2;m_2^2,m_1^2)
\end{equation}
\begin{equation}
B_{21}(q^2;m_1^2,m_2^2) = \left[B_{21} + 2B_1 + B_0\right](q^2;m_2^2,m_1^2)
\end{equation}
	or, using equation \ref{eqn-B1sym} to symmetrize this,
\begin{equation}
\left[B_{21} + B_1\right](q^2;m_1^2,m_2^2)
	= \left[B_{21} + B_1\right](q^2;m_2^2,m_1^2).
\end{equation}

The three--point integrals can be represented by a triangle with the 
vertices labelled by the external momenta and the sides labelled by the
masses of the particles in the loop.  The symmetry properties of the 
three--point integrals under permutations of $m_1$, $m_2$ and $m_3$
correspond to rotations and inversions of the 
triangle.  $C_0$ obeys the following symmetries:
\begin{eqnarray}
C_0(p_1^2, p_2^2, p^2; m_1^2, m_2^2, m_3^2) &=&
	C_0(p_2^2, p_1^2, p^2; m_3^2, m_2^2, m_1^2) \nonumber \\
= C_0(p^2, p_1^2, p_2^2; m_3^2, m_1^2, m_2^2) &=& 
	C_0(p_1^2, p^2, p_2^2; m_2^2, m_1^2, m_3^2) \nonumber \\
= C_0(p_2^2, p^2, p_1^2; m_2^2, m_3^2, m_1^2) &=& 
	C_0(p^2, p_2^2, p_1^2; m_1^2, m_3^2, m_2^2).
\end{eqnarray}
The same symmetries are obeyed by $C_{24}$.  The symmetries of the 
remaining three--point integrals are more complicated.  They are as 
follows.  The symmetries of $C_{11}$ are,
\begin{eqnarray}
C_{11}(p_1^2, p_2^2, p^2; m_1^2, m_2^2, m_3^2) 
	&=& [-C_{12} - C_0](p_2^2, p_1^2, p^2; m_3^2, m_2^2, m_1^2)
	\\
	&=& [-C_{11} + C_{12} - C_0](p^2, p_1^2, p_2^2; m_3^2, m_1^2, m_2^2)
	\\
	&=& [-C_{11} + C_{12} - C_0](p_1^2, p^2, p_2^2; m_2^2, m_1^2, m_3^2)
	\\
	&=& [-C_{12} - C_0](p_2^2, p^2, p_1^2; m_2^2, m_3^2, m_1^2)
	\\
	&=& C_{11} (p^2, p_2^2, p_1^2; m_1^2, m_3^2, m_2^2).
\end{eqnarray}
The symmetries of $C_{12}$ are,
\begin{eqnarray}
C_{12}(p_1^2, p_2^2, p^2; m_1^2, m_2^2, m_3^2)
	&=& [-C_{11} - C_0](p_2^2, p_1^2, p^2; m_3^2, m_2^2, m_1^2)
	\\
	&=& [-C_{11} - C_0](p^2, p_1^2, p_2^2; m_3^2, m_1^2, m_2^2)
	\\
	&=& C_{12}(p_1^2, p^2, p_2^2; m_2^2, m_1^2, m_3^2)
	\\
	&=& [C_{11} - C_{12}](p_2^2, p^2, p_1^2; m_2^2, m_3^2, m_1^2)
	\\
	&=& [C_{11} - C_{12}](p^2, p_2^2, p_1^2; m_1^2, m_3^2, m_2^2).
\end{eqnarray}
The symmetries of $C_{21}$ are,
\begin{eqnarray}
C_{21}(p_1^2, p_2^2, p^2; m_1^2, m_2^2, m_3^2)
	&=& [C_{22} + 2C_{12} + C_0]
		(p_2^2, p_1^2, p^2; m_3^2, m_2^2, m_1^2) \\
	&=& [C_{21} + C_{22} - 2C_{23} + 2C_{11} - 2C_{12} + C_0]
		\nonumber \\
		& &
		(p^2, p_1^2, p_2^2; m_3^2, m_1^2, m_2^2) \\
	&=& [C_{21} + C_{22} - 2C_{23} + 2C_{11} - 2C_{12} + C_0]
		\nonumber \\
		& & 
		(p_1^2, p^2, p_2^2; m_2^2, m_1^2, m_3^2) \\
	&=& [C_{22} + 2C_{12} + C_0]
		(p_2^2, p^2, p_1^2; m_2^2, m_3^2, m_1^2) 
		\label{eqn-C21symE}  \\
	&=& C_{21}(p^2, p_2^2, p_1^2; m_1^2, m_3^2, m_2^2).
		\label{eqn-C21symF}
\end{eqnarray}
The symmetries of $C_{22}$ are,
	%
\begin{eqnarray}
C_{22}(p_1^2, p_2^2, p^2; m_1^2, m_2^2, m_3^2)
	&=& [C_{21} + 2C_{11} + C_0]
		(p_2^2, p_1^2, p^2; m_3^2, m_2^2, m_1^2) \\
	&=& [C_{21} + 2C_{11} + C_0]
		(p^2, p_1^2, p_2^2; m_3^2, m_1^2, m_2^2) \\
	&=& C_{22}(p_1^2, p^2, p_2^2; m_2^2, m_1^2, m_3^2) \\
	&=& [C_{21} + C_{22} - 2C_{23}]
		(p_2^2, p^2, p_1^2; m_2^2, m_3^2, m_1^2) \\
	&=& [C_{21} + C_{22} - 2C_{23}]
		(p^2, p_2^2, p_1^2; m_1^2, m_3^2, m_2^2).
\end{eqnarray}
Finally, the symmetries of $C_{23}$ are,
\begin{eqnarray}
C_{23}(p_1^2, p_2^2, p^2; m_1^2, m_2^2, m_3^2)
	&=& [C_{23} + C_{12} + C_{11} + C_0]
		(p_2^2, p_1^2, p^2; m_3^2, m_2^2, m_1^2) \\
	&=& [C_{21} - C_{23} + 2C_{11} - C_{12} + C_0]
		\nonumber \\
		& &
		(p^2, p_1^2, p_2^2; m_3^2, m_1^2, m_2^2) 
		\label{eqn-C23symC}  \\
	&=& [C_{22} - C_{23} - C_{12}]
		(p_1^2, p^2, p_2^2; m_2^2, m_1^2, m_3^2) \\
	&=& [C_{22} - C_{23} - C_{11} + C_{12}]
		(p_2^2, p^2, p_1^2; m_2^2, m_3^2, m_1^2) 
		\label{eqn-C23symE}  \\
	&=& [C_{21} - C_{23}]
		(p^2, p_2^2, p_1^2; m_1^2, m_3^2, m_2^2).
\end{eqnarray}

The symmetry relations of $B_0$, $B_1$, and all the $C$ integrals agree
with those in \cite{Irulegui}, except for equations 
\ref{eqn-C21symE} and \ref{eqn-C21symF} for $C_{21}$ which are 
interchanged in \cite{Irulegui}, 
and equations \ref{eqn-C23symC} and \ref{eqn-C23symE} for $C_{23}$
which are interchanged in \cite{Irulegui}.
Reference \cite{Irulegui} does not list symmetry relations for 
$B_{22}$ and $B_{21}$.

\section{Approximations for the one--loop integrals in certain limits}
\label{1loop-approx}

Expansions for the one--loop integrals in certain limits have been given
in \cite{PHI}.  For easy reference, I reproduce here the expansions 
which I have used in this thesis.

\subsection{$(m/M)^2 \equiv R \neq 0,\infty$}

The results given below are valid for arbitrary R as long as 
$m^2,M^2 \gg p_1^2, p_2^2, p^2$.  Note that for the corrections
to the process $Z \rightarrow b \bar{b}$, the external momenta
are $p_1^2 = p^2 = m_b^2$ and $p_2^2 = M_Z^2$.  These results are
relevant for the charged Higgs corrections to $Z \rightarrow b \bar{b}$,
where $m=m_t$ and $M$ is a charged Higgs mass.
\begin{eqnarray}
B_1(p^2;m^2,M^2) &=& -\frac{1}{2} \left[ \Delta - \log(M^2/\mu^2)
	+ \frac{3}{2} + \frac{1}{R-1} - \frac{R^2 \log R}{(R-1)^2} \right]
	\nonumber \\
	& & - \frac{p^2}{6M^2} \left[ \frac{2}{R-1} + \frac{9}{(R-1)^2}
	+ \frac{6}{(R-1)^3} - \frac{6R^2 \log R}{(R-1)^4} \right] 
\end{eqnarray}
\begin{eqnarray}
C_0(p_1^2,p_2^2,p^2;M^2,m^2,m^2) &=& \frac{1}{M^2}
	\left[ -\frac{1}{R-1} + \frac{\log R}{(R-1)^2} \right]  \\
C_{24}(p_1^2,p_2^2,p^2;M^2,m^2,m^2) &=& 
	\frac{1}{4} \left[ \Delta - \log (M^2/\mu^2) + \frac{1}{2}
	- \frac{1}{R-1} - \frac{R(R-2)\log R}{(R-1)^2} \right]
	\nonumber \\
	& & + \frac{(p_1^2 + p^2)}{24M^2} \left[ \frac{1}{R-1}
	- \frac{3}{(R-1)^2} - \frac{6}{(R-1)^3} + \frac{6R \log R}{(R-1)^4}
	\right]  \nonumber \\
	& & + \frac{p_2^2}{72M^2} \left[ \frac{2}{R-1} 
	- \frac{3}{(R-1)^2} + \frac{6}{(R-1)^3} 
		\right. \nonumber \\
		& & \left.
	- \frac{6 \log R}{(R-1)^4}
	\right].
\end{eqnarray}

There are also expansions of $C_0$ and $C_{24}$ in the limit that all three
particles in the loop are much heavier than the external particles,
\begin{equation}
C_0(0,0,0;m_1^2,m_2^2,m_3^2) = 
	-\frac{[m_1^2 m_2^2 \log (m_1^2/m_2^2) + m_2^2 m_3^2 \log (m_2^2/m_3^2)
		+ m_3^2 m_1^2 \log (m_3^2/m_1^2)]}
	{[m_1^2 - m_2^2][m_2^2 - m_3^2][m_1^2 - m_3^2]},
\end{equation}
which is negative for any internal masses $m_1$, $m_2$, and $m_3$.
It is also symmetric under permutations of $m_1$, $m_2$, and $m_3$.
$C_{24}$ is related to this by
\begin{equation}
C_{24}(0,0,0;m_1^2,m_2^2,m_3^2) = \frac{1}{4} \left[ B_0(0;m_2^2,m_3^2)
	+ m_1^2 C_0(0,0,0;m_1^2,m_2^2,m_3^2) + \frac{1}{2} \right]
\end{equation}
This equation is also symmetric under permutations of 
$m_1$, $m_2$, and $m_3$.
$B_0$ is given by,
\begin{equation}
B_0(0;m_2^2,m_3^2) = \Delta - \log (m_2^2/\mu^2) 
	- \frac{m_3^2}{m_2^2 - m_3^2} \log (m_2^2/m_3^2) + 1.
\end{equation}

\subsection{$m\simeq0$ $(R \equiv (m/M)^2 \simeq 0)$}

In expanding the two-- and three--point integrals that appear 
in the neutral
Higgs corrections to $Z \rightarrow b \bar{b}$, $m=m_b$ and $M$ 
is a neutral Higgs mass.  The external momenta
are $p_1^2 = p^2 = m_b^2$ and $p_2^2 = M_Z^2$, as before.
However, now one cannot use
the expansions given above, because $m^2 \ll p_2^2$.  The previous expansions
were derived for $m^2 \gg p_2^2$.
As pointed out in reference \cite{PHI}, one cannot simply take 
the limit $R \rightarrow 0$ in the formulas above, since some of the 
results will diverge.  The divergences are cut off by $p_2^2$.
The results given below were derived in \cite{PHI} for  
$M^2 \gg p_1^2, p_2^2, p^2 \gg m^2$.  However, they are still valid
for the case $p_1^2 = p^2 = m^2 = m_b^2$.  In the formulas below we
neglect $m_b$.
\begin{equation}
B_1(p_1^2;m^2,M^2) = -\frac{1}{2} \left[ \Delta - \log (M^2/\mu^2)
	+ \frac{1}{2} \right] + \mathcal{O}(p_1^2/M^2)
\end{equation}
\begin{eqnarray}
C_0(p_1^2,p_2^2,p^2;M^2,m^2,m^2) &=& -\frac{1}{M^2}[\log (-M^2/p_2^2) + 1]
	\nonumber \\
	&=& -\frac{1}{M^2}[\log (M^2/p_2^2) + 1 + i\pi] \\
C_{24}(p_1^2,p_2^2,p^2;M^2,m^2,m^2) &=& \frac{1}{4} \left[
	\Delta - \log (M^2/\mu^2) + \frac{3}{2}
	+ \frac{p_2^2}{3M^2} \log (-M^2/p_2^2)
	+ \frac{5p_2^2}{18M^2}\right]  \nonumber \\
	&=& \frac{1}{4} \left[
	\Delta - \log (M^2/\mu^2) + \frac{3}{2}
	+ \frac{p_2^2}{3M^2} [\log (M^2/p_2^2) + i\pi]
		\right. \nonumber \\
		& & \left.
	+ \frac{5p_2^2}{18M^2} \right]
\end{eqnarray}
\begin{eqnarray}
[C_{22} - C_{23}](p_1^2,p_2^2,p^2;M^2,m^2,m^2) &=& \frac{1}{6 M^2}
	\left[ \log (-M^2/p_2^2) - \frac{1}{6} \right]
	\nonumber \\
	&=& \frac{1}{6M^2} \left[ \log (M^2/p_2^2) + i\pi - \frac{1}{6} 
	\right].
\end{eqnarray}
Note the imaginary parts of the three--point functions above.  These
are a consequence of the fact that the internal $b$ quark lines in 
the diagram of figure \ref{fig:h0loops}(b) can be cut, 
yielding the on--shell decay 
$Z \rightarrow b \bar{b}$.

\chapter{Combining the LEP and SLD measurements of $A_b$}
\label{app:Abderiv}

$A_b$ is measured directly at SLD from the $b$ quark left--right 
forward--backward asymmetry.  This is possible because 
the electron and positron polarizations at SLD are nonzero
and the polarizations are known.
However, at LEP, the electron and positron beams are not
polarized longitudinally.  For this reason, $A_b$ can 
only be measured indirectly at LEP, in the form of the 
$b$ quark forward--backward asymmetry.  The forward--backward
asymmetry also depends on $A_e$.  We assume lepton universality,
so that $A_e = A_{\mu} = A_{\tau} \equiv A_l$.

In this appendix I describe the procedure used for combining the
LEP and SLD determinations of $A_b$.
The procedure is as follows.  First the lepton asymmetry $A_l$ is 
extracted from the LEP and SLD data.  In particular, the LEP 
measurements of the electron forward--backward asymmetry and the
tau polarization measurements are used.  The SLD measurement 
of $\sin^2\theta^{\rm lept}_{\rm eff}$ also yields a measurement of $A_l$.
Second, the combined LEP and SLD value of $A_l$ is used to 
extract $A_b$ from the LEP measurement of the $b$ quark 
forward--backward asymmetry.  Finally, the value of $A_b$
found indirectly from the LEP data in this way is combined with
the direct measurement of $A_b$ from SLD.

At LEP, $A_l$ is measured in three different ways.  First, it can
be found from the 
lepton forward--backward asymmetry, $A^{0,l}_{FB}$,
\begin{equation}
A^{0,l}_{FB} = \frac{3}{4} A_l^2.
\end{equation}
Solving this for $A_l$, we obtain the first LEP determination
of $A_l$,
\begin{equation}
A_l = \sqrt{\frac{4}{3} A^{0,l}_{FB}}.
\end{equation}
The second and third LEP measurements of $A_l$ come from the 
$\tau$ polarization measurement.  This allows a direct measurement 
of $A_e$ and of $A_{\tau}$.  Assuming lepton universality, these
are both equated to $A_l$.

At SLD, $A_l$ is measured directly and the result is 
expressed in terms of $\sin^2\theta^{\rm lept}_{\rm eff}$.
Both $A_l$ and 
$\sin^2\theta^{\rm lept}_{\rm eff}$ are defined in terms of the 
vector and axial--vector couplings of leptons to $Z$,
$g^V_l$ and $g^A_l$.  In particular,
\begin{eqnarray}
\sin^2\theta^{\rm lept}_{\rm eff} &\equiv& \frac{1}{4} 
	\left( 1 - \frac{g^V_l}{g^A_l} \right) \\
A_l &=& \frac{2 g^V_l g^A_l}{(g^V_l)^2 + (g^A_l)^2}.
\end{eqnarray}
Solving these for $A_l$ in terms of $\sin^2\theta^{\rm lept}_{\rm eff}$,
$A_l$ is,
\begin{equation}
A_l = \frac{2 (1 - 4 \sin^2\theta^{\rm lept}_{\rm eff})}
	{1 + (1 - 4 \sin^2\theta^{\rm lept}_{\rm eff})^2}.
\end{equation}

The three LEP values of $A_l$ and the one SLD value
are then combined.

The LEP value of $A_b$ is extracted from the $b$ quark 
forward--backward asymmetry, $A^{0,b}_{FB}$, using the combined
LEP and SLD measurement of $A_l$.  $A^{0,b}_{FB}$ is,
\begin{equation}
A^{0,b}_{FB} = \frac{3}{4} A_e A_b.
\end{equation}
Solving this for $A_b$, and again assuming lepton universality,
\begin{equation}
A_b = \frac{4}{3} \frac{A^{0,b}_{FB}}{A_l}.
\end{equation}
Then $A_b$ is found by inserting the value of $A^{0,b}_{FB}$ 
measured at LEP and the combined value of $A_l$ determined from LEP 
and SLD as described above.

At SLD, $A_b$ is measured directly.  To find the combined LEP 
and SLD value of $A_b$, the SLD measurement is combined with
the LEP value determined above.

The current measurements of the input parameters and the combined
value of $A_b$ are summarized in table \ref{table:Abdata}.

\begin{table}
\begin{center}
\begin{tabular}{|l|r|}
	\hline
$A^{0,l}_{FB}$ (LEP) & $0.01683 \pm 0.00096$ \\
$A_e$ (LEP $\tau$ polarization) & $0.1479 \pm 0.0051$ \\
$A_{\tau}$ (LEP $\tau$ polarization) & $0.1431 \pm 0.0045$ \\
	\hline
$\sin^2\theta^{\rm lept}_{\rm eff}$ (SLD) & $0.23109 \pm 0.00029$ \\
	\hline
	\hline
$A_l$ from LEP and SLD & $0.1489 \pm 0.0017$ \\
	\hline
	\hline
$A^{0,b}_{FB}$ (LEP) & $0.0991 \pm 0.0020$ \\
	\hline
$A_b$ from SLD & $0.908 \pm 0.027$ \\
	\hline
	\hline
$A_b$ from LEP and SLD & $0.895 \pm 0.016$ \\
	\hline
\end{tabular}
\end{center}
\caption[Combined LEP and SLD determination of $A_b$]
{Input data and results for the combined LEP and SLD 
determination of $A_b$.  Inputs are taken from reference 
\cite{Clare99}.}
\label{table:Abdata}
\end{table}

\chapter{SM parameters used in numerical calculations}
\label{app:SMinputs}

In this appendix we list the SM parameters that have been used in 
our numerical calculations of $R_b$ and $A_b$ in extended
Higgs sectors.  The parameters are listed in table 
\ref{table:SMinputs}.

\begin{table}
\begin{center}
\begin{tabular}{|cccc|}
	\hline
Parameter & Value & Source & Reference \\
	\hline
$\alpha^{-1}(M_Z)$ & 128.898 & SM fit & \cite{Eidelmann95} \\
$\sin^2\theta^{lept}_{eff}$ & 0.23157 & SM fit 
	& \cite{Karlen98}  \\
$M_W$ & 80.371 GeV & SM fit &  \cite{Karlen98}  \\
$M_Z$ & 91.1865 GeV & SM fit & \cite{Karlen98}  \\
$m_b$ & 3.0 GeV & $\bar{m}_b(M_Z)$ (${\overline{\rm MS}}$ mass) 
	& \cite{Fusaoka98}  \\
$m_t$ & 171.8 GeV & SM fit & \cite{Clare99}  \\
	\hline
$R_b^{SM}$ & 0.21587 & SM fit (from pull) & \cite{Clare99} \\
$A_b^{SM}$ & 0.935 & SM fit & \cite{Clare99}  \\
	\hline
\end{tabular}
\end{center}
\caption[SM parameters used in calculations]
{The SM parameters used in our numerical 
calculations.}
\label{table:SMinputs}
\end{table}

For all the input parameters except $m_b$, we have used the values
extracted from a fit of the precision electroweak data to the SM.
As shown in table \ref{table:fit_vs_measured}, the difference between
using the SM fit values and the measured values of the parameters 
is very small. 
 
\begin{table}
\begin{center}
\begin{tabular}{|cccc|}
	\hline
Parameter & SM fit & Measured & Fractional difference (\%) \\
	\hline
$\alpha^{-1}(M_Z)$ & 128.898 & 128.878 & 0.02\% \\
$\sin^2\theta^{lept}_{eff}$ & 0.23157 & 0.2321 & 0.2\%  \\
$M_W$ & 80.371 GeV & 80.448 GeV & 0.1\% \\
$M_Z$ & 91.1865 GeV & 91.1867 GeV & 0.0005\% \\
$m_t$ & 171.8 GeV & 174.3 GeV & 1.4\% \\
	\hline
\end{tabular}
\end{center}
\caption[Comparison of the SM fit and measured values of SM 
parameters]
{A comparison of the SM fit values of the SM parameters with 
their measured values, and the percent difference between the
two.  The SM fit values are the same as in table \ref{table:SMinputs},
and the measured values are from reference \cite{Clare99}.}
\label{table:fit_vs_measured}
\end{table}

The largest fractional difference between the 
SM fit and measured value occurs for $m_t$.  In order to show the
effect of this difference on $R_b$, we calculate the correction
to $R_b$ from charged Goldstone boson exchange in the SM using 
both values for $m_t$.  We obtain,
\begin{eqnarray}
\Delta R_b (m_t^{\mathrm{SM-fit}}) &=& -0.00174  \\
\Delta R_b (m_t^{\mathrm{measured}}) &=& -0.00180.
\end{eqnarray}
The difference in $R_b$ between the two choices of $m_t$ is 
$0.00006$.  This is negligible compared to the experimental 
uncertainty in $R_b$, which is $0.00073$. 
We thus ignore the uncertainty in $m_t$ in our numerical calculations.

For the $b$ quark mass, we use the ${\overline{\rm MS}}$ 
running mass evaluated
at $M_Z$, denoted $\bar{m}_b(M_Z)$.  As discussed in reference 
\cite{Murayama97}, it is appropriate to use $\bar{m}_b(M_Z)$ 
in calculations involving $b$ quarks at the scale $M_Z$, because
the running mass takes into account the large QCD corrections
to $m_b$ which are enhanced by $\log(M_Z/m_b)$.

For the SM fit value of $R_b$, we have used 
$R_b^{SM} = 0.21587$, instead of the oft--quoted value,
$R_b^{SM} = 0.2158$.  The old SM prediction for $R_b$ is no
longer precise enough now that the measured value of $R_b$ and
its experimental error are quoted to five decimal places.
Our value is extracted from reference 
\cite{Clare99}, which quotes the measured value of $R_b$, the 
experimental error, and the pull, which is the number of standard
deviations between the measured value and the SM fit.  Although the
SM fit value is not quoted in reference \cite{Clare99}, we can deduce 
it from the pull.  The numbers are summarized in table 
\ref{table:Rbpull}.  Note that a pull of 1.27 means that the 
measured value of $R_b$ is 1.27 $\sigma$ above the SM fit value.

\begin{table}
\begin{center}
\begin{tabular}{|cccc|}
	\hline
 & Measurement & Pull & SM value \\
	\hline
$R_b$ & $0.21680 \pm 0.00073$ & 1.27 & 0.21587 \\
	\hline
\end{tabular}
\end{center}
\caption[$R_b$ measurement and SM fit]
{The measured value of $R_b$, its pull from the SM fit, and the
corresponding SM fit value.  The pull is defined as the number 
of standard deviations by which the measured value 
differs from the SM fit value.  The numbers are from reference 
\cite{Clare99}.}
\label{table:Rbpull}
\end{table}

Finally, in table \ref{table:Rbsigmas} we list the values of $R_b$
that correspond to the 95\%, 99\%, and 99.9\% confidence levels below
the measured value.  These confidence levels correspond to the contours
in our exclusion plots in chapter \ref{sec:models}.

\begin{table}
\begin{center}
\begin{tabular}{|lcr|}
	\hline
Confidence level & Number of $\sigma$ & $R_b$ value \\
	\hline
95\% & 1.96 & 0.21537 \\
99\% & 2.58 & 0.21492 \\
99.9\% & 3.3 & 0.21439 \\
	\hline
\end{tabular}
\end{center}
\caption[Values of $R_b$ corresponding to various confidence levels]
{The values of $R_b$ that correspond to the 95\%, 99\%, and 99.9\%
confidence levels below the measured value.}
\label{table:Rbsigmas}
\end{table}
\chapter{Constraints from direct Higgs searches}
\label{app:dirsearches}

In this appendix I describe the searches for Higgs boson production at LEP and
the resulting limits on the masses of Higgs states.

\section{Charged Higgs searches}

The search for singly--charged Higgs bosons at LEP looks for charged Higgs 
boson pair production via $Z \to H^+ H^-$,
followed by decays to $\tau \nu_{\tau}$ or $q\bar{q}^{\prime}$.
The best bound on the charged Higgs mass from the data at 
$\sqrt{s} = 189$ GeV comes from the OPAL collaboration, which 
excludes charged Higgs bosons up to a mass of 68.7 GeV at the
95\% confidence level \cite{OPALH+99}.  The bound is independent
of the branching ratio $BR(H^{\pm} \to \tau \nu_{\tau})$, assuming
that $BR(H^{\pm} \to q\bar{q}^{\prime}) + 
BR(H^{\pm} \to \tau \nu_{\tau}) =1$.  
This condition is true as long as 
the charged Higgs boson can only decay into quarks and leptons.
In models with two or more singly--charged Higgs bosons, the heavier charged
Higgs boson can decay into the lighter charged Higgs boson 
and a neutral Higgs boson, if
the decay is kinematically allowed.  However, we will be using the 
mass bound of reference \cite{OPALH+99} to constrain the mass of the 
lighter charged Higgs boson only, so
this is not a concern.  Finally, the charged Higgs boson decay into $W^{\pm}$
and a neutral boson is not kinematically allowed.

The bound also depends on the production cross section of the 
charged Higgs boson pair.  In the analysis of reference \cite{OPALH+99}
it is assumed that the $ZH^+H^-$ coupling is that in the 2HDM,
\begin{equation}
g_{ZH^+H^-} = -\frac{e}{s_Wc_W} \left( \frac{1}{2} - s^2_W \right).
\end{equation}
This coupling, and hence the production cross section, is the same 
in a model containing multiple doublets and singlets, and in the
Georgi--Machacek models for $H_3^{\pm}$.  The bound is used for these
models in figures \ref{fig:mtwohdm} and \ref{fig:tripcustRb}.
However, the coupling is not the
same in the models containing doublets and triplets without 
${\rm SU}(2)_c$ 
symmetry.  In the models with one or two doublets and one real,
$Y=0$ triplet, the coupling is larger than in the 2HDM, and hence
the production cross section is larger.  Therefore in these models,
the charged Higgs mass bound from reference \cite{OPALH+99} is 
still valid, and is in fact a conservative bound.  The bound is used
in figure \ref{fig:D2T_0+} for the model with two doublets and one 
$Y=0$ triplet.  
In the models with one or two doublets and
one complex, $Y=2$ triplet, the coupling is smaller than in the 2HDM.
Hence the charged Higgs boson production cross section is smaller, and the 
mass bound from reference \cite{OPALH+99} is no longer valid.  This 
is the case in figure \ref{fig:D2T_2+}, for the model with two doublets
and one $Y=2$ triplet.

\section{Neutral Higgs searches}

The search for neutral Higgs bosons at LEP looks for neutral Higgs boson 
production in two ways.  In the search for the standard model Higgs boson,
LEP looks for the process $e^+e^- \to Z^* \to Z h^0$.  In the search 
for the neutral Higgs bosons of the Minimal Supersymmetric Model (MSSM),
LEP looks for the above process, in addition to the process
$e^+e^- \to Z^* \to h^0 A^0$.  The implications of these searches
for the neutral Higgs bosons in the general 2HDM are discussed in the following
sections.

\subsection{SM Higgs search}

In the SM, the dominant Higgs boson production 
mode at LEP energies is 
\begin{equation}
e^+e^- \to Z^* \to Z h^0,
\end{equation}
where $Z^*$ is an off--shell $Z$ boson and the final $Z$ and $h^0$ are
on--shell.

From the LEP data taken at a
center--of--mass energy of $\sqrt{s} = 189$ GeV,
the lower limits on the SM Higgs mass from the four LEP experiments 
are summarized in table \ref{table:SMHiggsbounds} \cite{Felcini99}.
The best bound comes from the DELPHI and L3 experiments; they find
$M_{h^0_{SM}} > 95.2$ GeV.

\begin{table}
\begin{center}
\begin{tabular}{|lr|}
	\hline
 & $M_{h^0}$ (GeV) \\
	\hline
ALEPH & 90.2  \\
DELPHI & 95.2  \\
L3 & 95.2  \\
OPAL & 91.0  \\
	\hline
\end{tabular}
\end{center}
\caption[Lower bound on the SM Higgs mass from LEP]
{The observed lower bounds on the SM Higgs mass from the 
four LEP experiments, from data taken at $\sqrt{s} = 189$ GeV.
Data from reference \cite{Felcini99}.}
\label{table:SMHiggsbounds}
\end{table}

If we make some simplifying assumptions, we can extract from this measurement
a lower bound on the mass of the CP--even neutral Higgs bosons in the 2HDM.
In the 2HDM, the $ZZh^0$ coupling is reduced from its SM value by a
factor of $\sin(\beta - \alpha)$.  Thus the cross section for $Zh^0$ 
production is reduced by a factor of $\sin^2(\beta - \alpha)$.
Also, for a fixed center--of--mass energy, the $Zh^0$ production 
cross section decreases with increasing Higgs mass.  

If we assume that the background and the $Zh^0$ detection efficiency
are fairly flat as a function of the Higgs mass, 
as suggested by figure 27 of reference \cite{Adam98},
then we can take 
the LEP lower bound on the SM Higgs mass as an upper bound on the 
$Zh^0$ production cross section.  We can then translate the LEP 
bound on the SM Higgs mass into a bound on $\sin^2(\beta - \alpha)$
as a function of the Higgs mass.  

In the SM, the tree--level cross section for $Zh^0$ production in 
$e^+e^-$ collisions is \cite{HHG},
\begin{equation}
\sigma_{SM}(e^+e^- \to Z h^0) = \frac{\pi \alpha^2 \lambda^{1/2}
	[\lambda + 12 s M_Z^2][1 + (1 - 4 s^2_W)^2]}
	{192 s^2 s^4_W c^4_W (s - M_Z^2)^2},
\end{equation}
where $s$ is the square of the center--of--mass energy, 
$s_W = \sin\theta_W$, $c_W = \cos\theta_W$, and
$\lambda$ is a kinematic factor,
\begin{equation}
\lambda = (s - M^2_{h^0} - M^2_Z)^2 - 4 M^2_{h^0} M^2_Z.
\end{equation}
In the 2HDM, the cross section is multiplied by a factor of
$\sin^2(\beta - \alpha)$,
\begin{equation}
\sigma(e^+e^- \to Z h^0) = 
	\sigma_{SM}(e^+e^- \to Z h^0) \sin^2(\beta - \alpha).
\end{equation}

\begin{figure}
\resizebox{\textwidth}{!}{\rotatebox{270}{\includegraphics{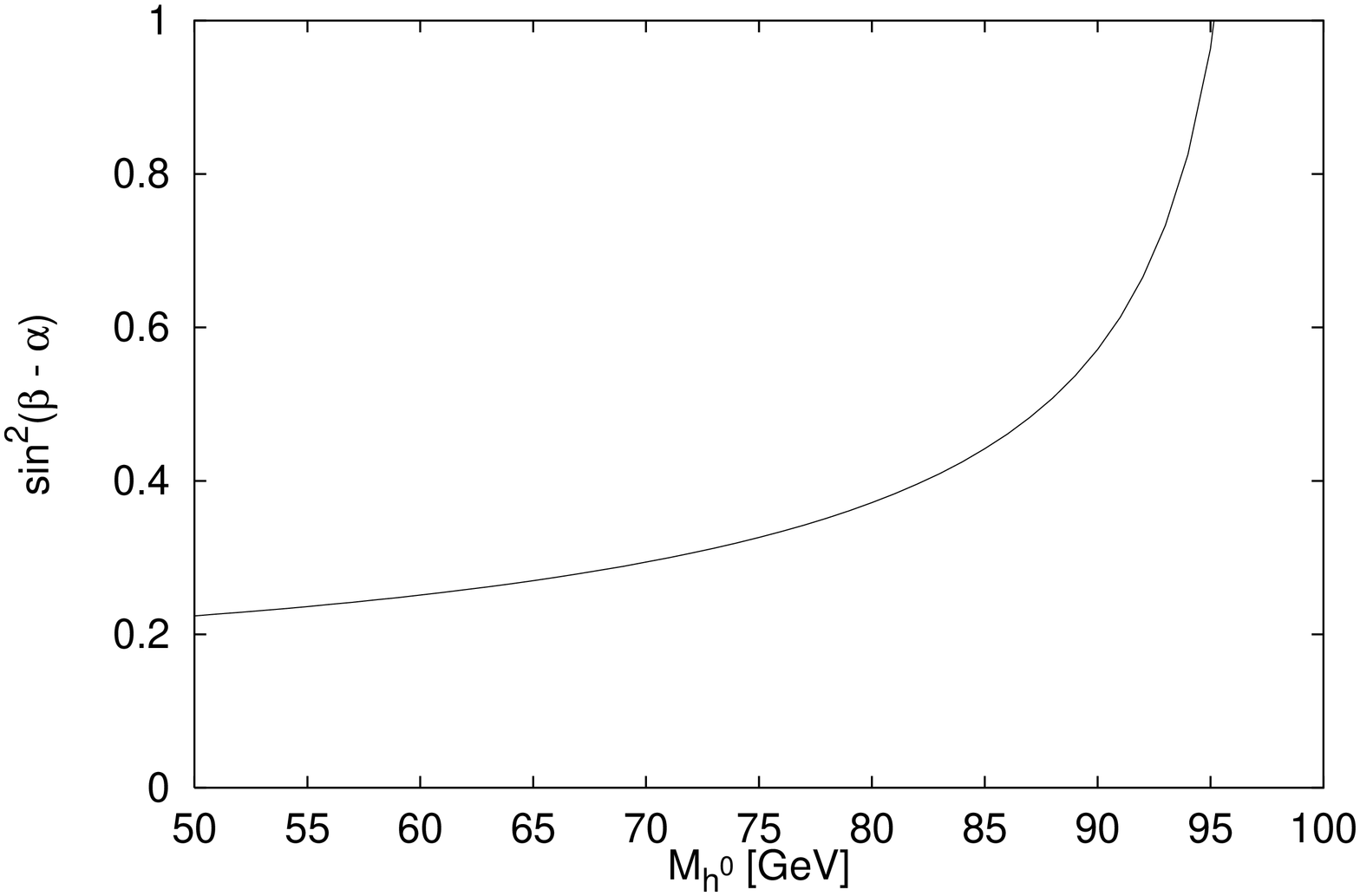}}}
\caption[Constraint on $\sin^2(\beta - \alpha)$ as a function of $M_{h^0}$
in the 2HDM]
{Constraint on $\sin^2(\beta - \alpha)$ as a function of $M_{h^0}$
in the 2HDM.  The area above the line is excluded.
We have assumed that the Higgs boson detection efficiency and
background are constant as a function of the Higgs mass.}
\label{fig:sinsqplot}
\end{figure}

The LEP bound on $M_{h^0}$ in the SM fixes 
$\sigma_{SM}(e^+e^- \to Z h^0)$.  If we then vary $M_{h^0}$
in the 2HDM cross section, we find a lower bound on the Higgs
mass as a function of $\sin^2(\beta - \alpha)$.
This bound is shown in figure
\ref{fig:sinsqplot}.  A similar analysis was done by Sopczak in 
reference \cite{Sopczak}, using the LEP data taken at 
center--of--mass energies between 161 and 172 GeV.

For $\sin^2(\beta - \alpha) = 1$, the bound on $M_{h^0}$ is the 
same as in the SM, $M_{h^0} > 95.2$ GeV.  This bound is 
used in figure \ref{fig:Rb0_0_no}. 
For $\sin^2(\beta - \alpha) = 1/2$, the bound on $M_{h^0}$ is 
$M_{h^0} > 87$ GeV.  This bound is used in figures 
\ref{fig:Ab0_2_no} and \ref{fig:Ab0_2_165}.

The decays of $h^0$ in the 2HDM differ from those of the SM Higgs boson
in one important way.
If the CP--odd state $A^0$ has less than half the mass of the $h^0$, 
then the decay mode $h^0 \to A^0 A^0$ becomes possible.  
Since the $h^0A^0A^0$ coupling
is typically of order unity while the $h^0b\bar{b}$ coupling is 
suppressed by a factor of $m_b/v_1$, $h^0 \to A^0 A^0$ could be the 
dominant decay of $h^0$.
In this case, the final state will contain six $b$ jets, since all 
three of the $A^0$ particles will decay dominantly to $b$ quarks.
This final state should be easy to detect in a dedicated search.

\subsection{2HDM Higgs search}

The search for neutral Higgs bosons at LEP is sensitive to the neutral 
Higgs bosons of the 2HDM.
However, 
the analysis of the LEP data is done in the context of the Higgs sector
of the Minimal Supersymmetric Model (MSSM), in which the Higgs sector 
is constrained by supersymmetry relations.  Fortunately, the Higgs
mass bounds in the MSSM can be reinterpreted as constraints on a general
2HDM.

In the MSSM, the dominant Higgs boson production
modes at LEP energies are,
\begin{equation}
e^+e^- \to Z^* \to Z h^0
\end{equation}
and
\begin{equation}
e^+e^- \to Z^* \to h^0 A^0,
\end{equation}
where again $Z^*$ is an off--shell $Z$ boson and the final--state particles
are on--shell.

From the LEP data taken at a
center--of--mass energy of $\sqrt{s} = 189$ GeV,
the lower limits on the MSSM Higgs masses from the four LEP experiments 
are summarized in table \ref{table:MSSMHiggsbounds} \cite{Felcini99}.
The best bound comes from the DELPHI experiment, which finds
$M_{h^0} > 83.5$ GeV and $M_{A^0} > 84.5$ GeV, for arbitrary 
$\tan\beta$.

\begin{table}
\begin{center}
\begin{tabular}{|lrr|}
	\hline
 & $M_{h^0}$ (GeV) & $M_{A^0}$ (GeV) \\
	\hline
ALEPH & 80.8 & 81.2 \\
DELPHI & 83.5 & 84.5 \\
L3 & 77.0 & 78.0 \\
OPAL & 74.8 & 76.5 \\
	\hline
\end{tabular}
\end{center}
\caption[Lower bounds on the MSSM Higgs masses from LEP]
{The observed lower bounds on the masses of the MSSM Higgs bosons
$h^0$ and $A^0$ from the 
four LEP experiments, from data taken at $\sqrt{s} = 189$ GeV.
Data from reference \cite{Felcini99}.}
\label{table:MSSMHiggsbounds}
\end{table}

As discussed in the previous section, the cross section for $Z h^0$
production is proportional to $\sin^2(\beta - \alpha)$ in a 2HDM.
The $Zh^0A^0$ coupling is proportional to $\cos(\beta - \alpha)$,
so the cross section for $h^0 A^0$ production is proportional to 
$\cos^2(\beta - \alpha)$.

The limits on the MSSM Higgs masses observed by the LEP experiments
actually depend on the value of $\tan\beta$.  For both the 
$h^0$ and $A^0$ searches, the bounds are weakest at large $\tan\beta$;
these are the values quoted in table \ref{table:MSSMHiggsbounds}.
In the MSSM, the parameters of the Higgs sector are correllated, so
that at large $\tan\beta$, $\cos(\beta - \alpha)$ is very close to one.
Thus the bounds quoted in table \ref{table:MSSMHiggsbounds} come
from the search for $h^0 A^0$ production.

As in the previous section, if we assume that the backgrounds and 
detection efficiency for $h^0$ and $A^0$ are independent of the 
Higgs masses, we can use the MSSM mass bounds to deduce bounds on
the $h^0$ and $A^0$ masses in the general 2HDM.

In the 2HDM, the tree--level cross section for $h^0 A^0$ production in 
$e^+e^-$ collisions is \cite{HHG},
\begin{equation}
\sigma(e^+e^- \to h^0 A^0) = \frac{g^2 f^2}{48 \pi}
	\left( \frac{8 s^4_W - 4 s^2_W + 1}{c^2_W} \right) 
	\frac{\kappa^3}{\sqrt{s}[(s - M_Z^2)^2 + \Gamma_Z^2 M_Z^2]}
\end{equation}
where $\Gamma_Z$ is the $Z$ decay width,
$\kappa$ is a kinematic factor,
\begin{equation}
\kappa = \frac{ \left[ (s - M^2_{h^0} - M^2_{A^0})^2 - 4 M^2_{h^0} M^2_{A^0} 
	\right]^{1/2}}
	{2\sqrt{s}}
\end{equation}
and $f$ is a coupling factor,
\begin{equation}
f = \frac{g}{2 c_W} \cos(\beta - \alpha).
\end{equation}

We can now extend the MSSM Higgs boson search results to the general
2HDM.
The LEP bound on $M_{h^0}$ and $M_{A^0}$ fixes the cross section with
$\cos^2(\beta - \alpha) = 1$.  With our assumptions, we can then
vary $M_{h^0}$ and $\cos^2(\beta - \alpha)$ and find lower bound 
on $M_{A^0}$.  In our analysis, we fix $\cos^2(\beta - \alpha)$, then
vary $M_{h^0}$ and find a lower bound on $M_{A^0}$.  
We do this for two values of $\cos^2(\beta - \alpha)$.

\begin{figure}
\resizebox{\textwidth}{!}{\rotatebox{270}{\includegraphics{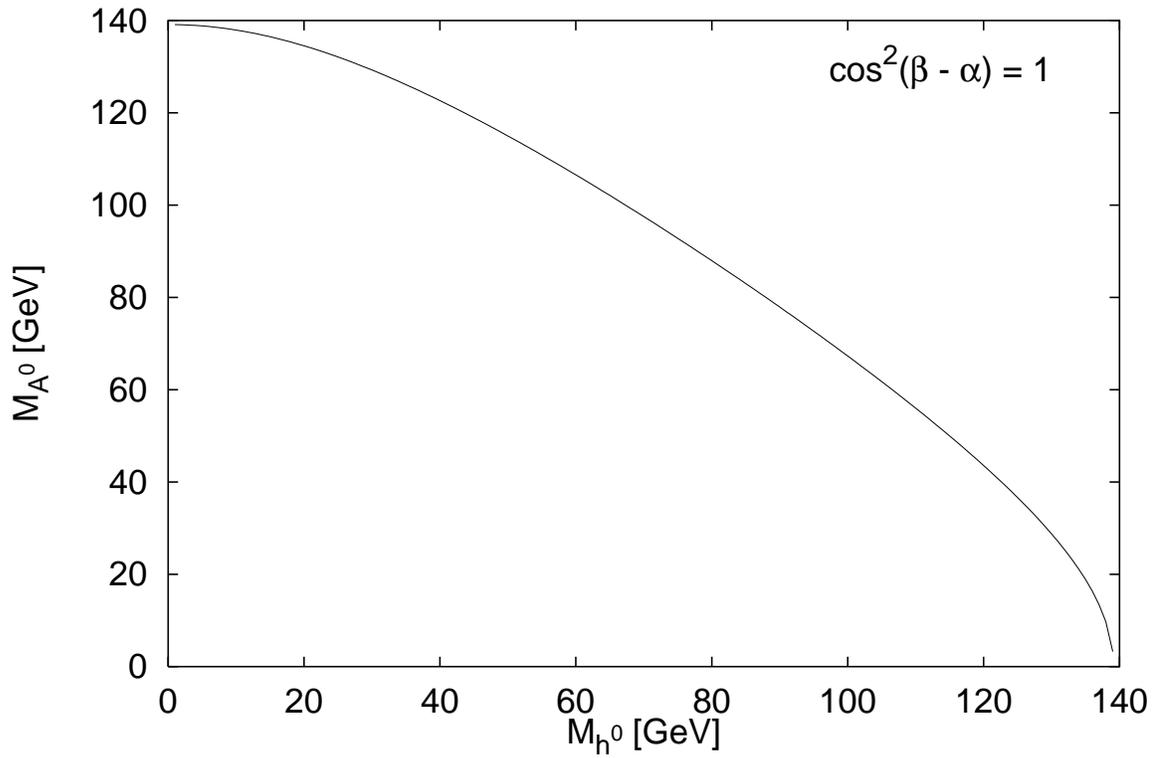}}}
\caption[Bounds on $M_{h^0}$ and $M_{A^0}$ in the 2HDM for 
$\cos^2(\beta - \alpha) = 1$]
{Bounds on $M_{h^0}$ and $M_{A^0}$ in the 2HDM for 
$\cos^2(\beta - \alpha) = 1$.  The area below the line is excluded.
We have assumed that the Higgs boson detection efficiency and
background are constant as a function of the Higgs masses.
}
\label{fig:ZhA_1plot}
\end{figure}

For $\cos^2(\beta - \alpha) = 1$, we find the bound shown in
figure \ref{fig:ZhA_1plot}.  As expected, this bound passes through
the MSSM point with $M_{h^0} = 83.5$ GeV and $M_{A^0} = 84.5$ GeV.
This bound is used in figures \ref{fig:Rb0_1_no} 
and \ref{fig:Rb0_1_165}.  Note also that for
$\cos^2(\beta - \alpha) = 0$, the bound shown in figure 
\ref{fig:ZhA_1plot} is the bound on the $A^0$ mass as a function of 
the mass of the heavier CP--even neutral Higgs boson, $H^0$.  This bound
is used in figure \ref{fig:Rb0_0_no}.

\begin{figure}
\resizebox{\textwidth}{!}{\rotatebox{270}{\includegraphics{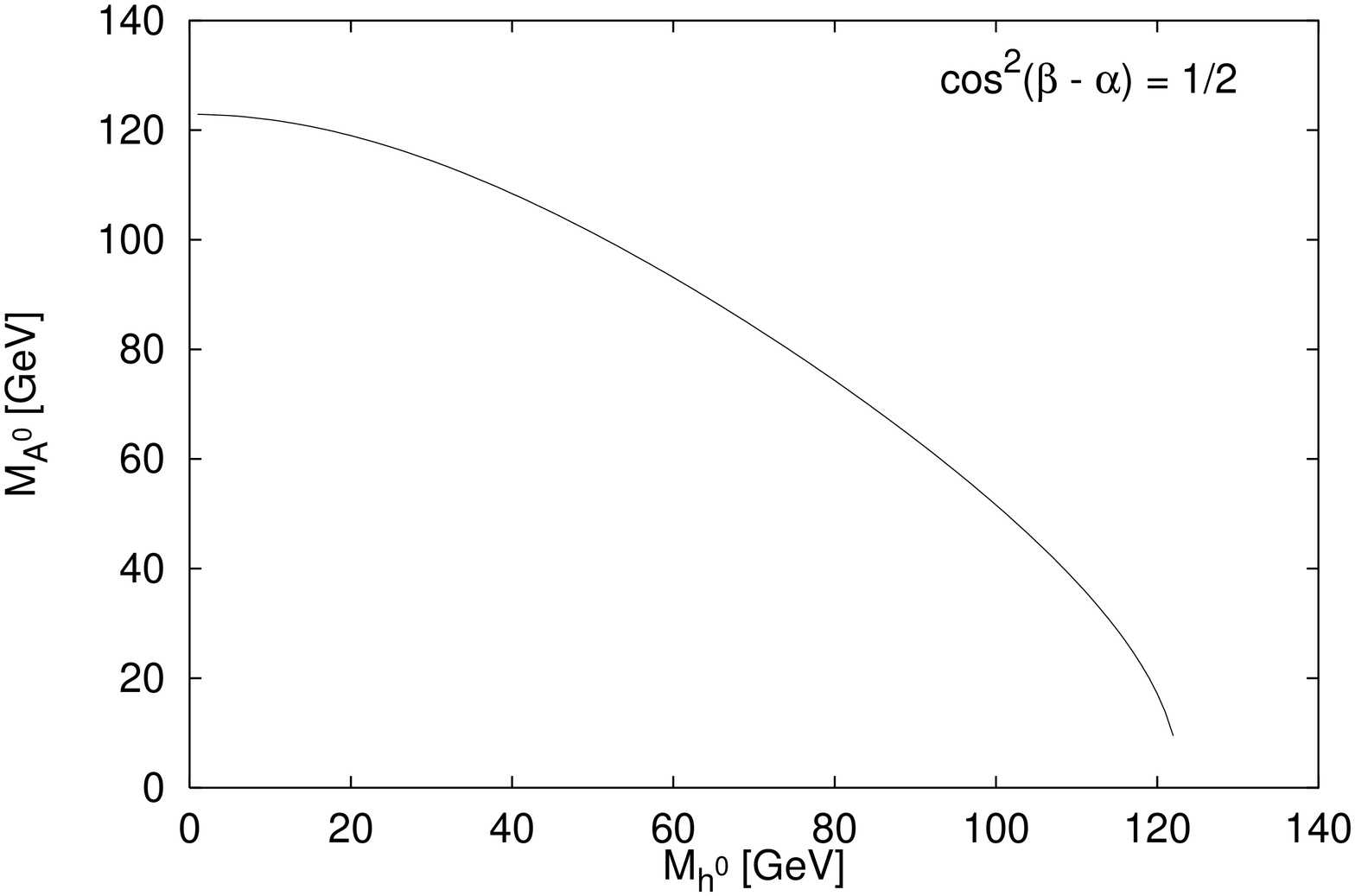}}}
\caption[Bounds on $M_{h^0}$ and $M_{A^0}$ in the 2HDM for 
$\cos^2(\beta - \alpha) = 1/2$]
{Bounds on $M_{h^0}$ and $M_{A^0}$ in the 2HDM for 
$\cos^2(\beta - \alpha) = 1/2$.  The area below the line is excluded.
We have assumed that the Higgs boson detection efficiency and
background are constant as a function of the Higgs masses.
}
\label{fig:ZhA_2plot}
\end{figure}

For $\cos^2(\beta - \alpha) = 1/2$, we find the bound shown in
figure \ref{fig:ZhA_2plot}.
This bound is used in figures \ref{fig:Ab0_2_no} and \ref{fig:Ab0_2_165}.

In making figures \ref{fig:ZhA_1plot} and \ref{fig:ZhA_2plot} we have
assumed that the backgrounds and detection efficiencies for $h^0$ and
$A^0$ are constant for all Higgs masses.  This is not a good assumption
for very light Higgs bosons (for example, below 10 GeV) because the Higgs
branching ratio to $b$ quarks will be suppressed by the non--negligible
$b$ quark mass.  The LEP searches look for Higgs boson decays to $b$ quarks 
in order to tag the events.  

Also, as discussed before, if the $A^0$ has less than half the mass 
of the $h^0$, the decay
mode $h^0 \to A^0 A^0$ becomes possible.  Since the $h^0A^0A^0$ coupling
is typically of order unity while the $h^0b\bar{b}$ coupling is 
suppressed by a factor of $m_b/v_1$, $h^0 \to A^0 A^0$ could be the 
dominant decay of $h^0$.
In this case, the final state will contain six $b$ jets, since all 
three of the $A^0$ particles will decay dominantly to $b$ quarks.
Again, one would think that this final state would be easy to detect.

In summary, the bounds presented in figures \ref{fig:ZhA_1plot} 
and \ref{fig:ZhA_2plot} are good for 
$M_{h^0}$ and $M_{A^0}$ near 80 GeV.  However, for either
$M_{h^0}$ or $M_{A^0}$ very small, the assumptions on which our 
derivation is based become less reliable.


\bibliographystyle{prsty}
\bibliography{intro,Hdirectsearch,paper,custodialSU2,loopints}

\end{document}